\documentclass[12pt,a4paper]{article}

\usepackage{amsfonts, amsmath, amssymb, amsthm, color, enumitem, float, mathrsfs, mathtools, multirow, nccmath, rotating, tikz, tocloft, subfig}

\usepackage{lmodern}[lmr]

\usepackage[colorlinks=true]{hyperref}
\hypersetup{
   colorlinks = true,
   linkcolor = blue,
   anchorcolor = blue,
   citecolor = blue,
   filecolor = blue,
   urlcolor = blue
}

\usepackage[hmargin=0.80in, vmargin=0.80in]{geometry}

\usepackage[round]{natbib}
\author{}
\title{}

\DeclareMathOperator*{\argmin}{\arg\!\min}

\DeclareMathOperator*{\plim}{p\!\lim}

\DeclareMathOperator*{\diag}{\normalfont\textrm{diag}}
\DeclareMathOperator*{\tr}{\normalfont\textrm{Tr}}

\newtheorem{lemma}{Lemma}

\newtheorem{remark}{Remark}
\newtheorem{theorem}{Theorem}
\newtheorem{assumption}{Assumption}

\renewcommand{\arraystretch}{1.2}

\newcommand\blfootnote[1]{
\begingroup
\renewcommand\thefootnote{}\footnote{#1}
\addtocounter{footnote}{-1}
\endgroup
}

\allowdisplaybreaks[4]

\begin{document}

\setlength{\abovedisplayskip}{7pt}
\setlength{\belowdisplayskip}{7pt}
 
\begin{titlepage}

\begin{center}
{\Large \bf Estimation and Inference for \\ Three-Dimensional Panel Data Models \blfootnote{ \\
An earlier version of this paper has been circulated and submitted to working paper series under the title ``Multi-Level Panel Data Models: Estimation and Empirical Analysis", which is the Working Paper \#4/22 available at https://ideas.repec.org/p/msh/ebswps/2022-4.html.

We thank the participants of \textit{2024 International Workshop on Econometrics with Its Application and Practice in Finance} at Nankai University for constructive suggestions and comments. Gao and Peng would like to acknowledge the Australian Research Council Discovery Projects Program for its financial support under Grant Numbers: DP200102769 \& DP210100476.  Liu's research is financially supported by the  National Natural Science Foundation of China under Grant No. 72203114.{\item $^{\ast}$University of North Texas, United States.  \\ $^\dag$Monash University, Australia. \\ $^\sharp$Nankai University, China.

}
}

} 
\medskip

{\sc Guohua Feng$^{\ast}$, Jiti Gao$^{\dag}$,  Fei Liu$^\sharp$ and Bin Peng$^{\dag}$}
\medskip

\bigskip\bigskip

\today

\bigskip

\begin{abstract}
Hierarchical panel data models have recently garnered significant attention. This study contributes to the relevant literature by introducing a novel three-dimensional (3D) hierarchical panel data model, which integrates panel regression with three sets of latent factor structures: one set of global factors and two sets of local factors. Instead of aggregating latent factors from various nodes, as seen in the literature of distributed principal component analysis (PCA), we propose an estimation approach capable of recovering the parameters of interest and disentangling latent factors at different levels and across different dimensions. We establish an asymptotic theory and provide a bootstrap procedure to obtain inference for the parameters of interest while accommodating various types of cross-sectional dependence and time series autocorrelation. Finally, we demonstrate the applicability of our framework by examining productivity convergence in manufacturing industries worldwide.

\medskip
\medskip


\medskip


\end{abstract}

\end{center}

\end{titlepage}

\section{Introduction}\label{Sec1}
\renewcommand{\theequation}{1.\arabic{equation}}
\setcounter{equation}{0}

In the past three decades or so, panel data research has undergone significant advancements, leveraging its capacity to extract richer insights from both cross-sectional and time dimensions. Within this domain, a notable strand of research (\citealp{Bai,FanLiaoWang}) has garnered increasing attention, with a particular focus on addressing cross-sectional dependence (CSD) in large $N$ and $T$ settings by accounting for correlations among observations across different entities within a panel. Specifically, strong CSD is often addressed through a factor structure approach involving the estimation of a low-rank representation, while weak CSD poses inference challenges related to estimating the asymptotic covariance matrix. Additionally, accounting for time series autocorrelation (TSA) is essential when analyzing data with a temporal dimension. To tackle both CSD and TSA jointly, various methods have been proposed in the relevant literature. For instance, \cite{goncalves_2011} suggests using the moving block bootstrap method, \cite{BAI2020} employ a thresholding method in conjunction with a heteroskedasticity and autocorrelation-consistent (HAC) covariance matrix estimator, and \cite{GPY2023} propose a dependent wild bootstrap procedure by extending the time series approach of \cite{shao2010} to a two--dimensional panel data framework.

Recent developments in panel data research, especially in contexts where observations exhibit hierarchical structures, have led to the emergence of the relevant literature on hierarchical panel data models (\citealp{Matyas, KSS2020, CYZ2022, Zhang2023, JLS2023, JLS2024}). This evolution reflects the recognition of the necessity to extend beyond traditional panel data models to capture dependencies at multiple levels of aggregation, such as individuals nested within groups or regions nested within countries. Consequently, the literature on hierarchical panel data models builds upon and extends methodologies developed in the cross-sectional dependence literature to accommodate these hierarchical structures, ultimately enabling researchers to more accurately model and analyze complex panel datasets.

Another pertinent literature is regression with incidental parameters, a concept pioneered by \cite{Neyman}. \cite{LANCASTER2000391} commends their seminal paper as a remarkable contribution, widely acknowledged for its influence within the statistical community. To see the challenges posed by incidental parameters in the context of 3 dimensional (3D) panels, we present the following table outlining the potential incidental parameters across different data structures:

\begin{table}[H]\centering
\begin{tabular}{ll}
\hline\hline
   Data Structure      & Possible Incidental Parameters \\
\multicolumn{1}{l|}{2 Dimensional Panel} &    (1). $\alpha_i$ (Individual effects); (2). $g_t$ (Time effects); \\
\multicolumn{1}{l|}{} & (3). $\alpha_i+g_t$ (Two way effects); (4). No effects \\ \multicolumn{1}{l|}{}& \\
\multicolumn{1}{l|}{3 Dimensional Panel} &    64 possible specifications (\citealp{LMS2021})              \\
\hline\hline       
\end{tabular}
\end{table}
\noindent While a 3D panel involves only one additional layer, it significantly increases complexity. To the best of our knowledge, no prior studies have attempted to integrate all potential specifications of incidental parameters into a single regression model. Hence, a unified framework is needed. Fortunately, incidental parameters can be integrated using a factor structure. For example, all four conceivable scenarios of a 2D panel can fit within a structure such as $ \pmb{\alpha}_i^\top \mathbf{g}_t$. This approach offers valuable insights into addressing the challenge of 3D panel data modeling.

With the above challenges in mind, the primary objective of this study is to contribute to the literature on hierarchical panel data models by introducing a panel data regression model with three sets of latent factor structures: one set of global factors and two sets of local factors. Compared with pure factor models, this model is capable of estimating the coefficient of a particular observed independent variable while accounting for both commonalities and individual differences across entities. In addition, in comparison with panel data regression models with only one or two sets of factors, the proposed model and estimation theory is capable to model dependencies that exist at multiple levels of aggregation. This model finds applications in various contexts. For example, in the context of modeling economic convergence, the latent global factors represent economic events affecting the growth of all countries and industries, while the two local factors represent events affecting specific countries or industries, respectively. In the context of modeling international trade, the latent global factors represent factors affecting the trade of all countries, while the two local factors represent events affecting importing or exporting countries, respectively.

To better understand the 3D factor structure, suppose that we have two groups of individuals:

\begin{eqnarray}
i \in [L]\quad\text{and}\quad j\in [N],
\end{eqnarray}
where $L$ and $N$ are two positive integers, and $[S]\coloneqq \{1,\ldots, S\}$ for any  positive integer $S$. In the literature of modeling empirical growth (e.g., \citealp{Rodrik2013}), $i$ represents industries, and $j$ denotes countries;  in the context of  bilateral trade (e.g., \citealp{CYZ2022}), $i$ and $j$ respectively represent importing countries and exporting countries; in the stochastic actor-oriented models (e.g., \citealp{KS2023}), $i$ refers to social groups and $j$ refers to individuals within each group; etc. In addition, we let $y_{ijt}$ denote a set of outcome variables:

\begin{eqnarray}
\{y_{ijt}\, | \, (i,j,t)\in[L]\times [N]\times [T]\}
\end{eqnarray}
which is indexed by both $i$ and $j$, and also varies over each time period $t$. The definition of $y_{ijt}$ should be self-evident in the aforementioned studies, so we omit these details here.  Throughout this paper, we assume that each layer of information is influenced by distinct factors, enabling us to extract essential insights through careful data processing. Our approach diverges in that we aim to disentangle factors layer by layer and across different blocks within each layer, rather than consolidating factors from disparate nodes.

For a clearer grasp of the 3D factor structure, we visually represent the three sets of latent factors ($\mathbf{f}_{t}$, $\mathbf{f}_{it}^\circ$, and $\mathbf{f}_{jt}^\bullet$) in Figure \ref{Fig1}, utilizing empirical growth data as a demonstration.

{\footnotesize

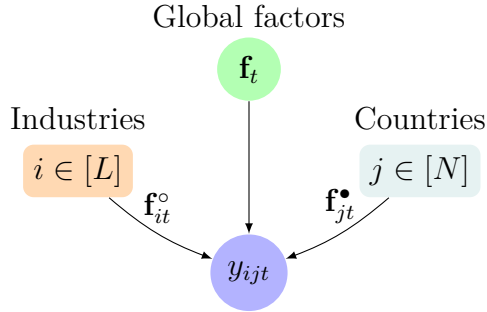
\begin{figure}[H]
\centering
\begin{tikzpicture}[scale=0.9]

\tikzstyle{annot} = [text width=10em, text centered]

\node[rectangle, rounded corners, minimum size = 6mm, fill=orange!30] (Input-1) at (0,0) {$i\in [L]$};

\node[rectangle, rounded corners, minimum size = 6mm, fill=teal!10] (Hidden-1) at (5,0) {$j\in [N]$};

\node[circle, minimum size = 6mm, fill=green!30] (Global-1) at (2.5,1.5) {$\mathbf{f}_t$};

\node[circle, minimum size = 6mm, fill=blue!30] (Global-2) at (2.5,-1.5) {$y_{ijt}$};

\draw[-latex] (Input-1) to [bend right=10] node[above] {$\mathbf{f}_{it}^\circ$} (Global-2);
\draw[-latex] (Hidden-1) to [bend left=10] node[above] {$\mathbf{f}_{jt}^\bullet$}  (Global-2);

\draw[-latex] (Global-1) -- (Global-2);

\node[annot,above of=Input-1, node distance=0.7cm] {Industries};
\node[annot,above of=Hidden-1, node distance=0.7cm] {Countries};
\node[annot,above of=Global-1, node distance=0.7cm] {Global factors};
\end{tikzpicture}
\caption{Different Factors of the Hierarchy}\label{Fig1}
\end{figure}}
\noindent Specifically, $\mathbf{f}_{t}$, $\mathbf{f}_{it}^\circ$ and $\mathbf{f}_{jt}^\bullet$ represent $\ell\times 1$, $\ell_i^\circ\times 1$, and $\ell_j^\bullet\times 1$ vectors respectively, denoting the global, country-specific, and industry-specific factors or shocks driving $y_{ijt}$. Here, $\ell$, $\ell_i^\circ$, and $\ell_j^\bullet$ are non-negative fixed integers, subject to the constraint:

\begin{eqnarray}\label{m.cond1}
0\le\min_{i,j}(\ell+\ell_i^\circ+ \ell_j^\bullet)\le  \max_{i,j}(\ell+\ell_i^\circ+ \ell_j^\bullet)\le c,
\end{eqnarray}
where $c$ is a fixed positive constant. This condition implies that $\ell$, $\ell_i^\circ$, and $\ell_j^\bullet$ can potentially be 0. In an extreme scenario, if $y_{ijt}$ behaves entirely randomly like an idiosyncratic error, then $\ell=\ell_i^\circ=\ell_j^\bullet=0$ for all $(i,j)$ pairs. Figure \ref{Fig1} illustrates that $f_t$ influences every industry and country, while each country-specific (or industry-specific) shock can also impact a specific subset of countries or industries. Mathematically, for all $t\in [T]$, Figure 1 reveals the following mapping:

{\footnotesize
\begin{eqnarray}\label{eq3d}
\begin{pmatrix}
\pmb{\gamma}_{11}^{\top} & \pmb{\gamma}_{11}^{\circ\top} &  \cdots & \mathbf{0} & \pmb{\gamma}_{11}^{\bullet\top} &  \cdots & \mathbf{0} \\
\vdots    & \vdots & \ddots & \vdots   & \vdots & \vdots & \vdots \\ 
\pmb{\gamma}_{L1}^{\top} &  \mathbf{0} & \cdots & \pmb{\gamma}_{L1}^{\circ\top}    &  \pmb{\gamma}_{L1}^{\bullet\top}   & \cdots &  \mathbf{0}\\
\vdots &  &  \vdots & & & \vdots &    \\
\pmb{\gamma}_{1N}^{\top} & \pmb{\gamma}_{1N}^{\circ\top}& \cdots &  \mathbf{0}  &\mathbf{0} & \cdots & \pmb{\gamma}_{1N}^{\bullet\top} \\
\vdots  & \vdots & \ddots & \vdots       & \vdots & \vdots & \vdots \\ 
\pmb{\gamma}_{LN}^{\top} & \mathbf{0} & \cdots & \pmb{\gamma}_{LN}^{\circ\top}  & \mathbf{0} & \cdots & \pmb{\gamma}_{LN}^{\bullet\top} \\
\end{pmatrix}
\begin{pmatrix}
\mathbf{f}_t \\
\mathbf{f}_{1t}^\circ \\
\vdots\\
\mathbf{f}_{Lt}^\circ \\
\mathbf{f}_{1t}^\bullet \\
\vdots\\
\mathbf{f}_{Nt}^\bullet 
\end{pmatrix}  \mapsto \begin{pmatrix}
y_{11t}  \\
\vdots \\ 
y_{L1t} \\
\vdots \\
y_{1Nt}  \\
\vdots \\ 
y_{LNt} 
\end{pmatrix},
\end{eqnarray}
where $\pmb{\gamma}_{ij}$,  $\pmb{\gamma}_{ij}^\circ$, and $\pmb{\gamma}_{ij}^\bullet$} are factor loadings indicating how each factor influences $y_{ijt}$ at each time period $t$. Figure \ref{Fig1}, along with the restriction \eqref{m.cond1} and the mapping \eqref{eq3d}, together entail the following observations:

\begin{itemize}[leftmargin=*, itemsep=0.5pt, parsep=0.5pt, topsep=0.6pt]
\item[1.] Figure \ref{Fig1}, the restriction \eqref{m.cond1}, and the mapping \eqref{eq3d} encompass all possible combinations of incidental parameters.

\item[2.] The global shock $\mathbf{f}_t$ (if exists) affects all countries and industries, while the country and industry shocks ($\mathbf{f}_{it}^\circ$ and $\mathbf{f}_{jt}^\bullet$) respectively impact a specific subset of countries or industries.

\item[3.] From the a signal-to-noise ratio point of view, $\mathbf{f}_t$ can be recovered by aggregating all available information, whereas $\mathbf{f}_{it}^\circ$ and $\mathbf{f}_{jt}^\bullet$ can be identified block by block.

\item[4.] The dimensions $i$ and $j$ exhibit symmetry, despite potential differences in the values of $L$ and $N$.
\end{itemize}

Finally, to accommodate potential endogeneity, we posit that the unobservable factors exert influence on regressors, subject to less restrictive conditions (refer to Remark \ref{rm1} and Assumption \ref{AS1} for more details). Further elaboration on this aspect will be provided when detailing the model setup and associated conditions in Section \ref{Sec2}. Our objective is twofold: to estimate the coefficients of these endogenous regressors and to discern factors across various levels or blocks of the hierarchy. 

Concluding this section, we summarize our contributions as follows. In addition to the introduction of a panel data regression model featuring three sets of latent factor structures, this study also provides additional contributions in the following areas:

\begin{itemize}[leftmargin=*, itemsep=0.5pt, parsep=0.5pt, topsep=0.6pt]
\item[1.]  We develop an estimation method to recover the parameters of interest and unobservable factors across levels and blocks of the hierarchy, and establish its asymptotic properties. The unified hierarchy structure in a 3D framework covers the total 64 possibilities of incidental parameters which have been the centre of a wide range of applications since \cite{Neyman}.

\item[2.] The main model and its estimation method discussed in Section 2 below are developed for the case where both cross--sectional dependence and serial correlation are allowed for the error terms. By contrast, the existing literature (see, \cite{JLS2024}, for example) focuses on the case where the error components are independent and identically distributed (i.i.d.).

\item[3.] We propose a bootstrap procedure to obtain inferences for the parameters of interest, accommodating various types of cross--sectional dependence (CSD) and time series autocorrelation (TSA) within the data generating process of the hierarchy.

\item[4.] To validate our theoretical findings, we conduct extensive simulations and analyze real data examples. In the empirical study, we specifically utilize data from manufacturing industries at the ISIC two-digit level to examine the twin hypotheses of conditional and unconditional convergence for manufacturing industries across countries.
\end{itemize}

The rest of the paper is organized as follows. Section \ref{Sec2} presents our model and methodology.   An asymptotic theory is established accordingly for each step involved in different estimation procedures. Section \ref{Sec3} conducts extensive numerical studies to examine the theoretical findings. Specifically, Section \ref{Sec3} examines the theoretical findings using extensive simulations, while Section \ref{Sec4} uses a set of data from manufacturing industries at the ISIC two-digit level to examine the twin hypotheses of conditional and unconditional-convergence for manufacturing industries across countries.  Section \ref{Sec5} concludes. In the online supplementary Appendices A and B, Appendix A lists the necessary tables and pictures for Section 4. Appendix \ref{App.2} provides two detailed numerical implementations; Appendix \ref{App_bias} proposes a Jackknife based bias correction method; Appendix \ref{AppB1} includes the necessary preliminary lemmas; Appendix \ref{AppB2} presents the detailed proofs; some additional estimation results of the empirical study are reported in Appendix B.

Before proceeding further, it is convenient to introduce some notations that will be repeatedly used throughout the article. Vectors and matrices are always written in bold font. For a matrix $\mathbf{A}$, $\|\mathbf{A}\|$ and $\|\mathbf{A}\|_2$  denote the Frobenius norm and the spectral norm of $\mathbf{A}$, respectively, and $\mathbf{A}^\top$ stands for the transpose of $\mathbf{A}$. Provided that $\mathbf{A}$ has full column rank, let $\mathbf{M}_{\mathbf{A}}=\mathbf{I}- \mathbf{P}_{\mathbf{A}}$ with $\mathbf{P}_{\mathbf{A}} =\mathbf{A}(\mathbf{A}^\top \mathbf{A})^{-1}\mathbf{A}^\top$. For two block wise matrices $\mathbf{A}=\{ \mathbf{a}_{ij}\}$ and $\mathbf{B}=\{ \mathbf{b}_{ij}\}$, provided $\mathbf{a}_{ij}\mathbf{b}_{ij}$ is well defined, we let $\circ$ define the block wise Hadamard product, i.e., $\mathbf{A}\circ \mathbf{B} =\{ \mathbf{a}_{ij}\mathbf{b}_{ij}\}$. For each $\mathbf{a}_{ij}$, let $\sum_{i,j} \mathbf{a}_{ij}=\sum_{i=1}^L\sum_{j=1}^N \mathbf{a}_{ij} $.
Also, we let $\diag\{\mathbf{A}, \mathbf{B} \}$ return a diagonal matrix with $\mathbf{A}$ and $\mathbf{B}$ on the main diagonal. For a vector $\mathbf{b}$, let $\|\mathbf{b}\|_1$ define the $L^1$ norm. For two scalars $m$ and $n$, $m\wedge n=\min\{m, n\}$, $m\vee n=\max\{m, n\}$. For two random variables $a$ and $b$, $a\asymp b$ stands for $a=O_P(b)$ and $b=O_P(a)$. $I(\cdot)$ represents the conventional indicator function, and  ``$\to_P$" and ``$\to_D$" stand for convergence in probability and convergence in distribution, respectively.  $E^*$ and $\text{Pr}^*$ stand for the expectation and probability induced by the bootstrap procedure.

 \section{Model, Assumptions and Estimation Method}\label{Sec2}

\renewcommand{\theequation}{2.\arabic{equation}}
\setcounter{equation}{0}

In this section, we will introduce our panel data regression model featuring three sets of latent factor structures, outline a method for estimating the model, and establish its asymptotic properties. Before delving into the specifics of our model, we wish to emphasize two crucial points. Firstly, our main focus is on regressing $y_{ijt}$ on a $d\times 1$ vector $\mathbf{x}_{ijt}$, where $d$ is fixed. Secondly, as discussed in the introduction, we hypothesize that the unobservable factors impact the regressors to address potential endogeneity. With this consideration, our panel data regression model with three sets of latent factor structures is formulated as follows:

\begin{eqnarray}\label{m.md1}
\begin{pmatrix}
1& -\pmb{\beta}_{ij}^\top \\
0&\mathbf{I}_d 
\end{pmatrix}
\begin{pmatrix}
y_{ijt}  \\ \mathbf{x}_{ijt}
\end{pmatrix} =\begin{pmatrix}
\pmb{\gamma}_{ij}^\top \\
\pmb{\phi}_{ij}^\top
\end{pmatrix} \mathbf{f}_t +\begin{pmatrix}
\pmb{\gamma}_{ij}^{\circ\top} \\
\pmb{\phi}_{ij}^{\circ\top}
\end{pmatrix} \mathbf{f}_{it}^\circ +\begin{pmatrix}
\pmb{\gamma}_{ij}^{\bullet\top} \\
\pmb{\phi}_{ij}^{\bullet\top}
\end{pmatrix} \mathbf{f}_{jt}^\bullet + \begin{pmatrix}
\varepsilon_{ijt} \\
\mathbf{v}_{ijt}
\end{pmatrix},
\end{eqnarray}
where $\pmb{\beta}_{ij}$ and every quantity on the right hand side of \eqref{m.md1} are unknown, and $(i,j,t)\in [L]\times [N]\times [T]$ with $(L,N,T)$ can all be large. 

Our goal is to infer $\pmb{\beta}_{ij}$, and to recover the spaces spanned by the three sets of unobservable factors. For the time being, we assume that $\ell$, $\ell_i^\circ$'s and $\ell_j^\bullet$'s are known, and we will address their estimation in Section \ref{Sec2.3}. The setup connects with the existing works, such as \cite{Ando}, \cite{KSS2020} and \cite{JLS2023}, by addressing many challenges of the relevant literature within one framework.  Meanwhile, it is worth mentioning that the paper by \cite{JLS2024} has an almost identical setup as our model \eqref{m.md1}, and they offer a set of tests on the heterogeneous coefficients which are driven by the independent and identically distributed (i.i.d.) error components attached to the coefficients. In this paper, we pay particular attention to the case where the random errors of the main model are allowed to be both cross--sectionally dependent and serially correlated. 

We now rewrite the main equation (i.e., the upper equation) of \eqref{m.md1} in vector form as follows:

\begin{eqnarray}\label{m.md2}
\mathbf{Y}_{ij\centerdot} &=&\mathbf{X}_{ij\centerdot}  \pmb{\beta}_{ij} + \mathbf{F}\pmb{\gamma}_{ij}  + \mathbf{F}_i^\circ \pmb{\gamma}_{ij}^\circ + \mathbf{F}_j^\bullet  \pmb{\gamma}_{ij}^\bullet+\pmb{\mathcal{E}}_{ij\centerdot}  ,
\end{eqnarray}
where {\small $\mathbf{Y}_{ij\centerdot} =(y_{ij1},\ldots, y_{ijT})^\top$, $\mathbf{X}_{ij\centerdot} =(\mathbf{x}_{ij1},\ldots, \mathbf{x}_{ijT})^\top$, $\pmb{\mathcal{E}}_{ij\centerdot} =(\varepsilon_{ij1},\ldots, \varepsilon_{ijT})^\top$,
$\mathbf{F} =(\mathbf{f}_1,\ldots, \mathbf{f}_T)^\top$}, $\mathbf{F}_i^\circ =(\mathbf{f}_{i1}^\circ,\ldots, \mathbf{f}_{iT}^\circ)^\top$, and $\mathbf{F}_j^\bullet =(\mathbf{f}_{j1}^\bullet,\ldots, \mathbf{f}_{jT}^\bullet)^\top$.

\begin{remark}\label{rm1}
We refrain from expressing $\mathbf{x}_{ijt}$ in matrix form at this stage, as we aim to minimize constraints on $\pmb{\phi}_{ij}$, $\pmb{\phi}_{ij}^\circ$ and $\pmb{\phi}_{ij}^\bullet$ as much as possible. These parameters may even exhibit characteristics akin to weak factor loadings, as observed in \cite{Yamagata2023}. If the focus is about the structure of $\mathbf{x}_{ijt}$ only, we refer the interested reader to \cite{JLS2023} for a comprehensive investigation on the factor structure without regressors.
\end{remark}

According to \eqref{m.md1} and \eqref{m.md2}, it is reasonable to assume that for $\forall (i,j)$

\begin{eqnarray}\label{def.Fstar}
\mathbf{F}_{ij}^*=(\mathbf{F}, \mathbf{F}_i^\circ,\mathbf{F}_j^\bullet)
\end{eqnarray}
has full column rank. Otherwise, one can always reorganize the unobservable factors to achieve a representation with full column rank. 

\begin{remark}\label{rm2}
If $\mathbf{F}_{ij}^*$ is observable, we can immediately rewrite \eqref{m.md2} as 
\begin{eqnarray}
\mathbf{M}_{\mathbf{F}_{ij}^*}\mathbf{Y}_{ij\centerdot}=\mathbf{M}_{\mathbf{F}_{ij}^*}\mathbf{X}_{ij\centerdot}  \pmb{\beta}_{ij} +\mathbf{M}_{\mathbf{F}_{ij}^*}\pmb{\mathcal{E}}_{ij\centerdot}  ,
\end{eqnarray}
which enables consistent estimation for $\pmb{\beta}_{ij}$ using the OLS method. For the case with unobservable $\mathbf{F}_{ij}^*$, a very intuitive idea following a typical two dimensional panel data approach (such as \citealp{Ando}) is to consider an objective function in the following form:

\begin{eqnarray}\label{obj0}
(\mathbf{Y}_{ij\centerdot}-\mathbf{X}_{ij\centerdot}  \mathbf{b}_{ij} )^\top \mathbf{M}_{\mathbf{C}_{ij}} (\mathbf{Y}_{ij\centerdot}-\mathbf{X}_{ij\centerdot}  \mathbf{b}_{ij} ),
\end{eqnarray}
where each $\mathbf{C}_{ij} = (\mathbf{C}, \mathbf{C}_{i}^\circ, \mathbf{C}_{j}^\bullet)$ is a generic matrix satisfying $\frac{1}{T}\mathbf{C}_{ij}^\top  \mathbf{C}_{ij} =\mathbf{I}_{\ell+\ell_i^\circ+\ell_i^\bullet}$. One would hope to obtain the estimates of $\pmb{\beta}_{ij}$ and $\mathbf{F}_{ij}^*$ by combining the quantity of \eqref{obj0} over $(i,j)$. However, the restriction  $\frac{1}{T}\mathbf{C}_{ij}^\top \mathbf{C}_{ij} =\mathbf{I}_{\ell+\ell_i^\circ+\ell_i^\bullet}$ simply cannot be fulfilled practically. To illustrate this, suppose we consider a fixed $i$. Then $(\mathbf{C}, \mathbf{C}_{i}^\circ)$ allow us to generate a set of $\mathbf{C}_{|i}^\bullet =\{\mathbf{C}_{j}^\bullet\}$. Once moving on to a different $i^*$, $(\mathbf{C}, \mathbf{C}_{i^*}^\circ)$ will generate another set of $\mathbf{C}_{|i^*}^\bullet =\{\mathbf{C}_{j}^\bullet\}$. In general, $\mathbf{C}_{|i}^\bullet $ and $\mathbf{C}_{|i^*}^\bullet $ are not necessarily the same unless (1): $(L\vee N)\ll T$; and (2): $\{\mathbf{C}_{i}^\circ\}$ and $\{\mathbf{C}_{j}^\bullet\}$ are generated from two orthogonal spaces from the perspective of vector multiplication. However, it rules out the most common restriction, such as $L \asymp N \asymp T$.
\end{remark}

To tackle the problem raised in Remark \ref{rm2}, we put our thoughts in a nutshell below. For simplicity, suppose that as $T\rightarrow \infty$

\begin{eqnarray}\label{condff}
\max_{i,j}\left\|\frac{1}{T}\mathbf{F}_{ij}^{*\top}\mathbf{F}_{ij}^*-\mathbf{I}_{\ell+\ell_i^\circ+\ell_j^\bullet}\right\|=o_P(1),
\end{eqnarray}
which does not lose any generality, as it is only a matter of rotating matrices in view of the factor structure admitting the following expression:

\begin{eqnarray}
\mathbf{F}\pmb{\gamma}_{ij}  + \mathbf{F}_i^\circ \pmb{\gamma}_{ij}^\circ + \mathbf{F}_j^\bullet  \pmb{\gamma}_{ij}^\bullet =\mathbf{F}_{ij}^* \pmb{\gamma}_{ij}^*
\end{eqnarray}
in which $\pmb{\gamma}_{ij}^* = (\pmb{\gamma}_{ij}^\top, \pmb{\gamma}_{ij}^{\circ\top}, \pmb{\gamma}_{ij}^{\bullet\top} )^\top$. Lemme \ref{LemmaT2} of the online Appendix A shows the feasibility of \eqref{condff} when taking $\max$ over $(i,j)$. By \eqref{condff}, simple algebra shows that

\begin{eqnarray}\label{clue1}
\max_{i,j}\left\|\mathbf{M}_{\mathbf{F}_{ij}^*} - \left(\mathbf{I}_T-\frac{1}{T}\mathbf{F}\mathbf{F}^\top -\frac{1}{T}\mathbf{F}_i^\circ\mathbf{F}_i^{\circ \top} - \frac{1}{T}\mathbf{F}_j^\bullet\mathbf{F}_j^{\bullet\top}\right)\right\|=o_P(1),
\end{eqnarray}
in which the representation $\mathbf{I}_T-\frac{1}{T}\mathbf{F}\mathbf{F}^\top -\frac{1}{T}\mathbf{F}_i^\circ\mathbf{F}_i^{\circ \top} - \frac{1}{T}\mathbf{F}_j^\bullet\mathbf{F}_j^{\bullet\top}$ automatically separates $\mathbf{F}$, $\mathbf{F}_i^\circ$, and $\mathbf{F}_j^\bullet$, and also serves as a projection matrix asymptotically. Thus, \eqref{clue1} sheds light on how to recover $\mathbf{F}$, $\mathbf{F}_i^\circ$, and $\mathbf{F}_j^\bullet$ within a single framework. We are now ready to proceed.
 
\subsection{Estimation Procedure}\label{sec2.1}

We start by introducing a set of new symbols. Let $\mathbf{C}$, $\mathbf{C}_i^\circ$, and $\mathbf{C}_j^\bullet$ be some generic matrices each sharing the same dimensions with $\mathbf{F}$, $\mathbf{F}_i^\circ$, and $\mathbf{F}_j^\bullet$ respectively. We define

\begin{eqnarray}\label{defCs}
\mathbf{C}_{\centerdot \centerdot} = (\mathbf{C}, \mathbf{C}^\circ, \mathbf{C}^\bullet)^\top,\quad \mathbf{C}^\circ = (\mathbf{C}_1^\circ,\ldots, \mathbf{C}_L^\circ),\quad \mathbf{C}^\bullet = (\mathbf{C}_1^\bullet,\ldots, \mathbf{C}_N^\bullet ) .
\end{eqnarray}
Similar to $\mathbf{C}^\circ$ and $\mathbf{C}^\bullet$, we define $\mathbf{F}^\circ = (\mathbf{F}_1^\circ,\ldots, \mathbf{F}_L^\circ)$ and $ \mathbf{F}^\bullet = (\mathbf{F}_1^\bullet,\ldots, \mathbf{F}_N^\bullet ).$ Also, let 

\[
\mathbf{b}_{\centerdot\centerdot} = (\mathbf{b}_{11},\ldots,\mathbf{b}_{1N},\ldots \ldots, \mathbf{b}_{LN})^\top
\] be a generic matrix with dimensions matching $\pmb{\beta}_{\centerdot\centerdot} = (\pmb{\beta}_{11},\ldots, \pmb{\beta}_{1N},\ldots\ldots,\pmb{\beta}_{LN})^\top$.

Using these symbols, we introduce the following objective function:

\begin{eqnarray}\label{eq3}
Q(\mathbf{b}_{\centerdot\centerdot} ,\mathbf{C}_{\centerdot \centerdot}) &=& \frac{1}{LNT}\sum_{i=1}^L\sum_{j=1}^N(\mathbf{Y}_{ij\centerdot} -\mathbf{X}_{ij\centerdot } \mathbf{b}_{ij})^\top \mathbf{C}_{ij}^\dag (\mathbf{Y}_{ij\centerdot} -\mathbf{X}_{ij\centerdot }\mathbf{b}_{ij}) ,
\end{eqnarray}
where $\mathbf{C}_{ij}^\dag = \mathbf{M}_{(\mathbf{C}, \mathbf{C}_i^\circ)}+\mathbf{M}_{(\mathbf{C}, \mathbf{C}_j^\bullet)}$ and the following conditions hold:

\begin{eqnarray}\label{conC}
\frac{1}{T}\mathbf{C}^\top \mathbf{C}  = \mathbf{I}_{\ell},\quad \frac{1}{T}\mathbf{C}_i^{\circ\top} \mathbf{C}_i^\circ  = \mathbf{I}_{\ell_i^\circ},\quad \frac{1}{T}\mathbf{C}_j^{\bullet\top} \mathbf{C}_j^\bullet  = \mathbf{I}_{\ell_j^\bullet},\quad \mathbf{C}^\top (\mathbf{C}_i^\circ, \mathbf{C}_j^\bullet)= \mathbf{0}_{\ell\times (\ell_i^\circ +\ell_i^\bullet)}.
\end{eqnarray}

\begin{remark}\label{RM3}

We stress a few key points below.

\begin{enumerate}[leftmargin=*, itemsep=0.5pt, parsep=0.5pt, topsep=0.6pt]

\item One may regard either $\mathbf{M}_{(\mathbf{C}, \mathbf{C}_i^\circ)}$ or $\mathbf{M}_{(\mathbf{C}, \mathbf{C}_j^\bullet)}$ as a projection matrix to be used to recover the global factors with slightly over-specified number of factors. The fact is reflected in the following operation:

\begin{eqnarray*}
\mathbf{F}_{ij}^\dag\mathbf{F}_{ij}^* = (\mathbf{M}_{(\mathbf{F},\mathbf{F}_i^\circ)}+\mathbf{M}_{(\mathbf{F},\mathbf{F}_j^\bullet)})(\mathbf{F},\mathbf{F}_i^\circ,\mathbf{F}_j^\bullet)=(\mathbf{0},\mathbf{M}_{(\mathbf{F},\mathbf{F}_j^\bullet)}\mathbf{F}_i^\circ,\mathbf{M}_{(\mathbf{F},\mathbf{F}_i^\circ)}\mathbf{F}_j^\bullet),
\end{eqnarray*}
where $\mathbf{F}_{ij}^\dag= \mathbf{M}_{(\mathbf{F},\mathbf{F}_i^\circ)}+\mathbf{M}_{(\mathbf{F},\mathbf{F}_j^\bullet)}$. By Assumption \ref{AS1}.3, it is then easy to show that

\begin{eqnarray*}
   \max_{i,j}\frac{1}{\sqrt{T}}\|(\mathbf{F}_{ij}^\dag\mathbf{F}_{ij}^*) - (\mathbf{0}, \mathbf{F}_i^\circ,  \mathbf{F}_j^\bullet)\|=o_P(1),
\end{eqnarray*}
so a structure like $\mathbf{F}_{ij}^\dag$ only projects out the global factor, and keeps the local factors in the system as residuals. It reveals one may be able to recover global and local factors sequentially due to different signal strength.

\item The first three conditions of \eqref{conC} are standard in the literature. The last condition essentially requires $\mathbf{C} \perp \mathbf{C}^\circ $ and $\mathbf{C} \perp \mathbf{C}^\bullet $. Rather than labeling it as a constraint, it merely dictates the sequence of calculations when minimizing $Q(\mathbf{b}_{\centerdot\centerdot} ,\mathbf{C}_{\centerdot \centerdot})$.

Consider $\mathbf{C}_i^\circ$ as an example. By \eqref{conC}, we can write
\[
\mathbf{C}_{ij}^\dag=\mathbf{M}_{(\mathbf{C}, \mathbf{C}_i^\circ)}+\mathbf{M}_{(\mathbf{C}, \mathbf{C}_j^\bullet)}=2\mathbf{I}_T-2\mathbf{P}_{\mathbf{C}}-\mathbf{P}_{\mathbf{C}_i^\circ}-\mathbf{P}_{\mathbf{C}_j^\bullet},
\] 
so $\mathbf{C}_{ij}^\dag$ maintains a structure similar to that in \eqref{clue1}. Minimizing $Q(\mathbf{b}_{\centerdot\centerdot} ,\mathbf{C}_{\centerdot \centerdot}) $ with respect to $\mathbf{C}_i^\circ$ is equivalent to maximizing the following quantity:

\begin{eqnarray}\label{eq4}
&&\max_{\mathbf{C}_i^\circ} \tr \left(\sum_{j=1}^N(\mathbf{Y}_{ij\centerdot} -\mathbf{X}_{ij\centerdot } \mathbf{b}_{ij})^\top \mathbf{C}_i^\circ\mathbf{C}_i^{\circ\top} (\mathbf{Y}_{ij\centerdot} -\mathbf{X}_{ij\centerdot } \mathbf{b}_{ij}) \right)\nonumber \\ 
&=&\max_{\mathbf{C}_i^\circ}\tr \left(\mathbf{C}_i^{\circ\top} \mathbf{M}_{\mathbf{C}} \sum_{j=1}^N(\mathbf{Y}_{ij\centerdot} -\mathbf{X}_{ij\centerdot } \mathbf{b}_{ij})(\mathbf{Y}_{ij\centerdot} -\mathbf{X}_{ij\centerdot } \mathbf{b}_{ij})^\top \mathbf{M}_{\mathbf{C}} \mathbf{C}_i^{\circ}\right) ,
\end{eqnarray}
where the equality holds because of $\mathbf{C}\perp \mathbf{C}_i^\circ$. Equation \eqref{eq4} implies that in order to obtain the estimates of $\mathbf{F}_{i}^\circ$ and $\mathbf{F}_{j}^\bullet$, one should project out the estimate of $\mathbf{F}$ firstly, which is automatically taken into account in view of \eqref{eq3} and \eqref{eq4}.  Now, how to minimize global and local factors sequentially becomes even clearer.

\item Additionally, since

\[
\lambda_{\min}(\mathbf{C}_{ij}^\dag)\ge \lambda_{\min} (\mathbf{M}_{(\mathbf{C}, \mathbf{C}_i^\circ)})+\lambda_{\min}(\mathbf{M}_{(\mathbf{C}, \mathbf{C}_j^\bullet)})= 0,
\]
we have $Q(\mathbf{b}_{\centerdot\centerdot} ,\mathbf{C}_{\centerdot \centerdot})\ge 0$ to ensure the non-negativity of the objective function.
\end{enumerate}
\end{remark}

Having explained the above points, we outline our estimation procedure below.

\bigskip

\hrule
\begin{itemize}[leftmargin=*, itemsep=0.5pt, parsep=0.5pt, topsep=0.6pt]
\item[] \textbf{Step 1} Perform the following optimization:

\begin{eqnarray}\label{est1}
(\widehat{\mathbf{b}}_{\centerdot\centerdot}, \widehat{\mathbf{C}}_{\centerdot \centerdot})=\argmin Q(\mathbf{b}_{\centerdot \centerdot},\mathbf{C}_{\centerdot \centerdot}) ,
\end{eqnarray}
where $\widehat{\mathbf{b}}_{\centerdot\centerdot} =(\widehat{\mathbf{b}}_{11},\ldots, \widehat{\mathbf{b}}_{LN})^\top$ and $\widehat{\mathbf{C}}_{\centerdot \centerdot} = (\widehat{\mathbf{C}},\widehat{\mathbf{C}}_1^\circ,\ldots, \widehat{\mathbf{C}}_L^\circ, \widehat{\mathbf{C}}_1^\bullet,\ldots, \widehat{\mathbf{C}}_N^\bullet)^\top$. Here, \eqref{est1} can be decomposed as follows:

\begin{itemize}[leftmargin=*, itemsep=0.5pt, parsep=0.5pt, topsep=0.6pt]
\item[1.] $\widehat{\mathbf{b}}_{ij} =  (\mathbf{X}_{ij\centerdot}^\top \widehat{\mathbf{C}}_{ij}^\dag \mathbf{X}_{ij\centerdot} )^{-1} \mathbf{X}_{ij\centerdot}^\top \widehat{\mathbf{C}}_{ij}^\dag \mathbf{Y}_{ij\centerdot},$ where $\widehat{\mathbf{C}}_{ij}^\dag= 2\mathbf{I}_T-2\mathbf{P}_{\widehat{\mathbf{C}}}-\mathbf{P}_{\widehat{\mathbf{C}}_i^\circ}-\mathbf{P}_{\widehat{\mathbf{C}}_j^\bullet}$;

\item[2.] $\widehat{\mathbf{C}}\widehat{\mathbf{V}} =\widehat{\pmb{\Sigma}} \widehat{\mathbf{C}}$, where $\widehat{\mathbf{V}}$ includes the largest $\ell$ eigenvalues of $\widehat{\pmb{\Sigma}} $, and 

\begin{eqnarray*}
\widehat{\pmb{\Sigma}} = \frac{1}{LNT}\sum_{i=1}^L\sum_{j=1}^N(\mathbf{Y}_{ij\centerdot} -\mathbf{X}_{ij\centerdot } \widehat{\mathbf{b}}_{ij} )(\mathbf{Y}_{ij\centerdot} -\mathbf{X}_{ij\centerdot } \widehat{\mathbf{b}}_{ij})^\top;
\end{eqnarray*}

\item[3.] For $\forall i\in [L]$, $\widehat{\mathbf{C}}_i^\circ\widehat{\mathbf{V}}_i^\circ =\widehat{\pmb{\Sigma}}_i^\circ \widehat{\mathbf{C}}_i^\circ$, where $\widehat{\mathbf{V}}_i^\circ$ includes the largest $\ell_i^\circ$ eigenvalues of $\widehat{\pmb{\Sigma}}_i^\circ$, and 

\begin{eqnarray*}
\widehat{\pmb{\Sigma}}_i^\circ = \frac{1}{NT} \sum_{j=1}^N \mathbf{M}_{\widehat{\mathbf{C}}}(\mathbf{Y}_{ij\centerdot} -\mathbf{X}_{ij\centerdot } \widehat{\mathbf{b}}_{ij})(\mathbf{Y}_{ij\centerdot} -\mathbf{X}_{ij\centerdot } \widehat{\mathbf{b}}_{ij})^\top\mathbf{M}_{\widehat{\mathbf{C}}};
\end{eqnarray*}

\item[4.] For $\forall j\in [N]$, $\widehat{\mathbf{C}}_j^\bullet\widehat{\mathbf{V}}_j^\bullet =\widehat{\pmb{\Sigma}}_j^\bullet \widehat{\mathbf{C}}_j^\bullet$, where $\widehat{\mathbf{V}}_j^\bullet$ includes the largest $\ell_j^\bullet$ eigenvalues of $\widehat{\pmb{\Sigma}}_j^\bullet$, and 

\begin{eqnarray*}
\widehat{\pmb{\Sigma}}_j^\bullet = \frac{1}{LT} \sum_{i=1}^L\mathbf{M}_{\widehat{\mathbf{C}}}(\mathbf{Y}_{ij\centerdot} -\mathbf{X}_{ij\centerdot } \widehat{\mathbf{b}}_{ij})(\mathbf{Y}_{ij\centerdot} -\mathbf{X}_{ij\centerdot } \widehat{\mathbf{b}}_{ij})^\top\mathbf{M}_{\widehat{\mathbf{C}}}.
\end{eqnarray*}
\end{itemize}

\item[] \textbf{Step 2} Finally, update the estimate of $\pmb{\beta}_{ij}$ by

\begin{eqnarray}\label{est2}
\widetilde{\mathbf{b}}_{ij} = (\mathbf{X}_{ij\centerdot}^\top \mathbf{M}_{\widehat{\mathbf{C}}_{ij}} \mathbf{X}_{ij\centerdot} )^{-1} \mathbf{X}_{ij\centerdot}^\top  \mathbf{M}_{\widehat{\mathbf{C}}_{ij}}  \mathbf{Y}_{ij\centerdot},
\end{eqnarray}
where $\widehat{\mathbf{C}}_{ij} =(\widehat{\mathbf{C}}, \widehat{\mathbf{C}}_i^\circ, \widehat{\mathbf{C}}_j^\bullet)$.
\end{itemize}
\hrule

\bigskip

The detailed numerical implementation in provided in Appendix \ref{App.2}. In the next subsection, we propose the above estimation procedure, and establish the corresponding asymptotic properties.

\subsection{Main Assumptions}\label{sec2.2}

In the following, we first provide some notation and mathematical symbols which will be repeatedly used in Assumptions and theoretical development of the paper. Key assumptions with their justifications will follow.

\noindent \textbf{Notation}:

\noindent  Regressors: $\mathbf{X}_{i \centerdot \centerdot } =( \mathbf{X}_{i1\centerdot} ,\ldots, \mathbf{X}_{iN\centerdot} ) $ and $\mathbf{X}_{\centerdot j\centerdot } =( \mathbf{X}_{1j\centerdot} ,\ldots, \mathbf{X}_{Lj\centerdot} ) $;

\noindent  Errors: $\pmb{\mathcal{E}}_{\centerdot \centerdot \centerdot}=(\pmb{\mathcal{E}}_{1\centerdot\centerdot} ,\ldots,\pmb{\mathcal{E}}_{L\centerdot\centerdot} )^\top$,  $\pmb{\mathcal{E}}_{i\centerdot\centerdot} = ( \pmb{\mathcal{E}}_{i1\centerdot}, \ldots, \pmb{\mathcal{E}}_{iN\centerdot})$, $\pmb{\mathcal{E}}_{\centerdot j \centerdot} =( \pmb{\mathcal{E}}_{1j\centerdot}, \ldots, \pmb{\mathcal{E}}_{Lj\centerdot})$, $\mathbf{V}_{\centerdot\centerdot\centerdot}=(\mathbf{V}_{1\centerdot\centerdot},\cdots, \mathbf{V}_{L\centerdot\centerdot})^\top$,  $\mathbf{V}_{i\centerdot\centerdot}=(\mathbf{V}_{i1\centerdot},\ldots,\mathbf{V}_{iN\centerdot})$, $\mathbf{V}_{\centerdot j\centerdot}=(\mathbf{V}_{1j\centerdot},\ldots,\mathbf{V}_{Lj\centerdot})$, and $\mathbf{V}_{ij\centerdot}=(\mathbf{v}_{ij1},\ldots, \mathbf{v}_{ijT})^\top$;

\noindent  Loadings: $\pmb{\phi}_{ij}^* = (\pmb{\phi}_{ij}^\top, \pmb{\phi}_{ij}^{\circ \top}, \pmb{\phi}_{ij}^{\bullet \top})^\top$, $\pmb{\gamma}_{ij}^* = (\pmb{\gamma}_{ij}^\top, \pmb{\gamma}_{ij}^{\circ \top}, \pmb{\gamma}_{ij}^{\bullet \top})^\top$,   $\pmb{\Gamma}_{i\centerdot}^\circ =(\pmb{\gamma}_{i1}^\circ,\ldots, \pmb{\gamma}_{iN}^\circ  )^\top$, and $\pmb{\Gamma}_{\centerdot j}^\bullet = (\pmb{\gamma}_{1j}^\bullet,\ldots, \pmb{\gamma}_{Lj}^\bullet )^\top$.

\begin{assumption}\label{AS1}
\item  

\begin{itemize}[leftmargin=*, itemsep=0.5pt, parsep=0.5pt, topsep=0.6pt]
\item[1.] For the error components, we impose the following conditions:

\begin{itemize}[leftmargin=*, itemsep=0.5pt, parsep=0.5pt, topsep=0.6pt]
\item[(a)] $\|\pmb{\mathcal{E}}_{\centerdot \centerdot \centerdot}\|_2=O_P(\sqrt{LN}\vee \sqrt{T})$, and $\left(\frac{ \max_i\| \pmb{\mathcal{E}}_{i\centerdot\centerdot}\|_2}{\sqrt{NT}}+\frac{ \max_j\| \pmb{\mathcal{E}}_{\centerdot j\centerdot}\|_2}{\sqrt{LT}} \right)\cdot\log(L\vee N\vee T)=o_P(1)$;

\item[(b)] $ |\frac{1}{LNT} \sum_{i,j}\pmb{\mathcal{E}}_{ij\centerdot}^\top   \mathbf{F}_{ij}^* \pmb{\gamma}_{ij}^* | =O_P ( \frac{1}{\sqrt{LN \wedge T}} )$, and $\left| \frac{1}{LNT} \sum_{i,j}\pmb{\mathcal{E}}_{ij\centerdot}^\top   \mathbf{X}_{ij\centerdot}\pmb{\beta}_{ij} \right| =O_P (\frac{1}{\sqrt{LN \wedge T}})$'

\item[(c)] $\|\pmb{\mathbf{V}}_{\centerdot \centerdot \centerdot}\|_2=O_P(\sqrt{LN}\vee \sqrt{T})$, $\|\pmb{\mathbf{V}}_{i \centerdot \centerdot}\|_2=O_P(\sqrt{N}\vee \sqrt{T})$, and $\|\pmb{\mathbf{V}}_{\centerdot j\centerdot}\|_2=O_P(\sqrt{L}\vee \sqrt{T})$,  for $\forall (i,j)$.
\end{itemize}

\item[2.] For the loadings, we impose the following conditions:

\begin{itemize}[leftmargin=*, itemsep=0.5pt, parsep=0.5pt, topsep=0.6pt]
\item[(a)] $\frac{1}{LN}\sum_{i,j} (\pmb{\gamma}_{ij}^*\pmb{\gamma}_{ij}^{*\top}-E[\pmb{\gamma}_{ij}^*\pmb{\gamma}_{ij}^{*\top}] )=o_P(1)$ in which 

\begin{eqnarray*}
0<\min_{i,j} \lambda_{\min}(E[\pmb{\gamma}_{ij}^*\pmb{\gamma}_{ij}^{*\top}] )\le \max_{i,j} \lambda_{\max}(E[\pmb{\gamma}_{ij}^*\pmb{\gamma}_{ij}^{*\top}] )<\infty;
\end{eqnarray*}

\item[(b)] $\max_i \|\frac{1}{N} \pmb{\Gamma}_{i\centerdot}^{\circ \top}\pmb{\Gamma}_{i\centerdot}^\circ  -\pmb{\Sigma}_{\pmb{\gamma}, i}^\circ  \| =o_P (1) $ in which

\begin{eqnarray*}
0<\min_i \lambda_{\min}(\pmb{\Sigma}_{\pmb{\gamma}, i}^\circ)\le \max_i \lambda_{\max}(\pmb{\Sigma}_{\pmb{\gamma}, i}^\circ)<\infty;
\end{eqnarray*}

\item[(c)] $\max_j \|\frac{1}{L} \pmb{\Gamma}_{\centerdot j}^{\bullet \top}\pmb{\Gamma}_{\centerdot j}^\bullet  -\pmb{\Sigma}_{\pmb{\gamma}, j}^\bullet \| =o_P (1)$ in which

\begin{eqnarray*}
0<\min_j \lambda_{\min}(\pmb{\Sigma}_{\pmb{\gamma}, j}^\bullet)\le \max_j \lambda_{\max}(\pmb{\Sigma}_{\pmb{\gamma}, j}^\bullet)<\infty.
\end{eqnarray*}
\end{itemize}

\item[3.] For the factors, we impose the following conditions:

\begin{itemize}[leftmargin=*, itemsep=0.5pt, parsep=0.5pt, topsep=0.6pt]
\item[(a)] $\log(L\vee N\vee T)\cdot\max_{i,j}\| \frac{1}{T} \mathbf{F}_{ij}^{*\top} \mathbf{F}_{ij}^* -\mathbf{I}_{\ell+\ell_i^\circ+\ell_j^\bullet}\| =o_P (1)$;

\item[(b)] $\|\mathbf{F}^\circ\|_2 =O_P(\sqrt{L}\vee \sqrt{T})$ and $\|\mathbf{F}^\bullet\|_2 =O_P(\sqrt{N}\vee \sqrt{T})$.

\end{itemize}

\item[4.] For the regressors, we impose the following conditions:

\begin{itemize}[leftmargin=*, itemsep=0.5pt, parsep=0.5pt, topsep=0.6pt]
\item[(a)] $\max_i\frac{1}{NT} \| \mathbf{X}_{i\centerdot\centerdot}\|^2 =O_P(1)$;

\item[(b)] $\max_j\frac{1}{LT}  \| \mathbf{X}_{\centerdot j\centerdot}\|^2 =O_P(1)$.
\end{itemize}

\item[5.] Suppose that the following conditions are satisfied with probability approaching 1:
\[\inf_{\mathbf{C}_{\centerdot\centerdot}} \lambda_{\min}\left(\frac{1}{T}\mathbf{D}_{\centerdot \centerdot}\right)>0,
\] 
where $\mathbf{D}_{\centerdot \centerdot} = \diag\{\mathbf{D}_{11},\ldots, \mathbf{D}_{LN}\}$, $\mathbf{D}_{ij, 1} = \mathbf{X}_{ij\centerdot}^\top\mathbf{C}_{ij}^\dag  \mathbf{X}_{ij\centerdot} $, $\mathbf{D}_{ij, 2} = \pmb{\gamma}_{ij} \otimes (\mathbf{C}_{ij}^\dag  \mathbf{X}_{ij\centerdot})$, and $\mathbf{D}_{ij} =\mathbf{D}_{ij, 1} -\mathbf{D}_{ij, 2}^\top (E[\pmb{\gamma}_{ij} \pmb{\gamma}_{ij}^{\top}] \otimes \mathbf{I}_T)^{-1}\mathbf{D}_{ij, 2}$.
\end{itemize}
\end{assumption}

Assumption \ref{AS1} imposes a set of regular conditions. All of the conditions involving $\max_i$, $\max_j$ and $\max_{ij}$ can be easily justified in the same fashion as in Lemma \ref{LemmaT2} in Appendix A below. However, as we cannot define some type of mixing conditions along with the dimension $i$ or $j$, so we state them as they are via a set of high level assumptions. 
Assumption \ref{AS1}.1.(a) specifies the spectral norm orders of random errors, which can be satisfied by various examples discussed in the supplementary Appendix S.2 of \cite{Moon2017}. In particular, it can be justified by i.i.d. $\varepsilon_{ijt}$ with zero mean and uniformly bounded fourth moment $E[\varepsilon_{ijt}^4]$. Following the arguments in Lemma S.2.1 of \cite{Moon2017}, we can demonstrate that Assumption \ref{AS1}.1.(a) also holds for the MA($\infty$) process: $\varepsilon_{ijt} = \sum_{\tau=0}^\infty a_{ij\tau}e_{ij,t-\tau}$, where $e_{ijt}$ are i.i.d. with zero mean and uniformly bounded fourth moment, and the MA coefficients ${a_{ij\tau}}$ satisfy certain uniform boundedness conditions.

Assumption \ref{AS1}.1.(b) is met by strictly exogenous and i.i.d. or $\alpha$-mixing errors. Assumption \ref{AS1}.2 regulates the behaviour of factor loadings, which can be justified by identically distributed factor loadings with weak cross--sectional dependence. Assumption \ref{AS1}.3 imposes conditions on the factors, which can be easily satisfied. For instance, Assumption \ref{AS1}.3.(a) can be met by  $\mathbf{f}_t$, $\mathbf{f}_{it}^\circ$, and $\mathbf{f}_{jt}^\bullet$ that are independently generated from a multivariate standard normal distribution. In Assumption \ref{AS1}.3.(b), the required orders of spectral norm of $\mathbf{F}^\circ$ and $\mathbf{F}^\bullet$ can be fulfilled by the i.i.d. or MA($\infty$) processes, similar to those discussed for Assumption \ref{AS1}.1.

As explained in Remark \ref{rm1}, Assumption \ref{AS1}.4 actually does not impose too many conditions on variables used to generate $\mathbf{x}_{ijt}$. For instance, one may have weak factor signal through the factor loadings associated with $\mathbf{x}_{ijt}$. If that is the case, one may add certain factor structure to $\mathbf{x}_{ijt}$. In short, we focus on the current estimation approach in Section \ref{Sec2}.  
We comment on Assumption \ref{AS1}.5 as follows. This assumption only includes the factor loadings of the global factors, so it is similar to Assumption A of \cite{Bai}, but with a lightly over specified number of factors that kicks in via $\mathbf{C}_{ij}^\dag= \mathbf{M}_{(\mathbf{C}, \mathbf{C}_i^\circ)}+\mathbf{M}_{(\mathbf{C}, \mathbf{C}_j^\bullet)}$ by the construction. This is indeed achievable as argued in \cite{Moon}. As we have not imposed too many conditions on the components of $\mathbf{x}_{ijt}$ up to this point, we state Assumption \ref{AS1}.5 as it is. In a very extreme case in which $\mathbf{x}_{ijt}\equiv \mathbf{v}_{ijt}$, this condition reduces to $\min_{i,j}\lambda_{\min}(\mathbf{V}_{ij\centerdot}^\top\mathbf{V}_{ij\centerdot}/T)$  which can be realized under a set of additional conditions on $\mathbf{v}_{ijt}$  and in connection with Lemma \ref{LemmaT2} in Appendix A below.  

\begin{assumption}\label{AS2}

\item 

\begin{itemize}[leftmargin=*, itemsep=0.5pt, parsep=0.5pt, topsep=0.6pt]
\item[1.] Suppose that $\{ \pmb{\zeta}_{ijt} : = ( \varepsilon_{ijt}, \mathbf{v}_{ijt}^\top)^\top\}$ are independent of the other variables, and $E[\pmb{\zeta}_{ijt} ]=\mathbf{0}$ and $ E[\varepsilon_{ijt} | \mathbf{v}_{ijt}]=0$, and for $\forall (i,j)$, $ \{\pmb{\zeta}_{ijt}  |  t\in [T]\}$ are strictly $\alpha$-mixing process with the $\alpha$-mixing coefficient 

\begin{eqnarray*}
\alpha_{ij}(t) = \sup_{A\in \mathcal{F}_{ij,-\infty}^0, B\in  \mathcal{F}_{ij,t}^\infty} |P(A)P(B) -P(AB)|
\end{eqnarray*}
satisfying $ \max_{i,j}\sum_{t=1}^\infty  [\alpha_{ij}(t)]^{\kappa_{ij}/(4+\kappa_{ij})} < \infty$ with $\kappa_{ij}>0$ and  $\max_{i,j} E\|\pmb{\zeta}_{ij1}\|^{2+\kappa_{ij}} <\infty$, where $\mathcal{F}_{ij,-\infty}^0$ and $\mathcal{F}_{ij,t}^\infty$ are the $\sigma$-algebras generated by $\{\pmb{\zeta}_{ijs}: s \leq 0\}$ and $\pmb{\zeta}_{ijs}: s \geq t\}$, respectively. Finally, there exits $\kappa^*\in (0, \min_{i,j}\kappa_{ij})$ such that $LN/T^{1+\kappa^*/4}\to 0$.

\item[2.] $\max_{i,j} (\|\pmb{\gamma}_{ij}^\circ\|^2+\|\pmb{\gamma}_{ij}^\bullet\|^2)=O_P(\log(LN))$.

\item[3.] The minimum eigenvalue of $\pmb{\Sigma}_{\mathbf{v},ij}=\lim_{T\to\infty}\frac{1}{T}E[\mathbf{V}_{ij\centerdot}^\top \mathbf{V}_{ij\centerdot}]$ is positive and uniformly bounded away from zero for all $i$ and $j$. 
\end{itemize}

\end{assumption}

The mixing condition of Assumption \ref{AS2}.1 is necessary to derive Lemma \ref{LemmaT2}, which is then further required to derive Lemma \ref{M.LM2} and the asymptotic distribution. The requirements about $\kappa_{ij}$'s impose conditions on the existence of some moments of $\pmb{\zeta}_{ijt}$, which can be easily fulfilled, e.g., $\pmb{\zeta}_{ijt}$ follows a normal distribution. Assumption \ref{AS2}.2 is standard.

\subsection{Asymptotic Properties}\label{sec2.3}

Under the main assumptions, we are now ready to establish the consistency and asymptotic normality results.

\begin{lemma}\label{M.LM1}
For {\normalfont \textbf{Step 1}}, under Assumption \ref{AS1}, as $(L,N,T)\to (\infty,\infty,\infty)$, we have

\begin{itemize}[leftmargin=*, itemsep=0.5pt, parsep=0.5pt, topsep=0.6pt]
\item[1.] $\frac{1}{\sqrt{LN}}\| \pmb{\beta}_{\centerdot \centerdot} -\widehat{\mathbf{b}}_{\centerdot \centerdot} \|=o_P(1)$;

\item[2.]  $\|\mathbf{P}_{\mathbf{F}} -\mathbf{P}_{\widehat{\mathbf{C}}} \|=O_P\left( \frac{1}{\sqrt{LN}}\|\pmb{\beta}_{\centerdot \centerdot}-\widehat{\mathbf{b}}_{\centerdot\centerdot}\| +\frac{1}{\sqrt{L\wedge N\wedge T}}\right)$.
\end{itemize}
\end{lemma}

The establishment of Lemma \ref{M.LM1} does not require an excessive number of conditions for the variables involved in \eqref{m.md1}. It shows that we can recover $\pmb{\beta}_{\centerdot \centerdot}$ as a whole. As expected, the global factor $\mathbf{F}$ can be estimated firstly by putting everything together.  

As shown in \eqref{eq3d} and \eqref{eq4}, in order to recover $\mathbf{F}_i^\circ$ and $\mathbf{F}_j^\bullet$, we need to divide the data into different blocks as presented in the estimation procedure, so $\widehat{\mathbf{b}}_{ij}$'s are automatically partitioned into blocks accordingly. As a consequence, we can no longer use the overall consistency of $\widehat{\mathbf{b}}_{\centerdot \centerdot}$ derived in Lemma \ref{M.LM1}. To proceed, we further impose Assumption \ref{AS2}, and summarize the results associated with $\mathbf{F}_i^\circ$ and $\mathbf{F}_j^\bullet$ in Lemma \ref{M.LM2} below.

\begin{lemma}\label{M.LM2}
For {\normalfont \textbf{Step 1}}, under Assumptions \ref{AS1} and \ref{AS2}, as $(L,N,T)\to (\infty,\infty,\infty)$,

\begin{itemize}[leftmargin=*, itemsep=0.5pt, parsep=0.5pt, topsep=0.6pt]
\item[1.] $\max_{i, j}\| \pmb{\beta}_{ij}-\widehat{\mathbf{b}}_{ij}\|=o_P(1)$;
 
\item[2.] $\max_i\|\mathbf{P}_{ \mathbf{F}_i^\circ} -\mathbf{P}_{ \widehat{\mathbf{C}}_i^\circ}\|=O_P (\max_{i,j} \|\pmb{\beta}_{ij}-\widetilde{\mathbf{b}}_{ij}\| + \frac{\sqrt{\log(LN)}}{\sqrt{N \wedge T}}  +\frac{1}{\sqrt{L}}+ \frac{\max_i\|\pmb{\mathcal{E}}_{i\centerdot\centerdot}\|_2}{\sqrt{N T}} +\frac{\max_i \| \mathbf{F}_i^{\circ\top} \mathbf{F}\|_2 }{T}  )$;

\item[3.] $\max_j\|\mathbf{P}_{ \mathbf{F}_j^\bullet} -\mathbf{P}_{ \widehat{\mathbf{C}}_j^\bullet}\|=O_P (\max_{i,j} \|\pmb{\beta}_{ij}-\widehat{\mathbf{b}}_{ij}\| + \frac{\sqrt{\log(LN)}}{\sqrt{L \wedge T}} +\frac{1}{\sqrt{N}} + \frac{\max_j\|\pmb{\mathcal{E}}_{\centerdot j\centerdot}\|_2}{\sqrt{L T}}+\frac{\max_j\| \mathbf{F}_j^{\bullet\top} \mathbf{F}\|_2}{T} )$.
\end{itemize}
\end{lemma}

Lemma \ref{M.LM2} justifies the validity of our approach across all blocks involved in \textbf{Step 1}. With this, we have successfully recovered all key components of model \eqref{m.md2}, thus concluding our examination of \textbf{Step 1} in the estimation procedure. 

Now, we proceed to Step 2, where we update the estimator for $\pmb{\beta}_{ij}$. Additionally, we consider two mean group estimators to infer the average values of  $\pmb{\beta}_{ij}$ on two cross-sectional dimensions, respectively. For $\forall (i,j)$, define
\begin{equation*}
\widetilde{\mathbf{b}}_{i\centerdot }=\frac{1}{N}\sum_{j=1}^N \widetilde{\mathbf{b}}_{ij}, \quad \widetilde{\mathbf{b}}_{\centerdot j}=\frac{1}{L}\sum_{i=1}^L \widetilde{\mathbf{b}}_{ij}.
\end{equation*} 

To facilitate the theoretical development, we introduce some additional notation. 
Let $\mathcal{R}_{ij}^{*}=(\mathcal{R}_{F}, \mathcal{R}_{F,i}^\circ, \mathcal{R}_{F,j}^\bullet)$, where  $\mathcal{R}_F = \sum_{ij}(\mathbf{F}_{i}^{\circ}\pmb{\gamma}_{ij}^\circ+\mathbf{F}_{j}^{\bullet}\pmb{\gamma}_{ij}^\bullet)\pmb{\gamma}_{ij}^\top (\pmb{\Gamma}_{\centerdot \centerdot }^\top \pmb{\Gamma}_{\centerdot \centerdot })^{-1}$, $\mathcal{R}_{F,i}^\circ=\sum_{j=1}^N(\mathbf{F}\pmb{\gamma}_{ij}+\mathbf{F}_{j}^{\bullet}\pmb{\gamma}_{ij}^\bullet)\pmb{\gamma}_{ij}^{\circ\top} (\pmb{\Gamma}_{i\centerdot  }^{\circ\top} \pmb{\Gamma}_{i\centerdot  }^\circ)^{-1}$ and $\mathcal{R}_{F,j}^\bullet = \sum_{i=1}^L(\mathbf{F}\pmb{\gamma}_{ij}+\mathbf{F}_{i}^{\circ}\pmb{\gamma}_{ij}^\circ)\pmb{\gamma}_{ij}^{\bullet\top} (\pmb{\Gamma}_{\centerdot j }^{\bullet\top} \pmb{\Gamma}_{\centerdot j }^\bullet)^{-1}$.

\begin{theorem}\label{M.Thm1}
Let  Assumptions \ref{AS1} and \ref{AS2} hold.  As $(L,N,T)\to (\infty,\infty,\infty)$, 
\begin{itemize}
\item[(1)]for $\forall (i,j)$, if  $\sqrt{T}/(L\wedge N) \to 0$,
\begin{eqnarray*}
\sqrt{T} (\widetilde{\mathbf{b}}_{ij} - \pmb{\beta}_{ij} ) \to_D N(\mathbf{0}, \pmb{\Sigma}_{\mathbf{v},ij}^{-1}\pmb{\Sigma}_{\mathbf{v},\varepsilon,ij}\pmb{\Sigma}_{\mathbf{v},ij}^{-1}),
\end{eqnarray*}
where $\pmb{\Sigma}_{\mathbf{v},\varepsilon,ij} =\lim_{T\to\infty}\frac{1}{T}E[\mathbf{V}_{ij\centerdot}^\top\pmb{\mathcal{E}}_{ij\centerdot } \pmb{\mathcal{E}}_{ij\centerdot }^\top \mathbf{V}_{ij\centerdot}]$.

\item[(2)]for $\forall i$, if  $\sqrt{T}/(L\wedge N) \to 0$ and $\sqrt{N}/(L\wedge T) \to 0$,
\begin{eqnarray*}
\sqrt{NT} (\widetilde{\mathbf{b}}_{i\centerdot} -  \pmb{\beta}_{i\centerdot} -\mathcal{B}_{i\centerdot} ) \to_D N(\mathbf{0}, \pmb{\Sigma}_{\mathbf{b},i\centerdot}),
\end{eqnarray*}
where $\mathcal{B}_{i\centerdot}=\frac{1}{NT}\sum_{j=1}^N\pmb{\Sigma}_{\mathbf{v},ij}^{-1} \pmb{\phi}_{ij}^{*\top} \mathcal{R}_{ij}^{*\top} \mathbf{M}_{\widehat{\mathbf{C}}_{ij}} \mathcal{R}_{ij}^* \pmb{\gamma}_{ij}^*$ is the bias term with the probability order of $O_P(\frac{1}{L\wedge N \wedge T})$, and
$\pmb{\Sigma}_{\mathbf{b},i\centerdot}=\lim_{N,T\to\infty} \frac{1}{NT}\sum_{j_1,j_2=1}^N \pmb{\Sigma}_{\mathbf{v},ij_1}^{-1} E[\mathbf{V}_{ij_1\centerdot}^\top\pmb{\mathcal{E}}_{ij_1\centerdot } \pmb{\mathcal{E}}_{ij_2\centerdot }^\top \mathbf{V}_{ij_2\centerdot}]\pmb{\Sigma}_{\mathbf{v},ij_2}^{-1}$.

\item[(3)]for $\forall j$, if  $\sqrt{T}/(L\wedge N) \to 0$ and $\sqrt{L}/(N\wedge T) \to 0$,
\begin{eqnarray*}
\sqrt{LT} (\widetilde{\mathbf{b}}_{\centerdot j} - \pmb{\beta}_{\centerdot j}-\mathcal{B}_{\centerdot j} ) \to_D N(\mathbf{0}, \pmb{\Sigma}_{\mathbf{b},\centerdot j}),
\end{eqnarray*}
where  $\mathcal{B}_{\centerdot j}=\frac{1}{LT}\sum_{i=1}^L\pmb{\Sigma}_{\mathbf{v},ij}^{-1} \pmb{\phi}_{ij}^{*\top} \mathcal{R}_{ij}^{*\top} \mathbf{M}_{\widehat{\mathbf{C}}_{ij}} \mathcal{R}_{ij}^* \pmb{\gamma}_{ij}^*$ is the bias term with the probability order of $O_P(\frac{1}{L\wedge N \wedge T})$, and $\pmb{\Sigma}_{\mathbf{b},\centerdot j}=\lim_{L,T\to\infty} \frac{1}{LT}\sum_{i_1,i_2=1}^L \pmb{\Sigma}_{\mathbf{v},i_1j}^{-1} E[\mathbf{V}_{i_1j\centerdot}^\top\pmb{\mathcal{E}}_{i_2j\centerdot } \pmb{\mathcal{E}}_{i_2j\centerdot }^\top \mathbf{V}_{i_2j\centerdot}]\pmb{\Sigma}_{\mathbf{v},i_2j}^{-1}$.
\end{itemize}

\end{theorem}

Theorem \ref{M.Thm1}.(1) provides the asymptotic distribution for each $\widetilde{\mathbf{b}}_{ij}$. Meanwhile, Theorem \ref{M.Thm1}.(2-3) establish the asymptotic normality for the mean group estimators, which incorporate bias terms to achieve the optimal rates of convergence  ($\sqrt{NT}$ and $\sqrt{LT}$, respectively). The presence of bias terms arises due to the hierarchical factor structure, which can be addressed using conventional methods such as the analytical approach or the Jackknife procedure \citep[see][for example]{Chen2021}. In Appendix \ref{App_bias} below, we propose a Jackknife method to remove the bias terms present in the theorem. 

To conduct practical inference, it is essential to deal with the time series correlation involved in the term $\pmb{\Sigma}_{\mathbf{v},\varepsilon,ij}$. Therefore, we now focus on Theorem  \ref{M.Thm1}.(1) and then develop a new dependent wild bootstrap procedure by considerably extending \cite{GPY2023} for the standard time--series and cross-sectional panel data setting to the following setting where the double--index $(i,j)$ involved can be both cross--sectional. For the second and third parts of Theorem  \ref{M.Thm1}, the corresponding bootstrap procedure involves a two--step approach in each case. The first step is a bias correction procedure as discussed in Appendix \ref{App_bias} below, and the second step develops a bootstrap procedure for each case in almost the same way as proposed below.

\bigskip

\noindent \textbf{Bootstrap Procedure:}
\hrule
\begin{itemize}[leftmargin=*, itemsep=0.5pt, parsep=0.5pt, topsep=0.6pt]
\item[(i).] Let $\pmb{\xi}=(\xi_{1},\ldots, \xi_{T})^{\top}$ be an
$m$-dependent time series for each bootstrap draw, and let the components of $\pmb{\xi}$ satisfy that 

\begin{eqnarray*}
E[\xi_{t}]=0,\quad E|\xi_{t}|^{2} =1,\quad E|\xi_{t}|^{2+\kappa_{ij}/2
}<\infty,\quad E[\xi_{t}\xi_{s}]=a\left(  \frac{t-s}{m} \right)  , 
\end{eqnarray*}
where $m\to\infty$, $\kappa_{ij}$ is a constant and defined in Assumption \ref{AS2}, and $a(\cdot)$ is a symmetric kernel defined on $[-1,1]$ satisfying that $a(0)=1$ and $K_{a}(x)=\int_{\mathbb{R}} a(u)e^{-iux}du\ge0$ for $x\in\mathbb{R}$. 

\item[(ii).] For $\forall (i,j)$, construct a new set of dependent
variables by $\mathbf{Y}_{ij\centerdot}^{*} = \mathbf{X}_{ij\centerdot} \widehat{\mathbf{b}}_{ij}+ \widehat{\pmb{\mathcal{E}}}_{ij\centerdot}\circ\pmb{\xi}$, where $\widehat{\pmb{\mathcal{E}}}_{ij\centerdot} =  \mathbf{Y}_{ij\centerdot} -\mathbf{X}_{ij\centerdot} \widehat{\mathbf{b}}_{ij}$. Accordingly, the new estimate  of $\pmb{\beta}_{ij}$ is obtained using 

\begin{eqnarray*}
\widetilde{\mathbf{b}}_{ij}^*=\left(\mathbf{X}_{ij\centerdot}^\top \mathbf{M}_{\widehat{\mathbf{C}}_{ij}} \mathbf{X}_{ij\centerdot}\right)^{-1} \mathbf{X}_{ij\centerdot}^\top \mathbf{M}_{\widehat{\mathbf{C}}_{ij}} \mathbf{Y}_{ij\centerdot}^*,
\end{eqnarray*}

\item[(iii).] We repeat the above procedure $\mathcal{L}$ times.
\end{itemize}
\hrule

\bigskip

\begin{theorem}\label{M.Thm2}
Let  Assumptions \ref{AS1} and \ref{AS2} hold. Assume further that $\frac{m}{\sqrt{Th}} \to 0$ and $mT^{2/\underline{\kappa}_{ij}-1}\rightarrow 0$, where $\underline{\kappa}_{ij}=\min(\kappa^\ast_{ij},4)$, $\kappa^\ast_{ij}=1/ (\frac{1}{2+\kappa_{ij}}+\frac{\kappa_{ij}}{2(4+\kappa_{ij})} )$, and $\kappa_{ij}$ is defined in Assumption \ref{AS2}.
As $(L,N,T)\to (\infty,\infty,\infty)$, for $\forall (i,j)$

\begin{eqnarray*}
\sup_{w\in \mathbb{R}}\left| \text{\normalfont Pr}^* (\sqrt{T}(\widetilde{\mathbf{b}}_{ij}^* -\widetilde{\mathbf{b}}_{ij})\le w) -\Pr (\sqrt{T}(\widetilde{\mathbf{b}}_{ij}-\pmb{\beta}_{ij})\le w)\right| =o_P(1).
\end{eqnarray*}
\end{theorem}

Theorem \ref{M.Thm2} provides a procedure to recover the asymptotic distribution for each given $(i,j)$. It is worth mentioning that the DWB method is initially introduced in \cite{shao2010} in the context of time series analysis, wherein a comprehensive comparison between the DWB and some existing bootstrap methods can be found.

\subsection{Selection of the Numbers of Factors}\label{Sec2.3}

To conclude our theoretical investigation, we provide a procedure to select the numbers of factors. For notational simplicity, we define $\pmb{\ell}^\circ =(\ell_1^\circ,\ldots, \ell_L^\circ)^\top$ and $\pmb{\ell}^\bullet =(\ell_1^\bullet,\ldots, \ell_N^\bullet)^\top.$ Thus, our goal is to estimate $\ell$, $\pmb{\ell}^\circ$, and $\pmb{\ell}^\bullet$.

First, we emphasize that we can still get consistent estimation of $\widehat{\mathbf{b}}_{\centerdot\centerdot}$ by prescribing a $d_{\max}$ to be the numbers of global, country and industry factors. As long as $d_{\max}\ge (\ell \vee (\max_i \ell_i^\circ) \vee (\max_i\ell_j^\bullet))$, the estimation procedure of Section \ref{sec2.1}, and the results developed in Section \ref{sec2.2} still hold true. In connection with the estimation procedure of Section \ref{sec2.1}, we provide the following steps to estimate $\ell$ firstly, and then estimate the elements of $\pmb{\ell}^\circ$ and $\pmb{\ell}^\bullet$ sequentially.

\bigskip

\hrule

\begin{itemize}[leftmargin=*, itemsep=0.5pt, parsep=0.5pt, topsep=0.6pt]
\item[] \textbf{Step 1*} Prescribe a $d_{\max}$ to be the numbers of global, country and industry factors, and implement the estimation procedure of Section \ref{sec2.1} to get $\widetilde{\mathbf{b}}_{\centerdot \centerdot} =(\widetilde{\mathbf{b}}_{11},\ldots, \widetilde{\mathbf{b}}_{1N},\ldots, \widetilde{\mathbf{b}}_{LN})^\top$. 

\item[] \textbf{Step 2*} Using $\widetilde{\mathbf{b}}_{\centerdot \centerdot}$, we calculate $\widehat{\pmb{\Sigma}}$ as in Section \ref{sec2.1}, and estimate the number of global factors as follows:

\begin{eqnarray}\label{EQ2.7}
\widehat{\ell} =\argmin_{0\le s \le d_{\max} }  \Big\{\frac{\widehat{\lambda}_{s+1}}{\widehat{\lambda}_s }\cdot I(\widehat{\lambda}_s \ge \omega)+I(\widehat{\lambda}_s < \omega)  \Big\},
\end{eqnarray}
where $\omega =[\log (L\vee N\vee T)]^{-1}$, $\widehat{\lambda}_0 =1$ is a mock eigenvalue, and $\widehat{\lambda}_s$ stands for the $s^{th}$ largest eigenvalue of $\widehat{\pmb{\Sigma}}$. Record the eigenvectors corresponding to the largest $\widehat{\ell}$ eigenvalues in $\widehat{\mathbf{C}}$.

\item[]\textbf{Step 3*} Having obtained $\widetilde{\mathbf{b}}_{\centerdot \centerdot}$ and $\widehat{\mathbf{C}}$, we sequentially estimate $\ell_i^\circ$'s and $\ell_j^\bullet$'s as follows:

\begin{eqnarray}\label{EQ2.9}
\widehat{\ell}_i^\circ &=&\argmin_{0\le s \le d_{\max} } \Big\{ \frac{\widehat{\lambda}_{i,s+1}^\circ}{\widehat{\lambda}_{i,s}^\circ} \cdot I(\widehat{\lambda}_{i, s}^\circ \ge \omega)+ I(\widehat{\lambda}_{i, s}^\circ < \omega)\Big\},\nonumber \\
\widehat{\ell}_j^\bullet &=&\argmin_{0\le s \le d_{\max} }  \Big\{ \frac{\widehat{\lambda}_{j,s+1}^\bullet}{\widehat{\lambda}_{j,s}^\bullet} \cdot I(\widehat{\lambda}_{j, s}^\bullet \ge \omega)+ I(\widehat{\lambda}_{j, s}^\bullet< \omega)\Big\},
\end{eqnarray}
where $\widehat{\lambda}_{i,0}^\circ=\widehat{\lambda}_{j,0}^\bullet=1$ are mock eigenvalues, and $\widehat{\lambda}_{i,s}^\circ$ and $\widehat{\lambda}_{j,s}^\bullet$ stand for the $s$-th largest eigenvalues of $\widehat{\pmb{\Sigma}}_i^\circ $ and $\widehat{\pmb{\Sigma}}_j^\bullet$ respectively. 
\end{itemize}
\hrule

\bigskip

The estimators in \eqref{EQ2.7} and \eqref{EQ2.9} can be considered as extensions of \cite{LY12}. However, as pointed out by \cite{LY12}, it remains unresolved to bound  the ratio associated with the eigenvalues which converge to 0 from below. To bypass this unresolved issue, we introduce a data driven tuning parameter $\omega$. The idea is that although it is challenging to study a ratio with a denominator converging to 0, we can discard this ratio and construct a V-shape curve by employing the indicator function in \eqref{EQ2.7} and \eqref{EQ2.9}, respectively.  

\begin{theorem} \label{Theorem2.3}
Under Assumptions \ref{AS1} and \ref{AS2}, as $(L,N,T)\to (\infty,\infty,\infty)$,  $\Pr(\widehat{\ell} =\ell, \widehat{\pmb{\ell}}^\circ= \pmb{\ell}^\circ, \widehat{\pmb{\ell}}^\bullet= \pmb{\ell}^\bullet)\to 1,$ where $\widehat{\pmb{\ell}}^\circ =(\widehat{\ell}_1^\circ,\ldots, \widehat{\ell}_L^\circ)^\top $ and $\widehat{\pmb{\ell}}^\bullet=(\widehat{\ell}_1^\bullet,\ldots, \widehat{\ell}_L^\bullet)^\top $.
\end{theorem} 

Up to this point, we have successfully recovered all the unknown quantities as mentioned in the beginning of Section \ref{Sec2}. In the next section, we examine the above theoretical results using extensive simulations.

\begin{remark}
To conclude this section, we note that practically it is not guaranteed that each country (or industry) has the same number of individuals, so it brings certain unbalanceness to the data set. Formally, $\forall i\in [L]$, we may have $j\in [N_i]$, which is exactly the same as our empirical study of Section \ref{Sec4}. After carefully checking the proofs, we can see that as long as $\min_i N_i\to \infty$ and $\frac{T}{\min_i N_i^2}\to 0$, all the aforementioned results are still valid with certain modifications.
\end{remark}

\section{Simulation Studies}\label{Sec3}

\renewcommand{\theequation}{3.\arabic{equation}}
\setcounter{equation}{0}

In this section, we conduct extensive numerical studies to examine the theoretical findings. Specifically, we provide simulation studies, and the main data generating process (DGP) follows \eqref{m.md1}. In what follows, we provide details about each component involved.

To capture heterogeneity of the coefficients, we generate them by $\pmb{\beta}_{ij} =(0.5+i/L, 0.5+j/N)^\top$ for $(i,j)\in [L]\times[N]$, which implies $d=2$. For factors and loadings, we first generate the numbers of factors. Specifically, we let $\ell =2$, and let $\ell_i^\circ$ and $ \ell_j^\bullet$ be randomly selected from the set $\{ 0,1,2\}$ with equal probability attached to each possible outcome. The design is to show that we allow $\ell_i^\circ$ and $ \ell_j^\bullet$ to be 0, which is meaningful for practical analysis.  Therefore, for each generated dataset, only $\ell$ is fixed. The values of $\pmb{\ell}^\circ$ and $\pmb{\ell}^\bullet$ vary at each simulation replication.

Sequentially, we generate factors and loadings, and let them be independently drawn from some normal distributions: $\mathbf{f}_t\sim N(\mathbf{0},\mathbf{I}_\ell)$,  $\mathbf{f}_{it}^\circ\sim N(\mathbf{0},2\cdot\mathbf{I}_{\ell_i^\circ})$ if $\ell_i^\circ>0$, $\mathbf{f}_{jt}^\bullet\sim N(\mathbf{0}, 2\cdot\mathbf{I}_{\ell_j^\bullet})$ if $\ell_j^\bullet>0$, $\pmb{\gamma}_{ij}\sim N(\mathbf{1},\mathbf{I}_\ell)$, $\pmb{\gamma}_{ij}^\circ\sim N(\mathbf{0},\mathbf{I}_{\ell_i^\circ})$ if $\ell_i^\circ>0$, $\pmb{\gamma}_{ij}^\bullet\sim N(-\mathbf{1},\mathbf{I}_{\ell_j^\bullet})$ if $\ell_j^\bullet>0$, $\pmb{\phi}_{ij, s}\sim N(\mathbf{1},\mathbf{I}_\ell)$, $\pmb{\phi}_{ij, s}^\circ\sim N(\mathbf{0},\mathbf{I}_{\ell_i^\circ})$ if $\ell_i^\circ>0$, $\pmb{\phi}_{ij, s}^\bullet\sim N(-\mathbf{1},\mathbf{I}_{\ell_j^\bullet})$ if $\ell_j^\bullet>0$, where $s\in [d]$, $\pmb{\phi}_{ij, s}$, $\pmb{\phi}_{ij, s}^\circ$, and $\pmb{\phi}_{ij, s}^\bullet$ stand for the $s^{th}$ columns of $\pmb{\phi}_{ij}$, $\pmb{\phi}_{ij}^\circ$, and $\pmb{\phi}_{ij}^\bullet$ respectively. We note that normal distributions are not really necessary here, as the choice of some other distributions doesn't affect the finite-sample performance. To present the main steps for the simulation and save the space, we do not further explore alternative options.

We generate residuals as follows:

\begin{eqnarray*}
&&\pmb{\varepsilon}_{\centerdot \centerdot t} = \rho_\varepsilon\pmb{\varepsilon}_{\centerdot \centerdot t-1} +0.5\cdot\pmb{\Sigma}_\varepsilon^{1/2} \cdot N(\mathbf{0}, \mathbf{I}_{LN})^\top,\quad \mathbf{v}_{\centerdot \centerdot t,s} = \rho_v \mathbf{v}_{\centerdot \centerdot t-1,s} +\pmb{\Sigma}_v^{1/2}\cdot N(\mathbf{0}, \mathbf{I}_{LN})^\top,
\end{eqnarray*}
where $\rho_\varepsilon=\rho_v =0.1$,  $\pmb{\Sigma}_\varepsilon=\pmb{\Sigma}_v  = \{ 0.2^{\|(i_1,j_1)-(i_2,j_2)\|}\}_{LN\times LN}$ with $(i,j)\in [L]\times[N]$, and

\begin{eqnarray*}
&&\pmb{\varepsilon}_{\centerdot \centerdot t} = (\pmb{\varepsilon}_{11t},\ldots, \pmb{\varepsilon}_{1Nt},\ldots\ldots, \pmb{\varepsilon}_{LNt}),\quad \mathbf{v}_{\centerdot \centerdot t,s} = (\mathbf{v}_{11t,s},\ldots, \mathbf{v}_{1Nt,s},\ldots\ldots, \mathbf{v}_{LNt,s}),
\end{eqnarray*}
in which $\mathbf{v}_{ijt,s}$ is the $s^{th}$ element of $\mathbf{v}_{ijt}$, and  $s\in [d]$. Based on the above DGP, we have weak serial correlation and cross-sectional dependence among $(\varepsilon_{ijt}, \mathbf{v}_{ijt}^\top)^\top$.

Finally, we assemble $y_{ijt}$'s and $\mathbf{x}_{ijt}$'s as in \eqref{m.md1}. For each dataset, we carefully calculate all the unknown quantities following the numerical implementation of Appendix \ref{App.2}. 

\medskip

To evaluate the performance of our proposed method, we repeat the above procedure $R=1000$ times, and define several criteria. For notational simplicity, the subindex $_k$ always stands for the corresponding quantities obtained at the $k^{th}$ replication in what follows. That said, we  use root mean square errors (RMSE) to measure our estimates about $\pmb{\beta}_{\centerdot\centerdot}$, $\mathbf{F}$, $\mathbf{F}^\circ$, and $\mathbf{F}^\bullet$:

\begin{eqnarray}\label{RMSE}
\text{RMSE}_{\pmb{\beta}} &=& \left\{\frac{1}{R}\sum_{k=1}^R\frac{1}{LN}\|\widetilde{\mathbf{b}}_{\centerdot\centerdot,k}-  \pmb{\beta}_{\centerdot\centerdot}\|^2 \right\}^{1/2},\nonumber \\
\text{RMSE}_{\mathbf{F}} &=&  \left\{\frac{1}{R}\sum_{k=1}^R \|\mathbf{P}_{\widehat{\mathbf{C}}_k}- \mathbf{P}_{\mathbf{F}_k}\|^2 \right\}^{1/2},\nonumber \\
\text{RMSE}_{\mathbf{F}^\circ} &=&  \left\{\frac{1}{R}\sum_{k=1}^R\frac{1}{L}\sum_{i=1}^L\|\mathbf{P}_{\widehat{\mathbf{C}}_{i,k}^\circ}- \mathbf{P}_{\mathbf{F}_{i,k}^\circ}\|^2 \right\}^{1/2},\nonumber \\
\text{RMSE}_{\mathbf{F}^\bullet} &=&  \left\{\frac{1}{R}\sum_{k=1}^R\frac{1}{N}\sum_{j=1}^N\|\mathbf{P}_{\widehat{\mathbf{C}}_{j,k}^\bullet}- \mathbf{P}_{\mathbf{F}_{j,k}^\bullet}\|^2 \right\}^{1/2},
\end{eqnarray}
where we let $\mathbf{P}_{\widehat{\mathbf{C}}_k} =\mathbf{0}$ if $\widehat{\ell}=0$, and similarly, we let $\mathbf{P}_{\widehat{\mathbf{C}}_{i,k}^\circ}=\mathbf{0}$, $\mathbf{P}_{\mathbf{F}_{i,k}^\circ}=\mathbf{0}$, $\mathbf{P}_{\widehat{\mathbf{C}}_{j,k}^\bullet}=\mathbf{0}$, and $\mathbf{P}_{\mathbf{F}_{j,k}^\bullet}=\mathbf{0}$ if $\widehat{\ell}_{i,k}^\circ =0$, $\ell_{i,k}^\circ =0$, $\widehat{\ell}_{j,k}^\bullet =0$, and $\ell_{j,k}^\bullet =0$ respectively. For the purpose of comparison, we estimate $\pmb{\beta}_{\centerdot\centerdot}$, $\mathbf{F}$, $\mathbf{F}^\circ$ and $\mathbf{F}^\bullet$ by assuming that $\ell$, $\pmb{\ell}^\circ$, and $\pmb{\ell}^\bullet$ are known, and calculate $\text{RMSE}_{\pmb{\beta}}^*$, $\text{RMSE}_{\mathbf{F}}^*$, $\text{RMSE}_{\mathbf{F}^{\circ}}^*$, and $\text{RMSE}_{\mathbf{F}^{\bullet}}^*$ as in \eqref{RMSE}.

To evaluate the estimation about the numbers of factors, we define the following rates:

\begin{eqnarray}\label{Rate}
\text{Rate}_\ell &=& \frac{1}{R}\sum_{k=1}^RI(\widehat{\ell}_k=\ell),\quad
\text{Rate}_{\ell}^- = \frac{1}{R}\sum_{k=1}^RI(\widehat{\ell}_k<\ell),\quad
\text{Rate}_{\ell}^+ = \frac{1}{R}\sum_{k=1}^RI(\widehat{\ell}_k>\ell),\nonumber \\
\text{Rate}_{\pmb{\ell}^\circ} &=& \frac{1}{RL}\sum_{k=1}^R \sum_{i=1}^LI(\widehat{\ell}_{i,k}^\circ=\ell_{i,k}^\circ),\quad \text{Rate}_{\pmb{\ell}^\bullet} = \frac{1}{RN}\sum_{k=1}^R \sum_{j=1}^NI(\widehat{\ell}_{j,k}^\bullet=\ell_{j,k}^\bullet),\nonumber \\
\text{Rate}_{\pmb{\ell}^\circ}^- &=& \frac{1}{RL}\sum_{k=1}^R \sum_{i=1}^LI(\widehat{\ell}_{i,k}^\circ<\ell_{i,k}^\circ),\quad
\text{Rate}_{\pmb{\ell}^\bullet}^- = \frac{1}{RN}\sum_{k=1}^R \sum_{j=1}^NI(\widehat{\ell}_{j,k}^\bullet<\ell_{j,k}^\bullet),\nonumber \\
\text{Rate}_{\pmb{\ell}^\circ}^+ &=& \frac{1}{RL}\sum_{k=1}^R \sum_{i=1}^LI(\widehat{\ell}_{i,k}^\circ >\ell_{i,k}^\circ),\quad
\text{Rate}_{\pmb{\ell}^\bullet}^+ = \frac{1}{RN}\sum_{k=1}^R \sum_{j=1}^NI(\widehat{\ell}_{j,k}^\bullet >\ell_{j,k}^\bullet).
\end{eqnarray}
Apparently, $\text{Rate}_\ell$, $\text{Rate}_{\pmb{\ell}^\circ}$, and $\text{Rate}_{\pmb{\ell}^\bullet}$ measure the percentages of correctly identifying the numbers of global, country and industry factors; $\text{Rate}_{\ell}^-$, $\text{Rate}_{\pmb{\ell}^\circ}^-$, and $\text{Rate}_{\pmb{\ell}^\bullet}^-$ measure the percentages of under estimation; $\text{Rate}_{\ell}^+$, $\text{Rate}_{\pmb{\ell}^\circ}^+$, and $\text{Rate}_{\pmb{\ell}^\bullet}^+$ measure the percentages of over estimation.

Finally, we evaluate the coverage rates (CR) of the bootstrap procedure. For $s\in [d]$,

\begin{eqnarray}
\text{CR}  = \frac{1}{dRLN}\sum_{s=1}^d\sum_{k=1}^R\sum_{i=1}^L\sum_{j=1}^N I(\widetilde{\mathbf{b}}_{ij,s,k} -\pmb{\beta}_{ij,s}\in \text{CI}_{ij,s,k})
\end{eqnarray}
where $\pmb{\beta}_{ij,s}$ stands for the $s^{th}$ element of $\pmb{\beta}_{ij}$,  $\text{CI}_{ij,s,k}$ stands for the 95\% confidence interval obtained from the bootstrap draws associated with $\widetilde{\mathbf{b}}_{ij,s,k}^*-\widetilde{\mathbf{b}}_{ij,s,k}$, and $\widetilde{\mathbf{b}}_{ij,s,k}$ and $\widetilde{\mathbf{b}}_{ij,s,k}^*$  respectively stand for the $s^{th}$ elements of $\widetilde{\mathbf{b}}_{ij}$ and $\widetilde{\mathbf{b}}_{ij}^*$ obtained at the $k^{th}$ simulation replication.

We summarize results in Tables \ref{TB1}, \ref{TB2}, and \ref{TB3}. As $i$ and $j$ are symmetric in a sense, we alter $N$ and $T$ only in the following tables, while keeping $L=60$ for simplicity. We now discuss each table in details. In Table \ref{TB1}, as expected, as either $N$ or $T$ increases, we have improved probability of correctly identifying $\ell$, $\pmb{\ell}^\circ$, and $\pmb{\ell}^\bullet$. Even for the case $(L, N, T)=(60,60,60)$, we are still able to identify different numbers of factors with probabilities greater than 70\%. In Table \ref{TB2}, as expected, $\text{RMSE}_{\pmb{\beta}}^*$,  $\text{RMSE}_{\mathbf{F}}^*$, $\text{RMSE}_{\mathbf{F}^{\circ}}^*$, and $\text{RMSE}_{\mathbf{F}^{\bullet}}^*$ outperform $\text{RMSE}_{\pmb{\beta}} $, $\text{RMSE}_{\mathbf{F}}$, $\text{RMSE}_{\mathbf{F}^{\circ}} $, and $\text{RMSE}_{\mathbf{F}^{\bullet}} $ respectively. However, the differences are not very significant, and gradually vanish as the sample size increases. This is not surprising, as we have better probabilities to identify different types of factors with large sample size. As a result, it is less likely to have misspecified factor structures as the sample size increases. Table \ref{TB3} reports the coverage rates of the bootstrap procedure. Overall, the numbers are very close to the nominal rate 95\%, and even for $T=60$. To conclude, as explained in Remark \ref{rm2}, we emphasize that the above results show our method works well for the case with $L\asymp N\asymp T$.

{\small

\begin{table}[ht]
\small\centering
\renewcommand{\arraystretch}{1}
\setlength{\tabcolsep}{4pt}
\caption{Selection of Factor Numbers}\label{TB1}
\begin{tabular}{rrrccccccccc}
\hline\hline
$N$   & $L$   & $T$   & $\text{Rate}_\ell$  & $\text{Rate}_\ell^-$ & $\text{Rate}_\ell^+$ & $\text{Rate}_{\pmb{\ell}^\circ}$ & $\text{Rate}_{\pmb{\ell}^\circ}^-$ & $\text{Rate}_{\pmb{\ell}^\circ}^+$ & $\text{Rate}_{\pmb{\ell}^\bullet}$ & $\text{Rate}_{\pmb{\ell}^\bullet}^-$ & $\text{Rate}_{\pmb{\ell}^\bullet}^+$ \\
60 & 60 & 60 & 0.702 & 0.000 & 0.298 & 0.709 & 0.219 & 0.072 & 0.764 & 0.165 & 0.071 \\
 &  & 120 & 0.956 & 0.000 & 0.044 & 0.902 & 0.063 & 0.035 & 0.941 & 0.029 & 0.029 \\
 &  & 180 & 0.990 & 0.000 & 0.010 & 0.937 & 0.036 & 0.026 & 0.973 & 0.008 & 0.019 \\
 & 120 & 60 & 0.716 & 0.000 & 0.284 & 0.703 & 0.219 & 0.077 & 0.768 & 0.212 & 0.020 \\
 &  & 120 & 0.946 & 0.000 & 0.054 & 0.887 & 0.072 & 0.042 & 0.954 & 0.043 & 0.003 \\
 &  & 180 & 1.000 & 0.000 & 0.000 & 0.937 & 0.033 & 0.030 & 1.000  & 0.000 & 0.000\\
    \hline\hline
\end{tabular}
\end{table}

\begin{table}[ht]
\small\centering
\renewcommand{\arraystretch}{1}
\setlength{\tabcolsep}{4pt}
\caption{Root Mean Square Errors}\label{TB2}
\begin{tabular}{rrrcccccccc}
\hline\hline
$N$   & $L$   & $T$   & $\text{RMSE}_{\pmb{\beta}}^*$ & $\text{RMSE}_{\mathbf{F}}^*$ & $\text{RMSE}_{\mathbf{F}^{\circ}}^*$ & $\text{RMSE}_{\mathbf{F}^{\bullet}}^*$ & $\text{RMSE}_{\pmb{\beta}} $ & $\text{RMSE}_{\mathbf{F}}$ & $\text{RMSE}_{\mathbf{F}^{\circ}} $ & $\text{RMSE}_{\mathbf{F}^{\bullet}} $ \\
60 & 60 & 60 & 0.270 & 0.207 & 0.694 & 0.332 & 0.335 & 0.582 & 0.842 & 0.587 \\
 &  & 120 & 0.210 & 0.169 & 0.574 & 0.261 & 0.236 & 0.268 & 0.619 & 0.352 \\
 &  & 180 & 0.191 & 0.159 & 0.533 & 0.233 & 0.209 & 0.189 & 0.559 & 0.283 \\
 & 120 & 60 & 0.277 & 0.197 & 0.726 & 0.298 & 0.310 & 0.569 & 0.872 & 0.558 \\
 &  & 120 & 0.224 & 0.165 & 0.617 & 0.225 & 0.230 & 0.284 & 0.664 & 0.309 \\
 &  & 180 & 0.201 & 0.154 & 0.566 & 0.195 & 0.203 & 0.154 & 0.584 & 0.196 \\
    \hline\hline   
\end{tabular}
\end{table}

}

{\footnotesize

\begin{table}[ht]
\small\centering
\caption{Coverage Rates}\label{TB3}
\begin{tabular}{rrrr }
\hline\hline
$N$   & $L$   & $T$ & CR  \\
60 & 60 & 60 & 0.958  \\
 &  & 120 & 0.930  \\
 &  & 180 & 0.948  \\
 & 120 & 60 & 0.989  \\
 &  & 120 & 0.943  \\
 &  & 180 & 0.950  \\
 \hline\hline
\end{tabular}
\end{table}

}

\section{An Empirical Study}\label{Sec4}

\renewcommand{\theequation}{4.\arabic{equation}}
\setcounter{equation}{0}

The analysis of economic and productivity convergence stands as a compelling area of inquiry within economics, offering an ideal application for our hierarchical panel data model as outlined in Section \ref{Sec2}. Specifically, starting with the seminal study by \cite{Baumol}, numerous studies have been devoted to testing whether income or productivity of poorer economies are converging to those of richer economies. As \cite{Durlauf2003} puts it, ``\textit{Few issues in empirical growth economics have received as much attention as the question of whether countries exhibit convergence}". A main technique employed by these studies is ``cross-country growth regressions", where aggregate- or industry-level cross-country data are used to regress the average growth rates of per capita income (or labour productivity) over a long period on the initial level of income per capita (or labour productivity) and some additional control variables\footnote{In this study we follow \cite{Rodrik2013} and define labour productivity of an industry as the industry's real value added divided by its number of employees.  As is well known, the value added of an industry, also referred to as gross domestic product (GDP)-by-industry, is the contribution of a private industry or government sector to overall GDP (The U.S. Bureau of Economic Analysis, 2006, available at \url{https://www.bea.gov/help/faq/184}). The definitions of labour productivity and value added, therefore, imply that an industry's labour productivity can be considered as the industry's ``GDP per capita", which in turn implies that neoclassical growth models predict not only conditional convergence in GDP per capita among economies but also industry-level convergence in labour productivity among economies.}. A negative and significant coefficient on the initial conditions is taken to be evident in favour of $\beta $-convergence. For excellent surveys of cross-country convergence studies, see \cite{Durlauf2003} for example. Despite the increasing availability of disaggregated data at industry level, the hierarchical structure of these data, to the best of our knowledge, has rarely been explored.

\subsection{Data}\label{Sec4.1}

We begin our empirical analysis by introducing the data. For labor productivity (or real value added per employee), we follow \cite{Rodrik2013} and use the dataset of UNIDO's INDSTAT2, which provides data on value added (in nominal U.S dollars) and employment for 23  manufacturing industries at the ISIC two-digit level for a large number of countries. With this dataset, real value added can be computed by deflating the nominal value added by the US producer price index, and labor productivity can then be obtained by further dividing real value added by employment (i.e., number of employees). Growth in labor productivity is then measured as percentage change in labor productivity.

Our control variables include a wide range of factors that have been found to be important for assessing convergence. These include human capital (as measured by school enrollment), investment price, trade openness and terms of trade, institutions (measured by civil liberties),  and government consumption share. We refer interested readers to \cite{Martin} for detailed definitions of a wide range of variables.  A summary of the dataset is given in Table \ref{TabEmp1} and Table \ref{TabEmp2}. It should be noted that since geographical factors are usually time-invariant, they will be captured by the factor structures and thus are not included in the control variables. When measuring dependent and independent variables, we follow \cite{Salimans}, and treat them differently. Specifically, the dependent variable is measured as a five-year moving average of economic growth, while all explanatory variables are measured at the beginning of each five year period. Due to data availability, our sample starts at 1963 and ends at 2018. For the same reason, the number of countries varies across industries.

\subsection{Estimation Results}\label{Sec4.2}

We now investigate conditional convergence for the manufacturing industries, which can be done by estimating equation \eqref{m.md2} where all components (i.e., initial productivity, control variables and three types of factors) are included. The factor selection is achieved using the method that is described in Section \ref{Sec2.3}. 

For the global factor, we identify one factor that affects the growth in labour productivity of each country in every manufacturing industry and it accounts for 22.50\% of the variations in $y_{ijt}$.  Panels A and B of Table \ref{TabEmp3} present the estimated numbers of industry and country factors. As evident from the table, the number of factors varies among industries and countries, ranging from 1 to 10. These factors collectively explain 56.65\% of $y_{ijt}$'s variations. Overall, the factor structure  can account for approximately 80\% of the variations in $y_{ijt}$, indicating a superior goodness of fit for our model.

Figures \ref{Boxplot_1} and \ref{Boxplot_2}  provide the boxplots of country-specific estimated coefficients for each explanatory variable across 23 industries.  As evident in Figure \ref{Boxplot_1}, the estimated partial effects from the initial productivity are generally negative with an average value of -0.157. 
 This suggests that when country characteristics are controlled, initial labour productivity is generally negatively related to the subsequent rate of growth in labour productivity. In other words, conditional convergence in labour productivity exists for the manufacturing industry. This finding is in line with that of \cite{Rodrik2013} who, by applying a fixed effects panel data model to the UNIDO's INDSTAT dataset, also finds conditional convergence in labour productivity for the total manufacturing industry. It is also consistent with the income convergence literature (\citealp{Martin}) that finds that once country characteristics are controlled for, the coefficient on initial income becomes negative and statistically significant. To show the disparity  in conditional convergence of productivity among countries, we present the country-specific estimated partial effects for initial productivity within the food and beverages industry in Table \ref{TabEmp5} as an example. Despite the consistent negative estimates across all countries, there exists significant divergence in their magnitudes. For instance,  coefficients for some countries such as Albania, Croatia, and Malawi exceed -0.25, whereas the effects are substantially lower in others like Pakistan and Peru, or even insignificant as in Jordan and Syria. This finding demonstrates the importance of employing heterogeneous panel data in the examination of productivity convergence. Moreover, such country-specific variations are also present in the estimates for other industries as well as the estimated partial effects for control variables. 
 

In order to confirm our results regarding conditional convergence in labour productivity, we conduct two robustness checks. First, we re-estimate the model for the following two subperiods: 1973-2018 and 1983-2018. The results are provided in Figure \ref{Boxplot_Robust}. It is evident that the estimated partial effects for initial productivity generally remain negative, confirming the conditional convergence of productivity.  Secondly, we follow \cite{Rodrik2013} and exclude OCED countries from our sample of countries. The results, which are also presented in Figure \ref{Boxplot_Robust}, show that our findings of conditional convergence is very robust to the exclusion of OECD countries. 
 
To summarize this section, we have examined the  hypothesis of the conditional convergence of productivity for manufacturing industries across countries. The empirical results presented in this section suggest that there is strong and consistent evidence of convergence once factors that affect steady-state levels of labour productivity are controlled for. As comparison, we also examine the unconditional convergence of productivity by excluding the control variables. We provide the estimation results and discussion on such issue in Appendix A.

\section{Conclusion}\label{Sec5}

The hierarchy of large datasets has garnered considerable attention recently (e.g., \citealp{CYZ2022}, \citealp{JLS2023}, \cite{JLS2024}, \citealp{Zhang2023}, along with numerous real data examples collected in \citealp{Matyas}). In this study, we contribute to the literature by proposing a panel data regression model with three sets of latent factor structures, which, to the best of our knowledge, has not been extensively explored in the existing literature. Rather than consolidating factors from various nodes, as seen in the literature of distributed PCA (\citealp{Fanetal2019}), we propose an estimation method to recover the parameters of interest by peeling off factors layer by layer and across different blocks within certain layers. We establish estimation theory and asymptotic properties accordingly. Additionally, we present a bootstrap procedure to infer the parameters of interest, allowing for different types of cross-sectional dependence (CSD) and time-series autocorrelation (TSA) in the data generating process (DGP). Finally, we validate our theoretical findings using extensive simulated and real data examples. In our empirical study, we utilize data from manufacturing industries at the ISIC two-digit level to examine the twin hypotheses of conditional and unconditional convergence for manufacturing industries across countries.

{\footnotesize
\bibliography{Refs.bib}

@article{Moon2017,
  title={Dynamic linear panel regression models with interactive fixed effects},
  author={Moon, Hyungsik Roger and Weidner, Martin},
  journal={Econometric Theory},
  volume={33},
  number={1},
  pages={158--195},
  year={2017},
  publisher={Cambridge University Press}
}

@article{McLeish1975,
  title={A maximal inequality and dependent strong laws},
  author={McLeish, Don L},
  journal={Annals of probability},
  volume={3},
  number={5},
  pages={829--839},
  year={1975},
  publisher={Institute of Mathematical Statistics}
}

@article{Neyman,
 author = {J. Neyman and Elizabeth L. Scott},
 journal = {Econometrica},
 number = {1},
 pages = {1--32},
 title = {Consistent Estimates Based on Partially Consistent Observations},
 volume = {16},
 year = {1948}
}

@article{Zhang2023,
author = {Bo Zhang and Guangming Pan and Qiwei Yao and Wang Zhou},
title = {Factor Modeling for Clustering High-Dimensional Time Series},
journal = {Journal of the American Statistical Association},
volume = {0},
number = {0},
pages = {1-12},
year = {2023}
}

@article{Yamagata2023,
author = {Yoshimasa Uematsu and Takashi Yamagata},
title = {Inference in Sparsity-Induced Weak Factor Models},
journal = {Journal of Business \& Economic Statistics},
volume = {41},
number = {1},
pages = {126-139},
year = {2023}
}

@article{Fanetal2019,
author = {Jianqing Fan and Dong Wang and Kaizheng Wang and Ziwei Zhu},
title = {{Distributed estimation of principal eigenspaces}},
volume = {47},
journal = {Annals of Statistics},
number = {6},
pages = {3009 -- 3031},
year = {2019}
}

@article{Hansen1992,
  title={Consistent covariance matrix estimation for dependent heterogeneous processes},
  author={Hansen, Bruce E},
  journal={Econometrica},
  volume = {60},
  number = {4},
  pages={967--972},
  year={1992}
}

@book{Bosq1996,
  title={Nonparametric Statistics for Stochastic Processes: Estimation and Prediction},
  author={Bosq, D.},
  isbn={9781468404890},
  lccn={96013588},
  series={Lecture Notes in Statistics},
  year={2012},
  publisher={Springer New York}
}

@article{SY1996,
  title={Weak convergence for weighted empirical processes of dependent sequences},
  author={Shao, Qi-Man and Yu, Hao},
  journal={Annals of Probability},
  volume={24}, 
  number={4}, 
  pages={2098-2127},
  year={1996}
}

@article{GPY2023,
author = {Jiti Gao and Bin Peng and Yayi Yan},
title = {Higher-Order Expansions and Inference for Panel Data Models},
journal = {Journal of the American Statistical Association},
volume = {0},
number = {0},
pages = {1-12},
year = {2023}
}

@ARTICLE {FanLiaoWang,
    author  = "J. Fan and Y. Liao and W. Wang",
    title   = "{Projected principal component analysis in factor models}",
    journal = "Annals of Statistics",
    year    = "2016",
    volume  = "44",
    number  = "1",
    pages   = "219-254"
}

@article{CYZ2022,
author = {Rong Chen and Dan Yang and Cun-Hui Zhang},
title = {Factor Models for High-Dimensional Tensor Time Series},
journal = {Journal of the American Statistical Association},
volume = {117},
number = {537},
pages = {94-116},
year = {2022}
}

@BOOK{Gao,
        author = {J. Gao},
        title = {Nonlinear Time Series: Semiparametric and Nonparametric Methods},
        year = {2007},
        publisher = {Chapman and Hall},
        edition = {}
        }

@article{goncalves_2011,
title={THE MOVING BLOCKS BOOTSTRAP FOR PANEL LINEAR REGRESSION MODELS WITH INDIVIDUAL FIXED EFFECTS},
volume={27},
number={5},
journal={Econometric Theory},
author={Gon{\c c}alves, S.},
year={2011},
pages={1048-1082}}

@article{BAI2020,
title = {Standard errors for panel data models with unknown clusters},
journal = {Journal of Econometrics},
year = {2020},
author = {Jushan Bai and Sung Hoon Choi and Yuan Liao},
 pages   = "forthcoming"
}

@UNPUBLISHED{JLS2023,
    author = "S. Jin and X. Lu and L. Su",
    title  = "Three-Dimensional Factor Models with Global and Local Factors",
    note   = "\url{https://teacher.econ.cuhk.edu.hk/~xunlu/}",
    year   = "2024"
}

@UNPUBLISHED{JLS2024,
    author = "S. Jin and X. Lu and L. Su",
    title  = "Three-Dimensional Heterogeneous Panel Data Models with Multi-level Interactive Fixed Effects",
    note   = "\url{https://teacher.econ.cuhk.edu.hk/~xunlu/}",
    year   = "2024"
}

@ARTICLE {shao2010,
    author    = "Shao, Xiaofeng",
    title     = "The dependent wild bootstrap",
    journal   = "Journal of the American Statistical Association",
    year      = "2010",
    volume    = "105",
    number    = "489",
    pages     = "218-235",
}

@article{KSS2020,
title = {Estimation and inference for multi-dimensional heterogeneous panel datasets with hierarchical multi-factor error structure},
journal = {Journal of Econometrics},
volume = {220},
number = {2},
pages = {504-531},
year = {2021},
author = {George Kapetanios and Laura Serlenga and Yongcheol Shin}
}

@ARTICLE {Ando,
    author  = "T. Ando and J. Bai",
    title   = "Clustering Huge Number of Financial Time Series: A Panel Data Approach With High-Dimensional Predictors and Factor Structures",
    journal = "Journal of the American Statistical Association",
    year    = "2017",
    volume  = "112",
    number  = "519",
    pages   = "1182-1198"
}

@ARTICLE {Bai,
    author  = "J. Bai",
    title   = "Panel Data Models with Interactive Fixed Effects",
    journal = "Econometrica",
    year    = "2009",
    volume  = "77",
    number  = "4",
    pages   = "1229-1279"
}

@ARTICLE {Moon,
    author  = "H. R. Moon and M. Weidner",
    title   = "Linear Regression for Panel with Unknown Number of Factors as Interactive Fixed Effects",
    journal = "Econometrica",
    year    = "2015",
    volume  = "83",
    number  = "4",
    pages   = "1543-1579"
}

@ARTICLE {LY12,
    author  = "C. Lam and Q. Yao",
    title   = "Factor modeling for high-dimensional time series: Inference for the number of factors",
    journal = "Annals of Statistics",
    year    = "2012",
    volume  = "40",
    number  = "2",
    pages   = "694-726"
}

@article{Chen2021,
title = {Nonlinear factor models for network and panel data},
journal = {Journal of Econometrics},
volume = {220},
number = {2},
pages = {296-324},
year = {2021},
author  = "Mingli Chen and Iv\'an Fern\'andez-Val and Martin Weidner",
}

@article{LANCASTER2000391,
title = {The incidental parameter problem since 1948},
journal = {Journal of Econometrics},
volume = {95},
number = {2},
pages = {391--413},
year = {2000},
author = {Tony Lancaster}
}

@article{KS2023,
    author = {Koskinen, Johan and Snijders, Tom A B},
    title = "{Multilevel longitudinal analysis of social networks}",
    journal = {Journal of the Royal Statistical Society Series A: Statistics in Society},
    volume = {186},
    number = {3},
    pages = {376-400},
    year = {2023}
}

@ARTICLE {LMS2021,
    author  = "X. Lu and K. Miao and L. Su",
    title   = "Determination of Different Types of Fixed Effects in Three-Dimensional Panels",
    journal = "Econometrics Reviews",
    year    = "2021",
    volume = {40},
number = {9},
pages = {867-898},
}

@BOOK {Matyas,
    author    = "{Matyas et al.}",
    title     = "The Econometrics of Multi-dimensional Panels: Theory and Applications",
    publisher = "Springer, Chamg",
    year      = "2017"
}

@BOOK {GL2013,
    author    = "G. H. Golub and C. F. {Van Loan}",
    title     = "Matrix Computations (4th Edition)",
    publisher = "The Johns Hopkins University Press",
    year      = "2013"
}

@article{Rodrik2013,
    author = {Rodrik, Dani},
    title = "Unconditional Convergence in Manufacturing",
    journal = {Quarterly Journal of Economics},
    volume = {128},
    number = {1},
    pages = {165-204},
    year = {2013},
}

@article{Baumol,
 author = {William J. Baumol},
 journal = {American Economic Review},
 number = {5},
 pages = {1072-1085},
 publisher = {American Economic Association},
 title = {Productivity Growth, Convergence, and Welfare: {W}hat the Long-Run Data Show},
 volume = {76},
 year = {1986}
}

@article{Durlauf2003,
  author={S. N. Durlauf},
  title={The Convergence Hypothesis After 10 Years},
  journal={Revista Economica de Castilla-La Mancha},
  year={2003},
  volume={2},
  pages={55-74}
}

@ARTICLE {Martin,
    author  = "X. Sala-I-Martin and G. Doppelhofer and R. I. Miller",
    title   = "Determinants of Long-Term Growth: A Bayesian Averaging of Classical Estimates (BACE) Approach",
    journal = "American Economic Review",
    year    = "2004",
    volume  = "94",
    number  = "4",
    pages   = "813-835"
}

@ARTICLE {Salimans,
    author  = "T. Salimans",
    title   = "Variable selection and functional form uncertainty in cross-country growth regressions",
    journal = "Journal of Econometrics",
    year    = "2012",
    volume  = "171",
    number  = "2",
    pages   = "267-280"
}
}

\renewcommand{\theequation}{A.\arabic{equation}}
\renewcommand{\thesection}{A.\arabic{section}}
\renewcommand{\thefigure}{A.\arabic{figure}}
\renewcommand{\thetable}{A.\arabic{table}}
\renewcommand{\thelemma}{A.\arabic{lemma}}
\renewcommand{\theremark}{A.\arabic{remark}}
\renewcommand{\thecorollary}{A.\arabic{corollary}}

\setcounter{equation}{0}
\setcounter{lemma}{0}
\setcounter{section}{0}
\setcounter{table}{0}
\setcounter{figure}{0}
\setcounter{remark}{0}
\setcounter{corollary}{0}
\setcounter{assumption}{0}

{\small
 
\section*{Appendix A}

This appendix provides the main tables listed in Section \ref{Sec4} as well as some extra estimation results for the empirical study discussed in Section \ref{Sec4}.  

Tables \ref{TabEmp1} and  \ref{TabEmp2}, and Figures \ref{Boxplot_1} and \ref{Boxplot_2} present some empirical results for conditional convergence. Figures \ref{Boxplot_1} and \ref{Boxplot_2} provide the boxplots of country-specific estimated coefficients for each explanatory variable across 23 industries. As shown in Figure \ref{Boxplot_1}, the majority of the estimated coefficients associated with initial productivity exhibit values below zero, with a limited number of exceptions, while their distribution varies across industries. For the control variables, the estimated coefficients are distributed around zero and they exhibit more evident heterogeneity in different industries. 

We then examine the unconditional convergence of labour productivity by excluding the control variables that are employed in Section \ref{Sec4}. We re-estimate the numbers of factors using the method proposed in Section \ref{Sec2.3} and report the results for industry- and country-specific factors in Table \ref{TabApp1}. The values are ranging from 1 to 9 for different industries and countries. In total, the industry- and country-specific factors can account for 65.77\% of $y_{ijt}$'s variations. In addition, we identify one global factor that can explains 27.88\% of the variations in $y_{ijt}$. For the estimated coefficients for initial productivity, we present their distributions for each industry in Figure \ref{Boxplot_App}. As can be seen from this figure, most estimated effects of initial productivity on the subsequent growth of productivity are negative and these manufacturing industries exhibit evident difference in the distributions. Furthermore, we calculate the average value of initial productivity's effects using different samples in Table \ref{TabApp2}. The results also indicate a negative relationship between the initial productivity and the subsequent growth rate, which is robust to various sample selections.  Additionally, we report the country-specific estimates and the associated 95\% confidence intervals for the initial productivity in the food and beverages industry in Table \ref{TabApp2}. It is evident to see that  the estimated effects are significant and negative for most countries.  

To summarize, we can confirm that both unconditional and conditional convergences of productivity exist for most countries after controlling unrecoverable heterogeneity properly.

\begin{table}[H]
\small
\begin{center}
\caption{\textbf{Summary statistics of the dataset}. This table provides the means and standard deviations for the dependent variable and regressors. The sample period is from 1963 to 2018. }\label{TabEmp1}
\begin{tabular}{lccc}
\hline\hline
Variable Name                    & Abbreviation & Mean   & Std   \\
Growth in Labor   productivity & GLP  & 0.038  & 0.483 \\
Initial productivity           & IniP & 7.919  & 2.678 \\
Investment price               & IP   & 0.278  & 0.213 \\
Government consumption share   & GCS  & 0.207  & 0.120 \\
Openness measure               & Open & -0.029 & 0.119 \\
Secondary school enrolment     & SSE  & 0.534  & 0.313 \\
Civil liberties                & CL   & 4.353  & 1.797\\
\hline\hline
\end{tabular}
\end{center}
\end{table}

\begin{table}[H]
\begin{center}
\caption{\textbf{The number  of countries for  each industry.} 
The dataset comprises 23 manufacturing industries at the 2-digit level of the International Standard Industrial Classification of All Economic Activities (ISIC). The sample period is from 1963 to 2018.}
\label{TabEmp2}
\begin{tabular}{lcc}
\hline\hline
Industry Name                                    & Abbreviation & No. of countries \\
Food and beverages                               & FB           & 71               \\
Tobacco products                                 & TP           & 65               \\
Textiles                                         & TE           & 71               \\
Wearing apparel, fur                             & WAF          & 69               \\
Leather, leather products and footwear           & LLF          & 53               \\
Wood products (excl. furniture)                  & WP           & 71               \\
Paper and paper products                         & PPP          & 68               \\
Printing and publishing                          & PP           & 70               \\
Coke, refined petroleum products, nuclear   fuel & CRN          & 61               \\
Chemicals and chemical products                  & CCP          & 71               \\
Rubber and plastics products                     & RPP          & 69               \\
Non-metallic mineral products                    & NMP          & 70               \\
Basic metals                                     & BM           & 68               \\
Fabricated metal products                        & FMP          & 70               \\
Machinery and equipment n.e.c.                   & ME           & 67               \\
Office, accounting and computing   machinery     & OACM         & 42               \\
Electrical machinery and apparatus               & EMA          & 68               \\
Radio, television and communication   equipment  & RTCE         & 31               \\
Medical, precision and optical   instruments     & MPOI         & 54               \\
Motor vehicles, trailers, semi-trailers          & MVTS         & 68               \\
Other transport equipment                        & OTE          & 46               \\
Furniture, manufacturing n.e.c.                  & FM           & 71               \\
Recycling                                        & RC           & 23              \\
\hline\hline
\end{tabular}
\end{center}
\end{table}

\setlength{\tabcolsep}{2.5pt}
\begin{table}[H]
\footnotesize
\caption{\textbf{Estimation of industry- and country-specific factors.}
Panels A and B present the estimated numbers of industry-specific and country-specific factors, respectively, along with the total proportion of variance explained by these two groups of factors.  The sample period is from 1963 to 2018.
}\label{TabEmp3}
\begin{tabular}{ccccccccc}
 \hline\hline
\multicolumn{2}{c}{Panel A}             &              & \multicolumn{6}{c}{Panel B}                                                          \\
\cline{1-2}\cline{4-9}
Industry        & No. of Factors& & Country & No. of Factors & Country & No. of Factors & Country & No. of Factors \\
FB          & 4               &  & ALB       & 1              & GTM       & 1              & PAK       & 5              \\
TP          & 5               &  & ARM       & 5              & HND       & 5              & PAN       & 2              \\
TE          & 1               &  & AZE       & 1              & HRV       & 4              & PER       & 5              \\
WAF         & 1               &  & BDI       & 3              & IND       & 5              & PHL       & 5              \\
LLF         & 1               &  & BEN       & 5              & IRN       & 5              & POL       & 4              \\
WP          & 1               &  & BFA       & 1              & JAM       & 1              & PRY       & 5              \\
PPP         & 5               &  & BGD       & 2              & JOR       & 3              & RUS       & 5              \\
PP          & 5               &  & BGR       & 5              & KAZ       & 5              & SDN       & 1              \\
CRN         & 8               &  & BLR       & 1              & KEN       & 1              & SEN       & 6              \\
CCP         & 5               &  & BRA       & 5              & KGZ       & 5              & SLV       & 2              \\
RPP         & 3               &  & BWA       & 1              & KHM       & 1              & SYR       & 4              \\
NMP         & 1               &  & CAF       & 1              & LAO       & 5              & THA       & 1              \\
BM          & 10              &  & CIV       & 1              & LBN       & 2              & TTO       & 2              \\
FMP         & 1               &  & CMR       & 5              & LKA       & 5              & TUN       & 1              \\
ME          & 4               &  & COL       & 3              & MDA       & 1              & TUR       & 1              \\
OACM        & 8               &  & DOM       & 3              & MDG       & 3              & TZA       & 5              \\
EMA         & 5               &  & DZA       & 6              & MEX       & 5              & UGA       & 4              \\
RTCE        & 4               &  & ECU       & 1              & MNG       & 9              & UKR       & 5              \\
MPOI        & 2               &  & EGY       & 1              & MOZ       & 5              & URY       & 1              \\
MVTS        & 8               &  & ETH       & 2              & MWI       & 5              & VEN       & 5              \\
OTE         & 1               &  & GAB       & 5              & MYS       & 2              & YEM       & 4              \\
FM          & 1               &  & GEO       & 1              & NIC       & 4              & ZAF       & 5              \\
RC          & 8               &  & GHA       & 3              & NPL       & 3              & ZWE       & 6              \\
            &                 &  & GMB       & 7              & OMN       & 5              &           &     \\
\hline
                  \% of Var($y_{jit}$)  &            &               &  &           &                &     &     & 56.65\%  \\
                  \hline\hline
\end{tabular}
\end{table}

\setlength{\tabcolsep}{2.5pt}
\begin{table}[H]
\footnotesize
\caption{\textbf{Estimation of industry- and country-specific factors (without control variables).}
Panels A and B of this table present the estimated numbers of industry-specific and country-specific factors, respectively, along with the total proportion of variance explained by these two groups of factors.  The sample period is from 1963 to 2018.
}\label{TabApp1}
\begin{tabular}{ccccccccc}
 \hline\hline
\multicolumn{2}{c}{Panel A}             &              & \multicolumn{6}{c}{Panel B}                                                          \\
\cline{1-2}\cline{4-9}
Industry        & No. of Factors& & Country & No. of Factors & Country & No. of Factors & Country & No. of Factors \\
FB                   & 1              &  & ALB       & 1              & GTM       & 1              & PAK       & 1              \\
TP                   & 5              &  & ARM       & 8              & HND       & 4              & PAN       & 2              \\
TE                   & 1              &  & AZE       & 2              & HRV       & 3              & PER       & 3              \\
WAF                  & 4              &  & BDI       & 3              & IND       & 1              & PHL       & 1              \\
LLF                  & 1              &  & BEN       & 2              & IRN       & 1              & POL       & 4              \\
WP                   & 2              &  & BFA       & 1              & JAM       & 1              & PRY       & 5              \\
PPP                  & 5              &  & BGD       & 2              & JOR       & 4              & RUS       & 1              \\
PP                   & 3              &  & BGR       & 1              & KAZ       & 3              & SDN       & 1              \\
CRN                  & 8              &  & BLR       & 1              & KEN       & 2              & SEN       & 6              \\
CCP                  & 1              &  & BRA       & 4              & KGZ       & 3              & SLV       & 2              \\
RPP                  & 3              &  & BWA       & 1              & KHM       & 5              & SYR       & 6              \\
NMP                  & 1              &  & CAF       & 1              & LAO       & 1              & THA       & 1              \\
BM                   & 9              &  & CIV       & 1              & LBN       & 2              & TTO       & 2              \\
FMP                  & 2              &  & CMR       & 5              & LKA       & 5              & TUN       & 1              \\
ME                   & 4              &  & COL       & 1              & MDA       & 2              & TUR       & 1              \\
OACM                 & 8              &  & DOM       & 6              & MDG       & 4              & TZA       & 3              \\
EMA                  & 5              &  & DZA       & 1              & MEX       & 2              & UGA       & 5              \\
RTCE                 & 4              &  & ECU       & 1              & MNG       & 5              & UKR       & 5              \\
MPOI                 & 2              &  & EGY       & 2              & MOZ       & 2              & URY       & 1              \\
MVTS                 & 9              &  & ETH       & 2              & MWI       & 3              & VEN       & 5              \\
OTE                  & 1              &  & GAB       & 2              & MYS       & 2              & YEM       & 5              \\
FM                   & 1              &  & GEO       & 1              & NIC       & 6              & ZAF       & 2              \\
RC                   & 7              &  & GHA       & 4              & NPL       & 3              & ZWE       & 3              \\
                     &                &  & GMB       & 1              & OMN       & 5              &           &        \\
\hline
                  \% of Var($y_{jit}$)  &            &               &  &           &                &     &     & 65.77\%  \\
                  \hline\hline
\end{tabular}
\end{table}

\begin{table}[H]
\footnotesize
\begin{center}
\caption{\textbf{The means of estimated coefficients.} }\label{TabApp2}
\begin{tabular}{lcrr}
\hline\hline
 Sample                                   &  Variable    & $\widehat{\beta}$ &  CI   \\
1963-2018                 & IniP & -0.107             & (-0.109, -0.101) \\
1973-2018                 & IniP & -0.119             & (-0.121, -0.113) \\
1983-2018                 & IniP & -0.141            & (-0.143, -0.133) \\
1993-2018                 & IniP & -0.154             & (-0.162, -0.141) \\
1963-2018, excluding OECD & IniP & -0.109             & (-0.111, -0.103) \\
                                    \hline\hline
\end{tabular}
\end{center}
\end{table}

\setlength{\tabcolsep}{3pt}
\begin{table}[H]
\footnotesize
\begin{center}
\caption{\textbf{Estimated coefficients for IniP in the food and beverages industry.}
The three-letter ISO 3166-1 alpha-3 codes are used as the abbreviations for countries. For each country in the food and beverages industry, the estimated coefficient for IniP and its $95\%$ confidence interval  are reported. The sample period is from 1963 to 2018.
}\label{TabEmp5}
\begin{tabular}{crr crr crr}
\hline\hline
Country    & $\widehat{\beta}$ &   CI       &  Country   & $\widehat{\beta}$ &   CI        &  Country     & $\widehat{\beta}$ &    CI   \\
ALB & -0.267            & (-0.326, -0.209) & GTM & -0.132            & (-0.188, -0.079) & PAK & -0.124            & (-0.223, -0.025) \\
ARM & -0.191            & (-0.251, -0.120) & HND & -0.156            & (-0.228, -0.077) & PAN & -0.213            & (-0.324, -0.101) \\
AZE & -0.215            & (-0.246, -0.181) & HRV & -0.311            & (-0.597, -0.040) & PER & -0.092            & (-0.151, -0.036) \\
BDI & -0.165            & (-0.242, -0.097) & IND & -0.196            & (-0.306, -0.071) & PHL & -0.215            & (-0.304, -0.137) \\
BEN & -0.130            & (-0.219, -0.038) & IRN & -0.130            & (-0.229, -0.050) & POL & -0.127            & (-0.180, -0.073) \\
BFA & -0.168            & (-0.223, -0.114) & JAM & -0.192            & (-0.276, -0.122) & PRY & -0.139            & (-0.254, -0.019) \\
BGD & -0.133            & (-0.237, -0.035) & JOR & 0.006             & (-0.027, 0.044) & RUS & -0.136            & (-0.234, -0.043) \\
BGR & -0.150            & (-0.237, -0.044) & KAZ & -0.233            & (-0.288, -0.180) & SDN & -0.244            & (-0.387, -0.110) \\
BLR & -0.079            & (-0.190, 0.022) & KEN & -0.160            & (-0.223, -0.098) & SEN & -0.231            & (-0.350, -0.111) \\
BRA & -0.165            & (-0.274, -0.064) & KGZ & -0.258            & (-0.373, -0.152) & SLV & -0.063            & (-0.124, 0.007) \\
BWA & -0.123            & (-0.199, -0.040) & KHM & -0.220            & (-0.358, -0.073) & SYR & -0.003            & (-0.201, 0.198) \\
CAF & -0.079            & (-0.156, -0.006) & LAO & -0.150            & (-0.283, 0.007) & THA & 0.049             & (-0.093, 0.186) \\
CIV & -0.215            & (-0.356, -0.066) & LBN & 0.026             & (-0.041, 0.089) & TTO & -0.096            & (-0.199, 0.012) \\
CMR & -0.196            & (-0.279, -0.114) & LKA & -0.138            & (-0.255, -0.032) & TUN & -0.148            & (-0.266, -0.017) \\
COL & -0.203            & (-0.257, -0.142) & MDA & -0.144            & (-0.216, -0.062) & TUR & -0.214            & (-0.361, -0.045) \\
DOM & -0.262            & (-0.340, -0.169) & MDG & -0.188            & (-0.257, -0.110) & TZA & -0.163            & (-0.237, -0.082) \\
DZA & -0.143            & (-0.200, -0.093) & MEX & -0.181            & (-0.276, -0.069) & UGA & -0.154            & (-0.324, 0.004) \\
ECU & -0.167            & (-0.303, -0.024) & MNG & -0.213            & (-0.284, -0.149) & UKR & -0.241            & (-0.385, -0.081) \\
EGY & -0.128            & (-0.222, -0.026) & MOZ & -0.234            & (-0.355, -0.112) & URY & -0.228            & (-0.310, -0.143) \\
ETH & -0.167            & (-0.256, -0.074) & MWI & -0.289            & (-0.411, -0.153) & VEN & -0.160            & (-0.264, -0.059) \\
GAB & -0.234            & (-0.309, -0.157) & MYS & -0.187            & (-0.279, -0.085) & YEM & -0.206            & (-0.298, -0.112) \\
GEO & -0.039            & (-0.114, 0.042) & NIC & -0.273            & (-0.419, -0.123) & ZAF & -0.244            & (-0.397, -0.102) \\
GHA & -0.028            & (-0.354, 0.323) & NPL & -0.133            & (-0.196, -0.074 ) & ZWE & -0.003            & (-0.124, 0.108) \\
GMB & -0.121            & (-0.261, 0.008) & OMN & -0.094            & (-0.190, 0.003) &     &                   &     \\
\hline\hline
\end{tabular}
\end{center}
\end{table}
 
\begin{table}[H]
\footnotesize
\begin{center}
\caption{\textbf{Estimated coefficients for IniP in the food and beverages industry (without control variables).}
In this table, the three-letter ISO 3166-1 alpha-3 codes are used as the abbreviations for countries. For each country in the food and beverages industry, the estimated coefficient for IniP and its $95\%$ confidence interval (CI) are reported. The sample period is from 1963 to 2018.
}\label{TabApp3}
\begin{tabular}{crr crr crr}
\hline\hline
Country    & $\widehat{\beta}$ &   CI       &  Country   & $\widehat{\beta}$ &   CI        &  Country     & $\widehat{\beta}$ &    CI   \\
ALB & -0.257            & (-0.326, -0.191) & GTM & -0.096            & (-0.216, 0.025) & PAK & -0.167            & (-0.271, -0.059)\\
ARM & -0.072            & (-0.248, 0.110) & HND & -0.166            & (-0.235, -0.078) & PAN & -0.085            & (-0.135, -0.033)\\
AZE & -0.206            & (-0.442, 0.084) & HRV & -0.219            & (-0.406, -0.013) & PER & -0.071            & (-0.165, 0.022)\\
BDI & -0.096            & (-0.165, -0.027) & IND & -0.075            & (-0.164, 0.030) & PHL & -0.144            & (-0.216, -0.055)\\
BEN & -0.196            & (-0.233, -0.160) & IRN & -0.132            & (-0.241, -0.018) & POL & -0.088            & (-0.134, -0.041)\\
BFA & -0.081            & (-0.158, 0.010) & JAM & -0.100            & (-0.129, -0.002) & PRY & -0.074            & (-0.122, -0.025)\\
BGD & -0.097            & (-0.198, 0.014) & JOR & -0.181            & (-0.256, -0.095) & RUS & -0.095            & (-0.218, 0.022)\\
BGR & -0.109            & (-0.283, 0.081) & KAZ & -0.063            & (-0.172, 0.055) & SDN & -0.228            & (-0.362, -0.090)\\
BLR & -0.056            & (-0.161, 0.054) & KEN & -0.115            & (-0.244, 0.018) & SEN & -0.214            & (-0.296, -0.108)\\
BRA & -0.096            & (-0.156, -0.032) & KGZ & -0.073            & (-0.199, 0.073) & SLV & -0.077            & (-0.128, -0.027) \\
BWA & -0.112            & (-0.227, -0.012) & KHM & -0.080            & (-0.157, -0.001) & SYR & -0.069            & (-0.194, 0.068)\\
CAF & -0.134            & (-0.223, -0.043) & LAO & -0.072            & (-0.158, 0.036) & THA & -0.153            & (-0.299, -0.012)\\
CIV & -0.264            & (-0.399, -0.104) & LBN & 0.016             & (-0.048, 0.088) & TTO & -0.119            & (-0.196, -0.022)\\
CMR & -0.129            & (-0.187, -0.061) & LKA & -0.124            & (-0.221, -0.035) & TUN & -0.087            & (-0.124, -0.039)\\
COL & -0.089            & (-0.152, -0.025) & MDA & -0.072            & (-0.139, 0.004) & TUR & -0.126            & (-0.187, -0.056)\\
DOM & -0.216            & (-0.484, 0.078) & MDG & -0.154            & (-0.245, -0.056) & TZA & -0.106            & (-0.247, 0.021)\\
DZA & -0.117            & (-0.239, -0.005) & MEX & -0.152            & (-0.297, 0.008 ) & UGA & -0.065            & (-0.178, 0.046)  \\
ECU & -0.190            & (-0.274, -0.083) & MNG & -0.149            & (-0.244, -0.067) & UKR & -0.054            & (-0.125, 0.021)\\
EGY & -0.213            & (-0.346, -0.052) & MOZ & -0.283            & (-0.455, -0.079) & URY & -0.172            & (-0.275, -0.052)\\
ETH & -0.119            & (-0.200, -0.024) & MWI & -0.230            & (-0.364, -0.048) & VEN & -0.128            & (-0.261, 0.008)\\
GAB & -0.190            & (-0.260, -0.114) & MYS & -0.147            & (-0.205, -0.064) & YEM & -0.162            & (-0.274, -0.060)\\
GEO & -0.041            & (-0.129, 0.040) & NIC & -0.097            & (-0.220, 0.031) & ZAF & -0.102            & (-0.180, -0.018)\\
GHA & -0.102            & (-0.195, -0.015) & NPL & -0.070            & (-0.137, 0.008) & ZWE & 0.062             & (-0.103, 0.202)\\
GMB & -0.069            & (-0.152, 0.001) & OMN & -0.017            & (-0.093, 0.060) &     &                   &     \\
\hline\hline
\end{tabular}
\end{center}
\end{table}

\begin{figure}[H]
	\centering
	{\includegraphics[width=1\textwidth]{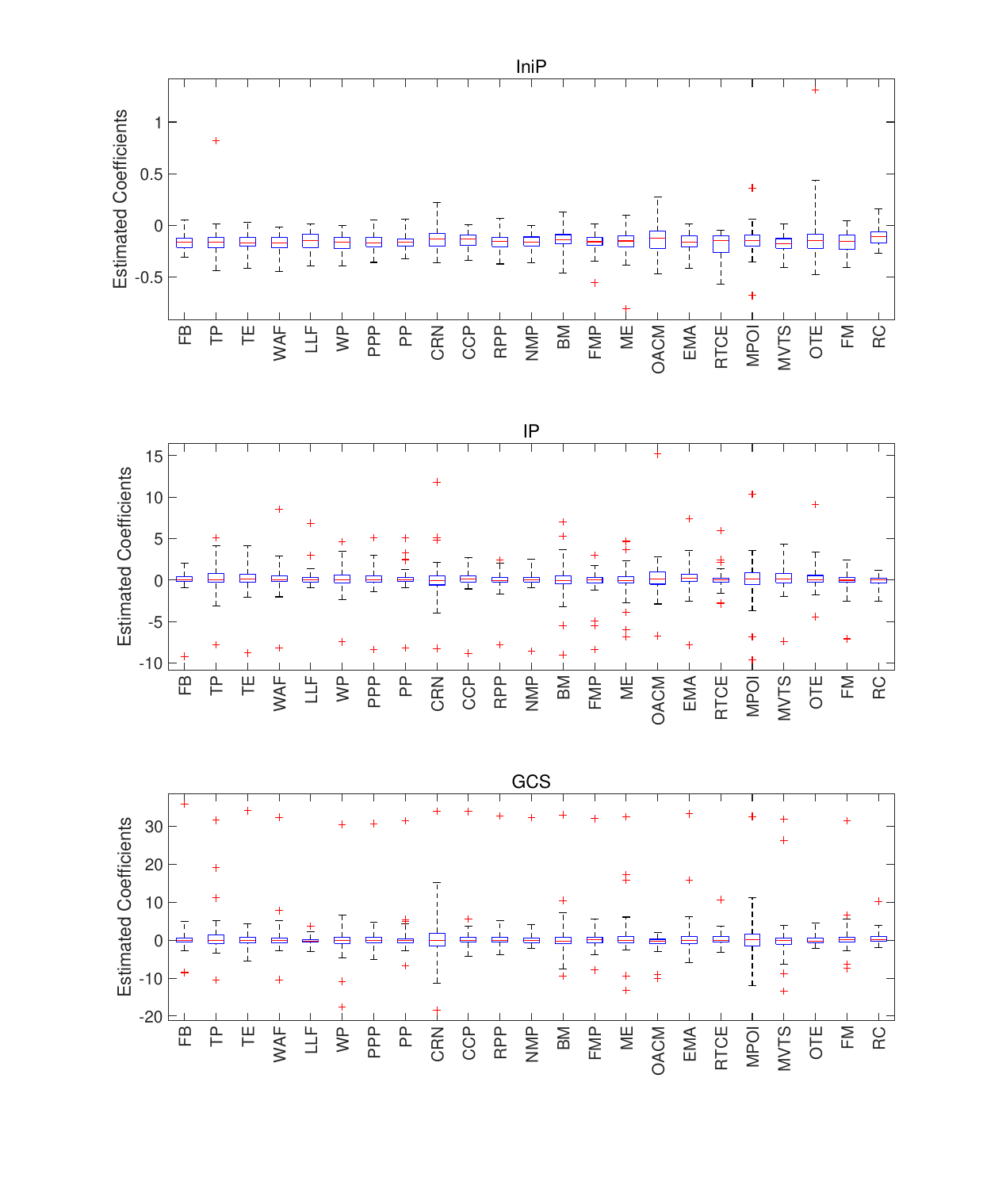}} 
	\vspace{-2cm}
	\caption{\textbf{Boxplots of estimated coefficients for IniP, IP and GCS.} The central mark, bottom edge and top edge of blue boxes indicate the median, 25th percentile and 75th percentile of the country-specific estimates, respectively, for each industry. The outliers are labelled by the red marker symbol.}
	\label{Boxplot_1}
\end{figure}

\begin{figure}[H]
	\centering
	{\includegraphics[width=1\textwidth]{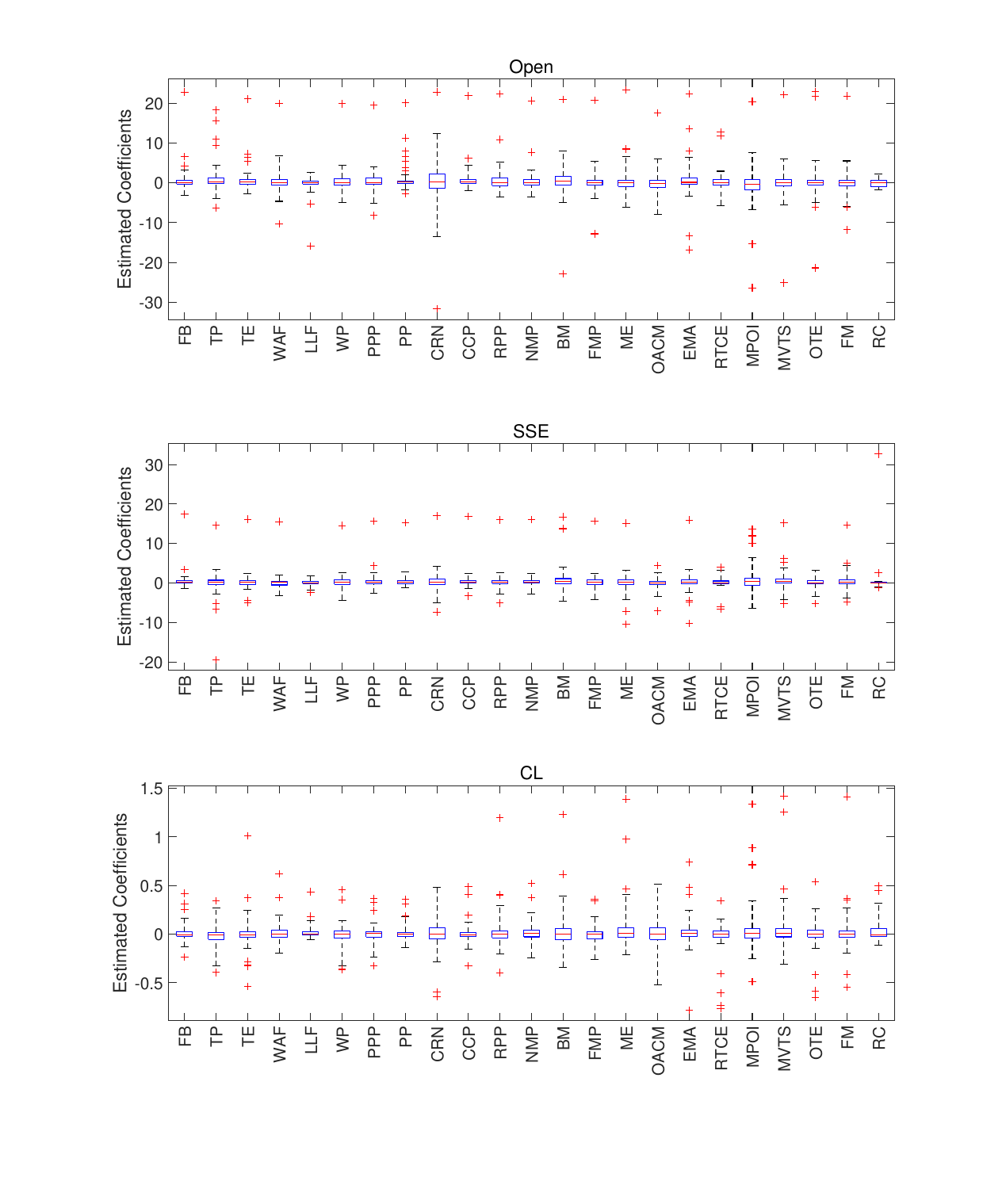}} 
	\vspace{-2cm}
	\caption{\textbf{Boxplots of estimated coefficients for Open, SSE and CL.} The central mark, bottom edge and top edge of blue boxes indicate the median, 25th percentile and 75th percentile of the country-specific estimates, respectively, for each industry. The outliers are labelled by the red marker symbol.}
	\label{Boxplot_2}
\end{figure}

\begin{figure}[H]
	\centering
	{\includegraphics[width=1\textwidth]{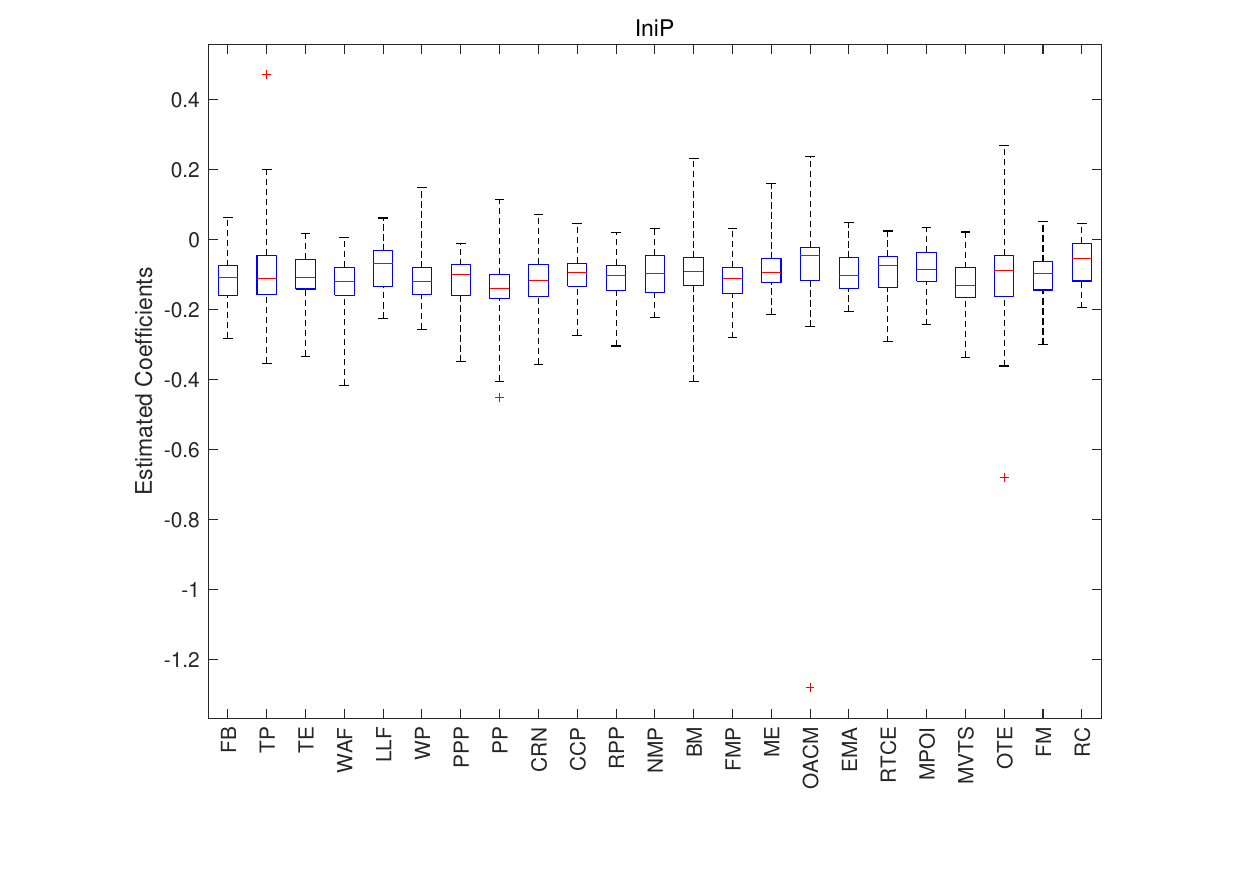}} 
	\vspace{-1cm}
	\caption{\textbf{Boxplots of estimated coefficients for IniP (without control variables).} The central mark, bottom edge and top edge of blue boxes indicate the median, 25th percentile and 75th percentile of the country-specific estimates, respectively, for each industry. The outliers are labelled by the red marker symbol.}
	\label{Boxplot_App}
\end{figure}

\begin{figure}[H]
	\centering
	{\includegraphics[width=1\textwidth]{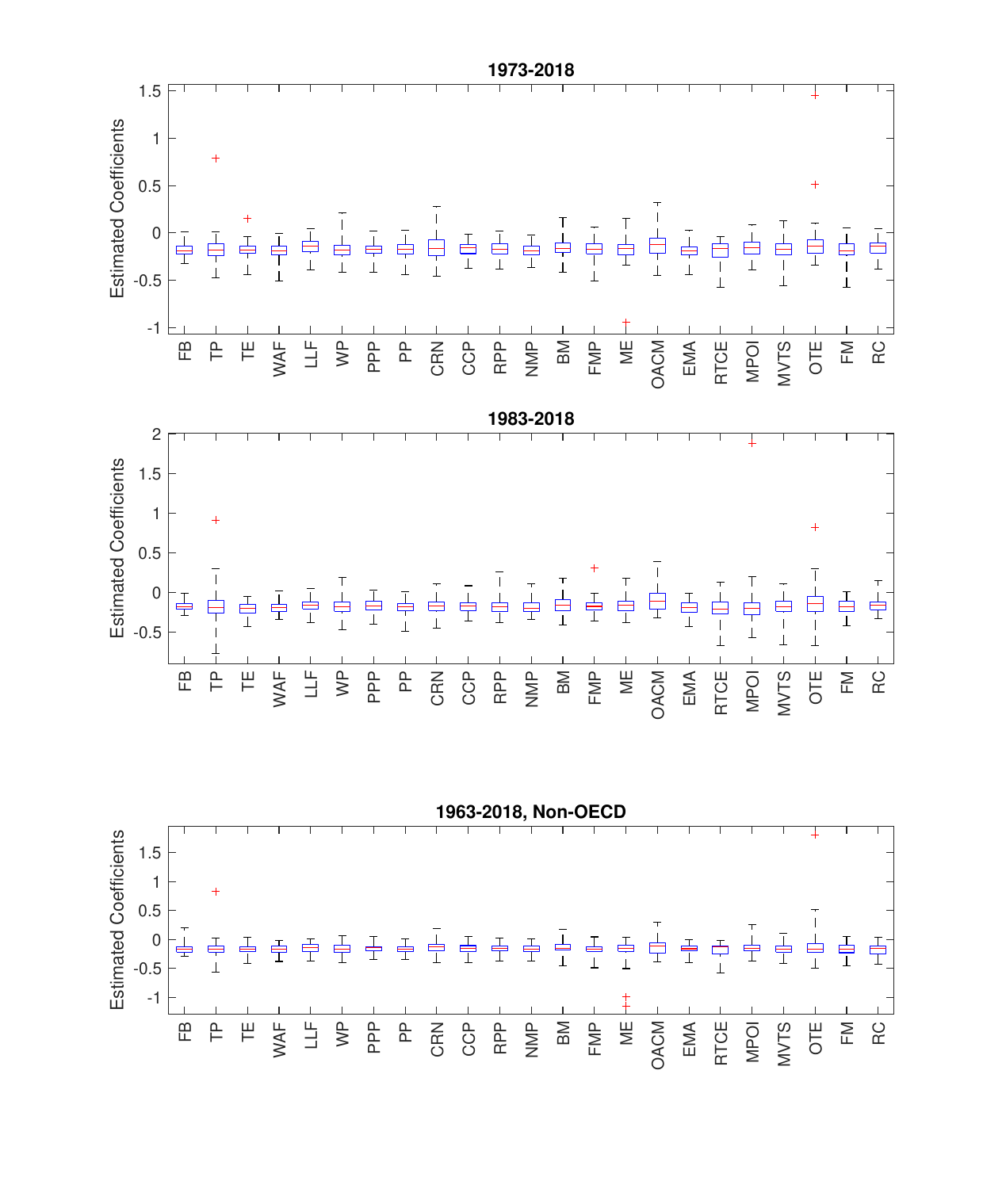}} 
	\vspace{-2cm}
	\caption{\textbf{Boxplots of estimated coefficients for IniP using different samples.} The central mark, bottom edge and top edge of blue boxes indicate the median, 25th percentile and 75th percentile of the country-specific estimates, respectively, for each industry. The outliers are labelled by the red marker symbol.}
	\label{Boxplot_Robust}
\end{figure}

\clearpage

{\small

\begin{center}
{\large \bf Online Supplementary Appendix B to \\``Estimation and Inference for Three-Dimensional Panel Data Models"}

\medskip

{\sc Guohua Feng$^{\ast}$, Jiti Gao$^{\dag}$,  Fei Liu$^\sharp$ and Bin Peng$^{\dag}$}

\medskip

$^{\ast}$University of North Texas,  $^\dag$Monash University and $^\sharp$Nankai University

\end{center}

\setcounter{page}{1}
\renewcommand{\theequation}{B.\arabic{equation}}
\renewcommand{\thesection}{B.\arabic{section}}
\renewcommand{\thefigure}{B.\arabic{figure}}
\renewcommand{\thetable}{B.\arabic{table}}
\renewcommand{\thelemma}{B.\arabic{lemma}}
\renewcommand{\theremark}{B.\arabic{remark}}
\renewcommand{\thecorollary}{B.\arabic{corollary}}

\setcounter{equation}{0}
\setcounter{lemma}{0}
\setcounter{section}{0}
\setcounter{table}{0}
\setcounter{figure}{0}
\setcounter{remark}{0}
\setcounter{corollary}{0}
\setcounter{assumption}{0}
 
\section*{Appendix B}

Appendix B is organized as follows. Appendix \ref{App.2} provides two detailed numerical implementations; Appendix \ref{App_bias} proposes a Jackknife procedure of bias correction for the mean group estimators; Appendix \ref{AppB1} includes the necessary preliminary lemmas; Appendix \ref{AppB2} presents the detailed proofs. 

\section{Numerical Implementation}\label{App.2}

In this appendix, we provide two numerical implementations. The first one is about the steps used to calculate simulation results, while the second one is specifically about the procedure of Section \ref{sec2.2}.

\bigskip

\noindent \textbf{1. Implementation about Simulation}
\hrule
\begin{itemize}[leftmargin=*, itemsep=0.5pt, parsep=0.5pt, topsep=0.6pt]
\item[1.] We prescribe $d_{\max}=5$ for instance. Let $\widehat{\ell} =d_{\max}$, $\widehat{\pmb{\ell}}^\circ = d_{\max}\cdot \mathbf{1}_L$, and $\widehat{\pmb{\ell}}^\bullet = d_{\max}\cdot \mathbf{1}_N$ to obtain $\widetilde{\mathbf{b}}_{\centerdot\centerdot}$ as in the second implementation of this section. 

\item[2.] Once receiving $\widetilde{\mathbf{b}}_{\centerdot\centerdot}$, we calculate $\widehat{\ell}$, $\widehat{\pmb{\ell}}^\circ$, and $\widehat{\pmb{\ell}}^\bullet$ as in Section \ref{Sec2.3}.

\item[3.] Using the estimates $\widehat{\ell}$, $\widehat{\pmb{\ell}}^\circ$, and $\widehat{\pmb{\ell}}^\bullet$, we obtain the estimate of $\pmb{\beta}_{\centerdot \centerdot}$ again as in the second implementation of this section, i.e., an updated $\widetilde{\mathbf{b}}_{\centerdot\centerdot}$. Simultaneously, we also obtain the estimates $\widehat{\mathbf{C}}$, $\widehat{\mathbf{C}}_i^\circ$ and $\widehat{\mathbf{C}}_j^\bullet$ which serve as the estimated $\mathbf{F}$, $\mathbf{F}_i^\circ$ and $\mathbf{F}_j^\bullet$ respectively.

\item[4.] With all estimates in hand, we implement the DWB procedure as in Section \ref{sec2.2} to calculate the coverage rates for all $\pmb{\beta}_{ij}$'s. For simplicity, we adopt the Bartlett kernel, i.e., $a(x) = 1-|x|$ if $|x|\le 1$, and let the bandwidth be $m=\lfloor 1.75  T^{1/3}\rfloor$.
\end{itemize}
\hrule

\bigskip
\bigskip

\noindent \textbf{2. Implementation about Section \ref{sec2.2}}
\hrule
\begin{itemize}[leftmargin=*, itemsep=0.5pt, parsep=0.5pt, topsep=0.6pt]

\item[1.] Randomly generate each element of $\widehat{\mathbf{C}}$ from standard normal distribution, and conduct SVD decomposition to update $\widehat{\mathbf{C}}$ and ensure $\frac{1}{T}\widehat{\mathbf{C}}^\top \widehat{\mathbf{C}}=\mathbf{I}$. Generate the elements of $\widehat{\mathbf{C}}_i^\circ$ and $\widehat{\mathbf{C}}_j^\bullet$ from standard normal distribution, and update $\widehat{\mathbf{C}}_i^\circ$ and $\widehat{\mathbf{C}}_j^\bullet$ via SVD decomposition for $\mathbf{M}_{\widehat{\mathbf{C}}}\widehat{\mathbf{C}}_i^\circ$ and $\mathbf{M}_{\widehat{\mathbf{C}}}\widehat{\mathbf{C}}_j^\bullet$.

\item[2.] Calculate $\widehat{\mathbf{b}}_{ij}$ using $(\widehat{\mathbf{C}}, \widehat{\mathbf{C}}_i^\circ, \widehat{\mathbf{C}}_j^\bullet)$ as in step (a) of \eqref{est1}. 

\item[3.] Based on $\widehat{\mathbf{b}}_{ij}$, update $(\widehat{\mathbf{C}}, \widehat{\mathbf{C}}_i^\circ, \widehat{\mathbf{C}}_j^\bullet)$ as in steps (b)-(d) of \eqref{est1}.

\item[4.] Calculate $\widetilde{\mathbf{b}}_{ij}$ using $(\widehat{\mathbf{C}}, \widehat{\mathbf{C}}_i^\circ, \widehat{\mathbf{C}}_j^\bullet)$ as in \eqref{est2}. 

\item[5.] Repeat steps (2)-(4) until reaching some convergence, say, $\|\widetilde{\mathbf{b}}_{\centerdot\centerdot}^{(n)}-\widetilde{\mathbf{b}}_{\centerdot\centerdot}^{(n-1)}\|/\sqrt{LN}<\epsilon$, where $\widetilde{\mathbf{b}}_{\centerdot\centerdot} =(\widetilde{\mathbf{b}}_{11},\ldots,  \widetilde{\mathbf{b}}_{LN})^\top$ and $\widetilde{\mathbf{b}}_{\centerdot\centerdot}^{(n)}$ stands for the value of $\widetilde{\mathbf{b}}_{\centerdot\centerdot}$ obtained in the $n^{th}$ replication.
\end{itemize}
\hrule

\section{Bias Correction}\label{App_bias}

In this appendix, we propose a Jackknife method to eliminate the bias in the CLT of mean group estimators $\widetilde{\mathbf{b}}_{i\centerdot }$ and $\widetilde{\mathbf{b}}_{\centerdot j}$. Since $\widetilde{\mathbf{b}}_{i\centerdot }$ and $\widetilde{\mathbf{b}}_{\centerdot j}$ have similar structures, we only focus on the bias correction for $\widetilde{\mathbf{b}}_{i\centerdot }$. 
Here, we introduce some new notation to reorganize the bias terms according to their probability orders. Specifically, we define the following mathematical symbols:\\{\footnotesize $\widetilde{\mathcal{R}}_{F,ij}^{\circ}=(\widetilde{\mathcal{R}}_{F,1}^{\circ}, \widetilde{\mathcal{R}}_{F,2,i}^{\circ}, \widetilde{\mathcal{R}}_{F,3,j}^{\circ})$ and $\widetilde{\mathcal{R}}_{F,ij}^{\bullet}=(\widetilde{\mathcal{R}}_{F,1}^{\bullet}, \widetilde{\mathcal{R}}_{F,2,i}^{\bullet}, \widetilde{\mathcal{R}}_{F,3,j}^{\bullet})$ with $\widetilde{\mathcal{R}}_{F,1}^{\circ}= \sum_{ij}\mathbf{F}_{i}^{\circ}\pmb{\gamma}_{ij}^\circ\pmb{\gamma}_{ij}^\top (\pmb{\Gamma}_{\centerdot \centerdot }^\top \pmb{\Gamma}_{\centerdot \centerdot })^{-1}$,\,
$\widetilde{\mathcal{R}}_{F,2,i}^\circ=\sum_{j=1}^N\widetilde{\mathcal{R}}_{F,1}^{\circ}\pmb{\gamma}_{ij}\pmb{\gamma}_{ij}^{\circ\top} (\pmb{\Gamma}_{i\centerdot  }^{\circ\top} \pmb{\Gamma}_{i\centerdot  }^\circ)^{-1}$,\,
$\widetilde{\mathcal{R}}_{F,3,j}^\circ=\sum_{i=1}^L\big(\widetilde{\mathcal{R}}_{F,1}^{\circ}\pmb{\gamma}_{ij}+\mathbf{F}_{i}^{\circ}\pmb{\gamma}_{ij}^\circ\big)\pmb{\gamma}_{ij}^{\bullet\top} (\pmb{\Gamma}_{i\centerdot  }^{\bullet\top} \pmb{\Gamma}_{i\centerdot }^\bullet)^{-1}$, \,
$\widetilde{\mathcal{R}}_{F,1}^{\bullet}= \sum_{ij}\mathbf{F}_{j}^{\bullet}\pmb{\gamma}_{ij}^\bullet\pmb{\gamma}_{ij}^\top (\pmb{\Gamma}_{\centerdot \centerdot }^\top \pmb{\Gamma}_{\centerdot \centerdot })^{-1}$,\,
$\widetilde{\mathcal{R}}_{F,2,i}^\bullet=\sum_{j=1}^N\big(\widetilde{\mathcal{R}}_{F,1}^{\bullet}\pmb{\gamma}_{ij}+\mathbf{F}_{j}^{\bullet}\pmb{\gamma}_{ij}^\bullet\big)\pmb{\gamma}_{ij}^{\circ\top} (\pmb{\Gamma}_{i\centerdot  }^{\circ\top} \pmb{\Gamma}_{i\centerdot  }^\circ)^{-1}$, and 
$\widetilde{\mathcal{R}}_{F,3,j}^\bullet=\sum_{i=1}^L\widetilde{\mathcal{R}}_{F,1}^{\bullet}\pmb{\gamma}_{ij}\pmb{\gamma}_{ij}^{\bullet\top} (\pmb{\Gamma}_{i\centerdot  }^{\bullet\top} \pmb{\Gamma}_{i\centerdot }^\bullet)^{-1}$.}

Drawing upon the arguments used in the proof of Theorem \ref{M.Thm1}.(2), we can further demonstrate that
\begin{eqnarray}\label{jackknife1}
\sqrt{NT} (\widetilde{\mathbf{b}}_{i\centerdot} -  \pmb{\beta}_{i\centerdot}  )&=&\frac{\sqrt{NT}}{L}\mathcal{A}_{i\centerdot,1}+\sqrt{\frac{T}{N}}\mathcal{A}_{i\centerdot,2}+\sqrt{\frac{T}{L}}\mathcal{A}_{i\centerdot,3}+\frac{1}{\sqrt{NT}}\sum_{j=1}^N\pmb{\Sigma}_{\mathbf{v},ij}^{-1} \mathbf{V}_{ij\centerdot}^\top\pmb{\mathcal{E}}_{ij\centerdot }+o_P(1),
\end{eqnarray}
where $\mathcal{A}_{i\centerdot,1}$, $\mathcal{A}_{i\centerdot,2}$ and $\mathcal{A}_{i\centerdot,3}$ denote the probability limits of the bias terms: 
\begin{eqnarray*}
\mathcal{A}_{i\centerdot,1}&=&\plim_{L,N,T\rightarrow\infty}\frac{L}{T}\sum_{j=1}^N\pmb{\Sigma}_{\mathbf{v},ij}^{-1} \pmb{\phi}_{ij}^{\ast\top} \widetilde{\mathcal{R}}_{F,ij}^{\circ\top} \mathbf{M}_{\mathbf{F}_{ij}^\ast} \widetilde{\mathcal{R}}_{F,ij}^\circ \pmb{\gamma}_{ij}^\ast,
\nonumber\\
\mathcal{A}_{i\centerdot,2}&=&\plim_{L,N,T\rightarrow\infty}\frac{N}{T}\sum_{j=1}^N\pmb{\Sigma}_{\mathbf{v},ij}^{-1} \pmb{\phi}_{ij}^{\ast\top} \widetilde{\mathcal{R}}_{F,ij}^{\bullet\top} \mathbf{M}_{\mathbf{F}_{ij}^\ast} \widetilde{\mathcal{R}}_{F,ij}^\bullet \pmb{\gamma}_{ij}^\ast,
\nonumber\\
\mathcal{A}_{i\centerdot,3}&=&\plim_{L,N,T\rightarrow\infty}\frac{\sqrt{LN}}{T}\sum_{j=1}^N\pmb{\Sigma}_{\mathbf{v},ij}^{-1}\Big( \pmb{\phi}_{ij}^{\ast\top} \widetilde{\mathcal{R}}_{F,ij}^{\bullet\top} \mathbf{M}_{\mathbf{F}_{ij}^\ast} \widetilde{\mathcal{R}}_{F,ij}^\circ \pmb{\gamma}_{ij}^\ast+\pmb{\phi}_{ij}^{\ast\top} \widetilde{\mathcal{R}}_{F,ij}^{\circ\top} \mathbf{M}_{\mathbf{F}_{ij}^\ast} \widetilde{\mathcal{R}}_{F,ij}^\bullet \pmb{\gamma}_{ij}^\ast\Big).
\end{eqnarray*}

We develop a split-sample bias correction method based on equation \eqref{jackknife1}. For notational simplicity, assume both $L$ and $N$ are even numbers. We can then split the sample into two halves along the $i$- and $j$-dimensions, respectively. 
Specifically, define the sets $\mathcal{S}_{L/2,1}=\big\{i,1,\ldots,\frac{L}{2}\big\}$, $\mathcal{S}_{L/2,2}=\big\{i,\frac{L}{2}+1,\ldots,L\big\}$, $\mathcal{S}_{N/2,1}=\big\{1,\ldots,\frac{N}{2}\big\}$, $\mathcal{S}_{N/2,2}=\big\{\frac{N}{2}+1,\ldots,N\big\}$.
Using these sample splits, we construct four mean group estimators:  $\widetilde{\mathbf{b}}_{i\centerdot,1}$, $\widetilde{\mathbf{b}}_{i\centerdot,2}$, $\widetilde{\mathbf{b}}_{i\centerdot,3}$ and $\widetilde{\mathbf{b}}_{i\centerdot,4}$ using the observations indexed by $\mathcal{S}_{L/2,1}\times \mathcal{S}_{N/2,1}$, $\mathcal{S}_{L/2,2}\times \mathcal{S}_{N/2,1}$, $\mathcal{S}_{L/2,1}\times \mathcal{S}_{N/2,2}$ and $\mathcal{S}_{L/2,2}\times \mathcal{S}_{N/2,2}$, respectively. 
The average of these four estimators is denoted by $\overline{\mathbf{b}}_{i\centerdot}$ . Using similar arguments to those in the proof of Theorem  \ref{M.Thm1}.(2), we can show that
\begin{equation}\label{jackknife2}
\sqrt{NT} (\overline{\mathbf{b}}_{i\centerdot} -  \pmb{\beta}_{i\centerdot}  )=2\frac{\sqrt{NT}}{L}\mathcal{A}_{i\centerdot,1}+2\sqrt{\frac{T}{N}}\mathcal{A}_{i\centerdot,2}+2\sqrt{\frac{T}{L}}\mathcal{A}_{i\centerdot,3}+\frac{1}{\sqrt{NT}}\sum_{j=1}^{N}\pmb{\Sigma}_{\mathbf{v},ij}^{-1} \mathbf{V}_{ij\centerdot}^\top\pmb{\mathcal{E}}_{ij\centerdot }+o_P(1).
\end{equation}

Based on these split-sample results, we define the bias-corrected estimator as:  $\widetilde{\mathbf{b}}_{i\centerdot,bc}=2\widetilde{\mathbf{b}}_{i\centerdot}-\overline{\mathbf{b}}_{i\centerdot}$.  Substituting the expansions for $\widetilde{\mathbf{b}}_{i\centerdot}$ and $\overline{\mathbf{b}}_{i\centerdot}$ in \eqref{jackknife1} and \eqref{jackknife2}, it follows that
\begin{eqnarray*}
\sqrt{NT} (\widetilde{\mathbf{b}}_{i\centerdot,bc} -  \pmb{\beta}_{i\centerdot}  )&=&\frac{1}{\sqrt{NT}}\sum_{j=1}^N\pmb{\Sigma}_{\mathbf{v},ij}^{-1} \mathbf{V}_{ij\centerdot}^\top\pmb{\mathcal{E}}_{ij\centerdot }+o_P(1).
\end{eqnarray*}

This result shows that the bias correction procedure successfully removes the leading bias terms, leaving an estimator with an asymptotic normal distribution centered around the true parameter $\pmb{\beta}_{i\centerdot}$. In connection with \eqref{asymp_term} in the proof of Theorem \ref{M.Thm1}.(2), it further demonstrates that the Jackknife bias-corrected estimator  achieves the same asymptotic distribution as the original estimator, but without the bias present in the latter. 

\section{Preliminary Lemmas}\label{AppB1}

We first introduce some additional notation. 

\begin{enumerate}[leftmargin=*, itemsep=0.5pt, parsep=0.5pt, topsep=0.6pt]

\item[] Parameters: $\pmb{\beta}_{i\centerdot } =(\pmb{\beta}_{j1} ,\ldots, \pmb{\beta}_{iN})^\top$ and $\pmb{\beta}_{\centerdot j} =(\pmb{\beta}_{1j} ,\ldots, \pmb{\beta}_{Lj})^\top$;

\item[]  Regressors: $\mathbf{X}_{i \centerdot \centerdot } =( \mathbf{X}_{i1\centerdot} ,\ldots, \mathbf{X}_{iN\centerdot} ) $ and $\mathbf{X}_{\centerdot j\centerdot } =( \mathbf{X}_{1j\centerdot} ,\ldots, \mathbf{X}_{Lj\centerdot} ) $;

\item[]  Errors:  $\pmb{\mathcal{E}}_d =\diag\{\pmb{\mathcal{E}}_{1\centerdot\centerdot} ,\ldots, \pmb{\mathcal{E}}_{L\centerdot\centerdot} \}$;

\item[]  Loadings: $\pmb{\gamma}_{ij}^* = (\pmb{\gamma}_{ij}^\top, \pmb{\gamma}_{ij}^{\circ \top}, \pmb{\gamma}_{ij}^{\bullet \top})^\top$, $\pmb{\Gamma}_{\centerdot \centerdot} =(\pmb{\gamma}_{11},\ldots, \pmb{\gamma}_{1N},\ldots, \pmb{\gamma}_{L1},\ldots, \pmb{\gamma}_{LN})^\top$, \\
$\pmb{\Gamma}_d^{\circ} = \diag\{\pmb{\Gamma}_{1\centerdot}^\circ  ,\ldots, \pmb{\Gamma}_{L\centerdot}^\circ  \}$,  $\pmb{\Gamma}_{i\centerdot}^\circ =(\pmb{\gamma}_{i1}^\circ,\ldots, \pmb{\gamma}_{iN}^\circ  )^\top$, and $\pmb{\Gamma}_{\centerdot j}^\bullet = (\pmb{\gamma}_{1j}^\bullet,\ldots, \pmb{\gamma}_{Lj}^\bullet )^\top$;

\item[]  Some matrices: $\mathbf{P}_{ij}^{\circ \bullet} = 2\mathbf{P}_{\mathbf{C}} + \mathbf{P}_{\mathbf{C}_i^\circ} + \mathbf{P}_{\mathbf{C}_j^\bullet}$, and $\pmb{\mathcal{C}}^\circ =(\mathbf{P}_{\mathbf{C}_1^\circ},\ldots, \mathbf{P}_{\mathbf{C}_L^\circ})^\top$.
\end{enumerate}

We present some preliminary lemmas below, of which Lemmas \ref{LemmaT1}, \ref{LemmaT2}, and \ref{LemA1} are some generic results.

\begin{lemma}\label{LemmaT1}
Suppose that $E\left[\xi_{t}\right]=0$,  and $\{\xi_{t}\}$  is strictly $\alpha$-mixing process with the $\alpha$-mixing coefficient 

\begin{eqnarray*}
\alpha(t) = \sup_{A\in \mathcal{F}_{-\infty}^0, B\in  \mathcal{F}_t^\infty} |P(A)P(B) -P(AB)|
\end{eqnarray*}
satisfying $\sum_{t=1}^{T}\alpha(t)^{\delta/(4+\delta)}=O(1)$ for some $\delta>0$ and $E|\xi_{t} |^{2+\delta/2}<\infty$, where $\mathcal{F}_{-\infty}^0$ and $\mathcal{F}_{t}^\infty$ are the $\sigma$-algebras generated by $\{\xi_s: s \leq 0\}$ and $\{\xi_s: s \geq t\}$, respectively. Then we have
\begin{equation*}
E \left |\frac{1}{T}\sum_{t=1}^T \xi_{t}\right |^{2+\delta^\ast/2} \leq \frac{C}{T^{1+\delta^\ast/4}}E  |\xi_{t} |^{2+\delta^\ast/2} ,
\end{equation*} 
where $C$ is a constant and $0<\delta^\ast<\delta$.
\end{lemma}

\begin{lemma}\label{LemmaT2}
Suppose that $E\left[\xi_{it}\right]=0$, and for $\forall i$, $\{\xi_{it}\}$ is strictly $\alpha$-mixing process with the $\alpha$-mixing coefficient 

\begin{eqnarray*}
\alpha_i(t) = \sup_{A\in \mathcal{F}_{i,-\infty}^0, B\in  \mathcal{F}_{i,t}^\infty} |P(A)P(B) -P(AB)|
\end{eqnarray*}
satisfying $\max_i\sum_{t=1}^{T}\alpha_i(t)^{\delta_i/(4+\delta_i)}=O(1)$ with $\delta_i>0$ and  $\max_iE |\xi_{it} |^{2+\delta_i/2} <\infty$, where $\mathcal{F}_{i,-\infty}^0$ and $\mathcal{F}_{i,t}^\infty$ are the $\sigma$-algebras generated by $\{\xi_{is}: s \leq 0\}$ and $\{\xi_{is}: s \geq t\}$, respectively. Then we have for any given $\varepsilon>0$, there exists $\delta^\ast \in (0,\min_i\delta_i)$ such that $$\Pr\left(\max_{1\leq i\leq N}\left |\frac{1}{T}\sum_{t=1}^T\xi_{it}\right |\geq \varepsilon \right)=O\left(\frac{N}{T^{1+\delta^\ast/4}}\right).$$

\end{lemma}

Lemma \ref{LemmaT2} infers, even for the case $N\asymp T$ or $N/T\to \infty$ (for this lemma only, $N$ and $T$ are generic notations and could be different from those in the main model), we are still able to achieve $\max_{1\leq i\leq N}  |\frac{1}{T}\sum_{t=1}^T\xi_{it} |=o_P(1)$ under rather minor conditions.  Notably, Lemma \ref{LemmaT2} does not impose any restrictions on $i$ dimension and on the tail behavior of $\xi_{it}$, so it permits $\{\xi_{it}\}$ to be correlated cross-sectionally  and to have heavy tail behavior.


\begin{lemma}\label{LemA1}
Suppose that $\mathbf{A}$ and $\mathbf{A}+\mathbf{E}$ are $n\times n$ symmetric matrices and that $\mathbf{Q} = (\mathbf{Q}_1,\mathbf{Q}_2)$, where $\mathbf{Q}_1$ is $n\times r$ and $\mathbf{Q}_2$ is $n\times (n-r)$, is an orthogonal matrix such that $\normalfont\text{span}(\mathbf{Q}_1)$ is an invariant subspace for $\mathbf{A}$. Decompose $\mathbf{Q}^\top \mathbf{A}\mathbf{Q}$ and $\mathbf{Q}^\top \mathbf{E} \mathbf{Q}$ as $\mathbf{Q}^\top \mathbf{A} \mathbf{Q} = \diag(\mathbf{D}_1,\mathbf{D}_2)$ and $\mathbf{Q}^\top \mathbf{E} \mathbf{Q} =\{\mathbf{E}_{ij}\}_{2\times 2}$. Let $\mathrm{sep}( \mathbf{D}_1, \mathbf{D}_2) = \min_{\lambda_1\in \lambda(\mathbf{D}_1),\ \lambda_2\in \lambda(\mathbf{D}_2)} |\lambda_1 -\lambda_2|$. If $\mathrm{sep}(\mathbf{D}_1,\mathbf{D}_2) > 0$ and $\|\mathbf{E}\|_{2} \leq \mathrm{sep}(\mathbf{D}_1,\mathbf{D}_2)/5$, then there exists a $(n-r)\times r$ matrix $\mathbf{P}$ with $\|\mathbf{P} \|_{2} \leq 4 \|\mathbf{E}_{21}\|_2/\mathrm{sep}(\mathbf{D}_1,\mathbf{D}_2)$, such that the columns of $\mathbf{Q}_1^0 = (\mathbf{Q}_1 + \mathbf{Q}_2\mathbf{P})(\mathbf{I}_r+\mathbf{P}^\top \mathbf{P})^{-1/2}$ define an orthonormal basis for a subspace that is invariant for $\mathbf{A}+\mathbf{E}$.
\end{lemma}

\begin{lemma}\label{LM1}
Under Assumptions \ref{AS1}.1-\ref{AS1}.4, as $(L, N, T)\to (\infty,\infty,\infty)$,

\begin{enumerate}[leftmargin=*, itemsep=0.5pt, parsep=0.5pt, topsep=0.6pt]
\item $\sup_{\mathbf{C}_{\centerdot \centerdot}}\left|\frac{1}{LNT}\sum_{i=1}^L\sum_{j=1}^N\pmb{\mathcal{E}}_{ij\centerdot}^\top \mathbf{P}_{ij}^{\circ \bullet}\pmb{\mathcal{E}}_{ij\centerdot}\right|  =O_P\left(\frac{1}{LN\wedge T} + \frac{\max_i \|\pmb{\mathcal{E}}_{i\centerdot\centerdot} \|_2^2}{NT} +\frac{\max_j \|\pmb{\mathcal{E}}_{\centerdot j\centerdot} \|_2^2}{LT} \right)$;

\item $\sup_{\mathbf{C}_{\centerdot \centerdot}}\left| \frac{1}{LNT}\sum_{i,j}\pmb{\mathcal{E}}_{ij\centerdot}^\top \mathbf{C}_{ij}^\dag \mathbf{F}_{ij}^* \pmb{\gamma}_{ij}^* \right|=O_P\left(\frac{1}{\sqrt{LN \wedge T}} + \frac{\max_i \| \pmb{\mathcal{E}}_{i\centerdot \centerdot}\|_2}{\sqrt{NT}}+\frac{\max_j \| \pmb{\mathcal{E}}_{\centerdot j\centerdot}\|_2}{\sqrt{LT}}\right)$;

\item $\sup_{\mathbf{C}_{\centerdot \centerdot}} \left|\frac{1}{LNT}\sum_{i,j}\pmb{\mathcal{E}}_{ij\centerdot}^\top \mathbf{C}_{ij}^\dag \mathbf{X}_{ij\centerdot}\pmb{\beta}_{ij} \right|=O_P\left(\frac{1}{\sqrt{LN \wedge T}} + \frac{\max_i \| \pmb{\mathcal{E}}_{i\centerdot \centerdot}\|_2}{\sqrt{NT}}+\frac{\max_j \| \pmb{\mathcal{E}}_{\centerdot j\centerdot}\|_2}{\sqrt{LT}}\right)$;

\item $\max_{i,j}\|\mathbf{M}_{\mathbf{F}_{ij}^*} -(\mathbf{I}_T- \frac{1}{T}\mathbf{F}_{ij}^* \mathbf{F}_{ij}^{*\top} ) \| =o_P(1)$;

\item $\sup_{\mathbf{C}}\frac{1}{LNT}\sum_{i,j}\pmb{\gamma}_{ij}^{\circ\top}  \mathbf{F}_i^{\circ\top}   \mathbf{P}_{\mathbf{C}} \mathbf{F}_i^\circ  \pmb{\gamma}_{ij}^\circ = O_P\left(  \frac{1}{L\wedge T}\right)$; 

\item {$\sup_{\mathbf{C}_j^\bullet}\frac{1}{LNT}\sum_{i,j}\pmb{\gamma}_{ij}^{\circ\top}  \mathbf{F}_i^{\circ\top}   \mathbf{P}_{\mathbf{C}_j^\bullet} \mathbf{F}_i^\circ  \pmb{\gamma}_{ij}^\circ = O_P\left( \frac{1}{L\wedge T}\right)$; }

\item {$\sup_{\mathbf{C}}\frac{1}{LNT}\sum_{i,j}\pmb{\gamma}_{ij}^{\bullet\top}  \mathbf{F}_j^{\bullet\top}   \mathbf{P}_{\mathbf{C}} \mathbf{F}_j^\bullet  \pmb{\gamma}_{ij}^\bullet = O_P\left( \frac{1}{N\wedge T}\right)$; }

\item {$\sup_{\mathbf{C}_i^\circ}\frac{1}{LNT}\sum_{i,j}\pmb{\gamma}_{ij}^{\bullet\top}  \mathbf{F}_j^{\bullet\top}   \mathbf{P}_{\mathbf{C}_i^\circ} \mathbf{F}_j^\bullet  \pmb{\gamma}_{ij}^\bullet = O_P\left( \frac{1}{N\wedge T}\right)$. }
\end{enumerate}
\end{lemma}

\medskip

The following lemmas are necessary for Section \ref{Sec2.3}. To study the estimation of the numbers of factors, recall that we have

\begin{eqnarray*}
\widehat{\pmb{\Sigma}} = \frac{1}{LNT}\sum_{i=1}^L\sum_{j=1}^N(\mathbf{Y}_{ij\centerdot} -\mathbf{X}_{ij\centerdot } \widetilde{\mathbf{b}}_{ij} )(\mathbf{Y}_{ij\centerdot} -\mathbf{X}_{ij\centerdot } \widetilde{\mathbf{b}}_{ij})^\top.
\end{eqnarray*}
Additionally, we let  

\begin{eqnarray*}
\pmb{\Sigma} &\coloneqq& \frac{1}{LNT}\mathbf{F} \pmb{\Gamma}_{\centerdot \centerdot}^{\top}\pmb{\Gamma}_{\centerdot \centerdot} \mathbf{F}^{\top},\nonumber \\
\mathbf{H}^\sharp &=&(\mathbf{H}_1^\sharp,\ldots, \mathbf{H}_\ell^\sharp) \coloneqq \frac{1}{LNT}\pmb{\Gamma}_{\centerdot\centerdot}^{\top}\pmb{\Gamma}_{\centerdot\centerdot} \cdot \mathbf{F}^{\top}\widehat{\mathbf{C}}^\sharp \widehat{\mathbf{V}}^{\sharp -1},\nonumber \\
\lambda_{s} &\coloneqq& \frac{1}{T}\mathbf{H}_s^{\sharp\top} \mathbf{F}^{\top} \pmb{\Sigma} \mathbf{F} \mathbf{H}_s^\sharp \quad \text{for}\quad s\in[\ell],
\end{eqnarray*} 
where $\widehat{\mathbf{V}}^\sharp$ is the $\ell\times \ell$  principal diagonal of $\widehat{\mathbf{V}}$, and $\widehat{\mathbf{C}}^\sharp$ includes the eigenvectors corresponding to $\widehat{\mathbf{V}}^\sharp$. 
 
\begin{lemma} \label{LemA3}
Under Assumptions \ref{AS1} and \ref{AS2}, as $(L,N,T)\to (\infty, \infty,\infty)$, we have the following results:

\begin{enumerate}[leftmargin=*, itemsep=0.5pt, parsep=0.5pt, topsep=0.6pt]
\item $|\widehat{\lambda}_s  -\lambda_s| = O_P\left( \frac{1}{\sqrt{LN}}\|\pmb{\beta}_{\centerdot \centerdot}-\widetilde{\mathbf{b}}_{\centerdot\centerdot}\| +\frac{1}{\sqrt{L\wedge N\wedge T}}\right)$ for $s\in[\ell]$, 

\item $\widehat{\lambda}_s = O_P\left( \frac{1}{LN}\|\pmb{\beta}_{\centerdot \centerdot}-\widetilde{\mathbf{b}}_{\centerdot\centerdot}\|^2 +\frac{1}{ L\wedge N\wedge T}\right)$ for $s =\ell+1,\ldots, d_{\max}$.
\end{enumerate}
\end{lemma}

Building on Lemma \ref{LemA3}, we let

\begin{eqnarray*}
\widehat{\pmb{\Sigma}}_i^\circ &=& \frac{1}{NT} \sum_{j=1}^N\mathbf{M}_{\widehat{\mathbf{C}}}(\mathbf{Y}_{ij\centerdot} -\mathbf{X}_{ij\centerdot } \widetilde{\mathbf{b}}_{ij} )(\mathbf{Y}_{ij\centerdot} -\mathbf{X}_{ij\centerdot } \widetilde{\mathbf{b}}_{ij})^\top \mathbf{M}_{\widehat{\mathbf{C}}},\nonumber \\
\widehat{\pmb{\Sigma}}_j^\bullet &=& \frac{1}{LT} \mathbf{M}_{\widehat{\mathbf{C}}}\sum_{j=1}^N(\mathbf{Y}_{ij\centerdot} -\mathbf{X}_{ij\centerdot } \widetilde{\mathbf{b}}_{ij} )(\mathbf{Y}_{ij\centerdot} -\mathbf{X}_{ij\centerdot } \widetilde{\mathbf{b}}_{ij})^\top\mathbf{M}_{\widehat{\mathbf{C}}}.
\end{eqnarray*}
Additionally, we let  

\begin{eqnarray*}
\pmb{\Sigma}_i^\circ &\coloneqq& \frac{1}{NT}\mathbf{F}_i^\circ \pmb{\Gamma}_{i \centerdot}^{\circ\top}\pmb{\Gamma}_{i \centerdot}^\circ \mathbf{F}_i^{\circ\top},\nonumber \\
\mathbf{H}_i^{\circ\sharp} &=&(\mathbf{H}_{i,1}^{\circ\sharp},\ldots, \mathbf{H}_{i,\ell_i^\circ}^{\circ\sharp}) \coloneqq \frac{1}{NT}\pmb{\Gamma}_{i\centerdot}^{\circ\top}\pmb{\Gamma}_{i\centerdot}^\circ \cdot \mathbf{F}_i^{\circ\top}\widehat{\mathbf{C}}_i^{\circ\sharp }\widehat{\mathbf{V}}_i^{\circ\sharp -1},\nonumber \\
\lambda_{i,s}^\circ &\coloneqq& \frac{1}{T}\mathbf{H}_{i,s}^{\circ\sharp\top} \mathbf{F}_i^{\circ\top} \pmb{\Sigma}_i^\circ \mathbf{F}_i^\circ \mathbf{H}_{i,s}^{\circ\sharp} \quad \text{for}\quad s\in[\ell_i^\circ],\nonumber \\
\pmb{\Sigma}_j^\bullet &\coloneqq& \frac{1}{LT}\mathbf{F}_j^\bullet \pmb{\Gamma}_{ \centerdot j}^{\bullet\top}\pmb{\Gamma}_{\centerdot j}^\bullet \mathbf{F}_j^{\bullet\top},\nonumber \\
\mathbf{H}_j^{\bullet\sharp} &=&(\mathbf{H}_{j,1}^{\bullet\sharp},\ldots, \mathbf{H}_{j,\ell_j^\bullet}^{\bullet\sharp}) \coloneqq \frac{1}{LT}\pmb{\Gamma}_{\centerdot j}^{\bullet\top}\pmb{\Gamma}_{\centerdot j}^\bullet \cdot \mathbf{F}_j^{\bullet\top}\widehat{\mathbf{C}}_j^{\bullet\sharp }\widehat{\mathbf{V}}_j^{\bullet\sharp -1},\nonumber \\
\lambda_{j,s}^\bullet &\coloneqq& \frac{1}{T}\mathbf{H}_{j,s}^{\bullet\sharp\top} \mathbf{F}_j^{\bullet\top} \pmb{\Sigma}_j^\bullet \mathbf{F}_j^\bullet \mathbf{H}_{j,s}^{\bullet\sharp} \quad \text{for}\quad s\in[\ell_j^\bullet]
\end{eqnarray*} 
where $\widehat{\mathbf{V}}_i^{\circ\sharp}$ is the $\ell_i^\circ\times \ell_i^\circ$  principal diagonal of $\widehat{\mathbf{V}}_i^\circ$,  $\widehat{\mathbf{C}}^{\circ\sharp}$ includes the eigenvectors corresponding to $\widehat{\mathbf{V}}_i^{\circ\sharp}$, $\widehat{\mathbf{V}}_j^{\bullet\sharp}$ is the $\ell_j^\bullet\times \ell_j^\bullet$  principal diagonal of $\widehat{\mathbf{V}}_j^\bullet$,  $\widehat{\mathbf{C}}^{\bullet\sharp}$ includes the eigenvectors corresponding to $\widehat{\mathbf{V}}_j^{\bullet\sharp}$. 
 
Let 

\begin{eqnarray*}
\Delta^\circ&=&\max_{i,j} \|\pmb{\beta}_{ij}-\widetilde{\mathbf{b}}_{ij}\| + \frac{\sqrt{\log(LN)}}{\sqrt{N \wedge T}}  +\frac{1}{\sqrt{L}}+ \frac{\max_i\|\pmb{\mathcal{E}}_{i\centerdot\centerdot}\|_2}{\sqrt{N T}} +\frac{\max_i \| \mathbf{F}_i^{\circ\top} \mathbf{F}\|_2 }{T} ,\nonumber \\
\Delta^\bullet &=&  \max_{i,j} \|\pmb{\beta}_{ij}-\widehat{\mathbf{b}}_{ij}\| + \frac{\sqrt{\log(LN)}}{\sqrt{L \wedge T}} +\frac{1}{\sqrt{N}}+ \frac{\max_j\|\pmb{\mathcal{E}}_{\centerdot j\centerdot}\|_2}{\sqrt{L T}} +\frac{\max_j\| \mathbf{F}_j^{\bullet\top} \mathbf{F}\|_2}{T}  .
\end{eqnarray*}
Apparently, we have $ \Delta^\circ\to 0$ and $\max_j\Delta^\bullet\to 0$ in view of Assumption \ref{AS1}.1 and Assumption \ref{AS1}.3.

\begin{lemma}\label{LemA4}
Under Assumptions \ref{AS1} and \ref{AS2}, as $(L, N,T)\to (\infty, \infty,\infty)$, we have  

\begin{enumerate}[leftmargin=*, itemsep=0.5pt, parsep=0.5pt, topsep=0.6pt]
\item $\max_i|\widehat{\lambda}_{i,s}^\circ -\lambda_{i,s}^\circ| =O_P (\Delta^\circ )$ for $s\in [\ell_i^\circ]$; 

\item $\max_i\widehat{\lambda}_{i,s}^\circ =O_P ( \Delta^{\circ 2}  )$ for $s =\ell_i^\circ+1,\ldots, d_{\max}$;

\item $\max_j|\widehat{\lambda}_{j,s}^\bullet -\lambda_{j,s}^\bullet| =O_P (\Delta^\bullet   )$ for $s\in [\ell_j^\bullet]$; 

\item $\max_j\widehat{\lambda}_{j,s}^\bullet =O_P (\Delta^{\bullet 2} )$ for $s =\ell_j^\bullet+1,\ldots, d_{\max}$.
\end{enumerate}
\end{lemma}

\section{Proofs}\label{AppB2}

\noindent \textbf{Proof of Lemma \ref{LemmaT1}:}

Lemma \ref{LemmaT1} holds immediately by Theorem 4.1 of \cite{SY1996}. Therefore, its proof is omitted.   \hspace*{\fill}{$\blacksquare$}

\bigskip

\noindent \textbf{Proof of Lemma \ref{LemmaT2}:}

Write

\begin{eqnarray*}
\Pr\left(\max_{1\leq i\leq N}\left |\frac{1}{T}\sum_{t=1}^T\xi_{it}\right |\geq \varepsilon \right)&\leq &\sum_{i=1}^N \Pr\left(\left |\frac{1}{T}\sum_{t=1}^T\xi_{it}\right |\geq \varepsilon \right) \nonumber \\
&\leq & \sum_{i=1}^N\frac{E   |\frac{1}{T}\sum_{t=1}^T \xi_{it}  |^{2+\delta_i^\ast/2}}{\varepsilon^{2+\delta_i^\ast/2}}
\nonumber\\
&\leq &O(1)\frac{1}{T^{1+ \min_i\delta_i^\ast/4}} \sum_{i=1}^N E \left |\xi_{it}\right |^{2+\delta_i^\ast/2} \nonumber \\
&=& O(1) \frac{N}{T^{1+\delta^\ast/4}},
\end{eqnarray*}
where we let $\delta^\ast \coloneqq \min_i\delta_i^\ast$, the second inequality follows from Chebyshev's inequality, and the third inequality holds by Lemma \ref{LemmaT1}. 

The proof is therefore completed. \hspace*{\fill}{$\blacksquare$}

%
%

\bigskip

\noindent \textbf{Proof of Lemma \ref{LemA1}:}

The proof is given in Theorem 8.1.10 of \cite{GL2013}. \hspace*{\fill}{$\blacksquare$}

\bigskip

\noindent \textbf{Proof of Lemma \ref{LM1}:}

In what follows, we repeatedly use the fact that $\tr(\mathbf{A})\le \text{rank}(\mathbf{A})\cdot \|\mathbf{A} \|_2$, and shall not mention it again.

(1). By the definition of $\mathbf{P}_{ij}^{\circ \bullet} $ defined in Appendix A, we can write

\begin{eqnarray*}
&&\frac{1}{LNT}\sum_{i,j}\pmb{\mathcal{E}}_{ij\centerdot}^\top \mathbf{P}_{ij}^{\circ \bullet}  \pmb{\mathcal{E}}_{ij\centerdot} \nonumber \\
&=&\frac{2}{LNT}\sum_{i,j}\pmb{\mathcal{E}}_{ij\centerdot}^\top \mathbf{P}_{\mathbf{C}}  \pmb{\mathcal{E}}_{ij\centerdot}+\frac{1}{LNT}\sum_{i,j}\pmb{\mathcal{E}}_{ij\centerdot}^\top \mathbf{P}_{\mathbf{C}_i^\circ}  \pmb{\mathcal{E}}_{ij\centerdot}+\frac{1}{LNT}\sum_{i,j}\pmb{\mathcal{E}}_{ij\centerdot}^\top \mathbf{P}_{\mathbf{C}_j^\bullet}  \pmb{\mathcal{E}}_{ij\centerdot}.
\end{eqnarray*}
We then consider the terms of the second line one by one.

For $\frac{1}{LNT}\sum_{i,j}\pmb{\mathcal{E}}_{ij\centerdot}^\top \mathbf{P}_{\mathbf{C}}  \pmb{\mathcal{E}}_{ij\centerdot}$, write

\begin{eqnarray}\label{eprove}
\sup_{\mathbf{C}}\frac{1}{LNT}\sum_{i,j}\pmb{\mathcal{E}}_{ij\centerdot}^\top  \mathbf{P}_{\mathbf{C}} \pmb{\mathcal{E}}_{ij\centerdot} &=& \sup_{\mathbf{C}} \frac{1}{LNT}\tr \{  \mathbf{P}_{\mathbf{C}} \,\pmb{\mathcal{E}}_{\centerdot\centerdot\centerdot}^\top \pmb{\mathcal{E}}_{\centerdot\centerdot\centerdot} \} \nonumber \\
&\le & \frac{1}{LNT} \|\pmb{\mathcal{E}}_{\centerdot\centerdot\centerdot} \|_2^2=O_P\left(\frac{1}{LN\wedge T} \right),
\end{eqnarray}
where we have used Assumption \ref{AS1}.1.

For $\frac{1}{LNT}\sum_{i,j}\pmb{\mathcal{E}}_{ij\centerdot}^\top \mathbf{P}_{ \mathbf{C}_i^\circ}\pmb{\mathcal{E}}_{ij\centerdot}$, write

\begin{eqnarray*}
\sup_{\mathbf{C}^\circ}\frac{1}{LNT}\sum_{i,j}\pmb{\mathcal{E}}_{ij\centerdot}^\top \mathbf{P}_{\mathbf{C}_i^\circ}  \pmb{\mathcal{E}}_{ij\centerdot} &= &\sup_{\mathbf{C}^\circ}\frac{1}{LNT}\sum_{i=1}^L  \tr \{ \mathbf{P}_{\mathbf{C}_i^\circ}  \pmb{\mathcal{E}}_{i\centerdot\centerdot}\pmb{\mathcal{E}}_{i\centerdot\centerdot}^\top \}\nonumber \\
&\le &O(1)\frac{1}{LNT}\sum_{i=1}^L \|\pmb{\mathcal{E}}_{i\centerdot\centerdot} \|_2^2\nonumber \\
&=&O_P(1)\frac{\max_i \|\pmb{\mathcal{E}}_{i\centerdot\centerdot} \|_2^2}{NT} ,
\end{eqnarray*}
where the second equality follows from Assumption \ref{AS1}.1. By noting $i$ and $j$ are symmetric, we can similarly obtain that

\begin{eqnarray*}
\sup_{ \mathbf{C}^\bullet}\frac{1}{LNT}\sum_{i,j}\pmb{\mathcal{E}}_{ij\centerdot}^\top \mathbf{P}_{\mathbf{C}^\bullet_j}  \pmb{\mathcal{E}}_{ij\centerdot} =O_P(1)\frac{\max_j \|\pmb{\mathcal{E}}_{\centerdot j\centerdot} \|_2^2}{LT}.
\end{eqnarray*}

Putting everything together, the result follows.

\medskip

(2). Recall that we have defined $\mathbf{F}_{ij}^*$ and $ \pmb{\gamma}_{ij}^*$ in \eqref{def.Fstar} and in Appendix A respectively.  Write

\begin{eqnarray}\label{eqlm1.2}
&&\frac{1}{LNT}\sum_{i,j}\pmb{\mathcal{E}}_{ij\centerdot}^\top  \mathbf{C}_{ij}^\dag \mathbf{F}_{ij}^* \pmb{\gamma}_{ij}^* \nonumber \\
&=&\frac{1}{LNT}\sum_{i,j}\pmb{\mathcal{E}}_{ij\centerdot}^\top   \mathbf{F}_{ij}^* \pmb{\gamma}_{ij}^*- \frac{2}{LNT}\sum_{i,j}\pmb{\mathcal{E}}_{ij\centerdot}^\top \mathbf{P}_{\mathbf{C}} \mathbf{F}_{ij}^* \pmb{\gamma}_{ij}^*\nonumber \\
&&- \frac{1}{LNT}\sum_{i,j}\pmb{\mathcal{E}}_{ij\centerdot}^\top \mathbf{P}_{\mathbf{C}_i^\circ}  \mathbf{F}_{ij}^* \pmb{\gamma}_{ij}^* - \frac{1}{LNT}\sum_{i,j}\pmb{\mathcal{E}}_{ij\centerdot}^\top \mathbf{P}_{ \mathbf{C}_j^\bullet }  \mathbf{F}_{ij}^* \pmb{\gamma}_{ij}^*.
\end{eqnarray}
In what follows, we investigate the terms on the right hand side one by one. By Assumption \ref{AS1}.1, we immediately obtain that

\begin{eqnarray}\label{eqlm1.21}
\left|\frac{1}{LNT}\sum_{i,j}\pmb{\mathcal{E}}_{ij\centerdot}^\top   \mathbf{F}_{ij}^* \pmb{\gamma}_{ij}^*\right| =O_P\left( \frac{1}{\sqrt{LN \wedge T}}\right).
\end{eqnarray}
Thus, we only need to consider three terms on the right hand side of \eqref{eqlm1.2}.

We start from the term $\frac{1}{LNT}\sum_{i,j}\pmb{\mathcal{E}}_{ij\centerdot}^\top \mathbf{P}_{\mathbf{C}} \mathbf{F}_{ij}^* \pmb{\gamma}_{ij}^*$, which can be further decomposed as follows:

\begin{eqnarray*}
&&\frac{1}{LNT}\sum_{i,j}\pmb{\mathcal{E}}_{ij\centerdot}^\top \mathbf{P}_{\mathbf{C}} \mathbf{F}_{ij}^* \pmb{\gamma}_{ij}^*\nonumber \\
&=&\frac{1}{LNT}\sum_{i,j}\pmb{\mathcal{E}}_{ij\centerdot}^\top \mathbf{P}_{\mathbf{C}} \mathbf{F} \pmb{\gamma}_{ij} +\frac{1}{LNT}\sum_{i,j}\pmb{\mathcal{E}}_{ij\centerdot}^\top \mathbf{P}_{\mathbf{C}} \mathbf{F}_i^\circ \pmb{\gamma}_{ij}^\circ +\frac{1}{LNT}\sum_{i,j}\pmb{\mathcal{E}}_{ij\centerdot}^\top \mathbf{P}_{\mathbf{C}} \mathbf{F}_j^\bullet \pmb{\gamma}_{ij}^\bullet .
\end{eqnarray*}
 
For $\frac{1}{LNT}\sum_{i,j}\pmb{\mathcal{E}}_{ij\centerdot}^\top \mathbf{P}_{\mathbf{C}} \mathbf{F} \pmb{\gamma}_{ij}$, write

\begin{eqnarray*}
\sup_{\mathbf{C}}\left|\frac{1}{LNT}\sum_{i,j}\pmb{\mathcal{E}}_{ij\centerdot}^\top \mathbf{P}_{\mathbf{C}} \mathbf{F} \pmb{\gamma}_{ij} \right| &=&\sup_{\mathbf{C}}\frac{1}{LNT} |\tr \{ \mathbf{P}_{\mathbf{C}} \mathbf{F} \pmb{\Gamma}_{\centerdot \centerdot}^\top \pmb{\mathcal{E}}_{\centerdot\centerdot\centerdot} \}  | \nonumber \\
&\le &O(1)\frac{1}{LNT}\| \mathbf{F} \|_2\cdot \|\pmb{\Gamma}_{\centerdot \centerdot}\|_2
\cdot \| \pmb{\mathcal{E}}_{\centerdot\centerdot\centerdot}\|_2 \nonumber \\
&= &O_P(1)\frac{1}{LNT}  \sqrt{T}\cdot \sqrt{LN}\cdot \sqrt{LN\vee T}\nonumber \\
&\le &O_P(1)\frac{1}{\sqrt{LN \wedge T}},
\end{eqnarray*}
where $\pmb{\Gamma}_{\centerdot \centerdot}$ is defined in Appendix A, and the second equality follows from Assumptions \ref{AS1}.1-\ref{AS1}.3. 

For $\frac{1}{LNT}\sum_{i,j}\pmb{\mathcal{E}}_{ij\centerdot}^\top \mathbf{P}_{\mathbf{C}} \mathbf{F}_i^\circ \pmb{\gamma}_{ij}^\circ$, write

\begin{eqnarray}\label{f0eprove}
&&\sup_{\mathbf{C}} \left| \frac{1}{LNT}\sum_{i,j}\pmb{\mathcal{E}}_{ij\centerdot}^\top \mathbf{P}_{\mathbf{C}} \mathbf{F}_i^\circ \pmb{\gamma}_{ij}^\circ \right|\nonumber \\
&=& \sup_{\mathbf{C}} \frac{1}{LNT}  \left|\sum_{i=1}^L\tr \{   \mathbf{P}_{\mathbf{C}} \mathbf{F}_i^\circ \pmb{\Gamma}_{i\centerdot}^{\circ\top} \pmb{\mathcal{E}}_{i\centerdot\centerdot}^\top \} \right| \nonumber \\
&=& \sup_{\mathbf{C}} \frac{1}{LNT}  \left| \tr \{   \mathbf{P}_{\mathbf{C}} \mathbf{F}^\circ \pmb{\Gamma}_d^{\circ \top} \pmb{\mathcal{E}}_{\centerdot\centerdot\centerdot}^\top \} \right| \nonumber \\
&\le &O_P(1) \frac{1}{LNT}  \sqrt{L\vee T}\cdot \sqrt{N} \cdot \sqrt{LN\vee T}\nonumber \\
&=&O_P(1)\frac{1}{\sqrt{(L\wedge T)(LN\wedge T)}},
\end{eqnarray}
where $\pmb{\Gamma}_d^{\circ}$ is defined in Appendix A, and the first inequality follows from Assumptions \ref{AS1}.1 and \ref{AS1}.3 and the following development:

\begin{eqnarray*}
\| \pmb{\Gamma}_d^{\circ}\|_2 =\sqrt{\lambda_{\max}(\pmb{\Gamma}_d^{\circ \top}\pmb{\Gamma}_d^{\circ}) } \le \sqrt{\max_i \|\pmb{\Gamma}_{i\centerdot}^\circ  \|^2} =O_P(\sqrt{N}),
\end{eqnarray*}
in which the second equality follows from Assumption \ref{AS1}.2.

Similar to $\frac{1}{LNT}\sum_{i,j}\pmb{\mathcal{E}}_{ij\centerdot}^\top \mathbf{P}_{\mathbf{C}} \mathbf{F}_i^\circ \pmb{\gamma}_{ij}^\circ$ and by noting $i$ and $j$ are symmetric, we obtain that

\begin{eqnarray*}
\sup_{\mathbf{C}} \left|\frac{1}{LNT}\sum_{i,j}\pmb{\mathcal{E}}_{ij\centerdot}^\top \mathbf{P}_{\mathbf{C}} \mathbf{F}_j^\bullet  \pmb{\gamma}_{ij}^\bullet\right|=O_P(1)\frac{1}{\sqrt{(N\wedge T)(LN\wedge T)}}.
\end{eqnarray*}

Up to this point, we can conclude that

\begin{eqnarray}\label{eqlm1.22}
\sup_{\mathbf{C}} \left|\frac{1}{LNT}\sum_{i,j}\pmb{\mathcal{E}}_{ij\centerdot}^\top \mathbf{P}_{\mathbf{C}} \mathbf{F}_{ij}^* \pmb{\gamma}_{ij}^*\right|=O_P(1)\frac{1}{\sqrt{L N\wedge T}}.
\end{eqnarray}

\medskip

Next, we investigate $\frac{1}{LNT}\sum_{i,j}\pmb{\mathcal{E}}_{ij\centerdot}^\top \mathbf{P}_{ \mathbf{C}_i^\circ}   \mathbf{F}_{ij}^* \pmb{\gamma}_{ij}^*$, which can also be further decomposed as follows:

\begin{eqnarray*}
&&\frac{1}{LNT}\sum_{i,j}\pmb{\mathcal{E}}_{ij\centerdot}^\top \mathbf{P}_{\mathbf{C}_i^\circ}   \mathbf{F}_{ij}^* \pmb{\gamma}_{ij}^*\nonumber \\
&=& \frac{1}{LNT}\sum_{i,j}\pmb{\mathcal{E}}_{ij\centerdot}^\top \mathbf{P}_{\mathbf{C}_i^\circ} \mathbf{F} \pmb{\gamma}_{ij} +\frac{1}{LNT}\sum_{i,j}\pmb{\mathcal{E}}_{ij\centerdot}^\top \mathbf{P}_{\mathbf{C}_i^\circ} \mathbf{F}_i^\circ \pmb{\gamma}_{ij}^\circ +\frac{1}{LNT}\sum_{i,j}\pmb{\mathcal{E}}_{ij\centerdot}^\top \mathbf{P}_{\mathbf{C}_i^\circ}\mathbf{F}_j^\bullet \pmb{\gamma}_{ij}^\bullet .
\end{eqnarray*}

For $ \frac{1}{LNT}\sum_{i,j}\pmb{\mathcal{E}}_{ij\centerdot}^\top \mathbf{P}_{\mathbf{C}_i^\circ} \mathbf{F} \pmb{\gamma}_{ij}$, write

\begin{eqnarray*}
\sup_{ \mathbf{C}^\circ }\left| \frac{1}{LNT}\sum_{i,j}\pmb{\mathcal{E}}_{ij\centerdot}^\top \mathbf{P}_{\mathbf{C}_i^\circ} \mathbf{F} \pmb{\gamma}_{ij} \right| &=&  \sup_{\mathbf{C}^\circ }\frac{1}{LNT}\left|\sum_{i=1}^L\tr\{\mathbf{F} \pmb{\Gamma}_{i\centerdot}^\top \pmb{\mathcal{E}}_{i\centerdot\centerdot}^\top \mathbf{P}_{\mathbf{C}_i^\circ} \} \right|\nonumber \\
&=&  \sup_{\mathbf{C}^\circ }\frac{1}{LNT}\left|\tr\{\mathbf{F} \pmb{\Gamma}_{\centerdot \centerdot}^\top \pmb{\mathcal{E}}_{d}^\top \pmb{\mathcal{C}}^\circ \} \right|\nonumber \\
&\le &O(1)\frac{1}{LNT}\|\mathbf{F} \|_2\| \pmb{\Gamma}_{\centerdot \centerdot}\|_2\| \pmb{\mathcal{E}}_{d}\|_2\| \pmb{\mathcal{C}}^\circ \|_2\nonumber \\
&\le &O_P(1)\frac{1}{LNT}\sqrt{T}\cdot \sqrt{LN}\cdot \max_i \| \pmb{\mathcal{E}}_{i\centerdot \centerdot}\|_2 \cdot \sqrt{L}\nonumber \\
&=&O_P(1)\frac{\max_i \| \pmb{\mathcal{E}}_{i\centerdot \centerdot}\|_2}{\sqrt{NT}},
\end{eqnarray*}
where $\pmb{\mathcal{C}}^\circ$ and $\pmb{\mathcal{E}}_{d}$ are defined in Appendix A, and the second inequality follows from Assumptions \ref{AS1}.1-\ref{AS1}.3 and the following development:

\begin{eqnarray}\label{eqcalC}
\|\pmb{\mathcal{C}}^\circ\|_2 =\sqrt{\lambda_{\max}\left(\sum_{i=1}^L \mathbf{P}_{\mathbf{C}_i^\circ} \right)}\le \sqrt{L}.
\end{eqnarray}

For $\frac{1}{LNT}\sum_{i,j}\pmb{\mathcal{E}}_{ij\centerdot}^\top \mathbf{P}_{\mathbf{C}_i^\circ} \mathbf{F}_i^\circ \pmb{\gamma}_{ij}^\circ $, we write

\begin{eqnarray*}
\sup_{\mathbf{C}^\circ}\left|\frac{1}{LNT}\sum_{i,j}\pmb{\mathcal{E}}_{ij\centerdot}^\top \mathbf{P}_{\mathbf{C}_i^\circ} \mathbf{F}_i^\circ \pmb{\gamma}_{ij}^\circ  \right| &=& \sup_{\mathbf{C}^\circ}\frac{1}{LNT} \left|\sum_{i=1}^L\tr\{\mathbf{P}_{\mathbf{C}_i^\circ} \mathbf{F}_i^\circ \pmb{\Gamma}_{i\centerdot}^{\circ \top} \pmb{\mathcal{E}}_{i\centerdot\centerdot}^\top \} \right|\nonumber \\
&\le &O_P(1)\frac{1}{NT}  \sqrt{T}\cdot \sqrt{N}\cdot \max_i \| \pmb{\mathcal{E}}_{i\centerdot \centerdot}\|_2 \nonumber \\
&\le &O_P(1) \frac{\max_i \| \pmb{\mathcal{E}}_{i\centerdot \centerdot}\|_2}{\sqrt{NT}},
\end{eqnarray*}
where  the first inequality follows from Assumptions \ref{AS1}.1-\ref{AS1}.3.

For $\frac{1}{LNT}\sum_{i,j}\pmb{\mathcal{E}}_{ij\centerdot}^\top \mathbf{P}_{ \mathbf{C}_i^\circ} \mathbf{F}_j^\bullet  \pmb{\gamma}_{ij}^\bullet$, write

\begin{eqnarray*}
&&\sup_{\mathbf{C}^\circ}\left|\frac{1}{LNT}\sum_{i,j}\pmb{\mathcal{E}}_{ij\centerdot}^\top \mathbf{P}_{ \mathbf{C}_i^\circ} \mathbf{F}_j^\bullet  \pmb{\gamma}_{ij}^\bullet\right| \nonumber \\
&= &\sup_{\mathbf{C}^\circ}\frac{1}{LNT}\left|\sum_{i=1}^L \tr \{ \mathbf{P}_{ \mathbf{C}_i^\circ}\mathbf{F}^\bullet \diag\{ \pmb{\Gamma}_{i\centerdot}^{\bullet\top}\} \pmb{\mathcal{E}}_{i\centerdot \centerdot}^\top \} \right|\nonumber \\
&\le &O_P(1)\frac{1}{NT} \sqrt{N\vee T} \cdot \frac{1}{L}\sum_{i=1}^L \|\diag\{ \pmb{\Gamma}_{i\centerdot}^{\bullet\top}\}\|_2 \cdot \max_i \| \pmb{\mathcal{E}}_{i\centerdot \centerdot}\|_2,\nonumber \\
&\le &O_P(1) \frac{1}{ \sqrt{N\wedge T}}\cdot \frac{\max_i \| \pmb{\mathcal{E}}_{i\centerdot \centerdot}\|_2}{\sqrt{NT}},
\end{eqnarray*}
where the first inequality follows from Assumptions \ref{AS1}.1 and \ref{AS1}.3, and the second inequality the following development:

\begin{eqnarray*}
\frac{1}{L}\sum_{i=1}^L \|\diag\{ \pmb{\Gamma}_{i\centerdot}^{\bullet\top}\}\|_2\le \max_j \frac{1}{L}\sum_{i=1}^L \|\gamma_{ij}^\bullet \|\le \max_j \frac{1}{\sqrt{L}}\|\pmb{\Gamma}_{\centerdot j}^{\bullet} \|=O_P(1)
\end{eqnarray*}
in which we have used Assumption \ref{AS1}.2.

Thus, we can conclude that

\begin{eqnarray}\label{eqlm1.23}
\sup_{ \mathbf{C}^\circ }\left|\frac{1}{LNT}\sum_{i,j}\pmb{\mathcal{E}}_{ij\centerdot}^\top \mathbf{P}_{ \mathbf{C}_i^\circ}   \mathbf{F}_{ij}^* \pmb{\gamma}_{ij}^*\right|=O_P(1) \frac{\max_i \| \pmb{\mathcal{E}}_{i\centerdot \centerdot}\|_2}{\sqrt{NT}}.
\end{eqnarray}

Similar to the study of $\frac{1}{LNT}\sum_{i,j}\pmb{\mathcal{E}}_{ij\centerdot}^\top \mathbf{P}_{ \mathbf{C}_i^\circ}   \mathbf{F}_{ij}^* \pmb{\gamma}_{ij}^*$ and by noting $i$ and $j$ being symmetric, we can obtain that

\begin{eqnarray}\label{eqlm1.24}
\sup_{ \mathbf{C}^\bullet }\left|\frac{1}{LNT}\sum_{i,j}\pmb{\mathcal{E}}_{ij\centerdot}^\top \mathbf{P}_{ \mathbf{C}_i^\bullet}   \mathbf{F}_{ij}^* \pmb{\gamma}_{ij}^*\right|=O_P(1) \frac{\max_j \| \pmb{\mathcal{E}}_{\centerdot j\centerdot}\|_2}{\sqrt{LT}}.
\end{eqnarray}
 
According to \eqref{eqlm1.2}, \eqref{eqlm1.21}, \eqref{eqlm1.22}, \eqref{eqlm1.23} and \eqref{eqlm1.24}, we finally obtain that

\begin{eqnarray*}
\sup_{\mathbf{C}_{\centerdot \centerdot}}\left| \frac{1}{LNT}\sum_{i,j}\pmb{\mathcal{E}}_{ij\centerdot}^\top \mathbf{C}_{ij}^\dag \mathbf{F}_{ij}^* \pmb{\gamma}_{ij}^* \right|=O_P \left(\frac{1}{\sqrt{LN \wedge T}} + \frac{\max_i \| \pmb{\mathcal{E}}_{i\centerdot \centerdot}\|_2}{\sqrt{NT}}+\frac{\max_j \| \pmb{\mathcal{E}}_{\centerdot j\centerdot}\|_2}{\sqrt{LT}}\right).
\end{eqnarray*}

\medskip

(3). Write

\begin{eqnarray}\label{eqlm1.3}
&&\frac{1}{LNT}\sum_{i,j}\pmb{\mathcal{E}}_{ij\centerdot}^\top \mathbf{C}_{ij}^\dag \mathbf{X}_{ij\centerdot}\pmb{\beta}_{ij} \nonumber \\
&=&\frac{1}{LNT}\sum_{i,j}\pmb{\mathcal{E}}_{ij\centerdot}^\top   \mathbf{X}_{ij\centerdot}\pmb{\beta}_{ij}  -\frac{2}{LNT}\sum_{i,j}\pmb{\mathcal{E}}_{ij\centerdot}^\top \mathbf{P}_{\mathbf{C} } \mathbf{X}_{ij\centerdot}\pmb{\beta}_{ij}\nonumber \\
&&-\frac{1}{LNT}\sum_{i,j}\pmb{\mathcal{E}}_{ij\centerdot}^\top \mathbf{P}_{ \mathbf{C}_i^\circ } \mathbf{X}_{ij\centerdot}\pmb{\beta}_{ij}-\frac{1}{LNT}\sum_{i,j}\pmb{\mathcal{E}}_{ij\centerdot}^\top \mathbf{P}_{ \mathbf{C}_j^\bullet } \mathbf{X}_{ij\centerdot}\pmb{\beta}_{ij}.
\end{eqnarray}
In what follows, we consider the terms on the right hand side of \eqref{eqlm1.3} one by one. By Assumption \ref{AS1}.1, we immediately obtain that

\begin{eqnarray}\label{eqlm1.31}
\left| \frac{1}{LNT}\sum_{i,j}\pmb{\mathcal{E}}_{ij\centerdot}^\top   \mathbf{X}_{ij\centerdot}\pmb{\beta}_{ij} \right| =O_P\left(\frac{1}{\sqrt{LN \wedge T}}\right).
\end{eqnarray}

Next, we consider $\frac{1}{LNT}\sum_{i,j}\pmb{\mathcal{E}}_{ij\centerdot}^\top \mathbf{P}_{\mathbf{C} } \mathbf{X}_{ij\centerdot}\pmb{\beta}_{ij}$.

\begin{eqnarray}\label{eqlm1.32}
&&\sup_{\mathbf{C}} \left| \frac{1}{LNT}\sum_{i,j}\pmb{\mathcal{E}}_{ij\centerdot}^\top \mathbf{P}_{\mathbf{C} } \mathbf{X}_{ij\centerdot}\pmb{\beta}_{ij}\right|\nonumber \\
&=& \sup_{\mathbf{C}} \frac{1}{LNT} |\tr\{\mathbf{P}_{\mathbf{C} } \mathbf{X}_{\centerdot\centerdot\centerdot}^\top\diag\{\pmb{\beta}_{\centerdot\centerdot}^\top \} \pmb{\mathcal{E}}_{\centerdot \centerdot \centerdot}\}  | \nonumber \\
&\le &  O(1) \frac{1}{LNT}\|\mathbf{X}_{\centerdot\centerdot\centerdot}\|_2 \cdot \| \pmb{\mathcal{E}}_{\centerdot \centerdot \centerdot}\|_2\nonumber\\
&\le &O_P(1) \frac{1}{LNT} \sqrt{LNT }\cdot \sqrt{LN \vee T}\nonumber \\
&=&O_P(1)\frac{1}{\sqrt{LN \wedge T}},
\end{eqnarray}
where the first inequality follows from the fact that $\|\diag\{\pmb{\beta}_{\centerdot\centerdot }\}\|_2<\infty$ due to $\pmb{\beta}_{ij}$'s being fixed parameters, and the second inequality follows from Assumptions \ref{AS1}.1 and \ref{AS1}.4. 

We then consider $\frac{1}{LNT}\sum_{i,j}\pmb{\mathcal{E}}_{ij\centerdot}^\top \mathbf{P}_{ \mathbf{C}_i^\circ } \mathbf{X}_{ij\centerdot}\pmb{\beta}_{ij}$.

\begin{eqnarray}\label{eqlm1.33}
&&\sup_{ \mathbf{C}^\circ } \left|\frac{1}{LNT}\sum_{i,j}\pmb{\mathcal{E}}_{ij\centerdot}^\top \mathbf{P}_{ \mathbf{C}_i^\circ } \mathbf{X}_{ij\centerdot}\pmb{\beta}_{ij} \right|\nonumber \\
&= & \sup_{ \mathbf{C}^\circ } \frac{1}{LNT}\left|\sum_{i=1}^L\tr\{ \mathbf{P}_{ \mathbf{C}_i^\circ } \mathbf{X}_{i\centerdot\centerdot}  \diag\{  \pmb{\beta}_{i\centerdot}^\top\}\pmb{\mathcal{E}}_{i\centerdot \centerdot}^\top\}\right|\nonumber \\
&\le &O_P(1)\frac{1}{NT}\sqrt{NT}\cdot \max_i \| \pmb{\mathcal{E}}_{i\centerdot \centerdot}\|_2\nonumber \\
&=&O_P\left(\frac{\max_i \| \pmb{\mathcal{E}}_{i\centerdot \centerdot}\|_2}{\sqrt{NT}} \right) ,
\end{eqnarray}
where $\pmb{\beta}_{i\centerdot }$ is defined in Appendix A, the first inequality follows from Assumptions \ref{AS1}.1 and \ref{AS1}.4 and the fact that $\|\diag\{\pmb{\beta}_{i\centerdot }^\top\}\|_2<\infty$ due to $\pmb{\beta}_{ij}$'s being fixed parameters.

Similar to the study for $\frac{1}{LNT}\sum_{i,j}\pmb{\mathcal{E}}_{ij\centerdot}^\top \mathbf{P}_{ \mathbf{C}_i^\circ } \mathbf{X}_{ij\centerdot}\pmb{\beta}_{ij} $ and by noting $i$ and $j$ being symmetric, we obtain that

\begin{eqnarray}\label{eqlm1.34}
\sup_{ \mathbf{C}^\bullet } \left|\frac{1}{LNT}\sum_{i,j}\pmb{\mathcal{E}}_{ij\centerdot}^\top \mathbf{P}_{ \mathbf{C}_j^\bullet } \mathbf{X}_{ij\centerdot}\pmb{\beta}_{ij} \right|=O_P(1)\frac{\max_j \| \pmb{\mathcal{E}}_{\centerdot j\centerdot}\|_2}{\sqrt{LT}}.
\end{eqnarray}

According to \eqref{eqlm1.3}, \eqref{eqlm1.31}, \eqref{eqlm1.32}, \eqref{eqlm1.33} and \eqref{eqlm1.34}, we obtain that

\begin{eqnarray*}
\sup_{\mathbf{C}_{\centerdot \centerdot}} \left|\frac{1}{LNT}\sum_{i,j}\pmb{\mathcal{E}}_{ij\centerdot}^\top \mathbf{C}_{ij}^\dag \mathbf{X}_{ij\centerdot}\pmb{\beta}_{ij} \right|=O_P \left(\frac{1}{\sqrt{LN \wedge T}} + \frac{\max_i \| \pmb{\mathcal{E}}_{i\centerdot \centerdot}\|_2}{\sqrt{NT}}+\frac{\max_j \| \pmb{\mathcal{E}}_{\centerdot j\centerdot}\|_2}{\sqrt{LT}}\right).
\end{eqnarray*}

\medskip

(4). Write

\begin{eqnarray*}
&&\max_{i,j}\left\|\mathbf{P}_{\mathbf{F}_{ij}^*} - \frac{1}{T}\mathbf{F}_{ij}^* \mathbf{F}_{ij}^{*\top}\right\| \nonumber \\
&= &\max_{i,j}\frac{1}{T}\left\|\mathbf{F}_{ij}^{*} \left[(\mathbf{F}_{ij}^{*\top}\mathbf{F}_{ij}^*/T)^{-1} -\mathbf{I}_{\ell +\ell_i^\circ +\ell_j^\bullet}\right] \mathbf{F}_{ij}^{*\top}\right\| \nonumber \\
&=&o_P(1),
\end{eqnarray*}
where the last step follows from Assumption \ref{AS1}.3.

\medskip

(5). Consider $\frac{1}{LNT}\sum_{i,j}\pmb{\gamma}_{ij}^{\circ\top}\mathbf{F}_{i}^{\circ\top}\mathbf{P}_{\mathbf{C}} \mathbf{F}_{i}^{\circ }\pmb{\gamma}_{ij}^{\circ}$, and write 

\begin{eqnarray*}
&&\sup_{\mathbf{C}}\frac{1}{LNT}\sum_{i,j}\pmb{\gamma}_{ij}^{\circ\top}  \mathbf{F}_i^{\circ\top}   \mathbf{P}_{\mathbf{C}} \mathbf{F}_i^\circ  \pmb{\gamma}_{ij}^\circ \nonumber \\
&=& \sup_{\mathbf{C}}\tr\left\{\frac{1}{LNT}\sum_{j=1}^N\mathbf{P}_{\mathbf{C}}  \mathbf{F}^\circ\diag\{ \pmb{\Gamma}_{\centerdot j}^{\circ}\}  \diag\{ \pmb{\Gamma}_{\centerdot j}^{\circ\top}\}^\top \mathbf{F}^{\circ\top} \right\}\nonumber \\
&\le &O(1) \frac{1}{LNT}\sum_{j=1}^N \|\diag\{ \pmb{\Gamma}_{\centerdot j}^{\circ}\}\|_2^2\cdot \| \mathbf{F}^\circ\|_2^2\nonumber \\
&= & O_P(1) \frac{1}{L\wedge T}
\end{eqnarray*}
in which we have again used Assumptions \ref{AS1}.2 and \ref{AS1}.3 and 

\begin{eqnarray*} 
\frac{1}{N}\sum_{j=1}^N\|\diag\{ \pmb{\Gamma}_{\centerdot j}^{\circ \top}\} \|_2^2\le \max_i \frac{1}{N}\sum_{j=1}^N \|\pmb{\gamma}_{i j}^\circ \|^2 =O_P(1)
\end{eqnarray*}
in which we have used Assumption \ref{AS1}.2.

\medskip

(6). Similar to the proof of (5), we can write 

\begin{eqnarray*}
&&\sup_{\mathbf{C}_j^\bullet}\frac{1}{LNT}\sum_{i,j}\pmb{\gamma}_{ij}^{\circ\top}  \mathbf{F}_i^{\circ\top}   \mathbf{P}_{\mathbf{C}_j^\bullet} \mathbf{F}_i^\circ  \pmb{\gamma}_{ij}^\circ \nonumber \\
&=&\sup_{\mathbf{C}_j^\bullet}\frac{1}{LNT}\sum_{j=1}^N\tr\left\{\mathbf{P}_{\mathbf{C}_j^\bullet}  \mathbf{F}^\circ\diag\{ \pmb{\Gamma}_{\centerdot j}^{\circ}\}  \diag\{ \pmb{\Gamma}_{\centerdot j}^{\circ\top}\}^\top \mathbf{F}^{\circ\top} \right\}\nonumber \\
&\le &O(1) \frac{1}{LNT}\sum_{j=1}^N \|\diag\{ \pmb{\Gamma}_{\centerdot j}^{\circ}\}\|_2^2\cdot \| \mathbf{F}^\circ\|_2^2\nonumber \\
&= & O_P(1) \frac{1}{L\wedge T}.
\end{eqnarray*}

\medskip

Equations (7)-(8) are obvious in view of the development of (5)-(6) as $i$ and $j$ dimensions are symmetric. The proof is now completed. \hspace*{\fill}{$\blacksquare$}

\bigskip

\noindent \textbf{Proof of Lemma \ref{M.LM1}:}

(1). Recall that we have defined a few notations in Assumption \ref{AS1}. Write

\begin{eqnarray*}
Q(\mathbf{b}_{\centerdot \centerdot}  ,\mathbf{C}_{\centerdot \centerdot}) &=&\frac{1}{LNT}\sum_{i,j}(\pmb{\beta}_{ij} -\mathbf{b}_{ij})^\top\mathbf{X}_{ij\centerdot}^\top \mathbf{C}_{ij}^\dag \mathbf{X}_{ij\centerdot}(\pmb{\beta}_{ij} -\mathbf{b}_{ij})\nonumber \\
&&+\frac{1}{LNT}\sum_{i,j}\pmb{\mathcal{E}}_{ij\centerdot}^\top  \mathbf{C}_{ij}^\dag \pmb{\mathcal{E}}_{ij\centerdot}\nonumber \\
&&+\frac{1}{LNT}\sum_{i,j}\pmb{\gamma}_{ij}^{*\top}\mathbf{F}_{ij}^{*\top}  \mathbf{C}_{ij}^\dag \mathbf{F}_{ij}^* \pmb{\gamma}_{ij}^* \nonumber \\
&&+\frac{2}{LNT}\sum_{i,j}(\pmb{\beta}_{ij} -\mathbf{b}_{ij})^\top\mathbf{X}_{ij\centerdot}^\top \mathbf{C}_{ij}^\dag \pmb{\mathcal{E}}_{ij\centerdot}\nonumber \\
&&+\frac{2}{LNT}\sum_{i,j}(\pmb{\beta}_{ij} -\mathbf{b}_{ij})^\top\mathbf{X}_{ij\centerdot}^\top \mathbf{C}_{ij}^\dag \mathbf{F}_{ij}^* \pmb{\gamma}_{ij}^* \nonumber \\
&&+\frac{2}{LNT}\sum_{i,j}\pmb{\mathcal{E}}_{ij\centerdot}^\top  \mathbf{C}_{ij}^\dag \mathbf{F}_{ij}^* \pmb{\gamma}_{ij}^* .
\end{eqnarray*}

{
By Lemma \ref{LM1} and Cauchy-Schwarz inequality, it is easy to know that

\begin{eqnarray*}
&&\frac{1}{LNT}\sum_{i,j}\pmb{\gamma}_{ij}^{*\top}\mathbf{F}_{ij}^{*\top}  \mathbf{C}_{ij}^\dag \mathbf{F}_{ij}^* \pmb{\gamma}_{ij}^{*} \nonumber \\
&=&\frac{1}{LNT}\sum_{i,j}\pmb{\gamma}_{ij}^{\top}\mathbf{F}^\top  \mathbf{C}_{ij}^\dag \mathbf{F} \pmb{\gamma}_{ij}+\frac{1}{LNT}\sum_{i,j}\pmb{\gamma}_{ij}^{\circ\top}\mathbf{F}_{i}^{\circ\top} \mathbf{F}_{i}^{\circ }\pmb{\gamma}_{ij}^{\circ}+\frac{1}{LNT}\sum_{i,j}\pmb{\gamma}_{ij}^{\circ\top}\mathbf{F}_{i}^{\circ\top}\mathbf{M}_{\mathbf{C}_i^\circ} \mathbf{F}_{i}^{\circ }\pmb{\gamma}_{ij}^{\circ}\notag \\
&&+\frac{1}{LNT}\sum_{i,j}\pmb{\gamma}_{ij}^{\bullet\top}\mathbf{F}_{j}^{\bullet\top} \mathbf{F}_{j}^{\bullet}\pmb{\gamma}_{ij}^{\bullet}+\frac{1}{LNT}\sum_{i,j}\pmb{\gamma}_{ij}^{\bullet\top}\mathbf{F}_{j}^{\bullet\top}\mathbf{M}_{\mathbf{C}_j^\bullet} \mathbf{F}_{j}^{\bullet}\pmb{\gamma}_{ij}^{\bullet}+O_P\left(\frac{1}{\sqrt{L\wedge N\wedge T}}\right),
\end{eqnarray*}
where $\mathbf{F}_{ij}^*$ is defined in \eqref{def.Fstar} and $\pmb{\gamma}_{ij}^{*} $ is defined in the beginning of Appendix A. Also, as shown in Remark \ref{RM3},

\begin{eqnarray*}
   \max_{i,j}\frac{1}{\sqrt{T}}\|(\mathbf{F}_{ij}^\dag\mathbf{F}_{ij}^*) - (\mathbf{0}, \mathbf{F}_i^\circ,  \mathbf{F}_j^\bullet)\|=o_P(1).
\end{eqnarray*}

We now start our investigation, and consider  

\begin{eqnarray*}
Q(\mathbf{b}_{\centerdot \centerdot}  ,\mathbf{C}_{\centerdot \centerdot}) -Q (\pmb{\beta}_{\centerdot\centerdot} , \mathbf{F}_{\centerdot \centerdot} ),
\end{eqnarray*}
where the definition of $\mathbf{F}_{\centerdot \centerdot}$ is obvious. Note that the terms   
\begin{eqnarray*}
\frac{1}{LNT}\sum_{i,j}\pmb{\mathcal{E}}_{ij\centerdot}^\top \pmb{\mathcal{E}}_{ij\centerdot} ,\quad \frac{1}{LNT}\sum_{i,j}\pmb{\gamma}_{ij}^{\circ\top}\mathbf{F}_{i}^{\circ\top} \mathbf{F}_{i}^{\circ }\pmb{\gamma}_{ij}^{\circ},\quad \frac{1}{LNT}\sum_{i,j}\pmb{\gamma}_{ij}^{\bullet\top}\mathbf{F}_{j}^{\bullet\top} \mathbf{F}_{j}^{\bullet}\pmb{\gamma}_{ij}^{\bullet}
\end{eqnarray*}
vanish automatically when taking difference. Thus, using Lemma \ref{LM1}, we immediately obtain that

\begin{eqnarray}\label{main.eq1}
&&Q(\mathbf{b}_{\centerdot \centerdot}  ,\mathbf{C}_{\centerdot \centerdot}) -Q (\pmb{\beta}_{\centerdot\centerdot} , \mathbf{F}_{\centerdot \centerdot})\nonumber \\
&=&\frac{1}{LNT}\sum_{i,j} (\pmb{\beta}_{ij} -\mathbf{b}_{ij})^\top\mathbf{X}_{ij\centerdot}^\top \mathbf{C}_{ij}^\dag \mathbf{X}_{ij\centerdot}(\pmb{\beta}_{ij} -\mathbf{b}_{ij})\nonumber\\
&&+\frac{1}{LNT}\sum_{i,j}\pmb{\gamma}_{ij}^{\top}\mathbf{F}^\top  \mathbf{C}_{ij}^\dag \mathbf{F} \pmb{\gamma}_{ij} \nonumber \\
&& +\frac{2}{LNT}\sum_{i,j}(\pmb{\beta}_{ij} -\mathbf{b}_{ij})^\top\mathbf{X}_{ij\centerdot }^\top \mathbf{C}_{ij}^\dag\mathbf{F}  \pmb{\gamma}_{ij}\nonumber \\
&& +\frac{2}{LNT}\sum_{i,j}(\pmb{\beta}_{ij} -\mathbf{b}_{ij})^\top\mathbf{X}_{ij\centerdot }^\top \mathbf{C}_{ij}^\dag( \mathbf{F}_i^\circ,  \mathbf{F}_j^\bullet) (\pmb{\gamma}_{ij}^{\circ\top},\pmb{\gamma}_{ij}^{\bullet\top})^\top\nonumber \\
&& +\frac{2}{LNT}\sum_{i,j}(\pmb{\beta}_{ij} -\mathbf{b}_{ij})^\top\mathbf{X}_{ij\centerdot }^\top \mathbf{C}_{ij}^\dag \pmb{\mathcal{E}}_{ij\centerdot} \notag \\
&&+\frac{1}{LNT}\sum_{i,j}\pmb{\gamma}_{ij}^{\circ\top}\mathbf{F}_{i}^{\circ\top}\mathbf{M}_{\mathbf{C}_i^\circ} \mathbf{F}_{i}^{\circ }\pmb{\gamma}_{ij}^{\circ}\notag \\
&&+\frac{1}{LNT}\sum_{i,j}\pmb{\gamma}_{ij}^{\bullet\top}\mathbf{F}_{j}^{\bullet\top}\mathbf{M}_{\mathbf{C}_j^\bullet} \mathbf{F}_{j}^{\bullet}\pmb{\gamma}_{ij}^{\bullet}+o_P(1).
\end{eqnarray}

We now focus on the right hand side of \eqref{main.eq1}, and consider the following two cases:

\begin{eqnarray*}
\text{Case 1:}\   \frac{1}{\sqrt{LN}}\| \pmb{\beta}_{ \centerdot \centerdot} -\mathbf{b}_{ \centerdot \centerdot}\|\le c \quad \text{and}\quad \text{Case 2:} \  \frac{1}{\sqrt{LN}}\| \pmb{\beta}_{ \centerdot \centerdot} -\mathbf{b}_{ \centerdot \centerdot}\|> c ,
\end{eqnarray*}
where $c $ is a large positive constant. Note that for Case 1, using Lemma \ref{LM1} and Assumption \ref{AS1}, equation \eqref{main.eq1} can further be simplified as follows:

\begin{eqnarray}\label{EQA.2}
&&Q(\mathbf{b}_{\centerdot \centerdot}  ,\mathbf{C}_{\centerdot \centerdot}) -Q (\pmb{\beta}_{\centerdot\centerdot} , \mathbf{F}_{\centerdot \centerdot})\nonumber \\
&=&\frac{1}{LNT}\sum_{i,j} (\pmb{\beta}_{ij} -\mathbf{b}_{ij})^\top\mathbf{X}_{ij\centerdot}^\top \mathbf{C}_{ij}^\dag \mathbf{X}_{ij\centerdot}(\pmb{\beta}_{ij} -\mathbf{b}_{ij})\nonumber\\
&&+\frac{1}{LNT}\sum_{i,j}\pmb{\gamma}_{ij}^{\top}\mathbf{F}^\top  \mathbf{C}_{ij}^\dag \mathbf{F} \pmb{\gamma}_{ij} \nonumber \\
&& +\frac{2}{LNT}\sum_{i,j}(\pmb{\beta}_{ij} -\mathbf{b}_{ij})^\top\mathbf{X}_{ij\centerdot }^\top \mathbf{C}_{ij}^\dag\mathbf{F}  \pmb{\gamma}_{ij}  \notag \\
&&+\frac{1}{LNT}\sum_{i,j}\pmb{\gamma}_{ij}^{\circ\top}\mathbf{F}_{i}^{\circ\top}\mathbf{M}_{\mathbf{C}_i^\circ} \mathbf{F}_{i}^{\circ }\pmb{\gamma}_{ij}^{\circ}\notag \\
&&+\frac{1}{LNT}\sum_{i,j}\pmb{\gamma}_{ij}^{\bullet\top}\mathbf{F}_{j}^{\bullet\top}\mathbf{M}_{\mathbf{C}_j^\bullet} \mathbf{F}_{j}^{\bullet}\pmb{\gamma}_{ij}^{\bullet}+o_P(1) \nonumber \\
&=&\frac{1}{LNT}\sum_{i,j}(\pmb{\beta}_{ij} -\mathbf{b}_{ij})^\top \mathbf{D}_{ij,1}(\pmb{\beta}_{ij} -\mathbf{b}_{ij}) \notag \\
&&+\frac{1}{LNT}\sum_{i,j}\pmb{\eta}_{ij}^\top [E[\pmb{\gamma}_{ij}^*\pmb{\gamma}_{ij}^{*\top}] \otimes \mathbf{I}_T]  \pmb{\eta}_{ij}  \nonumber \\
&&+\frac{2}{LNT}\sum_{i,j}(\pmb{\beta}_{ij} -\mathbf{b}_{ij})^\top \mathbf{D}_{ij,2}^\top \pmb{\eta}_{ij} \nonumber \\
&&+\frac{1}{LNT}\sum_{i,j}\pmb{\gamma}_{ij}^{\circ\top}\mathbf{F}_{i}^{\circ\top}\mathbf{M}_{\mathbf{C}_i^\circ} \mathbf{F}_{i}^{\circ }\pmb{\gamma}_{ij}^{\circ}\notag \\
&&+\frac{1}{LNT}\sum_{i,j}\pmb{\gamma}_{ij}^{\bullet\top}\mathbf{F}_{j}^{\bullet\top}\mathbf{M}_{\mathbf{C}_j^\bullet} \mathbf{F}_{j}^{\bullet}\pmb{\gamma}_{ij}^{\bullet}+o_P(1),
\end{eqnarray}
where $\pmb{\eta}_{ij}=\text{vec}(\mathbf{C}_{ij}^\dag \mathbf{F} )$, and the second equality follows from Assumption \ref{AS1}.2. Following the development on page 1265 of \cite{Bai} and using \eqref{EQA.2} and Assumption \ref{AS1}.2, it is easy to see that $ \frac{1}{\sqrt{LN}}\| \pmb{\beta}_{ \centerdot \centerdot} -\widehat{\mathbf{b}}_{ \centerdot \centerdot}\|=o_P(1)$ if we show that  Case 2 does not hold, which is exactly what we are about to do.

\medskip

For Case 2, we write \eqref{main.eq1} as follows:

\begin{eqnarray}\label{EQA.3}
&&Q(\mathbf{b}_{\centerdot \centerdot}  ,\mathbf{C}_{\centerdot \centerdot}) -Q (\pmb{\beta}_{\centerdot\centerdot} , \mathbf{F}_{\centerdot \centerdot})\nonumber \\
&=& \frac{1}{LNT}\sum_{i,j}[(\pmb{\beta}_{ij} -\mathbf{b}_{ij})^\top \mathbf{D}_{ij} (\pmb{\beta}_{ij} -\mathbf{b}_{ij}) +  \pmb{\theta}_{ij}^\top \mathbf{B}_{ij}\pmb{\theta}_{ij} ] \nonumber \\
&&+\frac{2}{LNT}\sum_{i,j}(\pmb{\beta}_{ij} -\mathbf{b}_{ij})^\top\mathbf{X}_{ij\centerdot }^\top \mathbf{C}_{ij}^\dag( \mathbf{F}_i^\circ,  \mathbf{F}_j^\bullet) (\pmb{\gamma}_{ij}^{\circ\top},\pmb{\gamma}_{ij}^{\bullet\top})^\top \notag \\
&& +\frac{2}{LNT}\sum_{i,j}(\pmb{\beta}_{ij} -\mathbf{b}_{ij})^\top\mathbf{X}_{ij\centerdot }^\top \mathbf{C}_{ij}^\dag\pmb{\mathcal{E}}_{ij\centerdot}  \notag \\
&&+\frac{1}{LNT}\sum_{i,j}\pmb{\gamma}_{ij}^{\circ\top}\mathbf{F}_{i}^{\circ\top}\mathbf{M}_{\mathbf{C}_i^\circ} \mathbf{F}_{i}^{\circ }\pmb{\gamma}_{ij}^{\circ}\notag \\
&&+\frac{1}{LNT}\sum_{i,j}\pmb{\gamma}_{ij}^{\bullet\top}\mathbf{F}_{j}^{\bullet\top}\mathbf{M}_{\mathbf{C}_j^\bullet} \mathbf{F}_{j}^{\bullet}\pmb{\gamma}_{ij}^{\bullet}+o_P(1)\notag \\
&\ge &a_0 c^2 +\frac{1}{LN}\sum_{i,j} \pmb{\theta}_{ij}^\top \mathbf{B}_{ij}\pmb{\theta}_{ij} \notag \\
&&+\frac{2}{LNT}\sum_{i,j}(\pmb{\beta}_{ij} -\mathbf{b}_{ij})^\top\mathbf{X}_{ij\centerdot }^\top \mathbf{C}_{ij}^\dag( \mathbf{F}_i^\circ,  \mathbf{F}_j^\bullet) (\pmb{\gamma}_{ij}^{\circ\top},\pmb{\gamma}_{ij}^{\bullet\top})^\top \notag\\
&&+\frac{2}{LNT}\sum_{i,j}(\pmb{\beta}_{ij} -\mathbf{b}_{ij})^\top\mathbf{X}_{ij\centerdot }^\top\mathbf{C}_{ij}^\dag\pmb{\mathcal{E}}_{ij\centerdot} \notag \\
&&+\frac{1}{LNT}\sum_{i,j}\pmb{\gamma}_{ij}^{\circ\top}\mathbf{F}_{i}^{\circ\top}\mathbf{M}_{\mathbf{C}_i^\circ} \mathbf{F}_{i}^{\circ }\pmb{\gamma}_{ij}^{\circ}\notag \\
&&+\frac{1}{LNT}\sum_{i,j}\pmb{\gamma}_{ij}^{\bullet\top}\mathbf{F}_{j}^{\bullet\top}\mathbf{M}_{\mathbf{C}_j^\bullet} \mathbf{F}_{j}^{\bullet}\pmb{\gamma}_{ij}^{\bullet}+o_P(1) ,
\end{eqnarray}
where $\pmb{\theta}_{ij}$  and $\mathbf{B}_{ij}$ are defined in the same fashion as that on  page 1265 of \cite{Bai}, $a_0$ is a positive constant by Assumption \ref{AS1}. Apparently,  Case 2 cannot hold by comparing the right hand sides of \eqref{main.eq1} and \eqref{EQA.3}. The proof of the first result is now completed. 

}
\medskip

(2). By \eqref{eq4}, we write

\begin{eqnarray}\label{EQA.4}
&&\widehat{\mathbf{C}} \widehat{\mathbf{V}} = \frac{1}{LNT}\sum_{i,j}\mathbf{X}_{ij\centerdot} (\pmb{\beta}_{ij}-\widehat{\mathbf{b}}_{ij})(\pmb{\beta}_{ij}-\widehat{\mathbf{b}}_{ij})^\top  \mathbf{X}_{ij\centerdot}^\top  \widehat{\mathbf{C}} \nonumber \\
&&+\frac{1}{LNT}\sum_{i,j}\mathbf{X}_{ij\centerdot}(\pmb{\beta}_{ij}-\widehat{\mathbf{b}}_{ij}) \pmb{\gamma}_{ij}^{\top} \mathbf{F}^{\top} \widehat{\mathbf{C}}   +  \frac{1}{LNT}\sum_{i,j} \mathbf{F}  \pmb{\gamma}_{ij} (\pmb{\beta}_{ij}-\widehat{\mathbf{b}}_{ij})^\top \mathbf{X}_{ij\centerdot}^\top \widehat{\mathbf{C}}  \nonumber\\
&&+\frac{1}{LNT}\sum_{i,j}\mathbf{X}_{ij\centerdot} (\pmb{\beta}_{ij}-\widehat{\mathbf{b}}_{ij})  \pmb{\gamma}_{ij}^{\circ\top}  \mathbf{F}_i^{\circ\top} \widehat{\mathbf{C}}    +  \frac{1}{LNT}\sum_{i,j} \mathbf{F}_i^\circ \pmb{\gamma}_{ij}^\circ  (\pmb{\beta}_{ij}-\widehat{\mathbf{b}}_{ij})^\top \mathbf{X}_{ij\centerdot}^\top \widehat{\mathbf{C}}  \nonumber\\
&&+\frac{1}{LNT}\sum_{i,j}\mathbf{X}_{ij\centerdot} (\pmb{\beta}_{ij}-\widehat{\mathbf{b}}_{ij}) \pmb{\gamma}_{ij}^{\bullet\top} \mathbf{F}_j^{\bullet\top} \widehat{\mathbf{C}}    +  \frac{1}{LNT}\sum_{i,j} \mathbf{F}_j^\bullet \pmb{\gamma}_{ij}^\bullet  (\pmb{\beta}_{ij}-\widehat{\mathbf{b}}_{ij})^\top \mathbf{X}_{ij\centerdot}^\top \widehat{\mathbf{C}}  \nonumber\\
&&+ \frac{1}{LNT}\sum_{i,j} \mathbf{X}_{ij\centerdot}(\pmb{\beta}_{ij}-\widehat{\mathbf{b}}_{ij})\pmb{\mathcal{E}}_{ij\centerdot}^\top \widehat{\mathbf{C}}  + \frac{1}{LNT}\sum_{i,j} \pmb{\mathcal{E}}_{ij\centerdot}(\pmb{\beta}_{ij}-\widehat{\mathbf{b}}_{ij})^\top \mathbf{X}_{ij\centerdot}^\top  \widehat{\mathbf{C}}  \nonumber\\
&&+ \frac{1}{LNT}\sum_{i,j} \mathbf{F} \pmb{\gamma}_{ij} \pmb{\gamma}_{ij}^{\top}  \mathbf{F}^{\top} \widehat{\mathbf{C}}+\frac{1}{LNT}\sum_{i,j} \mathbf{F} \pmb{\gamma}_{ij} \pmb{\gamma}_{ij}^{\circ\top} \mathbf{F}_i^{\circ\top} \widehat{\mathbf{C}}\nonumber\\
&&+ \frac{1}{LNT}\sum_{i,j}\mathbf{F}_i^\circ \pmb{\gamma}_{ij}^\circ \pmb{\gamma}_{ij}^\top \mathbf{F}^\top \widehat{\mathbf{C}} +\frac{1}{LNT}\sum_{i,j} \mathbf{F} \pmb{\gamma}_{ij} \pmb{\gamma}_{ij}^{\bullet\top}   \mathbf{F}_j^{\bullet\top} \widehat{\mathbf{C}} \nonumber \\
&&+\frac{1}{LNT}\sum_{i,j} \mathbf{F}_j^\bullet \pmb{\gamma}_{ij}^\bullet \pmb{\gamma}_{ij}^{\top} \mathbf{F}^\top  \widehat{\mathbf{C}} + \frac{1}{LNT}\sum_{i,j} \mathbf{F} \pmb{\gamma}_{ij} \pmb{\mathcal{E}}_{ij\centerdot}^\top \widehat{\mathbf{C}} \nonumber \\
&&+ \frac{1}{LNT}\sum_{i,j} \pmb{\mathcal{E}}_{ij\centerdot}\pmb{\gamma}_{ij} ^\top\mathbf{F}^\top  \widehat{\mathbf{C}}+\frac{1}{LNT}\sum_{i,j} \mathbf{F}_i^\circ  \pmb{\gamma}_{ij}^\circ \pmb{\gamma}_{ij}^{\circ\top}   \mathbf{F}_i^{\circ\top} \widehat{\mathbf{C}}  \nonumber \\
&& +\frac{1}{LNT}\sum_{i,j} \mathbf{F}_i^\circ  \pmb{\gamma}_{ij}^\circ  \pmb{\gamma}_{ij}^{\bullet\top} \mathbf{F}_j^{\bullet\top} \widehat{\mathbf{C}}+\frac{1}{LNT}\sum_{i,j} \mathbf{F}_j^\bullet \pmb{\gamma}_{ij}^\bullet  \pmb{\gamma}_{ij}^{\circ\top} \mathbf{F}_i^{\circ\top} \widehat{\mathbf{C}}  \nonumber\\
&& + \frac{1}{LNT}\sum_{i,j} \mathbf{F}_i^\circ  \pmb{\gamma}_{ij}^\circ  \pmb{\mathcal{E}}_{ij\centerdot}^\top   \widehat{\mathbf{C}} + \frac{1}{LNT}\sum_{i,j}\pmb{\mathcal{E}}_{ij\centerdot} \pmb{\gamma}_{ij}^{\circ\top}   \mathbf{F}_i^{\circ\top} \widehat{\mathbf{C}}  \nonumber \\
&&+\frac{1}{LNT}\sum_{i,j} \mathbf{F}_j^\bullet  \pmb{\gamma}_{ij}^\bullet  \pmb{\gamma}_{ij}^{\bullet\top}  \mathbf{F}_j^{\bullet\top} \widehat{\mathbf{C}}+ \frac{1}{LNT}\sum_{i,j} \mathbf{F}_j^\bullet \pmb{\gamma}_{ij}^\bullet  \pmb{\mathcal{E}}_{ij\centerdot}^\top \widehat{\mathbf{C}}  \nonumber \\
&& +\frac{1}{LNT}\sum_{i,j} \pmb{\mathcal{E}}_{ij\centerdot}  \pmb{\gamma}_{ij}^{\bullet\top}  \mathbf{F}_j^{\bullet\top} \widehat{\mathbf{C}} + \frac{1}{LNT}\sum_{i,j} \pmb{\mathcal{E}}_{ij\centerdot}\pmb{\mathcal{E}}_{ij\centerdot}^\top \widehat{\mathbf{C}}  \nonumber \\
&\coloneqq &\mathbf{J}_{1}+\cdots + \mathbf{J}_{25},
\end{eqnarray}
where the definitions of $ \mathbf{J}_{1}$ to $ \mathbf{J}_{25}$ are self evident.

For $ \mathbf{J}_{1}$, write

\begin{eqnarray*} 
\frac{1}{\sqrt{T}} \| \mathbf{J}_1 \|_2 &=&\frac{1}{\sqrt{T}}  \left\|  \frac{1}{LNT}\sum_{i,j}\mathbf{X}_{ij\centerdot} (\pmb{\beta}_{ij}-\widehat{\mathbf{b}}_{ij})(\pmb{\beta}_{ij}-\widehat{\mathbf{b}}_{ij})^\top  \mathbf{X}_{ij\centerdot}^\top  \widehat{\mathbf{C}}  \right\|_2 \nonumber \\
&\le &O(1)  \frac{1}{LNT}\sum_{i,j}\|\mathbf{X}_{ij\centerdot}(\pmb{\beta}_{ij}-\widehat{\mathbf{b}}_{ij})\| ^2\nonumber \\
&\le &O_P\left( \frac{1}{LN}\|\pmb{\beta}_{\centerdot \centerdot}-\widehat{\mathbf{b}}_{\centerdot\centerdot}\| ^2 \right),
\end{eqnarray*}
where the first inequality follows from the triangle inequality and the fact that $\frac{1}{T} \widehat{\mathbf{C}}^\top  \widehat{\mathbf{C}}= \mathbf{I}_\ell$, and the second inequality follows from Assumption \ref{AS1}.4.

For $ \mathbf{J}_2$, write

\begin{eqnarray*}
\frac{1}{\sqrt{T}} \| \mathbf{J}_2 \|_2 &=&\frac{1}{\sqrt{T}} \left\|\frac{1}{LNT}\sum_{i,j}\mathbf{X}_{ij\centerdot}(\pmb{\beta}_{ij}-\widehat{\mathbf{b}}_{ij}) \pmb{\gamma}_{ij}^{\top} \mathbf{F}^{\top} \widehat{\mathbf{C}}    \right\|_2\nonumber \\
&\le &O(1) \left\|\frac{1}{LNT}\sum_{i,j}\mathbf{X}_{ij\centerdot}(\pmb{\beta}_{ij}-\widehat{\mathbf{b}}_{ij}) \pmb{\gamma}_{ij}^{\top} \mathbf{F}^{\top} \right\|_2 \nonumber \\
&\le &O(1) \left\{\frac{1}{LNT}\sum_{i,j}\|\mathbf{X}_{ij\centerdot}  (\pmb{\beta}_{ij}-\widehat{\mathbf{b}}_{ij} )\| ^2 \right\}^{1/2}\left\{\frac{1}{LNT}\sum_{i,j}\| \mathbf{F} \pmb{\gamma}_{ij}\| ^2 \right\}^{1/2}\nonumber \\
&\le &O_P\left( \frac{1}{\sqrt{LN}}\|\pmb{\beta}_{\centerdot \centerdot}-\widehat{\mathbf{b}}_{\centerdot\centerdot}\|  \right),
\end{eqnarray*}
where the second inequality follows from Cauchy-Schwarz inequality, and third inequality follows from the development for $ \mathbf{J}_1$ and Assumptions \ref{AS1}.2-\ref{AS1}.4. Similarly, we obtain that

\begin{eqnarray*}
\frac{1}{\sqrt{T}} \| \mathbf{J}_3 \| =O_P\left( \frac{1}{\sqrt{LN}}\|\pmb{\beta}_{\centerdot \centerdot}-\widehat{\mathbf{b}}_{\centerdot\centerdot}\|  \right).
\end{eqnarray*}

For $ \mathbf{J}_4$, write

\begin{eqnarray*}
\frac{1}{\sqrt{T}} \| \mathbf{J}_4 \|_2 &=&\frac{1}{\sqrt{T}} \left\|\frac{1}{LNT}\sum_{i,j}\mathbf{X}_{ij\centerdot}(\pmb{\beta}_{ij}-\widehat{\mathbf{b}}_{ij})  \pmb{\gamma}_{ij}^{\circ\top}  \mathbf{F}_i^{\circ\top} \widehat{\mathbf{C}} \right\|_2\nonumber \\
&\le &O(1)\frac{1}{LNT} \left\|\sum_{j=1}^N\mathbf{X}_{\centerdot j \centerdot } \diag\{ \pmb{\beta}_{\centerdot j}^\top-\widehat{\mathbf{b}}_{\centerdot j}^\top\} \diag\{ \pmb{\Gamma}_{\centerdot j}^{\circ\top}\}^\top \mathbf{F}^{\circ\top}\right\|_2\nonumber \\
&\le &O(1)\frac{1}{LNT}\|\mathbf{F}^{\circ}\|_2\cdot \left\{\sum_{j=1}^N\|\mathbf{X}_{\centerdot j \centerdot } \diag\{ \pmb{\beta}_{\centerdot j}^\top-\widehat{\mathbf{b}}_{\centerdot j}^\top\}\|_2^2  \right\}^{1/2}\left\{\sum_{j=1}^N \| \diag\{ \pmb{\Gamma}_{\centerdot j}^{\circ \top}\} \|_2^2 \right\}^{1/2}\nonumber \\
&\le &O_P(1) \frac{1}{LNT}\sqrt{L\vee T}\cdot  \sqrt{T}\|\pmb{\beta}_{\centerdot \centerdot}-\widehat{\mathbf{b}}_{\centerdot\centerdot}\|\cdot \sqrt{N} \nonumber \\
&\le &O_P\left( \frac{1}{\sqrt{L\wedge T}} \cdot \frac{1}{\sqrt{LN}}\|\pmb{\beta}_{\centerdot \centerdot}-\widehat{\mathbf{b}}_{\centerdot \centerdot}\|\right)
\end{eqnarray*}
where the third inequality follows from the development of $\mathbf{J}_1$ and $\frac{1}{N}\sum_{j=1}^N\|\diag\{ \pmb{\Gamma}_{\centerdot j}^{\circ \top}\} \|_2^2 =O_P(1)$ as shown in the proof of Lemma \ref{LM1}.

Similar to the development of $ \mathbf{J}_4$, we obtain that

\begin{eqnarray*}
\frac{1}{\sqrt{T}} \| \mathbf{J}_5 \|_2  =O_P\left( \frac{1}{\sqrt{L\wedge T}} \cdot \frac{1}{\sqrt{LN}}\|\pmb{\beta}_{\centerdot \centerdot}-\widehat{\mathbf{b}}_{\centerdot \centerdot}\|\right).
\end{eqnarray*}

Similar to $ \mathbf{J}_4$ and $\mathbf{J}_5$ and by noting $i$ and $j$ being symmetric, we obtain that

\begin{eqnarray*}
\frac{1}{\sqrt{T}} \| \mathbf{J}_6 \|_2  &=&O_P\left( \frac{1}{\sqrt{N\wedge T}} \cdot \frac{1}{\sqrt{LN}}\|\pmb{\beta}_{\centerdot \centerdot}-\widehat{\mathbf{b}}_{\centerdot \centerdot}\|\right),\nonumber \\
\frac{1}{\sqrt{T}} \| \mathbf{J}_7 \|_2  &=&O_P\left( \frac{1}{\sqrt{N\wedge T}} \cdot \frac{1}{\sqrt{LN}}\|\pmb{\beta}_{\centerdot \centerdot}-\widehat{\mathbf{b}}_{\centerdot \centerdot}\|\right).
\end{eqnarray*}

Similar to the development of $\mathbf{J}_2$, we can obtain that

\begin{eqnarray*}
\frac{1}{\sqrt{T}} \|\mathbf{J}_8 \|_2&=& O_P\left( \frac{1}{\sqrt{LN}}\|\pmb{\beta}_{\centerdot\centerdot}-\widehat{\mathbf{b}}_{\centerdot\centerdot}\| \right),\nonumber \\
\frac{1}{\sqrt{T}} \|\mathbf{J}_9 \|_2&=& O_P\left( \frac{1}{\sqrt{LN}}\|\pmb{\beta}_{\centerdot\centerdot}-\widehat{\mathbf{b}}_{\centerdot\centerdot}\| \right).
\end{eqnarray*}

For $\mathbf{J}_{11}$, write

\begin{eqnarray*}
\frac{1}{\sqrt{T}} \|\mathbf{J}_{11} \|_2 &=& \frac{1}{\sqrt{T}} \left\|\frac{1}{LNT}\sum_{i,j} \mathbf{F} \pmb{\gamma}_{ij} \pmb{\gamma}_{ij}^{\circ\top}   \mathbf{F}_i^{\circ\top} \widehat{\mathbf{C}} \right\|_2\nonumber \\
&\le & \left\{\frac{1}{LNT}\sum_{i,j}\|  \mathbf{F} \pmb{\gamma}_{ij} \| ^2  \right\}^{1/2} \left\{\frac{1}{LNT}\sum_{i,j} \pmb{\gamma}_{ij}^{\circ\top}  \mathbf{F}_i^{\circ\top}   \mathbf{P}_{\widehat{\mathbf{C}}} \mathbf{F}_i^\circ  \pmb{\gamma}_{ij}^\circ \right\}^{1/2}\nonumber \\
&\le &O_P(1) \frac{1}{\sqrt{L\wedge T}} ,
\end{eqnarray*}
where the second inequality follows from Assumptions \ref{AS1}.2 and \ref{AS1}.3, and Lemma \ref{LM1}.

For $\mathbf{J}_{12}$, write

\begin{eqnarray}\label{mge_e19}
\frac{1}{\sqrt{T}} \|\mathbf{J}_{12} \|_2&=&   \frac{1}{\sqrt{T}} \left\|\frac{1}{LNT}\sum_{i,j} \mathbf{F}_i^\circ \pmb{\gamma}_{ij}^\circ  \pmb{\gamma}_{ij}^{\top} \mathbf{F}^{\top} \widehat{\mathbf{C}} \right\|_2\nonumber \\
&\le &O(1) \frac{1}{LNT}\sum_{j=1}^N\left\| \mathbf{F}^\circ \diag\{ \pmb{\Gamma}_{\centerdot j}^{^\circ\top} \} \pmb{\Gamma}_{\centerdot j} \mathbf{F}^{\top} \right\|_2\nonumber \\
&\le &O(1) \frac{1}{LNT} \| \mathbf{F}^\circ \|_2\left\{\sum_{j=1}^N\| \diag\{ \pmb{\Gamma}_{\centerdot j}^{^\circ\top} \}\|_2^2\right\}^{1/2}\left\{\sum_{j=1}^N\| \pmb{\Gamma}_{\centerdot j}\|_2^2\right\}^{1/2} \|\mathbf{F} \|_2\nonumber \\
&\le &O_P(1)\frac{1}{LT}\sqrt{L\vee T}\cdot\sqrt{L}\cdot\sqrt{T}\nonumber \\
&\le &O_P\left( \frac{1}{\sqrt{L\wedge T}}\right),
\end{eqnarray}
where the third inequality follows from Assumptions \ref{AS1}.2 and \ref{AS1}.3 and $\frac{1}{N}\sum_{j=1}^N\|\diag\{ \pmb{\Gamma}_{\centerdot j}^{\circ \top}\} \|_2^2 =O_P(1)$ as shown in the proof of Lemma \ref{LM1}.

By noting $i$ and $j$ being symmetric and following the development of $\mathbf{J}_{11}$ and $\mathbf{J}_{12}$, we obtain that

\begin{eqnarray}\label{mge_e20}
\frac{1}{\sqrt{T}} \|\mathbf{J}_{13} \|_2 &=&O_P\left( \frac{1}{\sqrt{N\wedge T}}\right),\nonumber\\
\frac{1}{\sqrt{T}} \|\mathbf{J}_{14} \|_2 &=&O_P\left( \frac{1}{\sqrt{N\wedge T}}\right).
\end{eqnarray}

For $\mathbf{J}_{15}$, write

\begin{eqnarray*}
\frac{1}{\sqrt{T}} \|\mathbf{J}_{15} \|_2 &=&  \frac{1}{\sqrt{T}} \left\|\frac{1}{LNT}\sum_{i,j}\mathbf{F} \pmb{\gamma}_{ij} \pmb{\mathcal{E}}_{ij\centerdot}^\top   \widehat{\mathbf{C}} \right\|_2\nonumber \\
&\le &  O(1)\frac{1}{LNT}  \| \mathbf{F} \pmb{\Gamma}_{\centerdot\centerdot}^{\top} \pmb{\mathcal{E}}_{\centerdot\centerdot\centerdot} \|_2\nonumber \\
&=&O_P\left(\frac{ 1}{\sqrt{LN\wedge T}} \right),
\end{eqnarray*}
where the last step follows from Assumptions \ref{AS1}.1, \ref{AS1}.2 and \ref{AS1}.3. 

Similar to the development of $\mathbf{J}_{15}$, we obtain that  

\begin{eqnarray*}
\frac{1}{\sqrt{T}}  \|\mathbf{J}_{16} \|_2 =O_P\left(\frac{1}{\sqrt{LN\wedge T}} \right).
\end{eqnarray*}

For $\mathbf{J}_{17}$, write

\begin{eqnarray*}
\frac{1}{\sqrt{T}}  \|\mathbf{J}_{17} \|_2 &=& \frac{1}{\sqrt{T}} \left\|\frac{1}{LNT}\sum_{i,j} \mathbf{F}_i^\circ  \pmb{\gamma}_{ij}^\circ  \pmb{\gamma}_{ij}^{^\circ\top}  \mathbf{F}_i^{^\circ\top} \widehat{\mathbf{C}}\right\|_2=O_P\left(\frac{1}{L\wedge T} \right),
\end{eqnarray*}
where the second equality follows from Lemma \ref{LM1}.

For $\mathbf{J}_{18}$, write

\begin{eqnarray*}
\frac{1}{\sqrt{T}}  \|\mathbf{J}_{18} \|_2 &=& \frac{1}{\sqrt{T}} \left\|\frac{1}{LNT}\sum_{i,j}  \mathbf{F}_i^\circ  \pmb{\gamma}_{ij}^\circ  \pmb{\gamma}_{ij}^{\bullet\top} \mathbf{F}_j^{\bullet\top} \widehat{\mathbf{C}}\right\|_2\nonumber \\
&\le &O(1)  \frac{1}{LNT} \| \mathbf{F}^\circ\cdot  \pmb{\Gamma}^{\circ\bullet}\cdot \mathbf{F}^{\bullet\top} \|_2\nonumber \\
&=&O_P(1)\frac{1}{\sqrt{(L\wedge T)(N\wedge T)}},
\end{eqnarray*}
where $ \pmb{\Gamma}^{\circ\bullet} =\{ \pmb{\gamma}_{ij}^\circ\pmb{\gamma}_{ij}^{\bullet\top}\}_{\|\pmb{\ell}^\circ\|_1\times\|\pmb{\ell}^\bullet\|_1}$, and the second equality follows from Assumption \ref{AS1}. 

Similar to the development of $\mathbf{J}_{18}$,

\begin{eqnarray*}
\frac{1}{\sqrt{T}}  \|\mathbf{J}_{19} \|_2=O_P(1)\frac{1}{\sqrt{(L\wedge T)(N\wedge T)}}.
\end{eqnarray*}

For $\mathbf{J}_{20}$, write

\begin{eqnarray*}
\frac{1}{\sqrt{T}} \|\mathbf{J}_{20} \|_2 &=&  \frac{1}{\sqrt{T}} \left\|\frac{1}{LNT}\sum_{i,j}\mathbf{F}_i^\circ  \pmb{\gamma}_{ij}^\circ \pmb{\mathcal{E}}_{ij\centerdot}^\top \widehat{\mathbf{C}}  \right\|_2 =O_P(1)\frac{1}{\sqrt{(L\wedge T)(LN\wedge T)}},
\end{eqnarray*}
where the last equality follows from \eqref{f0eprove}. 

Similar to the study of $\mathbf{J}_{20}$, we obtain that

\begin{eqnarray*}
\frac{1}{\sqrt{T}} \|\mathbf{J}_{21} \|_2 =O_P(1)\frac{1}{\sqrt{(L\wedge T)(LN\wedge T)}}.
\end{eqnarray*}

 Similar to $\mathbf{J}_{20}$ and $\mathbf{J}_{21}$ and by noting $i$ and $j$ being symmetric, we immediately obtain that

\begin{eqnarray*}
\frac{1}{\sqrt{T}} \|\mathbf{J}_{23} \|_2 &=&O_P(1)\frac{1}{\sqrt{(N\wedge T)(LN\wedge T)}},\nonumber \\
\frac{1}{\sqrt{T}} \|\mathbf{J}_{24} \|_2 &=&O_P(1)\frac{1}{\sqrt{(N\wedge T)(LN\wedge T)}}.
\end{eqnarray*}

Following the development of $\mathbf{J}_{17}$, we obtain that
 
 \begin{eqnarray*}
 \frac{1}{\sqrt{T}} \|\mathbf{J}_{22} \|_2=O_P\left(\frac{1}{N\wedge T} \right).
 \end{eqnarray*}

For $\mathbf{J}_{25}$, write

\begin{eqnarray*}
 \frac{1}{\sqrt{T}} \|\mathbf{J}_{25} \|_2 &=&  \frac{1}{\sqrt{T}}\left\| \frac{1}{LNT}\sum_{i,j} \pmb{\mathcal{E}}_{ij\centerdot} \pmb{\mathcal{E}}_{ij\centerdot}^\top \widehat{\mathbf{C}} \right\|_2 =O_P\left(\frac{1}{LN\wedge T} \right),
\end{eqnarray*}
where the last equality follows from \eqref{eprove}.

Thus, we can conclude that

\begin{eqnarray} \label{ff1}
\frac{1}{\sqrt{T}} \|\widehat{\mathbf{C}}  - \mathbf{F} \mathbf{H}   \|_2 &=&O_P\left( \frac{1}{\sqrt{LN}}\|\pmb{\beta}_{\centerdot \centerdot}-\widehat{\mathbf{b}}_{\centerdot\centerdot}\| +\frac{1}{\sqrt{L\wedge N\wedge T}}\right),
\end{eqnarray}
where $\mathbf{H} =\frac{1}{LNT} \pmb{\Gamma}_{\centerdot\centerdot}^{\top} \pmb{\Gamma}_{\centerdot\centerdot} \mathbf{F}^{\top} \widehat{\mathbf{C}} \widehat{\mathbf{V}}^{-1}$. Here, it is noteworthy that $\widehat{\mathbf{V}}$ converges to the eigenvalues of $\frac{1}{NT} \pmb{\Gamma}_{\centerdot\centerdot}^{\top} \pmb{\Gamma}_{\centerdot\centerdot} $ as we have $\frac{1}{T}\mathbf{F}^\top\mathbf{F}=\mathbf{I}_\ell+o_P(1)$ in Assumption \ref{AS1}.3. In connection with the fact that

\begin{eqnarray*}
\mathbf{P}_{\mathbf{F}} &=&(\mathbf{F}-\widehat{\mathbf{C}}\mathbf{H} ^{-1}+\widehat{\mathbf{C}}\mathbf{H} ^{-1})\nonumber\\
&&\cdot [(\mathbf{F}-\widehat{\mathbf{C}}\mathbf{H} ^{-1}+\widehat{\mathbf{C}}\mathbf{H} ^{-1})^\top (\mathbf{F}-\widehat{\mathbf{C}}\mathbf{H} ^{-1}+\widehat{\mathbf{C}}\mathbf{H} ^{-1})]^{-1} (\mathbf{F}-\widehat{\mathbf{C}}\mathbf{H} ^{-1}+\widehat{\mathbf{C}}\mathbf{H} ^{-1})^\top,
\end{eqnarray*}
the second result of Lemma \ref{M.LM1} follows immediately. \hspace*{\fill}{$\blacksquare$}

\bigskip

\noindent \textbf{Proof of Lemma \ref{M.LM2}:}

(1). Note that we can have the following expansion:

\begin{eqnarray*}
\widehat{\mathbf{b}}_{ij}-\pmb{\beta}_{ij}  &=& (\mathbf{X}_{ij\centerdot}^\top \widehat{\mathbf{C}}_{ij}^\dag \mathbf{X}_{ij\centerdot} )^{-1} \mathbf{X}_{ij\centerdot}^\top \widehat{\mathbf{C}}_{ij}^\dag \mathbf{F} \pmb{\gamma}_{ij}\nonumber \\
&&+(\mathbf{X}_{ij\centerdot}^\top\widehat{\mathbf{C}}_{ij}^\dag \mathbf{X}_{ij\centerdot} )^{-1} \mathbf{X}_{ij\centerdot}^\top \widehat{\mathbf{C}}_{ij}^\dag  \mathbf{F}_i^\circ \pmb{\gamma}_{ij}^\circ\nonumber \\
&&+(\mathbf{X}_{ij\centerdot}^\top \widehat{\mathbf{C}}_{ij}^\dag \mathbf{X}_{ij\centerdot} )^{-1} \mathbf{X}_{ij\centerdot}^\top \widehat{\mathbf{C}}_{ij}^\dag \mathbf{F}_j^\bullet \pmb{\gamma}_{ij}^\bullet\nonumber \\
&&+(\mathbf{X}_{ij\centerdot}^\top \widehat{\mathbf{C}}_{ij}^\dag \mathbf{X}_{ij\centerdot} )^{-1} \mathbf{X}_{ij\centerdot}^\top \widehat{\mathbf{C}}_{ij}^\dag \pmb{\mathcal{E}}_{ij\centerdot } .
\end{eqnarray*}

Our goal is to prove $\max_{i, j}\|\widehat{\mathbf{b}}_{ij}- \pmb{\beta}_{ij}\|=o_P(1).$ Suppose that there exits some $(i,j)$ such that $\|\widehat{\mathbf{b}}_{ij}- \pmb{\beta}_{ij}\|\ne o_P(1)$.

For the term $(\mathbf{X}_{ij\centerdot}^\top \widehat{\mathbf{C}}_{ij}^\dag \mathbf{X}_{ij\centerdot} )^{-1} \mathbf{X}_{ij\centerdot}^\top \widehat{\mathbf{C}}_{ij}^\dag \mathbf{F} \pmb{\gamma}_{ij}$, we write

\begin{eqnarray*}
&&\max_{i, j}\|(\mathbf{X}_{ij\centerdot}^\top \widehat{\mathbf{C}}_{ij}^\dag \mathbf{X}_{ij\centerdot} )^{-1} \mathbf{X}_{ij\centerdot}^\top \widehat{\mathbf{C}}_{ij}^\dag \mathbf{F} \pmb{\gamma}_{ij} \| \nonumber \\
&\le &\max_{i, j}\|(\mathbf{X}_{ij\centerdot}^\top \widehat{\mathbf{C}}_{ij}^\dag \mathbf{X}_{ij\centerdot} )^{-1} \mathbf{X}_{ij\centerdot}^\top \mathbf{P}_{\widehat{\mathbf{C}}_{i}^\circ} \mathbf{F} \pmb{\gamma}_{ij} \|\nonumber \\
&& +\max_{i, j}\|(\mathbf{X}_{ij\centerdot}^\top \widehat{\mathbf{C}}_{ij}^\dag \mathbf{X}_{ij\centerdot} )^{-1} \mathbf{X}_{ij\centerdot}^\top \mathbf{P}_{\widehat{\mathbf{C}}_{j}^\bullet} \mathbf{F} \pmb{\gamma}_{ij} \|+o_P(1)\nonumber \\
&\le &o_P(1),
\end{eqnarray*}
where the first inequality follows from $\frac{1}{\sqrt{T}}\|\mathbf{M}_{\widehat{\mathbf{C}}}\mathbf{F}\|=o_P(1)$ by the second result of Lemma \ref{M.LM1}, and the second result follows from the fourth condition of \eqref{conC} and \eqref{ff1}.

Next, suppose that $\mathbf{x}_{ijt}$ is a scalar for simplicity for now only. Then write

\begin{eqnarray*}
 \max_{i, j}\frac{1}{T} | \mathbf{X}_{ij\centerdot}^\top \widehat{\mathbf{C}}_{ij}^\dag \pmb{\mathcal{E}}_{ij\centerdot } |  &\le &\max_{i, j}\frac{1}{T} | \mathbf{X}_{ij\centerdot}^\top  \pmb{\mathcal{E}}_{ij\centerdot } |+\max_{i, j}\frac{1}{T} | \tr( (\mathbf{P}_{\widehat{\mathbf{C}}}+\mathbf{P}_{\widehat{\mathbf{C}}_i^\circ}  +\mathbf{P}_{\widehat{\mathbf{C}}_j^\bullet }) \pmb{\mathcal{E}}_{ij\centerdot } \mathbf{X}_{ij\centerdot}^\top)|\nonumber \\
  &\le &\max_{i, j}\frac{1}{T} | \mathbf{X}_{ij\centerdot}^\top  \pmb{\mathcal{E}}_{ij\centerdot } |+O(1)\max_{i, j}\frac{1}{T}\|\pmb{\mathcal{E}}_{ij\centerdot } \mathbf{X}_{ij\centerdot}^\top\|_2\nonumber \\
  &=&\max_{i, j}\frac{1}{T} | \mathbf{X}_{ij\centerdot}^\top  \pmb{\mathcal{E}}_{ij\centerdot } |+O(1)\max_{i, j}\frac{1}{T}\|\mathbf{X}_{ij\centerdot}^\top \pmb{\mathcal{E}}_{ij\centerdot } \|_2\nonumber \\
  &=&o_P(1),
\end{eqnarray*}
where the last equality follows from Lemma \ref{LemmaT2} and Assumption \ref{AS2} straightaway. Thus,

\begin{eqnarray*}
\| (\mathbf{X}_{ij\centerdot}^\top  \widehat{\mathbf{C}}_{ij}^\dag \mathbf{X}_{ij\centerdot} )^{-1} \mathbf{X}_{ij\centerdot}^\top  \widehat{\mathbf{C}}_{ij}^\dag \pmb{\mathcal{E}}_{ij\centerdot }\|=o_P(1).
\end{eqnarray*}

Therefore, if $\|\widehat{\mathbf{b}}_{ij}- \pmb{\beta}_{ij}\|\ne o_P(1)$, we conclude that

\begin{eqnarray*}
\|(\mathbf{X}_{ij\centerdot}^\top  \widehat{\mathbf{C}}_{ij}^\dag \mathbf{X}_{ij\centerdot} )^{-1} \mathbf{X}_{ij\centerdot}^\top  \widehat{\mathbf{C}}_{ij}^\dag ( \mathbf{F}_i^\circ \pmb{\gamma}_{ij}^\circ+ \mathbf{F}_j^\bullet \pmb{\gamma}_{ij}^\bullet)\|\ne o_P(1).
\end{eqnarray*}
Thus, it infers 

\begin{eqnarray*}
\frac{1}{\sqrt{T}}\| \widehat{\mathbf{C}}_{ij}^\dag ( \mathbf{F}_i^\circ ,\mathbf{F}_j^\bullet)\|\ne o_P(1).
\end{eqnarray*}
We now show it violates the first result of Lemma \ref{M.LM1} and \eqref{est1}. In view of Lemma \ref{M.LM1} and Assumption \ref{AS1}.4, we can always replace $\widehat{\mathbf{C}}_i^\circ$ and $\widehat{\mathbf{C}}_j^\bullet$ with $\mathbf{F}_i^\circ$ and $\mathbf{F}_j^\bullet$ to achieve

\begin{eqnarray*}
Q(\widehat{\mathbf{b}}_{\centerdot \centerdot}, \widehat{\mathbf{C}}_{\centerdot \centerdot|\mathbf{F}_i^\circ,\mathbf{F}_j^\bullet}^*)\le Q(\widehat{\mathbf{b}}_{\centerdot \centerdot}, \widehat{\mathbf{C}}_{\centerdot\centerdot}),
\end{eqnarray*}
where $\widehat{\mathbf{C}}_{\centerdot \centerdot|\mathbf{F}_i^\circ,\mathbf{F}_j^\bullet}^* $ is obtained by replacing $\widehat{\mathbf{C}}_i^\circ$ and $\widehat{\mathbf{C}}_j^\bullet$ with $\mathbf{F}_i^\circ$ and $\mathbf{F}_j^\bullet$. Thus, the result follows.

\medskip

(2)-(3). Note that by construction, $i$ and $j$ are symmetric, so we only need to prove one. Without loss of generality, we consider \textbf{Step 1}.d.  

For $\forall j\in [N]$, write

\begin{eqnarray}\label{Expan_J}
&&\widehat{\mathbf{C}}_j^\bullet \widehat{\mathbf{V}}_j^\bullet = \widehat{\pmb{\Sigma}}_j^\bullet \widehat{\mathbf{C}}_j^\bullet\nonumber \\
&=&  \frac{1}{LT}\sum_{i=1}^L \mathbf{M}_{\widehat{\mathbf{C}}}\mathbf{X}_{ij\centerdot} (\pmb{\beta}_{ij}-\widehat{\mathbf{b}}_{ij})(\pmb{\beta}_{ij}-\widehat{\mathbf{b}}_{ij})^\top  \mathbf{X}_{ij\centerdot}^\top \mathbf{M}_{\widehat{\mathbf{C}}} \widehat{\mathbf{C}}_j^\bullet \nonumber \\
&&+\frac{1}{LT}\sum_{i=1}^L\mathbf{M}_{\widehat{\mathbf{C}}}\mathbf{X}_{ij\centerdot}(\pmb{\beta}_{ij}-\widehat{\mathbf{b}}_{ij}) \pmb{\gamma}_{ij}^{\top} \mathbf{F}^{\top} \mathbf{M}_{\widehat{\mathbf{C}}} \widehat{\mathbf{C}}_j^\bullet +  \frac{1}{LT}\sum_{i=1}^L \mathbf{M}_{\widehat{\mathbf{C}}}\mathbf{F}  \pmb{\gamma}_{ij} (\pmb{\beta}_{ij}-\widehat{\mathbf{b}}_{ij})^\top \mathbf{X}_{ij\centerdot}^\top \mathbf{M}_{\widehat{\mathbf{C}}}  \widehat{\mathbf{C}}_j^\bullet \nonumber\\
&&+\frac{1}{LT}\sum_{i=1}^L \mathbf{M}_{\widehat{\mathbf{C}}}\mathbf{X}_{ij\centerdot} (\pmb{\beta}_{ij}-\widehat{\mathbf{b}}_{ij}) \pmb{\gamma}_{ij}^{\circ\top}   \mathbf{F}_i^{\circ\top}\mathbf{M}_{\widehat{\mathbf{C}}} \widehat{\mathbf{C}}_j^\bullet +  \frac{1}{LT}\sum_{i=1}^L\mathbf{M}_{\widehat{\mathbf{C}}}\mathbf{F}_i^\circ \pmb{\gamma}_{ij}^\circ  (\pmb{\beta}_{ij}-\widehat{\mathbf{b}}_{ij})^\top \mathbf{X}_{ij\centerdot}^\top \mathbf{M}_{\widehat{\mathbf{C}}}  \widehat{\mathbf{C}}_j^\bullet \nonumber\\
&&+\frac{1}{LT}\sum_{i=1}^L\mathbf{M}_{\widehat{\mathbf{C}}}\mathbf{X}_{ij\centerdot} (\pmb{\beta}_{ij}-\widehat{\mathbf{b}}_{ij}) \pmb{\gamma}_{ij}^{\bullet\top}   \mathbf{F}_j^{\bullet\top} \mathbf{M}_{\widehat{\mathbf{C}}} \widehat{\mathbf{C}}_j^\bullet  +  \frac{1}{LT}\sum_{i=1}^L\mathbf{M}_{\widehat{\mathbf{C}}} \mathbf{F}_j^\bullet \pmb{\gamma}_{ij}^\bullet  (\pmb{\beta}_{ij}-\widehat{\mathbf{b}}_{ij})^\top \mathbf{X}_{ij\centerdot}^\top \mathbf{M}_{\widehat{\mathbf{C}}} \widehat{\mathbf{C}}_j^\bullet \nonumber\\
&&+ \frac{1}{LT}\sum_{i=1}^L\mathbf{M}_{\widehat{\mathbf{C}}} \mathbf{X}_{ij\centerdot}(\pmb{\beta}_{ij}-\widehat{\mathbf{b}}_{ij})\pmb{\mathcal{E}}_{ij\centerdot}^\top \mathbf{M}_{\widehat{\mathbf{C}}} \widehat{\mathbf{C}}_j^\bullet + \frac{1}{LT}\sum_{i=1}^L\mathbf{M}_{\widehat{\mathbf{C}}} \pmb{\mathcal{E}}_{ij\centerdot}(\pmb{\beta}_{ij}-\widehat{\mathbf{b}}_{ij})^\top \mathbf{X}_{ij\centerdot}^\top \mathbf{M}_{\widehat{\mathbf{C}}} \widehat{\mathbf{C}}_j^\bullet \nonumber\\
&&+ \frac{1}{LT}\sum_{i=1}^L\mathbf{M}_{\widehat{\mathbf{C}}} \mathbf{F} \pmb{\gamma}_{ij} \pmb{\gamma}_{ij}^{\top}  \mathbf{F}^{\top} \mathbf{M}_{\widehat{\mathbf{C}}}\widehat{\mathbf{C}}_j^\bullet +\frac{1}{LT}\sum_{i=1}^L\mathbf{M}_{\widehat{\mathbf{C}}} \mathbf{F} \pmb{\gamma}_{ij} \pmb{\gamma}_{ij}^{\circ\top}\mathbf{F}_i^{\circ\top} \mathbf{M}_{\widehat{\mathbf{C}}} \widehat{\mathbf{C}}_j^\bullet\nonumber\\
&&+ \frac{1}{LT}\sum_{i=1}^L\mathbf{M}_{\widehat{\mathbf{C}}} \mathbf{F}_i^\circ \pmb{\gamma}_{ij}^\circ \pmb{\gamma}_{ij}^\top \mathbf{F}^\top \mathbf{M}_{\widehat{\mathbf{C}}} \widehat{\mathbf{C}}_j^\bullet +\frac{1}{LT}\sum_{i=1}^L\mathbf{M}_{\widehat{\mathbf{C}}} \mathbf{F} \pmb{\gamma}_{ij} \pmb{\gamma}_{ij}^{\bullet\top}  \mathbf{F}_j^{\bullet\top} \mathbf{M}_{\widehat{\mathbf{C}}}\widehat{\mathbf{C}}_j^\bullet\nonumber \\
&&+\frac{1}{LT}\sum_{i=1}^L\mathbf{M}_{\widehat{\mathbf{C}}} \mathbf{F}_j^\bullet \pmb{\gamma}_{ij}^\bullet \pmb{\gamma}_{ij}^{\top} \mathbf{F}^\top  \mathbf{M}_{\widehat{\mathbf{C}}}\widehat{\mathbf{C}}_j^\bullet + \frac{1}{LT}\sum_{i=1}^L\mathbf{M}_{\widehat{\mathbf{C}}} \mathbf{F} \pmb{\gamma}_{ij} \pmb{\mathcal{E}}_{ij\centerdot}^\top \mathbf{M}_{\widehat{\mathbf{C}}}\widehat{\mathbf{C}}_j^\bullet\nonumber \\
&&+ \frac{1}{LT}\sum_{i=1}^L\mathbf{M}_{\widehat{\mathbf{C}}} \pmb{\mathcal{E}}_{ij\centerdot}\pmb{\gamma}_{ij} ^\top\mathbf{F}^\top \mathbf{M}_{\widehat{\mathbf{C}}} \widehat{\mathbf{C}}_j^\bullet +\frac{1}{LT}\sum_{i=1}^L\mathbf{M}_{\widehat{\mathbf{C}}}  \mathbf{F}_i^\circ  \pmb{\gamma}_{ij}^\circ \pmb{\gamma}_{ij}^{\circ\top}  \mathbf{F}_i^{\circ\top} \mathbf{M}_{\widehat{\mathbf{C}}} \widehat{\mathbf{C}}_j^\bullet  \nonumber \\
&& +\frac{1}{LT}\sum_{i=1}^L\mathbf{M}_{\widehat{\mathbf{C}}} \mathbf{F}_i^\circ   \pmb{\gamma}_{ij}^\circ \pmb{\gamma}_{ij}^{\bullet\top} \mathbf{F}_j^{\bullet\top} \mathbf{M}_{\widehat{\mathbf{C}}}\widehat{\mathbf{C}}_j^\bullet +\frac{1}{LT}\sum_{i=1}^L\mathbf{M}_{\widehat{\mathbf{C}}} \mathbf{F}_j^\bullet  \pmb{\gamma}_{ij}^\bullet \pmb{\gamma}_{ij}^{\circ\top} \mathbf{F}_i^{\circ\top} \mathbf{M}_{\widehat{\mathbf{C}}}\widehat{\mathbf{C}}_j^\bullet \nonumber\\
&& + \frac{1}{LT}\sum_{i=1}^L \mathbf{M}_{\widehat{\mathbf{C}}} \mathbf{F}_i^\circ  \pmb{\gamma}_{ij}^\circ  \pmb{\mathcal{E}}_{ij\centerdot}^\top   \mathbf{M}_{\widehat{\mathbf{C}}}\widehat{\mathbf{C}}_j^\bullet + \frac{1}{LT}\sum_{i=1}^L\mathbf{M}_{\widehat{\mathbf{C}}} \pmb{\mathcal{E}}_{ij\centerdot} \pmb{\gamma}_{ij}^{\circ\top}  \mathbf{F}_i^{\circ\top} \mathbf{M}_{\widehat{\mathbf{C}}}\widehat{\mathbf{C}}_j^\bullet \nonumber \\
&&+\frac{1}{LT}\sum_{i=1}^L\mathbf{M}_{\widehat{\mathbf{C}}} \mathbf{F}_j^\bullet \pmb{\gamma}_{ij}^\bullet  \pmb{\gamma}_{ij}^{\bullet\top} \mathbf{F}_j^{\bullet\top} \mathbf{M}_{\widehat{\mathbf{C}}}\widehat{\mathbf{C}}_j^\bullet + \frac{1}{LT}\sum_{i=1}^L\mathbf{M}_{\widehat{\mathbf{C}}} \mathbf{F}_j^\bullet  \pmb{\gamma}_{ij}^\bullet  \pmb{\mathcal{E}}_{ij\centerdot}^\top \mathbf{M}_{\widehat{\mathbf{C}}}\widehat{\mathbf{C}}_j^\bullet \nonumber \\
&& +\frac{1}{LT}\sum_{i=1}^L\mathbf{M}_{\widehat{\mathbf{C}}} \pmb{\mathcal{E}}_{ij\centerdot} \pmb{\gamma}_{ij}^{\bullet\top} \mathbf{F}_j^{\bullet\top} \mathbf{M}_{\widehat{\mathbf{C}}} \widehat{\mathbf{C}}_j^\bullet + \frac{1}{LT}\sum_{i=1}^L\mathbf{M}_{\widehat{\mathbf{C}}} \pmb{\mathcal{E}}_{ij\centerdot}\pmb{\mathcal{E}}_{ij\centerdot}^\top \mathbf{M}_{\widehat{\mathbf{C}}} \widehat{\mathbf{C}}_j^\bullet \nonumber \\
&\coloneqq &\mathbf{W}_{j,1}+\cdots + \mathbf{W}_{j,25} ,
\end{eqnarray}
where the definitions of $\mathbf{W}_{j,1}$ to $\mathbf{W}_{j,25}$ should be obvious.

For $\mathbf{W}_{j,1}$, write

\begin{eqnarray}\label{nh1}
\max_j\frac{1}{\sqrt{T}} \|\mathbf{W}_{j,1}\|_2 &=&\max_j\frac{1}{\sqrt{T}}  \left\|  \frac{1}{LT}\sum_{i=1}^L \mathbf{M}_{\widehat{\mathbf{C}}}\mathbf{X}_{ij\centerdot} (\pmb{\beta}_{ij}-\widehat{\mathbf{b}}_{ij})(\pmb{\beta}_{ij}-\widehat{\mathbf{b}}_{ij})^\top  \mathbf{X}_{ij\centerdot}^\top \mathbf{M}_{\widehat{\mathbf{C}}} \widehat{\mathbf{C}}_j^\bullet \right\|_2 \nonumber \\
&\le &O_P(1) \max_j \frac{1}{LT}\sum_{i=1}^L\|\mathbf{X}_{ij\centerdot}\|^2  \|\pmb{\beta}_{ij}-\widehat{\mathbf{b}}_{ij}\| ^2\nonumber \\
&= &O_P ( \max_{i,j} \|\pmb{\beta}_{ij}-\widehat{\mathbf{b}}_{ij}\| ^2 ),
\end{eqnarray}
where the first inequality follows from the triangle inequality and the fact that $\frac{1}{T} \widehat{\mathbf{C}}_j^{\bullet\top} \widehat{\mathbf{C}}_j^\bullet= \mathbf{I}_{\ell_j^\bullet}$, and the second inequality follows from  Assumption \ref{AS1}.4.

For $\mathbf{W}_{j,2}$, write

\begin{eqnarray*}
\max_j\frac{1}{\sqrt{T}} \|\mathbf{W}_{j,2}\|_2 &=&\max_j\frac{1}{\sqrt{T}} \left\|\frac{1}{LT}\sum_{i=1}^L \mathbf{M}_{\widehat{\mathbf{C}}}\mathbf{X}_{ij\centerdot}(\pmb{\beta}_{ij}-\widehat{\mathbf{b}}_{ij}) \pmb{\gamma}_{ij}^{\top} \mathbf{F}^{\top} \mathbf{M}_{\widehat{\mathbf{C}}}\widehat{\mathbf{C}}_j^\bullet \right\|_2\nonumber \\
&\le &O_P(1) \max_j\left\|\frac{1}{LT}\sum_{i=1}^L\mathbf{M}_{\widehat{\mathbf{C}}}\mathbf{X}_{ij\centerdot}(\pmb{\beta}_{ij}-\widehat{\mathbf{b}}_{ij}) \pmb{\gamma}_{ij}^{\top} \mathbf{F}^{\top} \mathbf{M}_{\widehat{\mathbf{C}}}\right\|_2 \nonumber \\
&\le &O_P(1) \left\{\frac{1}{LT}\sum_{i=1}^L\|\mathbf{X}_{ij\centerdot} \|^2  \|\pmb{\beta}_{ij}-\widehat{\mathbf{b}}_{ij}\| ^2 \right\}^{1/2}\left\{\frac{1}{LT}\sum_{i=1}^L \| \mathbf{M}_{\widehat{\mathbf{C}}}\mathbf{F} \pmb{\gamma}_{ij}\| ^2 \right\}^{1/2}\nonumber \\
&= &O_P\left( \max_{i,j} \|\pmb{\beta}_{ij}-\widehat{\mathbf{b}}_{ij}\|\cdot \frac{1}{\sqrt{T}}\|\mathbf{M}_{\widehat{\mathbf{C}}}\mathbf{F}\| \right),
\end{eqnarray*}
where the second inequality follows from Cauchy-Schwarz inequality, and third inequality follows from the development for $\mathbf{W}_{j,1}$ and Assumptions \ref{AS1}.1 and \ref{AS1}.2. 

Similar to the development of $\mathbf{W}_{j,2}$, we obtain that

\begin{eqnarray*}
\max_j\frac{1}{\sqrt{T}} \|\mathbf{W}_{j,3}\|_2 =O_P\left( \max_{i,j} \|\pmb{\beta}_{ij}-\widehat{\mathbf{b}}_{ij}\|\cdot \frac{1}{\sqrt{T}}\|\mathbf{M}_{\widehat{\mathbf{C}}}\mathbf{F}\|\right).
\end{eqnarray*}

For $\mathbf{W}_{j,4}$, write

\begin{eqnarray*}
\max_j\frac{1}{\sqrt{T}} \| \mathbf{W}_{j,4} \|_2 &=&\max_j\frac{1}{\sqrt{T}} \left\|\frac{1}{LT}\sum_{i=1}^L\mathbf{M}_{\widehat{\mathbf{C}}}\mathbf{X}_{ij\centerdot}(\pmb{\beta}_{ij}-\widehat{\mathbf{b}}_{ij}) \pmb{\gamma}_{ij}^{\circ\top}  \mathbf{F}_i^{\circ\top} \mathbf{M}_{\widehat{\mathbf{C}}}\widehat{\mathbf{C}}_j^\bullet \right\|_2\nonumber \\
&\le &O(1)\max_j\frac{1}{LT} \|\mathbf{X}_{\centerdot j \centerdot } \diag\{ \pmb{\beta}_{\centerdot j}^\top-\widehat{\mathbf{b}}_{\centerdot j}^\top\} \diag\{ \pmb{\Gamma}_{\centerdot j}^{\circ\top}\}^\top \mathbf{F}^{\circ\top} \|_2\nonumber \\
&\le &O_P(1)\frac{1}{LT}\sqrt{LT}\cdot (\max_{i,j} \|\pmb{\beta}_{ij}-\widehat{\mathbf{b}}_{ij}\|)\cdot \sqrt{\log (LN)}\cdot\sqrt{L\vee T} \nonumber \\
&= &O_P\left( \frac{\sqrt{\log (LN)}}{\sqrt{L\wedge T}} \cdot \max_{i,j}\|\pmb{\beta}_{ij}-\widehat{\mathbf{b}}_{ij}\|\right),
\end{eqnarray*}
where the second inequality follows from Assumptions \ref{AS1}.3, \ref{AS1}.4 and \ref{AS2}.2. Similarly, we obtain that

\begin{eqnarray}\label{mge_13}
\max_j\frac{1}{\sqrt{T}} \| \mathbf{W}_{j,5} \|_2  =O_P\left( \frac{\sqrt{\log (LN)}}{\sqrt{L\wedge T}} \cdot \max_{i,j}\|\pmb{\beta}_{ij}-\widehat{\mathbf{b}}_{ij}\|\right).
\end{eqnarray}

Similar to the development of $\mathbf{W}_{j,2}$ and noting that $\max_j\frac{1}{\sqrt{T}}\|\mathbf{M}_{\widehat{\mathbf{C}}} \mathbf{F}_j^\bullet\|=O_P(1)$ by Lemma \ref{M.LM1} and Assumption \ref{AS1}.3, we obtain that

\begin{eqnarray*}
\max_j\frac{1}{\sqrt{T}} \|\mathbf{W}_{j,6}\|_2  &=&O_P( \max_{i,j} \|\pmb{\beta}_{ij}-\widehat{\mathbf{b}}_{ij}\|),\nonumber \\
\max_j\frac{1}{\sqrt{T}} \|\mathbf{W}_{j,7} \|_2  &=&O_P( \max_{i,j} \|\pmb{\beta}_{ij}-\widehat{\mathbf{b}}_{ij}\|).
\end{eqnarray*}

For $\mathbf{W}_{j,8}$, using the development similar to those for $\mathbf{W}_{j,4}$, we can write

\begin{eqnarray}\label{mge_e18}
\max_j\frac{1}{\sqrt{T}} \|\mathbf{W}_{j,8} \|_2 &=&  \max_j\frac{1}{\sqrt{T}} \left\|\frac{1}{LT}\sum_{i=1}^L\mathbf{M}_{\widehat{\mathbf{C}}} \mathbf{X}_{ij\centerdot} (\pmb{\beta}_{ij}-\widehat{\mathbf{b}}_{ij}) \pmb{\mathcal{E}}_{ij\centerdot}^\top  \mathbf{M}_{\widehat{\mathbf{C}}}\widehat{\mathbf{C}}_j^\bullet  \right\|_2\nonumber \\
&\le &O(1)  \max_j \frac{1}{L T} \|\mathbf{X}_{\centerdot j \centerdot } \diag\{ \pmb{\beta}_{\centerdot j}^\top-\widehat{\mathbf{b}}_{\centerdot j}^\top\} \pmb{\mathcal{E}}_{\centerdot j\centerdot}^\top\|_2\nonumber \\
& \le&O_P(1)\frac{1}{LT} \sqrt{LT}\cdot(\max_{i,j} \|\pmb{\beta}_{ij}-\widehat{\mathbf{b}}_{ij}\|)  \cdot \max_j\|\pmb{\mathcal{E}}_{\centerdot j\centerdot}\|_2\nonumber \\
&= &O_P\left( \frac{\max_{i,j} \|\pmb{\beta}_{ij}-\widehat{\mathbf{b}}_{ij}\| \cdot \max_j\|\pmb{\mathcal{E}}_{\centerdot j\centerdot}\|_2}{ \sqrt{LT}}\right),
\end{eqnarray}
where the second inequality follows from Assumptions \ref{AS1}.1 and \ref{AS1}.4. Similarly, we can obtain

\begin{eqnarray*}
\max_j\frac{1}{\sqrt{T}} \|\mathbf{W}_{j,9} \|_2 = O_P\left( \frac{\max_{i,j} \|\pmb{\beta}_{ij}-\widehat{\mathbf{b}}_{ij}\| \cdot \max_j\|\pmb{\mathcal{E}}_{\centerdot j\centerdot}\|_2}{ \sqrt{LT}}\right).
\end{eqnarray*}

For $\mathbf{W}_{j,10}$, write

\begin{eqnarray*}
\max_j\frac{1}{\sqrt{T}} \|\mathbf{W}_{j,10} \|_2 &=&\max_j\frac{1}{\sqrt{T}}\left\| \frac{1}{LT}\sum_{i=1}^L\mathbf{M}_{\widehat{\mathbf{C}}} \mathbf{F} \pmb{\gamma}_{ij} \pmb{\gamma}_{ij}^{\top} \mathbf{F}^\top  \mathbf{M}_{\widehat{\mathbf{C}}}\widehat{\mathbf{C}}_j^\bullet  \right\|_2
\nonumber \\
&\le &  O(1)\max_j\frac{1}{LT} \|\mathbf{M}_{\widehat{\mathbf{C}}} \mathbf{F} \pmb{\Gamma}_{\centerdot j}^\top \pmb{\Gamma}_{\centerdot j} \mathbf{F}^\top  \mathbf{M}_{\widehat{\mathbf{C}}}  \|_2\nonumber \\
&=&O_P(1)\frac{1}{T}\|\mathbf{M}_{\widehat{\mathbf{C}}} \mathbf{F}\|_2^2\nonumber \\
&=&O_P\left( \frac{1}{LN}\|\pmb{\beta}_{\centerdot \centerdot}-\widehat{\mathbf{b}}_{\centerdot \centerdot} \|^2+ \frac{1}{L\wedge N\wedge T}\right),
\end{eqnarray*}
where the second equality follows from Assumption \ref{AS1}.2, and the last step follows from \eqref{ff1}.

For $\mathbf{W}_{j,11}$, write

\begin{eqnarray*}
\max_j\frac{1}{\sqrt{T}} \|\mathbf{W}_{j,11} \|_2  &=&\max_j\frac{1}{\sqrt{T}}\left\| \frac{1}{LT}\sum_{i=1}^L\mathbf{M}_{\widehat{\mathbf{C}}} \mathbf{F} \pmb{\gamma}_{ij} \pmb{\gamma}_{ij}^{\circ\top} \mathbf{F}_i^{\circ\top}  \mathbf{M}_{\widehat{\mathbf{C}}}\widehat{\mathbf{C}}_j^\bullet \right\|_2\nonumber \\
&\le &O(1)\max_j\frac{1}{LT}\|\mathbf{M}_{\widehat{\mathbf{C}}} \mathbf{F} \pmb{\Gamma}_{\centerdot j}^\top \diag\{ \pmb{\Gamma}_{\centerdot j}^{\circ\top} \}^\top \mathbf{F}^{\circ\top}\|_2\nonumber \\
&\le &O_P(1)\frac{1}{\sqrt{LT}}\frac{1}{\sqrt{T}}\|\mathbf{M}_{\widehat{\mathbf{C}}} \mathbf{F}\|_2 \cdot \frac{1}{\sqrt{L}} \| \pmb{\Gamma}_{\centerdot j}\|_2\cdot \| \diag\{ \pmb{\Gamma}_{\centerdot j}^{\circ\top} \}\|_2\cdot \| \mathbf{F}^\circ\|_2\nonumber \\
&\le &O_P(1)\frac{\sqrt{\log(LN)}}{\sqrt{L\wedge T}}  \left( \frac{1}{\sqrt{LN}}\|\pmb{\beta}_{\centerdot \centerdot}-\widehat{\mathbf{b}}_{\centerdot \centerdot} \| + \frac{1}{\sqrt{L\wedge N\wedge T}}\right),
\end{eqnarray*}
where we have used \eqref{ff1} and Assumptions \ref{AS1}.2, \ref{AS1}.3, and \ref{AS2}.2. 

Similar to the development of $\mathbf{W}_{j,11}$, we can obtain that

\begin{eqnarray*}
\max_j\frac{1}{\sqrt{T}} \|\mathbf{W}_{j,12} \|_2 =O_P(1)\frac{\sqrt{\log(LN)}}{\sqrt{L\wedge T}}  \left( \frac{1}{\sqrt{LN}}\|\pmb{\beta}_{\centerdot \centerdot}-\widehat{\mathbf{b}}_{\centerdot \centerdot} \| + \frac{1}{\sqrt{L\wedge N\wedge T}}\right).
\end{eqnarray*}

Similar to the development of $\mathbf{W}_{j,10}$ and by noting $\max_j\frac{1}{\sqrt{T}}\|\mathbf{M}_{\widehat{\mathbf{C}}}  \mathbf{F}_j^\bullet\|_2=O_P(1)$ using Assumption \ref{AS1}.3 and \eqref{ff1}, we can also obtain that

\begin{eqnarray*}
\max_j\frac{1}{\sqrt{T}} \|\mathbf{W}_{j,13} \|_2&=&O_P\left(  \frac{1}{\sqrt{T}}\|\mathbf{M}_{\widehat{\mathbf{C}}} \mathbf{F}\|_2\right)=O_P\left( \frac{1}{\sqrt{LN}}\|\pmb{\beta}_{\centerdot \centerdot}-\widehat{\mathbf{b}}_{\centerdot \centerdot} \| + \frac{1}{\sqrt{L\wedge N\wedge T}} \right),\nonumber \\
\max_j\frac{1}{\sqrt{T}} \|\mathbf{W}_{j,14} \|_2&=&O_P\left(  \frac{1}{\sqrt{T}}\|\mathbf{M}_{\widehat{\mathbf{C}}} \mathbf{F}\|_2\right)=O_P\left( \frac{1}{\sqrt{LN}}\|\pmb{\beta}_{\centerdot \centerdot}-\widehat{\mathbf{b}}_{\centerdot \centerdot} \| + \frac{1}{\sqrt{L\wedge N\wedge T}} \right).  
\end{eqnarray*}

For $\mathbf{W}_{j,15}$, we write

\begin{eqnarray*}
\max_j\frac{1}{\sqrt{T}} \|\mathbf{W}_{j,15} \|_2 &=&  \max_j\frac{1}{\sqrt{T}} \left\|\frac{1}{LT}\sum_{i=1}^L\mathbf{M}_{\widehat{\mathbf{C}}} \mathbf{F} \pmb{\gamma}_{ij} \pmb{\mathcal{E}}_{ij\centerdot}^\top \mathbf{M}_{\widehat{\mathbf{C}}} \widehat{\mathbf{C}}_j^\bullet \right\|_2\nonumber \\
&\le &  O(1)\max_j\frac{1}{LT}  \|\mathbf{M}_{\widehat{\mathbf{C}}} \mathbf{F} \pmb{\Gamma}_{\centerdot j}^{\top} \pmb{\mathcal{E}}_{\centerdot j\centerdot}^\top  \|_2\nonumber \\
&=&O_P(1)\frac{\max_j\|\pmb{\mathcal{E}}_{\centerdot j\centerdot}\|_2}{\sqrt{L T}} \left( \frac{1}{\sqrt{LN}}\|\pmb{\beta}_{\centerdot \centerdot}-\widehat{\mathbf{b}}_{\centerdot \centerdot} \| + \frac{1}{\sqrt{L\wedge N\wedge T}}\right),
\end{eqnarray*}
where the second equality follows  Assumption \ref{AS1}.2 and \eqref{ff1}. 

Similar to the development of $\mathbf{W}_{j,15}$, we obtain that
 
\begin{eqnarray*}
\max_j\frac{1}{\sqrt{T}} \|\mathbf{W}_{j,16} \|_2 =O_P(1)\frac{\max_j\|\pmb{\mathcal{E}}_{\centerdot j\centerdot}\|_2}{\sqrt{L T}} \left( \frac{1}{\sqrt{LN}}\|\pmb{\beta}_{\centerdot \centerdot}-\widehat{\mathbf{b}}_{\centerdot \centerdot} \| + \frac{1}{\sqrt{L\wedge N\wedge T}}\right).
\end{eqnarray*}

Similar to the development of $\mathbf{J}_{16}$ and invoking Assumption \ref{AS2}.2, we get
 
\begin{eqnarray*}
\max_j\frac{1}{\sqrt{T}} \|\mathbf{W}_{j,17} \|_2 &=&O_P\left( \frac{\log (LN)}{L\wedge T}\right),\nonumber \\
\max_j\frac{1}{\sqrt{T}} \|\mathbf{W}_{j,18} \|_2 &=&O_P\left( \frac{\sqrt{\log (LN)}}{\sqrt{L\wedge T}}\right),\nonumber \\
\max_j\frac{1}{\sqrt{T}} \|\mathbf{W}_{j,19} \|_2 &=&O_P\left( \frac{\sqrt{\log (LN)}}{\sqrt{L\wedge T}}\right).
\end{eqnarray*}

By the development similar to $\mathbf{J}_{20}$ and invoking Assumption \ref{AS2}.2, we obtain that

\begin{eqnarray*}
\max_j\frac{1}{\sqrt{T}} \|\mathbf{W}_{j, 20} \|_2 =O_P\left( \frac{\sqrt{\log(LN)}}{\sqrt{L\wedge T}} \cdot \frac{\max_j\|\pmb{\mathcal{E}}_{\centerdot j\centerdot}\|_2}{\sqrt{L T}}  \right),\nonumber \\
\max_j\frac{1}{\sqrt{T}} \|\mathbf{W}_{j, 21} \|_2 =O_P\left( \frac{\sqrt{\log(LN)}}{\sqrt{L\wedge T}} \cdot \frac{\max_j\|\pmb{\mathcal{E}}_{\centerdot j\centerdot}\|_2}{\sqrt{L T}}  \right).
\end{eqnarray*}

Similar to the development of $\mathbf{W}_{j, 15}$ and by noting $\max_j\frac{1}{\sqrt{T}}\|\mathbf{M}_{\widehat{\mathbf{C}}}  \mathbf{F}_j^\bullet\|_2=O_P(1)$, we obtain that

\begin{eqnarray*}
\max_j\frac{1}{\sqrt{T}} \|\mathbf{W}_{j, 23} \|_2 &=&O_P(1) \frac{\max_j\|\pmb{\mathcal{E}}_{\centerdot j\centerdot}\|_2}{\sqrt{L T}},\nonumber \\
\max_j\frac{1}{\sqrt{T}} \|\mathbf{W}_{j, 24} \|_2 &=&O_P(1) \frac{\max_j\|\pmb{\mathcal{E}}_{\centerdot j\centerdot}\|_2}{\sqrt{L T}}.
\end{eqnarray*}

For $\mathbf{W}_{j, 25}$, we write

\begin{eqnarray}\label{mge_e27}
\max_j\frac{1}{\sqrt{T}} \|\mathbf{W}_{j, 25}\|_2&=& \max_j \frac{1}{\sqrt{T}} \left\| \frac{1}{LT}\sum_{i=1}^{L} \mathbf{M}_{\widehat{\mathbf{C}}} \pmb{\mathcal{E}}_{ij\centerdot} \pmb{\mathcal{E}}_{ij\centerdot}^\top\mathbf{M}_{\widehat{\mathbf{C}}} \widehat{\mathbf{C}}_j^\bullet\right\|_2 \nonumber \\
&\le &O(1)\max_j\frac{1}{LT}\| \pmb{\mathcal{E}}_{\centerdot j\centerdot}\pmb{\mathcal{E}}_{\centerdot j\centerdot}^\top\|_2=O_P\left( \left[ \frac{\max_j\|\pmb{\mathcal{E}}_{\centerdot j\centerdot}\|_2}{\sqrt{L T}}\right]^2\right),
\end{eqnarray}
where we have used Assumption \ref{AS1}.1.

Finally, we focus on $\mathbf{W}_{j,22}$.

\begin{eqnarray*}
&&\frac{1}{LT}\sum_{i=1}^{L}\mathbf{M}_{\widehat{\mathbf{C}}}\mathbf{F}_j^\bullet  \pmb{\gamma}_{ij}^\bullet \pmb{\gamma}_{ij}^{\bullet\top}\mathbf{F}_j^{\bullet\top} \mathbf{M}_{\widehat{\mathbf{C}}} \widehat{\mathbf{C}}_j^\bullet\nonumber \\
&=&\frac{1}{LT}\sum_{i=1}^{L}\mathbf{F}_j^\bullet  \pmb{\gamma}_{ij}^\bullet \pmb{\gamma}_{ij}^{\bullet\top}\mathbf{F}_j^{\bullet\top} \widehat{\mathbf{C}}_j^\bullet+\frac{1}{LT}\sum_{i=1}^{L}\mathbf{P}_{\widehat{\mathbf{C}}} \mathbf{F}_j^\bullet  \pmb{\gamma}_{ij}^\bullet \pmb{\gamma}_{ij}^{\bullet\top}\mathbf{F}_j^{\bullet\top} \mathbf{P}_{\widehat{\mathbf{C}}} \widehat{\mathbf{C}}_j^\bullet\nonumber \\
&&+\frac{1}{LT}\sum_{i=1}^{L} \mathbf{F}_j^\bullet  \pmb{\gamma}_{ij}^\bullet \pmb{\gamma}_{ij}^{\bullet\top}\mathbf{F}_j^{\bullet\top} \mathbf{P}_{\widehat{\mathbf{C}}}  \widehat{\mathbf{C}}_j^\bullet+\frac{1}{LT}\sum_{i=1}^{L}\mathbf{P}_{\widehat{\mathbf{C}}} \mathbf{F}_j^\bullet  \pmb{\gamma}_{ij}^\bullet \pmb{\gamma}_{ij}^{\bullet\top}\mathbf{F}_j^{\bullet\top} \widehat{\mathbf{C}}_j^\bullet.
\end{eqnarray*}

Note that

\begin{eqnarray*}
&&\max_j\frac{1}{LT}\left\|\sum_{i=1}^{L} \mathbf{F}_j^\bullet  \pmb{\gamma}_{ij}^\bullet \pmb{\gamma}_{ij}^{\bullet\top}\mathbf{F}_j^{\bullet\top} \mathbf{P}_{\widehat{\mathbf{C}}} \right\|_2\nonumber \\
&\le &\max_j\frac{1}{LT} \cdot \frac{1}{T}\left\|\sum_{i=1}^{L} \mathbf{F}_j^\bullet  \pmb{\gamma}_{ij}^\bullet \pmb{\gamma}_{ij}^{\bullet\top}\mathbf{F}_j^{\bullet\top} (\widehat{\mathbf{C}} - \mathbf{F}\mathbf{H})\widehat{\mathbf{C}}^\top \right\|_2\nonumber \\
&&+\max_j\frac{1}{LT} \cdot \frac{1}{T}\left\|\sum_{i=1}^{L} \mathbf{F}_j^\bullet  \pmb{\gamma}_{ij}^\bullet \pmb{\gamma}_{ij}^{\bullet\top}\mathbf{F}_j^{\bullet\top} \mathbf{F} \mathbf{H}\widehat{\mathbf{C}}^\top \right\|_2\nonumber \\
&\le &O_P(1)  \frac{1}{\sqrt{T}}\| \mathbf{C}  - \mathbf{F}\mathbf{H} \|_2+O_P(1) \max_j\frac{1}{T}\| \mathbf{F}_j^{\bullet\top} \mathbf{F}\|_2 ,
\end{eqnarray*}
where the second inequality follows from Assumptions \ref{AS1}.2 and \ref{AS1}.3. Similarly, it is easy to know that 

\begin{eqnarray*}
\frac{1}{LT}\left\|\sum_{i=1}^{L}\mathbf{P}_{\widehat{\mathbf{C}}}  \mathbf{F}_j^\bullet  \pmb{\gamma}_{ij}^\bullet \pmb{\gamma}_{ij}^{\bullet\top}\mathbf{F}_j^{\bullet\top} \mathbf{P}_{\widehat{\mathbf{C}}}  \right\|_2 &=&O_P\left( \frac{1}{T}\| \mathbf{C}  - \mathbf{F}\mathbf{H} \|_2^2 + \frac{1}{T^2}[\max_j\| \mathbf{F}_j^{\bullet\top} \mathbf{F}\|_2]^2\right).
\end{eqnarray*}

Therefore, we conclude that

\begin{eqnarray*}
&&\frac{1}{\sqrt{T}}\left\|\widehat{\mathbf{C}}_j^\bullet \mathbf{V}_{j}^\bullet - \frac{1}{LT}\sum_{i=1}^{L} \mathbf{F}_j^\bullet  \pmb{\gamma}_{ij}^\bullet \pmb{\gamma}_{ij}^{\bullet\top}\mathbf{F}_j^{\bullet\top} \widehat{\mathbf{C}}_j^\bullet\right\|_2 \nonumber \\
&=&O_P\left(\max_{i,j} \|\pmb{\beta}_{ij}-\widehat{\mathbf{b}}_{ij}\| + \frac{\sqrt{\log(LN)}}{\sqrt{L \wedge T}} +\frac{1}{\sqrt{N}}+ \frac{\max_j\|\pmb{\mathcal{E}}_{\centerdot j\centerdot}\|_2}{\sqrt{L T}} +\frac{\max_j\| \mathbf{F}_j^{\bullet\top} \mathbf{F}\|_2}{T} \right).
\end{eqnarray*}
The results then follow immediately. \hspace*{\fill}{$\blacksquare$}

\bigskip

\noindent \textbf{Proof of Theorem \ref{M.Thm1}:}

(1) We next investigate the asymptotic distribution.

\begin{eqnarray*}
\widetilde{\mathbf{b}}_{ij}= (\mathbf{X}_{ij\centerdot}^\top \mathbf{M}_{\widehat{\mathbf{C}}_{ij}} \mathbf{X}_{ij\centerdot} )^{-1} \mathbf{X}_{ij\centerdot}^\top \mathbf{M}_{\widehat{\mathbf{C}}_{ij}} \mathbf{Y}_{ij\centerdot},
\end{eqnarray*}
where $\widehat{\mathbf{C}}_{ij} =(\widehat{\mathbf{C}},\widehat{\mathbf{C}}_i^\circ,\widehat{\mathbf{C}}_j^\bullet)$. Note that

\begin{eqnarray}\label{bhatij}
\widetilde{\mathbf{b}}_{ij} &=& \pmb{\beta}_{ij}+(\mathbf{X}_{ij\centerdot}^\top \mathbf{M}_{\widehat{\mathbf{C}}_{ij}} \mathbf{X}_{ij\centerdot} )^{-1} \mathbf{X}_{ij\centerdot}^\top \mathbf{M}_{\widehat{\mathbf{C}}_{ij}} \mathbf{F} \pmb{\gamma}_{ij}\nonumber \\
&&+(\mathbf{X}_{ij\centerdot}^\top \mathbf{M}_{\widehat{\mathbf{C}}_{ij}} \mathbf{X}_{ij\centerdot} )^{-1} \mathbf{X}_{ij\centerdot}^\top \mathbf{M}_{\widehat{\mathbf{C}}_{ij}} \mathbf{F}_i^\circ \pmb{\gamma}_{ij}^\circ\nonumber \\
&&+(\mathbf{X}_{ij\centerdot}^\top \mathbf{M}_{\widehat{\mathbf{C}}_{ij}} \mathbf{X}_{ij\centerdot} )^{-1} \mathbf{X}_{ij\centerdot}^\top \mathbf{M}_{\widehat{\mathbf{C}}_{ij}} \mathbf{F}_j^\bullet \pmb{\gamma}_{ij}^\bullet\nonumber \\
&&+(\mathbf{X}_{ij\centerdot}^\top \mathbf{M}_{\widehat{\mathbf{C}}_{ij}} \mathbf{X}_{ij\centerdot} )^{-1} \mathbf{X}_{ij\centerdot}^\top \mathbf{M}_{\widehat{\mathbf{C}}_{ij}} \pmb{\mathcal{E}}_{ij\centerdot } .
\end{eqnarray}

We first note that in what follows, we shall repeatedly expand $\mathbf{M}_{\widehat{\mathbf{C}}_{ij}}\mathbf{F}_{ij}^*$ and $\widehat{\mathbf{b}}_{ij} - \pmb{\beta}_{ij}$ as in Lemma \ref{M.LM1} and Lemma \ref{M.LM2}. Therefore, if we can show the terms involved in expansions are $o_P(1)\frac{1}{\sqrt{T}} \|\mathbf{M}_{\widehat{\mathbf{C}}_{ij}}\mathbf{F}_{ij}^*\|_2$ or $o_P(1)(\widehat{\mathbf{b}}_{ij} - \pmb{\beta}_{ij})$, then they are negligible. The reason is that we can repeatedly expand these terms and they  will eventually attain a desired order.

We start with $\frac{1}{T}\mathbf{X}_{ij\centerdot}^\top \mathbf{M}_{\widehat{\mathbf{C}}_{ij}} \mathbf{F} \pmb{\gamma}_{ij}$, and consider 

\begin{eqnarray}
\frac{1}{T}\mathbf{X}_{ij\centerdot}^\top \mathbf{M}_{\widehat{\mathbf{C}}_{ij}} \mathbf{F} \pmb{\gamma}_{ij} &=&-\frac{1}{T}\mathbf{X}_{ij\centerdot}^\top \mathbf{M}_{\widehat{\mathbf{C}}_{ij}} (\widehat{\mathbf{C}}\mathbf{H}^{-1} -\mathbf{F}) \pmb{\gamma}_{ij} \nonumber \\
&=&-\frac{1}{T}\mathbf{X}_{ij\centerdot}^\top \mathbf{M}_{\widehat{\mathbf{C}}_{ij}} \sum_{s =1,\ne 10}^{25} \mathbf{J}_{s}\pmb{\Pi}\pmb{\gamma}_{ij} \nonumber \\
&\coloneqq &-(\mathbf{A}_{ij,1}+\cdots+\mathbf{A}_{ij,24}),
\end{eqnarray}
where $\mathbf{J}_{s}$'s are defined in the proof of Lemma \ref{M.LM1}, $\pmb{\Pi} = (\mathbf{F}^\top\widehat{\mathbf{C}}/T)^{-1}(\pmb{\Gamma}_{\centerdot \centerdot }^\top \pmb{\Gamma}_{\centerdot \centerdot }/(LN))^{-1}$, and the definitions of $\mathbf{A}_{ij,1},\ldots,\mathbf{A}_{ij,24}$ should be obvious. First, we show that

\begin{eqnarray}\label{Aijs1}
 \sum_{s=1}^9 \|\mathbf{A}_{ij,s} \|_2=o_P \left(\frac{1}{\sqrt{T}} \|\mathbf{M}_{\widehat{\mathbf{C}}_{ij}}\mathbf{F}_{ij}^*\|_2 +\frac{1}{\sqrt{T}} \right).
\end{eqnarray}
Take $ \mathbf{A}_{ij,2}$ as an example without loss of generality. Note that

\begin{eqnarray}\label{expandX}
\frac{1}{\sqrt{T}} \mathbf{M}_{\widehat{\mathbf{C}}_{ij}}\mathbf{X}_{ij\centerdot\centerdot}
&=&\frac{1}{\sqrt{T}} \mathbf{M}_{\widehat{\mathbf{C}}_{ij}}(\mathbf{F}_{ij}^*\pmb{\gamma}_{ij}^* + \mathbf{V}_{ij\centerdot})\nonumber \\
&=&\frac{1}{\sqrt{T}} \mathbf{M}_{\widehat{\mathbf{C}}_{ij}} (\widehat{\mathbf{C}}\mathbf{H}^{-1} -\mathbf{F}) +\frac{1}{\sqrt{T}} \mathbf{M}_{\widehat{\mathbf{C}}_{ij}} (\widehat{\mathbf{C}}_i^\circ\mathbf{H}_i^{\circ -1} -\mathbf{F}_i^\circ)\nonumber \\
&&+\frac{1}{\sqrt{T}} \mathbf{M}_{\widehat{\mathbf{C}}_{ij}} (\widehat{\mathbf{C}}_j^\bullet\mathbf{H}_j^{\bullet -1} -\mathbf{F}_j^\bullet) +\frac{1}{\sqrt{T}} \mathbf{M}_{\widehat{\mathbf{C}}_{ij}}  \mathbf{V}_{ij\centerdot},
\end{eqnarray}
where $\mathbf{H}_i^\circ =\frac{1}{NT} \pmb{\Gamma}_{i\centerdot}^{\circ\top} \pmb{\Gamma}_{i\centerdot}^\circ \mathbf{F}_i^{\circ\top} \widehat{\mathbf{C}}_i^\circ \widehat{\mathbf{V}}_i^{\circ -1}$ and $\mathbf{H}_j^\bullet =\frac{1}{LT} \pmb{\Gamma}_{j\centerdot}^{\bullet\top} \pmb{\Gamma}_{j\centerdot}^\bullet \mathbf{F}_j^{\bullet\top} \widehat{\mathbf{C}}_j^\bullet \widehat{\mathbf{V}}_j^{\bullet -1}$. Then we can write

\begin{eqnarray*}
\mathbf{A}_{ij,2}&=&\frac{1}{T}\mathbf{X}_{ij\centerdot}^\top \mathbf{M}_{\widehat{\mathbf{C}}_{ij}} \frac{1}{LNT}\sum_{i_2=1}^L\sum_{j_2=1}^{N}\mathbf{X}_{i_2j_2\centerdot}(\pmb{\beta}_{i_2j_2}-\widehat{\mathbf{b}}_{i_2j_2}) \pmb{\gamma}_{i_2j_2}^{\top} \mathbf{F}^{\top} \widehat{\mathbf{C}}     \pmb{\Pi}\pmb{\gamma}_{ij}\nonumber \\
&=&\frac{1}{T}\mathbf{X}_{ij\centerdot}^\top \mathbf{M}_{\widehat{\mathbf{C}}_{ij}} \frac{1}{LN}\sum_{i_2=1}^L\sum_{j_2=1}^{N}\mathbf{X}_{i_2j_2\centerdot} \pmb{\gamma}_{i_2j_2}^{\top} \left(\frac{\pmb{\Gamma}_{\centerdot \centerdot }^\top \pmb{\Gamma}_{\centerdot \centerdot }}{LN}\right)^{-1}\pmb{\gamma}_{ij} (\pmb{\beta}_{i_2j_2}-\widehat{\mathbf{b}}_{i_2j_2})\nonumber \\
&\le &O_P(1)\frac{1}{\sqrt{T}} \|\mathbf{M}_{\widehat{\mathbf{C}}_{ij}}\mathbf{F}_{ij}^*\|_2\cdot (\max_{i,j}\|\pmb{\beta}_{ij}-\widehat{\mathbf{b}}_{ij}\|) \nonumber \\
&&+\frac{1}{T}\mathbf{V}_{ij\centerdot}^\top \mathbf{M}_{\widehat{\mathbf{C}}_{ij}} \frac{1}{LN}\sum_{i_2=1}^L\sum_{j_2=1}^{N}\mathbf{X}_{i_2j_2\centerdot} \pmb{\gamma}_{i_2j_2}^{\top} \left(\frac{\pmb{\Gamma}_{\centerdot \centerdot }^\top \pmb{\Gamma}_{\centerdot \centerdot }}{LN}\right)^{-1}\pmb{\gamma}_{ij} (\pmb{\beta}_{i_2j_2}-\widehat{\mathbf{b}}_{i_2j_2})\nonumber \\
&\le &o_P \left(\frac{1}{\sqrt{T}} \|\mathbf{M}_{\widehat{\mathbf{C}}_{ij}}\mathbf{F}_{ij}^*\|_2 +\frac{1}{\sqrt{T}} \right)
\end{eqnarray*}
where the first inequality follows from \eqref{expandX} and Assumption \ref{AS1}, and the second inequality follows from $\max_{i,j}\|\pmb{\beta}_{ij}-\widehat{\mathbf{b}}_{ij}\|=o_P(1)$ by the third result of Lemma \ref{M.LM2}. The rest terms involved in $\sum_{s=1}^9 \|\mathbf{A}_{ij,s} \|_2$ can be proved similarly, so we omit the details. Thus, \eqref{Aijs1} is proved.

For $\mathbf{A}_{ij,10}$, write

\begin{eqnarray*}
\|\mathbf{A}_{ij,10}\|_2&=& \left\|\frac{1}{T}\mathbf{X}_{ij\centerdot}^\top \mathbf{M}_{\widehat{\mathbf{C}}_{ij}} \frac{1}{LNT}\sum_{i_2=1}^L\sum_{j_2=1}^{N}\mathbf{F}\pmb{\gamma}_{i_2j_2} \pmb{\gamma}_{i_2j_2}^{\circ\top} \mathbf{F}_{i_2}^{\circ\top}\widehat{\mathbf{C}}     \pmb{\Pi}\pmb{\gamma}_{ij}\right\|_2 \nonumber \\
&\le &O_P(1) \frac{1}{\sqrt{T}}\| \mathbf{M}_{\widehat{\mathbf{C}}_{ij}}(\mathbf{F}-\widehat{\mathbf{C}}\mathbf{H}^{-1})\|_2 \cdot \frac{1}{LNT}\left\| \sum_{i_2=1}^L\sum_{j_2=1}^{N}\pmb{\gamma}_{i_2j_2} \pmb{\gamma}_{i_2j_2}^{\circ\top} \mathbf{F}_{i_2}^{\circ\top}\widehat{\mathbf{C}}   \right\|_2\nonumber \\
&\le &o_P(1) \frac{1}{\sqrt{T}}\| \mathbf{M}_{\widehat{\mathbf{C}}_{ij}}(\mathbf{F}-\widehat{\mathbf{C}}\mathbf{H}^{-1})\|_2,
\end{eqnarray*}
where the second inequality follows from the development of $\mathbf{J}_{11}$.

For $\mathbf{A}_{ij,11}$, it is easy to show that

\begin{eqnarray*}
\| \mathbf{A}_{ij,11}\|_2 &=& o_P \left(\frac{1}{\sqrt{T}} \|\mathbf{M}_{\widehat{\mathbf{C}}_{ij}}\mathbf{F}_{ij}^*\|_2 +\frac{1}{\sqrt{T}} \right)
\end{eqnarray*}
following the development of $\mathbf{A}_{ij,2}$ and $\mathbf{J}_{12}$.

Similar to $\mathbf{A}_{ij,10}$ and $\mathbf{A}_{ij,11}$, we know that

\begin{eqnarray*}
\|\mathbf{A}_{ij,12}\|_2 &=&o_P(1) \frac{1}{\sqrt{T}}\| \mathbf{M}_{\widehat{\mathbf{C}}_{ij}}(\mathbf{F}-\widehat{\mathbf{C}}\mathbf{H}^{-1})\|_2,\nonumber\\
\|\mathbf{A}_{ij,13}\|_2 &=& o_P \left(\frac{1}{\sqrt{T}} \|\mathbf{M}_{\widehat{\mathbf{C}}_{ij}}\mathbf{F}_{ij}^*\|_2 +\frac{1}{\sqrt{T}} \right),\nonumber \\
\|\mathbf{A}_{ij,14}\|_2 &=&o_P(1) \frac{1}{\sqrt{T}}\| \mathbf{M}_{\widehat{\mathbf{C}}_{ij}}(\mathbf{F}-\widehat{\mathbf{C}}\mathbf{H}^{-1})\|_2,\nonumber\\
\|\mathbf{A}_{ij,15}\|_2 &=& o_P \left(\frac{1}{\sqrt{T}} \|\mathbf{M}_{\widehat{\mathbf{C}}_{ij}}\mathbf{F}_{ij}^*\|_2 +\frac{1}{\sqrt{T}} \right).
\end{eqnarray*}

Similar to the development of $\mathbf{J}_{17}$ to $\mathbf{J}_{25}$, it is easy to know that

\begin{eqnarray*}
\mathbf{A}_{ij,16} &=&O_P\left(\frac{1}{L\wedge T} \right),\nonumber \\
\mathbf{A}_{ij,17} &=&O_P\left(\frac{1}{\sqrt{L\wedge T} \sqrt{N\wedge T}} \right),\nonumber \\
\mathbf{A}_{ij,18} &=&O_P\left(\frac{1}{\sqrt{L\wedge T}\sqrt{N\wedge T}} \right),\nonumber \\
\mathbf{A}_{ij,19} &=&O_P\left(\frac{1}{L\wedge T} \right),\nonumber \\
\mathbf{A}_{ij,20} &=&O_P\left(\frac{1}{L\wedge T} \right),\nonumber \\
\mathbf{A}_{ij,21} &=&O_P\left(\frac{1}{N\wedge T} \right),\nonumber \\
\mathbf{A}_{ij,22} &=&O_P\left(\frac{1}{N\wedge T} \right),\nonumber \\
\mathbf{A}_{ij,23} &=&O_P\left(\frac{1}{N\wedge T} \right),\nonumber \\
\mathbf{A}_{ij,24} &=&O_P\left(\frac{1}{LN\wedge T} \right). 
\end{eqnarray*}

Therefore,

\begin{eqnarray*}
\frac{1}{T}\|\mathbf{X}_{ij\centerdot}^\top \mathbf{M}_{\widehat{\mathbf{C}}_{ij}} \mathbf{F} \pmb{\gamma}_{ij}\|_2= o_P \left(\frac{1}{\sqrt{T}} \|\mathbf{M}_{\widehat{\mathbf{C}}_{ij}}\mathbf{F}_{ij}^*\|_2 +\frac{1}{\sqrt{T}} \right).
\end{eqnarray*}
in view of the condition $\sqrt{T}/(L\wedge N) \to 0$ in the body of this theorem.

\medskip

Next, we consider $\frac{1}{T} \mathbf{X}_{ij\centerdot}^\top \mathbf{M}_{\widehat{\mathbf{C}}_{ij}} \mathbf{F}_j^\bullet \pmb{\gamma}_{ij}^\bullet$.

\begin{eqnarray}
\frac{1}{T} \mathbf{X}_{ij\centerdot}^\top \mathbf{M}_{\widehat{\mathbf{C}}_{ij}} \mathbf{F}_j^\bullet \pmb{\gamma}_{ij}^\bullet &=&-\frac{1}{T} \mathbf{X}_{ij\centerdot}^\top \mathbf{M}_{\widehat{\mathbf{C}}_{ij}} (\widehat{\mathbf{C}}_j^\bullet \mathbf{H}_j^{\bullet -1}- \mathbf{F}_j^\bullet)  \pmb{\gamma}_{ij}^\bullet \nonumber \\
&=&-\frac{1}{T}\mathbf{X}_{ij\centerdot}^\top \mathbf{M}_{\widehat{\mathbf{C}}_{ij}} \sum_{s =1,\ne 22}^{25} \mathbf{W}_{j,\ell}\pmb{\Pi}_j \pmb{\gamma}_{ij}^\bullet\nonumber \\
&:=&-(\mathbf{B}_{ij,1}+\cdots+\mathbf{B}_{ij, 24}),
\end{eqnarray}
where $\pmb{\Pi}_j = (\mathbf{F}_j^{\bullet\top} \widehat{\mathbf{C}}_j^\bullet/T)^{-1}(\pmb{\Gamma}_{\centerdot j}^{^\bullet\top} \pmb{\Gamma}_{\centerdot j }^\bullet/L)^{-1}$, and the definitions of $\mathbf{B}_{ij,1},\ldots,\mathbf{B}_{ij,24}$ should be obvious. In what follows, we consider these terms one by one.  

Similar to the development of $\mathbf{A}_{ij,1}$ to $\mathbf{A}_{ij,9}$, we have

\begin{eqnarray*}
 \sum_{s=1}^9 \|\mathbf{B}_{ij,s} \|_2=o_P \left(\frac{1}{\sqrt{T}} \|\mathbf{M}_{\widehat{\mathbf{C}}_{ij}}\mathbf{F}_{ij}^*\|_2 +\frac{1}{\sqrt{T}} \right).
\end{eqnarray*}

For $\mathbf{B}_{ij,10}$ to $\mathbf{B}_{ij,12}$, it is easy to know that

\begin{eqnarray*}
 \sum_{s=10}^{12} \|\mathbf{B}_{ij,s} \|_2 &=&o_P(1) \frac{1}{\sqrt{T}}\| \mathbf{M}_{\widehat{\mathbf{C}}_{ij}}\mathbf{F}\|_2
\end{eqnarray*}
by the development of $\mathbf{W}_{j,10}$ to $\mathbf{W}_{j,12}$.

For $\mathbf{B}_{ij,13}$, write

\begin{eqnarray*}
\mathbf{B}_{ij,13}&=&\frac{1}{T}\mathbf{X}_{ij\centerdot}^\top \mathbf{M}_{\widehat{\mathbf{C}}_{ij}} \frac{1}{LT}\sum_{i_2=1}^L \mathbf{M}_{\widehat{\mathbf{C}}}\mathbf{F}\pmb{\gamma}_{i_2j}  \pmb{\gamma}_{i_2j}^{\bullet\top} \mathbf{F}_j^{\bullet\top}\mathbf{M}_{\widehat{\mathbf{C}}}  \widehat{\mathbf{C}}_j^\bullet \pmb{\Pi}_j \pmb{\gamma}_{ij}^\bullet\nonumber \\
&=&\frac{1}{T}\mathbf{X}_{ij\centerdot}^\top \mathbf{M}_{\widehat{\mathbf{C}}_{ij}}\mathbf{F}  \pmb{\Gamma}_{\centerdot j}^{\top} \pmb{\Gamma}_{\centerdot j}^\bullet (\pmb{\Gamma}_{\centerdot j }^{\bullet\top} \pmb{\Gamma}_{\centerdot j}^\bullet)^{-1} \pmb{\gamma}_{ij}^\bullet.
\end{eqnarray*}
Therefore, the investigation of $\mathbf{B}_{ij,13}$ reduces to the study of $\frac{1}{T}\mathbf{X}_{ij\centerdot}^\top \mathbf{M}_{\widehat{\mathbf{C}}_{ij}} \mathbf{F} \pmb{\gamma}_{ij} $, which infers that

\begin{eqnarray*}
\|\mathbf{B}_{ij,13}\|_2 =o_P \left(\frac{1}{\sqrt{T}} \|\mathbf{M}_{\widehat{\mathbf{C}}_{ij}}\mathbf{F}_{ij}^*\|_2 +\frac{1}{\sqrt{T}} \right).
\end{eqnarray*}

Invoking \eqref{expandX}, it is easy to show that

\begin{eqnarray*}
\|\mathbf{B}_{ij,14}\|_2 =o_P \left(\frac{1}{\sqrt{T}} \|\mathbf{M}_{\widehat{\mathbf{C}}_{ij}}\mathbf{F}_{ij}^*\|_2 +\frac{1}{\sqrt{T}} \right).
\end{eqnarray*}

For $\mathbf{B}_{ij, 15}$ and $\mathbf{B}_{ij,16}$, following the development similar to $\mathbf{W}_{j,15}$ and $\mathbf{W}_{j,16}$, it is easy to know that

\begin{eqnarray*}
\mathbf{B}_{ij, 15} &=&o_P(1) \frac{1}{\sqrt{T}}\| \mathbf{M}_{\widehat{\mathbf{C}}}\mathbf{F}\|_2,\nonumber \\
\mathbf{B}_{ij, 16} &=&o_P(1) \frac{1}{\sqrt{T}}\| \mathbf{M}_{\widehat{\mathbf{C}}}\mathbf{F}\|_2.
\end{eqnarray*}

For $\mathbf{B}_{ij, 17}$ to $\mathbf{B}_{ij, 25}$, invoking \eqref{expandX} and following the development for $\mathbf{W}_{j, 17}$ to $\mathbf{W}_{j, 25}$, it is easy to know that

\begin{eqnarray*}
\sum_{s=17}^{25} \|\mathbf{B}_{ij,s} \|_2=o_P \left(\frac{1}{\sqrt{T}} \|\mathbf{M}_{\widehat{\mathbf{C}}_{ij}}\mathbf{F}_{ij}^*\|_2 +\frac{1}{\sqrt{T}} \right).
\end{eqnarray*}

Thus, we can conclude that
\begin{eqnarray}\label{mge_e11}
\frac{1}{T} \|\mathbf{X}_{ij\centerdot}^\top \mathbf{M}_{\widehat{\mathbf{C}}_{ij}} \mathbf{F}_j^\bullet \pmb{\gamma}_{ij}^\bullet\|=o_P \left(\frac{1}{\sqrt{T}} \|\mathbf{M}_{\widehat{\mathbf{C}}_{ij}}\mathbf{F}_{ij}^*\|_2 +\frac{1}{\sqrt{T}} \right).
\end{eqnarray}

Further, by noting $i$ and $j$ are symmetric, we immediately obtain that

\begin{eqnarray*}
\frac{1}{T} \|\mathbf{X}_{ij\centerdot}^\top \mathbf{M}_{\widehat{\mathbf{C}}_{ij}} \mathbf{F}_i^\circ \pmb{\gamma}_{ij}^\circ\|=o_P \left(\frac{1}{\sqrt{T}} \|\mathbf{M}_{\widehat{\mathbf{C}}_{ij}}\mathbf{F}_{ij}^*\|_2 +\frac{1}{\sqrt{T}} \right).
\end{eqnarray*}

Thus, when driving the asymptotic distribution, we only need to focus on the following term

\begin{eqnarray*}
\frac{1}{\sqrt{T}}\mathbf{X}_{ij\centerdot}^\top \mathbf{M}_{\widehat{\mathbf{C}}_{ij}} \pmb{\mathcal{E}}_{ij\centerdot },
\end{eqnarray*}
which in connection with \eqref{expandX} infers that we just need to focus on

\begin{eqnarray*}
\frac{1}{\sqrt{T}}\mathbf{V}_{ij\centerdot}^\top\pmb{\mathcal{E}}_{ij\centerdot }.
\end{eqnarray*}

By standard large block and small block technique (cf., \citealp{Gao}), we immediately obtain  that for $\forall (i,j)$,

\begin{eqnarray*}
\frac{1}{\sqrt{T}}\mathbf{V}_{ij\centerdot}^\top\pmb{\mathcal{E}}_{ij\centerdot }\to_D N(\mathbf{0}, \pmb{\Sigma}_{\mathbf{v},\varepsilon,ij}),
\end{eqnarray*}
where $\pmb{\Sigma}_{\mathbf{v},\varepsilon,ij} =\lim_{T\to\infty}\frac{1}{T}E[\mathbf{V}_{ij\centerdot}^\top\pmb{\mathcal{E}}_{ij\centerdot } \pmb{\mathcal{E}}_{ij\centerdot }^\top \mathbf{V}_{ij\centerdot}]$. In connection with the fact that for $\forall (i,j)$,

\begin{eqnarray*}
\frac{1}{T}\mathbf{X}_{ij\centerdot}^\top \mathbf{M}_{\widehat{\mathbf{C}}_{ij}}\mathbf{X}_{ij\centerdot}\to_P\pmb{\Sigma}_{\mathbf{v},ij},
\end{eqnarray*}
the proof is now completed. 

(2) We now investigate the mean group estimators. For each $i$, recall that we have defined $\widetilde{\mathbf{b}}_{i\centerdot }=\frac{1}{N}\sum_{j=1}^N \widetilde{\mathbf{b}}_{ij}$ and $\boldsymbol{\beta}_{i\centerdot}=\frac{1}{N}\sum_{j=1}^N \boldsymbol{\beta}_{ij}$. Additionally,  we have
$\mathbf{X}_{ij\centerdot}=\mathbf{F}_{ij}^*\pmb{\phi}_{ij}^*+\mathbf{V}_{ij\centerdot}$,
where $\mathbf{F}_{ij}^*=(\mathbf{F}, \mathbf{F}_i^\circ,\mathbf{F}_j^\bullet)$ and $\pmb{\phi}_{ij}^*=(\pmb{\phi}_{ij}^\top,\pmb{\phi}_{ij}^{\circ\top}, \pmb{\phi}_{ij}^{\bullet\top} )^\top$.

Then, we have the following expansion for $\widetilde{\mathbf{b}}_{i\centerdot}$:
\begin{eqnarray*}
\widetilde{\mathbf{b}}_{i\centerdot }&=&\frac{1}{N}\sum_{j=1}^N  (\mathbf{X}_{ij\centerdot}^\top \mathbf{M}_{\widehat{\mathbf{C}}_{ij}} \mathbf{X}_{ij\centerdot} )^{-1} \mathbf{X}_{ij\centerdot}^\top \mathbf{M}_{\widehat{\mathbf{C}}_{ij}} \mathbf{Y}_{ij\centerdot}
\nonumber\\
&=& \pmb{\beta}_{i\centerdot}+\frac{1}{N}\sum_{j=1}^N(\mathbf{X}_{ij\centerdot}^\top \mathbf{M}_{\widehat{\mathbf{C}}_{ij}} \mathbf{X}_{ij\centerdot} )^{-1}\pmb{\phi}_{ij}^{*\top} \mathbf{F}_{ij}^{*\top}\mathbf{M}_{\widehat{\mathbf{C}}_{ij}}\mathbf{F}_{ij}^* \pmb{\gamma}_{ij}^*
\nonumber \\
&&+\frac{1}{N}\sum_{j=1}^N(\mathbf{X}_{ij\centerdot}^\top \mathbf{M}_{\widehat{\mathbf{C}}_{ij}} \mathbf{X}_{ij\centerdot} )^{-1}\mathbf{V}_{ij\centerdot}^\top\mathbf{M}_{\widehat{\mathbf{C}}_{ij}}\mathbf{F}_{ij}^* \pmb{\gamma}_{ij}^*
\nonumber\\
&&+\frac{1}{N}\sum_{j=1}^N(\mathbf{X}_{ij\centerdot}^\top \mathbf{M}_{\widehat{\mathbf{C}}_{ij}} \mathbf{X}_{ij\centerdot} )^{-1}\pmb{\phi}_{ij}^{*\top} \mathbf{F}_{ij}^{*\top}\mathbf{M}_{\widehat{\mathbf{C}}_{ij}}\pmb{\mathcal{E}}_{ij\centerdot }
\nonumber\\
&&+\frac{1}{N}\sum_{j=1}^N(\mathbf{X}_{ij\centerdot}^\top \mathbf{M}_{\widehat{\mathbf{C}}_{ij}} \mathbf{X}_{ij\centerdot} )^{-1}\mathbf{V}_{ij\centerdot}^\top\mathbf{M}_{\widehat{\mathbf{C}}_{ij}}\pmb{\mathcal{E}}_{ij\centerdot }
\nonumber\\
&:=&\pmb{\beta}_{i\centerdot}+\mathbf{D}_{i,1}+\cdots+\mathbf{D}_{i,4},
\end{eqnarray*}
where $\pmb{\gamma}_{ij}^* = (\pmb{\gamma}_{ij}^\top, \pmb{\gamma}_{ij}^{\circ\top}, \pmb{\gamma}_{ij}^{\bullet\top} )^\top$ and the definitions of $\mathbf{D}_{i,1},\ldots,\mathbf{D}_{i,4}$ are obvious. In what follows, we study the these terms one by one. For the simplicity of  notation, we define $\widehat{\boldsymbol{\Xi}}_{ij}=T^{-1}\mathbf{X}_{ij\centerdot}^\top \mathbf{M}_{\widehat{\mathbf{C}}_{ij}} \mathbf{X}_{ij\centerdot}$.  

For $\mathbf{D}_{i,1}$, we have
\begin{eqnarray*}
\mathbf{D}_{i,1}&=&\frac{1}{NT}\sum_{j=1}^N\pmb{\Sigma}_{\mathbf{v},ij}^{-1} \pmb{\phi}_{ij}^{*\top} \mathbf{F}_{ij}^{*\top} \mathbf{M}_{\widehat{\mathbf{C}}_{ij}} \mathbf{F}_{ij}^* \pmb{\gamma}_{ij}^*
+\frac{1}{NT}\sum_{j=1}^N\big(\widehat{\boldsymbol{\Xi}}_{ij}^{-1}-\pmb{\Sigma}_{\mathbf{v},ij}^{-1} \big) \pmb{\phi}_{ij}^{*\top} \mathbf{F}_{ij}^{*\top} \mathbf{M}_{\widehat{\mathbf{C}}_{ij}} \mathbf{F}_{ij}^* \pmb{\gamma}_{ij}^*
\nonumber\\
&:=&\mathbf{D}_{i,1,1}+\mathbf{D}_{i,1,2}.
\end{eqnarray*}
To study the order of these terms, we can further write
\begin{eqnarray*}
\mathbf{F}_{ij}^{*\top} \mathbf{M}_{\widehat{\mathbf{C}}_{ij}} \mathbf{F}_{ij}^*
&=&\left(
\begin{array}{c }
\big(\mathbf{F}-\widehat{\mathbf{C}}\mathbf{H}^{-1}\big)^\top\\
\big(\mathbf{M}_{\widehat{\mathbf{C}}}\mathbf{F}_i^\circ-\widehat{\mathbf{C}}_i^\circ\mathbf{H}_i^{\circ -1}\big)^\top\\
\big(\mathbf{M}_{\widehat{\mathbf{C}}}\mathbf{F}_j^\bullet-\widehat{\mathbf{C}}_j^\bullet\mathbf{H}_j^{\bullet -1}\big)^\top\\
\end{array}
\right)\mathbf{M}_{\widehat{\mathbf{C}}_{ij}}\Big(\mathbf{F}-\widehat{\mathbf{C}}\mathbf{H}^{-1}, \mathbf{M}_{\widehat{\mathbf{C}}}\mathbf{F}_i^\circ-\widehat{\mathbf{C}}_i^\circ\mathbf{H}_i^{\circ -1}, \mathbf{M}_{\widehat{\mathbf{C}}}\mathbf{F}_j^\bullet-\widehat{\mathbf{C}}_j^\bullet\mathbf{H}_j^{\bullet -1}\Big)
\nonumber\\
&:=&\left(
\begin{array}{c c c }
\mathcal{M}_{ij,11},\cdots, \mathcal{M}_{ij,13}\\
\vdots,\vdots,\vdots\\
\mathcal{M}_{ij,31},\cdots, \mathcal{M}_{ij,33}\\
\end{array}\right),
\end{eqnarray*}
where the first equality holds by the orthogonality between $\widehat{\mathbf{C}}$, $\widehat{\mathbf{C}}_i^\circ$ and $\widehat{\mathbf{C}}_j^\bullet$,  $\mathbf{H}=\
\frac{1}{LNT}\pmb{\Gamma}_{\centerdot \centerdot}^\top \pmb{\Gamma}_{\centerdot \centerdot}\mathbf{F}^\top \widehat{\mathbf{C}}\widehat{\mathbf{V}}^{-1}$, $\mathbf{H}_i^\circ =\frac{1}{NT} \pmb{\Gamma}_{i\centerdot}^{\circ\top} \pmb{\Gamma}_{i\centerdot}^\circ \mathbf{F}_i^{\circ\top} \widehat{\mathbf{C}}_i^\circ \widehat{\mathbf{V}}_i^{\circ -1}$ and $\mathbf{H}_j^\bullet =\frac{1}{LT} \pmb{\Gamma}_{\centerdot j}^{\bullet\top} \pmb{\Gamma}_{\centerdot j}^\bullet \mathbf{F}_j^{\bullet\top} \widehat{\mathbf{C}}_j^\bullet \widehat{\mathbf{V}}_j^{\bullet -1}$. Here, $\mathcal{M}_{ij,11} =(\mathbf{F}-\widehat{\mathbf{C}}\mathbf{H}^{-1})^\top\mathbf{M}_{\widehat{\mathbf{C}}_{ij}}(\mathbf{F}-\widehat{\mathbf{C}}\mathbf{H}^{-1})$, and the other block matrices are defined in a similar manner.

Then, for the term with $\mathcal{M}_{ij,11}$, the expansion in \eqref{EQA.4} gives
\begin{eqnarray}
\frac{1}{NT}\sum_{j=1}^N\pmb{\Sigma}_{\mathbf{v},ij}^{-1}\pmb{\phi}_{ij}^\top \mathcal{M}_{ij,11} \pmb{\gamma}_{ij}
&=&\frac{1}{NT}\sum_{j=1}^N \pmb{\Sigma}_{\mathbf{v},ij}^{-1} \pmb{\phi}_{ij}^\top \Big(\sum_{s_1\neq 10}^{25}\mathbf{J}_{s_1}\pmb{\Pi}\Big)^\top
\mathbf{M}_{\widehat{\mathbf{C}}_{ij}} \Big(\sum_{s_2\neq 10}^{25}\mathbf{J}_{s_2}\pmb{\Pi}\Big)\pmb{\gamma}_{ij}
\nonumber\\
&:=&\sum_{s_1\neq 10}^{25}\sum_{s_2\neq 10}^{25}\mathcal{M}^\ast_{i,11, s_1s_2},
\end{eqnarray}
where $\pmb{\Pi} = (\mathbf{F}^\top\widehat{\mathbf{C}}/T)^{-1}(\pmb{\Gamma}_{\centerdot \centerdot }^\top \pmb{\Gamma}_{\centerdot \centerdot }/(LN))^{-1}$ and $\mathcal{M}^\ast_{i,11, s_1s_2}=\frac{1}{NT}\sum_{j=1}^N \pmb{\Sigma}_{\mathbf{v},ij}^{-1} \pmb{\phi}_{ij}^\top\pmb{\Pi}^\top\mathbf{J}_{s_1}^\top
\mathbf{M}_{\widehat{\mathbf{C}}_{ij}}\mathbf{J}_{s_2}\pmb{\Pi}\pmb{\gamma}_{ij}$.

For the term with $\mathbf{J}_1$, simple algebra yields that 
\begin{eqnarray*}
\sum_{s_2\neq 10}\mathcal{M}^\ast_{i,11, 1s_2}=O_P\left(\max_{j} \|\pmb{\beta}_{ij}-\widehat{\mathbf{b}}_{ij}\| ^2\cdot\frac{1}{\sqrt{T}}\|\mathbf{F}-\widehat{\mathbf{C}}\mathbf{H}^{-1}\|_2\right),
\end{eqnarray*}
which is negligible.
For the term with $\mathbf{J}_2$, 
\begin{eqnarray*}
\mathbf{J}_2
&=&\frac{1}{LNT}\sum_{i,j}\mathbf{F}\pmb{\phi}_{ij}(\pmb{\beta}_{ij}-\widehat{\mathbf{b}}_{ij}) \pmb{\gamma}_{ij}^{\top} \mathbf{F}^{\top} \widehat{\mathbf{C}}+\frac{1}{LNT}\sum_{i,j}\mathbf{F}_i^\circ\pmb{\phi}_{ij}^\circ(\pmb{\beta}_{ij}-\widehat{\mathbf{b}}_{ij}) \pmb{\gamma}_{ij}^{\top} \mathbf{F}^{\top} \widehat{\mathbf{C}}
\nonumber\\
&&+\frac{1}{LNT}\sum_{i,j}\mathbf{F}_j^\bullet\pmb{\phi}_{ij}^\bullet(\pmb{\beta}_{ij}-\widehat{\mathbf{b}}_{ij}) \pmb{\gamma}_{ij}^{\top} \mathbf{F}^{\top} \widehat{\mathbf{C}}
+\frac{1}{LNT}\sum_{i,j}\mathbf{V}_{ij\centerdot}(\pmb{\beta}_{ij}-\widehat{\mathbf{b}}_{ij}) \pmb{\gamma}_{ij}^{\top} \mathbf{F}^{\top} \widehat{\mathbf{C}}
\nonumber\\
&:=&\mathbf{J}_{2,1}+\cdots+\mathbf{J}_{2,4}.
\end{eqnarray*}
For $\mathbf{J}_{2,1}$, it is clear to see that 
\begin{equation}\label{mge_e5}
\frac{1}{NT}\sum_{j=1} \pmb{\Sigma}_{\mathbf{v},ij}^{-1} \pmb{\phi}_{ij}^\top \pmb{\Pi}^\top\mathbf{J}_{2,1}^\top
\mathbf{M}_{\widehat{\mathbf{C}}_{ij}} \Big(\sum_{s_2\neq 10}^{25}\mathbf{J}_{s_2}\pmb{\Pi}\Big)\pmb{\gamma}_{ij}=O_P\left(\max_{j} \|\pmb{\beta}_{ij}-\widehat{\mathbf{b}}_{ij}\| \cdot\frac{1}{T}\|\mathbf{F}-\widehat{\mathbf{C}}\mathbf{H}^{-1}\|^2_2\right).
\end{equation}

For the second term in $\mathbf{J}_2$,   let $\pmb{\Phi}_{\beta,j} =(\pmb{\gamma}_{1j}(\pmb{\beta}_{1j}-\widehat{\mathbf{b}}_{1j})^{\top} \pmb{\phi}_{1j}^{\circ\top}, \cdots, \pmb{\gamma}_{Lj}(\pmb{\beta}_{Lj}-\widehat{\mathbf{b}}_{Lj}) ^{\top}\pmb{\phi}_{Lj}^{\circ\top})^\top$. Then, we have
\begin{eqnarray}\label{mge_e14}
\frac{1}{LNT}\left\|\sum_{i,j}\mathbf{F}_i^\circ\pmb{\phi}_{ij}^\circ(\pmb{\beta}_{ij}-\widehat{\mathbf{b}}_{ij}) \pmb{\gamma}_{ij}^{\top} \mathbf{F}^{\top} \widehat{\mathbf{C}}\right\|_2
&=&\frac{1}{LNT}\left\|\sum_{j}\mathbf{F}^\circ\pmb{\Phi}_{\beta,j}\mathbf{F}^{\top} \widehat{\mathbf{C}}\right\|_2
\nonumber\\
&=& O_P\left(\frac{1}{\sqrt{L}}\left\|\mathbf{F}^\circ\right\|_2\max_{i,j}\|\pmb{\beta}_{ij}-\widehat{\mathbf{b}}_{ij}\|\right)
\nonumber\\
&=&O_P\left(\frac{\sqrt{L\vee T}}{\sqrt{L}}\max_{i,j}\|\pmb{\beta}_{ij}-\widehat{\mathbf{b}}_{ij}\|\right),
\end{eqnarray}
where the last equality holds by Assumption \ref{AS1}. It further yields
\begin{eqnarray*}
\frac{1}{NT}\sum_{j=1} \pmb{\Sigma}_{\mathbf{v},ij}^{-1} \pmb{\phi}_{ij}^\top \pmb{\Pi}^\top\mathbf{J}_{2,2}^\top
\mathbf{M}_{\widehat{\mathbf{C}}_{ij}} \Big(\sum_{s_2\neq 10}^{25}\mathbf{J}_{s_2}\pmb{\Pi}\Big)\pmb{\gamma}_{ij}=O_P\left(\frac{1}{\sqrt{L\wedge T}}\max_{i,j}\|\pmb{\beta}_{ij}-\widehat{\mathbf{b}}_{ij}\| \cdot\frac{1}{\sqrt{T}}\|\mathbf{F}-\widehat{\mathbf{C}}\mathbf{H}^{-1}\|_2\right).
\end{eqnarray*}

Analogously, the following result holds for $\mathbf{J}_{2,3}$:
\begin{eqnarray*}
\frac{1}{LNT}\left\|\sum_{i,j}\mathbf{F}_j^\bullet\pmb{\phi}_{ij}^\bullet(\pmb{\beta}_{ij}-\widehat{\mathbf{b}}_{ij}) \pmb{\gamma}_{ij}^{\top} \mathbf{F}^{\top} \widehat{\mathbf{C}}\right\|_2&=&O_P\left(\frac{\sqrt{N\vee T}}{\sqrt{N}}\max_{i,j}\|\pmb{\beta}_{ij}-\widehat{\mathbf{b}}_{ij}\|\right),
\end{eqnarray*}
which immediately gives
\begin{eqnarray*}
\frac{1}{NT}\sum_{j=1} \pmb{\Sigma}_{\mathbf{v},ij}^{-1} \pmb{\phi}_{ij}^\top \pmb{\Pi}^\top\mathbf{J}_{2,3}^\top
\mathbf{M}_{\widehat{\mathbf{C}}_{ij}} \Big(\sum_{s_2\neq 10}^{25}\mathbf{J}_{s_2}\pmb{\Pi}\Big)\pmb{\gamma}_{ij}=O_P\left(\frac{1}{\sqrt{N\wedge T}}\max_{i,j}\|\pmb{\beta}_{ij}-\widehat{\mathbf{b}}_{ij}\| \cdot\frac{1}{\sqrt{T}}\|\mathbf{F}-\widehat{\mathbf{C}}\mathbf{H}^{-1}\|_2\right).
\end{eqnarray*}

For $\mathbf{J}_{2,4}$, by Assumption \ref{AS1}, 
\begin{eqnarray}\label{mge_e10}
\|\mathbf{J}_{2,4}\|_2&=&\frac{1}{LNT}\left\|\mathbf{V}_{\centerdot\centerdot\centerdot}^\top\pmb{\Phi}^\dag_{\beta}\mathbf{F}^{\top} \widehat{\mathbf{C}}\right\|_2
\nonumber\\
&=& O_P\left(\frac{1}{\sqrt{LN}}\left\|\mathbf{V}_{\centerdot\centerdot\centerdot}\right\|_2\max_{i,j}\|\pmb{\beta}_{ij}-\widehat{\mathbf{b}}_{ij}\|\right)
\nonumber\\
&=&O_P\left(\frac{\sqrt{LN\vee T}}{\sqrt{LN}}\max_{i,j}\|\pmb{\beta}_{ij}-\widehat{\mathbf{b}}_{ij}\|\right),
\end{eqnarray} 
where $\pmb{\Phi}^\dag_{\beta} =(\pmb{\gamma}_{11}(\pmb{\beta}_{11}-\widehat{\mathbf{b}}_{11})^{\top} , \cdots, \pmb{\gamma}_{LN}(\pmb{\beta}_{LN}-\widehat{\mathbf{b}}_{LN}) ^{\top})^\top$. In summary, these results can jointly imply that $\sum_{s_2\neq 10}^{25}\mathcal{M}^\ast_{i,11, 2s_2}$ is negligible.

For $\mathbf{J}_3=\frac{1}{LNT}\sum_{i,j} \mathbf{F}  \pmb{\gamma}_{ij} (\pmb{\beta}_{ij}-\widehat{\mathbf{b}}_{ij})^\top \mathbf{X}_{ij\centerdot}^\top \widehat{\mathbf{C}}$, considering  $\mathbf{M}_{\widehat{\mathbf{C}}_{ij}}\mathbf{F}=\mathbf{M}_{\widehat{\mathbf{C}}_{ij}}(\mathbf{F}-\widehat{\mathbf{C}}\mathbf{H}^{-1})$, we have
\begin{equation}\label{mge_e6}
\sum_{s_2\neq 10}^{25}\mathcal{M}^\ast_{i,11, 3s_2}=O_P\left(\max_{i,j} \|\pmb{\beta}_{ij}-\widehat{\mathbf{b}}_{ij}\| \cdot\frac{1}{T}\|\mathbf{F}-\widehat{\mathbf{C}}\mathbf{H}^{-1}\|^2_2\right).
\end{equation}

For $\mathbf{J}_4$, we write
\begin{eqnarray*}
\mathbf{J}_4
&=&\frac{1}{LNT}\sum_{i,j}\mathbf{F}\pmb{\phi}_{ij}(\pmb{\beta}_{ij}-\widehat{\mathbf{b}}_{ij}) \pmb{\gamma}_{ij}^{\circ\top}  \mathbf{F}_i^{\circ\top}\widehat{\mathbf{C}}+\frac{1}{LNT}\sum_{i,j}\mathbf{F}_i^\circ\pmb{\phi}_{ij}^\circ(\pmb{\beta}_{ij}-\widehat{\mathbf{b}}_{ij}) \pmb{\gamma}_{ij}^{\circ\top}  \mathbf{F}_i^{\circ\top}\widehat{\mathbf{C}}
\nonumber\\
&&+\frac{1}{LNT}\sum_{i,j}\mathbf{F}_j^\bullet\pmb{\phi}_{ij}^\bullet(\pmb{\beta}_{ij}-\widehat{\mathbf{b}}_{ij}) \pmb{\gamma}_{ij}^{\circ\top}  \mathbf{F}_i^{\circ\top} \widehat{\mathbf{C}}
+\frac{1}{LNT}\sum_{i,j}\mathbf{V}_{ij\centerdot}(\pmb{\beta}_{ij}-\widehat{\mathbf{b}}_{ij}) \pmb{\gamma}_{ij}^{\circ\top}  \mathbf{F}_i^{\circ\top}\widehat{\mathbf{C}}
\nonumber\\
&:=&\mathbf{J}_{4,1}+\cdots+\mathbf{J}_{4,4}.
\end{eqnarray*} 
For $\mathbf{J}_{4,1}$, using analogous arguments to those in \eqref{mge_e5} and \eqref{mge_e6}, we can readily obtain
\begin{equation}\label{mge_e7}
\frac{1}{NT}\sum_{j=1} \pmb{\Sigma}_{\mathbf{v},ij}^{-1} \pmb{\phi}_{ij}^\top \pmb{\Pi}^\top\mathbf{J}_{4,1}^\top
\mathbf{M}_{\widehat{\mathbf{C}}_{ij}} \Big(\sum_{s_2\neq 10}^{25}\mathbf{J}_{s_2}\pmb{\Pi}\Big)\pmb{\gamma}_{ij}=O_P\left(\max_{j} \|\pmb{\beta}_{ij}-\widehat{\mathbf{b}}_{ij}\| \cdot\frac{1}{T}\|\mathbf{F}-\widehat{\mathbf{C}}\mathbf{H}^{-1}\|^2_2\right).
\end{equation}
Define $\pmb{\Phi}^\ast_{\beta,j} =(T^{-1}\widehat{\mathbf{C}}^\top \mathbf{F}_1^{\circ}\pmb{\gamma}^\circ_{1j}(\pmb{\beta}_{1j}-\widehat{\mathbf{b}}_{1j})^{\top} \pmb{\phi}_{1j}^{\circ\top}, \cdots, T^{-1}\widehat{\mathbf{C}}^\top \mathbf{F}_L^{\circ}\pmb{\gamma}^\circ_{Lj}(\pmb{\beta}_{Lj}-\widehat{\mathbf{b}}_{Lj}) ^{\top}\pmb{\phi}_{Lj}^{\circ\top})^\top$. By Assumption \ref{AS1},
\begin{eqnarray*}
\frac{1}{LNT}\left\|\sum_{i,j}\mathbf{F}_i^\circ\pmb{\phi}_{ij}^\circ(\pmb{\beta}_{ij}-\widehat{\mathbf{b}}_{ij}) \pmb{\gamma}_{ij}^{\circ\top}  \mathbf{F}_i^{\circ\top}\widehat{\mathbf{C}}\right\|_2
&=&\frac{1}{LN}\left\|\sum_{j}\mathbf{F}^\circ\pmb{\Phi}_{\beta,j}^\ast\right\|_2
\nonumber\\
&=& O_P\left(\frac{1}{\sqrt{L}}\left\|\mathbf{F}^\circ\right\|_2\max_{i,j}\|\pmb{\beta}_{ij}-\widehat{\mathbf{b}}_{ij}\|\right)
\nonumber\\
&=&O_P\left(\frac{\sqrt{L\vee T}}{\sqrt{L}}\max_{i,j}\|\pmb{\beta}_{ij}-\widehat{\mathbf{b}}_{ij}\|\right).
\end{eqnarray*} 
Therefore, it holds for $\mathbf{J}_{4,2}$ that
\begin{eqnarray}\label{mge_e8}
\frac{1}{NT}\sum_{j=1}^N \pmb{\Sigma}_{\mathbf{v},ij}^{-1} \pmb{\phi}_{ij}^\top \pmb{\Pi}^\top\mathbf{J}_{4,2}^\top
\mathbf{M}_{\widehat{\mathbf{C}}_{ij}} \Big(\sum_{s_2\neq 10}^{25}\mathbf{J}_{s_2}\pmb{\Pi}\Big)\pmb{\gamma}_{ij}=O_P\left(\frac{1}{\sqrt{L\wedge T}}\max_{i,j}\|\pmb{\beta}_{ij}-\widehat{\mathbf{b}}_{ij}\| \cdot\frac{1}{\sqrt{T}}\|\mathbf{F}-\widehat{\mathbf{C}}\mathbf{H}^{-1}\|_2\right).
\nonumber\\
\end{eqnarray}
Analogously, for $\mathbf{J}_{4,3}$ and $\mathbf{J}_{4,4}$, 
\begin{eqnarray}\label{mge_e9}
\frac{1}{NT}\sum_{j=1}^N \pmb{\Sigma}_{\mathbf{v},ij}^{-1} \pmb{\phi}_{ij}^\top \pmb{\Pi}^\top\mathbf{J}_{4,3}^\top
\mathbf{M}_{\widehat{\mathbf{C}}_{ij}} \Big(\sum_{s_2\neq 10}^{25}\mathbf{J}_{s_2}\pmb{\Pi}\Big)\pmb{\gamma}_{ij}&=&O_P\left(\frac{1}{\sqrt{N\wedge T}}\max_{i,j}\|\pmb{\beta}_{ij}-\widehat{\mathbf{b}}_{ij}\| \cdot\frac{1}{\sqrt{T}}\|\mathbf{F}-\widehat{\mathbf{C}}\mathbf{H}^{-1}\|_2\right),
\nonumber\\
\frac{1}{NT}\sum_{j=1}^N \pmb{\Sigma}_{\mathbf{v},ij}^{-1} \pmb{\phi}_{ij}^\top \pmb{\Pi}^\top\mathbf{J}_{4,L}^\top
\mathbf{M}_{\widehat{\mathbf{C}}_{ij}} \Big(\sum_{s_2\neq 10}^{25}\mathbf{J}_{s_2}\pmb{\Pi}\Big)\pmb{\gamma}_{ij}&=&O_P\left(\frac{1}{\sqrt{L\wedge T}}\max_{i,j}\|\pmb{\beta}_{ij}-\widehat{\mathbf{b}}_{ij}\| \cdot\frac{1}{\sqrt{T}}\|\mathbf{F}-\widehat{\mathbf{C}}\mathbf{H}^{-1}\|_2\right).
\nonumber\\
\end{eqnarray}
In summary of \eqref{mge_e7}, \eqref{mge_e8}, and \eqref{mge_e9}, 

\begin{eqnarray*}
\sum_{s_2\neq 10}^{25}\mathcal{M}^\ast_{i,11, 4s_2}=O_P\left(\frac{1}{\sqrt{L\wedge N \wedge T}}\max_{i,j}\|\pmb{\beta}_{ij}-\widehat{\mathbf{b}}_{ij}\| \cdot\frac{1}{\sqrt{T}}\|\mathbf{F}-\widehat{\mathbf{C}}\mathbf{H}^{-1}\|_2\right),
\end{eqnarray*}
Analogously, we can show that $\sum_{s_2\neq 10}^{25}\mathcal{M}^\ast_{i,11, 5s_2}$, $\sum_{s_2\neq 10}^{25}\mathcal{M}^\ast_{i,11, 6s_2}$ and $\sum_{s_2\neq 10}^{25}\mathcal{M}^\ast_{i,11, 7s_2}$ are all negligible. For $\mathbf{J}_8$ and $\mathbf{J}_9$, together with the condition $\|\pmb{\mathcal{E}}_{\centerdot \centerdot\centerdot}\|_2=O_P(LN\wedge T)$, similar arguments to those in \eqref{mge_e10} can be applied here to show that
\begin{eqnarray*}
\sum_{s_2\neq 10}^{25}\mathcal{M}^\ast_{i,11, 8s_2}=O_P\left(\frac{1}{\sqrt{L N \wedge T}}\max_{i,j}\|\pmb{\beta}_{ij}-\widehat{\mathbf{b}}_{ij}\| \cdot\frac{1}{\sqrt{T}}\|\mathbf{F}-\widehat{\mathbf{C}}\mathbf{H}^{-1}\|_2\right),
\end{eqnarray*}
and 
\begin{eqnarray*}
\sum_{s_2\neq 10}^{25}\mathcal{M}^\ast_{i,11, 9s_2}=O_P\left(\frac{1}{\sqrt{L N \wedge T}}\max_{i,j}\|\pmb{\beta}_{ij}-\widehat{\mathbf{b}}_{ij}\| \cdot\frac{1}{\sqrt{T}}\|\mathbf{F}-\widehat{\mathbf{C}}\mathbf{H}^{-1}\|_2\right).
\end{eqnarray*}
 Let $\pmb{\Phi}_{\gamma,j}=(\pmb{\gamma}_{1j} \pmb{\gamma}_{1j}^{\circ\top},\cdots, \pmb{\gamma}_{Lj} \pmb{\gamma}_{Lj}^{\circ\top}  )^\top$. For $\mathbf{J}_{11}$, it follows Assumption \ref{AS1} that
\begin{eqnarray*}
\frac{1}{LN}\left\|\sum_{i,j}  \pmb{\gamma}_{ij} \pmb{\gamma}_{ij}^{\circ\top} \mathbf{F}_i^{\circ\top}\right\|_2&\leq & \frac{1}{LN}\sum_{j} \left\| \pmb{\Phi}_{\gamma,j}\right\|\cdot\left\|\mathbf{F}^{\circ}\right\|_2
\nonumber\\
&=&O_P\left(\frac{\sqrt{L\vee T}}{\sqrt{L}}\right).
\end{eqnarray*}
Together with the fact $\mathbf{M}_{\widehat{\mathbf{C}}_{ij}}\mathbf{F}=\mathbf{M}_{\widehat{\mathbf{C}}_{ij}}(\mathbf{F}-\widehat{\mathbf{C}}\mathbf{H}^{-1})$, it yields 
\begin{eqnarray*}
\sum_{s_2\neq 10}^{25}\mathcal{M}^\ast_{i,11, 11s_2}=O_P\left(\frac{1}{\sqrt{L  \wedge T}} \cdot\frac{1}{T}\|\mathbf{F}-\widehat{\mathbf{C}}\mathbf{H}^{-1}\|^2_2\right).
\end{eqnarray*}
Using analogous arguments, we can obtain 
\begin{eqnarray*}
\sum_{s_2\neq 10}^{25}\mathcal{M}^\ast_{i,11, 13s_2}=O_P\left(\frac{1}{\sqrt{N  \wedge T}} \cdot\frac{1}{T}\|\mathbf{F}-\widehat{\mathbf{C}}\mathbf{H}^{-1}\|^2_2\right).
\end{eqnarray*}
For $\mathbf{J}_{15}$ and $\mathbf{J}_{16}$, since $\left\| \pmb{\mathcal{E}}_{\centerdot\centerdot\centerdot}\right\|_2=O_P(\sqrt{LN\vee T})$,
\begin{eqnarray}
\frac{1}{LN}\left\|\sum_{i,j} \pmb{\mathcal{E}}_{ij\centerdot} \pmb{\gamma}_{ij} ^\top\right\|_2&\leq & \frac{1}{LN}\left\| \pmb{\mathcal{E}}_{\centerdot\centerdot\centerdot}\right\|_2\left\|\pmb{\Gamma}_{\centerdot\centerdot}\right\|_2
\nonumber\\
&=&O_P\left(\frac{\sqrt{LN\vee T}}{\sqrt{LN}}\right),
\end{eqnarray}
which immediately yields
\begin{eqnarray*}
\sum_{s_2\neq 10}^{25}\mathcal{M}^\ast_{i,11, 15s_2}=O_P\left(\frac{1}{\sqrt{LN  \wedge T}} \cdot\frac{1}{T}\|\mathbf{F}-\widehat{\mathbf{C}}\mathbf{H}^{-1}\|^2_2\right),
\end{eqnarray*}
and 
\begin{eqnarray*}
\sum_{s_2\neq 10}^{25}\mathcal{M}^\ast_{i,11, 16s_2}=O_P\left(\frac{1}{\sqrt{LN  \wedge T}} \cdot\frac{1}{T}\|\mathbf{F}-\widehat{\mathbf{C}}\mathbf{H}^{-1}\|^2_2\right).
\end{eqnarray*}
Analogously,  we can show that $\sum_{s_2\neq 10}^{25}\mathcal{M}^\ast_{i,11, 20s_2}$, $\sum_{s_2\neq 10}^{25}\mathcal{M}^\ast_{i,11, 21s_2}$, $\sum_{s_2\neq 10}^{25}\mathcal{M}^\ast_{i,11, 23s_2}$, $\sum_{s_2\neq 10}^{25}\mathcal{M}^\ast_{i,11, 24s_2}$ and $\sum_{s_2\neq 10}^{25}\mathcal{M}^\ast_{i,11, 25s_2}$ are all negligible. Let $\pmb{\Phi}^\ast_{\gamma,j}=\big( \pmb{\gamma}_{1j}^{\circ} \pmb{\gamma}_{1j}^{\bullet\top},\cdots, \pmb{\gamma}_{Lj}^{\circ} \pmb{\gamma}_{Lj}^{\bullet\top}\big)^\top$ and $\pmb{\Phi}^\ast_{\gamma}=(\pmb{\Phi}^\ast_{\gamma,1},\cdots, \pmb{\Phi}^\ast_{\gamma,N})$. For $\mathbf{J}_{17}$, it holds by Assumption \ref{AS1} that
\begin{eqnarray*}
\frac{1}{LNT}\left\|\sum_{i,j} \mathbf{F}_j^\bullet \pmb{\gamma}_{ij}^\bullet  \pmb{\gamma}_{ij}^{\circ\top} \mathbf{F}_i^{\circ\top}\right\|_2&=&\frac{1}{LNT}\left\|\mathbf{F}^\bullet\pmb{\Phi}^\ast_{\gamma}\mathbf{F}^{\circ\top}\right\|_2
\nonumber\\
&\leq& \frac{1}{LNT}\left\|\mathbf{F}^\bullet\right\|_2\left\|\mathbf{F}^\circ\right\|_2\left\|\pmb{\Phi}^\ast_{\gamma}\right\|_2
\nonumber\\
&=&O_P\left(\frac{1}{\sqrt{L  \wedge T}\sqrt{N  \wedge T}}\right).
\end{eqnarray*}
Therefore, 
\begin{equation}\label{mge_e17}
\sum_{s_2\neq 10}^{25}\mathcal{M}^\ast_{i,11, 17s_2}=O_P\left(\frac{1}{\sqrt{L  \wedge T}\sqrt{N  \wedge T}}\cdot\frac{1}{\sqrt{T}}\|\mathbf{F}-\widehat{\mathbf{C}}\mathbf{H}^{-1}\|_2 \right).
\end{equation}
Analogously to \eqref{mge_e17}, $\sum_{s_2\neq 10}^{25}\mathcal{M}^\ast_{i,11, 18s_2}$, $\sum_{s_2\neq 10}^{25}\mathcal{M}^\ast_{i,11, 19s_2}$ and $\sum_{s_2\neq 10}^{25}\mathcal{M}^\ast_{i,11, 22s_2}$ are negligible.

For the terms with $\mathbf{J}_{12}$ and $\mathbf{J}_{14}$, by \eqref{mge_e19} and \eqref{mge_e20}, 
\begin{eqnarray*}
\sum_{s_2\neq 10}^{25}\mathcal{M}^\ast_{i,11, 12s_2}&=&O_P\left(\frac{1}{\sqrt{L  \wedge T}}\cdot\frac{1}{\sqrt{T}}\|\mathbf{F}-\widehat{\mathbf{C}}\mathbf{H}^{-1}\|_2 \right),
\nonumber\\
\sum_{s_2\neq 10}^{25}\mathcal{M}^\ast_{i,11, 14s_2}&=&O_P\left(\frac{1}{\sqrt{N  \wedge T}}\cdot\frac{1}{\sqrt{T}}\|\mathbf{F}-\widehat{\mathbf{C}}\mathbf{H}^{-1}\|_2 \right).
\end{eqnarray*}
Following from the established results, we have
\begin{eqnarray}\label{mge_e12}
\frac{1}{NT}\sum_{j=1}^N\pmb{\Sigma}_{\mathbf{v},ij}^{-1}\pmb{\phi}_{ij}^\top \mathcal{M}_{ij,11} \pmb{\gamma}_{ij}
&=&\sum_{s_1= 12,14}\sum_{s_2= 12,14}\mathcal{M}^\ast_{i,11, s_1s_2}+o_P\left(\frac{1}{\sqrt{NT}}\right),
\nonumber\\
&=& \frac{1}{NT}\sum_{j=1}^N\pmb{\Sigma}_{\mathbf{v},ij}^{-1}\pmb{\phi}_{ij}^\top \mathcal{R}_F^\top \mathbf{M}_{\widehat{\mathbf{C}}_{ij}} \mathcal{R}_F \pmb{\gamma}_{ij}+o_P\left(\frac{1}{\sqrt{NT}}\right)
,
\end{eqnarray}
where $\mathcal{R}_F = \sum_{ij}(\mathbf{F}_{i}^{\circ}\pmb{\gamma}_{ij}^\circ+\mathbf{F}_{j}^{\bullet}\pmb{\gamma}_{ij}^\bullet)\pmb{\gamma}_{ij}^\top (\pmb{\Gamma}_{\centerdot \centerdot }^\top \pmb{\Gamma}_{\centerdot \centerdot })^{-1}$, and  
\begin{eqnarray}\label{mge_e23}
\frac{1}{NT}\left\|\sum_{j=1}^N\pmb{\Sigma}_{\mathbf{v},ij}^{-1}\pmb{\phi}_{ij}^\top \mathcal{R}_F^\top \mathbf{M}_{\widehat{\mathbf{C}}_{ij}} \mathcal{R}_F \pmb{\gamma}_{ij}\right\|=O_P\left(\frac{1}{L\wedge N \wedge T}\right).
\end{eqnarray}

We now proceed to investigate the remaining terms in $\mathbf{D}_{i,1,1}$.
Noteworthily, $\mathcal{M}_{ij,22}$, $\mathcal{M}_{ij,33}$, $\mathcal{M}_{ij,12}$,  $\mathcal{M}_{ij,21}$, $\mathcal{M}_{ij,13}$, $\mathcal{M}_{ij,31}$, $\mathcal{M}_{ij,23}$ and $\mathcal{M}_{ij,32}$ have similar structures. To avoid repetitive and tedious calculations, we only provide detailed derivations for $\mathcal{M}_{ij,13}$. For the term with $\mathcal{M}_{ij,13}$,
By \eqref{EQA.4} and \eqref{Expan_J},
\begin{eqnarray*}
\frac{1}{NT}\sum_{j=1}^N\pmb{\Sigma}_{\mathbf{v},ij}^{-1}\pmb{\phi}_{ij}^{\bullet\top} \mathcal{M}_{ij,31} \pmb{\gamma}_{ij}
&=&\frac{1}{NT}\sum_{j=1}^N\pmb{\Sigma}_{\mathbf{v},ij}^{-1}\pmb{\phi}_{ij}^{\bullet\top}\Big(\mathbf{M}_{\widehat{\mathbf{C}}}\mathbf{F}_j^{\bullet}-\widehat{\mathbf{C}}_j^\bullet \mathbf{H}_j^{\bullet -1}\Big)^\top \mathbf{M}_{\widehat{\mathbf{C}}_{ij}} (\mathbf{F}-\widehat{\mathbf{C}}\mathbf{H}^{-1})\pmb{\gamma}_{ij}
\nonumber\\
&=&\frac{1}{NT}\sum_{j=1}^N\pmb{\Sigma}_{\mathbf{v},ij}^{-1}\pmb{\phi}_{ij}^{\bullet\top}\Big(\sum_{s_1 =1,\ne 22}^{25} \mathbf{W}_{j,s_1}\pmb{\Pi}_j\Big)^\top \mathbf{M}_{\widehat{\mathbf{C}}_{ij}} \Big(\sum_{s_2=1,\neq 10}^{25}\mathbf{J}_{s_2}\pmb{\Pi}\Big) \pmb{\gamma}_{ij}
\nonumber\\
&:=&\sum_{s_1=1,\neq 22}^{25}\sum_{s_2=1,\neq 10}^{25}\mathcal{M}^\ast_{i,31, s_1s_2},
\end{eqnarray*}
where $\mathcal{M}^\ast_{i,31, s_1s_2}=\frac{1}{NT}\sum_{j=1}^N\pmb{\Sigma}_{\mathbf{v},ij}^{-1}\pmb{\phi}_{ij}^{\bullet\top}\pmb{\Pi}_j^\top \mathbf{W}_{j,s_1}^\top\mathbf{M}_{\widehat{\mathbf{C}}_{ij}} \mathbf{J}_{s_2}\pmb{\Pi}\pmb{\gamma}_{ij}$, $\mathbf{H}_j^\bullet =(LT)^{-1} \pmb{\Gamma}_{j\centerdot}^{\bullet\top} \pmb{\Gamma}_{j\centerdot}^\bullet \mathbf{F}_j^{\bullet\top} \widehat{\mathbf{C}}_j^\bullet \widehat{\mathbf{V}}_j^{\bullet -1}$ and $\pmb{\Pi}_j = (\mathbf{F}_j^{\bullet\top} \widehat{\mathbf{C}}_j^\bullet/T)^{-1}(\pmb{\Gamma}_{\centerdot j}^{^\bullet\top} \pmb{\Gamma}_{\centerdot j }^\bullet/L)^{-1}$. 

Additionally, directly applying the results for $\mathbf{W}_{j,1}, \cdots, \mathbf{W}_{j,5}$ in the equations from \eqref{nh1} to \eqref{mge_13}, we obtain that $\sum_{s_2=1,\neq 10}^{25}\mathcal{M}^\ast_{i,31, 1s_2},\cdots, \sum_{s_2=1,\neq 10}^{25}\mathcal{M}^\ast_{i,31, 5s_2}$ are all negligible.  For the term with $\mathbf{W}_{j,6}=\frac{1}{LT}\sum_{i=1}^L\mathbf{M}_{\widehat{\mathbf{C}}}\mathbf{X}_{ij\centerdot} (\pmb{\beta}_{ij}-\widehat{\mathbf{b}}_{ij}) \pmb{\gamma}_{ij}^{\bullet\top}   \mathbf{F}_j^{\bullet\top} \mathbf{M}_{\widehat{\mathbf{C}}} \widehat{\mathbf{C}}_j^\bullet$, we write
\begin{eqnarray*}
\mathbf{W}_{j,6}&=&\frac{1}{LT}\sum_{i=1}^L\mathbf{M}_{\widehat{\mathbf{C}}}\mathbf{F}\pmb{\phi}_{ij} (\pmb{\beta}_{ij}-\widehat{\mathbf{b}}_{ij}) \pmb{\gamma}_{ij}^{\bullet\top}   \mathbf{F}_j^{\bullet\top} \mathbf{M}_{\widehat{\mathbf{C}}} \widehat{\mathbf{C}}_j^\bullet
\nonumber\\
&&+\frac{1}{LT}\sum_{i=1}^L\mathbf{M}_{\widehat{\mathbf{C}}}\mathbf{F}_i^\circ\pmb{\phi}_{ij}^\circ (\pmb{\beta}_{ij}-\widehat{\mathbf{b}}_{ij}) \pmb{\gamma}_{ij}^{\bullet\top}   \mathbf{F}_j^{\bullet\top} \mathbf{M}_{\widehat{\mathbf{C}}} \widehat{\mathbf{C}}_j^\bullet
\nonumber\\
&&+\frac{1}{LT}\sum_{i=1}^L\mathbf{M}_{\widehat{\mathbf{C}}}\mathbf{F}_j^\bullet\pmb{\phi}_{ij}^\bullet (\pmb{\beta}_{ij}-\widehat{\mathbf{b}}_{ij}) \pmb{\gamma}_{ij}^{\bullet\top}   \mathbf{F}_j^{\bullet\top} \mathbf{M}_{\widehat{\mathbf{C}}} \widehat{\mathbf{C}}_j^\bullet
\nonumber\\
&&+\frac{1}{LT}\sum_{i=1}^L\mathbf{M}_{\widehat{\mathbf{C}}}\mathbf{V}_{ij\centerdot} (\pmb{\beta}_{ij}-\widehat{\mathbf{b}}_{ij}) \pmb{\gamma}_{ij}^{\bullet\top}   \mathbf{F}_j^{\bullet\top} \mathbf{M}_{\widehat{\mathbf{C}}} \widehat{\mathbf{C}}_j^\bullet
\nonumber\\
&=&\mathbf{W}_{j,6,1}+\cdots+\mathbf{W}_{j,6,4}.
\end{eqnarray*}
For $\mathbf{W}_{j,6,1}$, it is clear to see that 
\begin{eqnarray*}
\mathbf{W}_{j,6,1}=\frac{1}{LT}\sum_{i=1}^L\mathbf{M}_{\widehat{\mathbf{C}}}(\mathbf{F}-\widehat{\mathbf{C}}\mathbf{H}^{-1})\pmb{\phi}_{ij} (\pmb{\beta}_{ij}-\widehat{\mathbf{b}}_{ij}) \pmb{\gamma}_{ij}^{\bullet\top}   \mathbf{F}_j^{\bullet\top} \mathbf{M}_{\widehat{\mathbf{C}}} \widehat{\mathbf{C}}_j^\bullet,
\end{eqnarray*}
and therefore
\begin{eqnarray*}
\frac{1}{NT}\left\|\sum_{i_2,j}\pmb{\Sigma}_{\mathbf{v},i_1j}^{-1}\pmb{\phi}_{i_1j}^{\bullet\top}\pmb{\Pi}_j^\top \mathbf{W}_{j,6,1}^\top  \mathbf{M}_{\widehat{\mathbf{C}}_{i_1j}} (\mathbf{F}-\widehat{\mathbf{C}}\mathbf{H}^{-1})\pmb{\gamma}_{i_1j}\right\|=O_P\left(\max_{j} \|\pmb{\beta}_{ij}-\widehat{\mathbf{b}}_{ij}\| \cdot\frac{1}{T}\|\mathbf{F}-\widehat{\mathbf{C}}\mathbf{H}^{-1}\|^2_2\right).
\end{eqnarray*}
Similarly to \eqref{mge_e14}, we have
\begin{eqnarray}
\frac{1}{LN}\left\|\sum_{i}\mathbf{M}_{\widehat{\mathbf{C}}}\mathbf{F}_i^\circ\pmb{\phi}_{ij}^\circ(\pmb{\beta}_{ij}-\widehat{\mathbf{b}}_{ij}) \pmb{\gamma}_{ij}^{\bullet\top} \right\|_2
&=&O_P\left(\frac{\sqrt{L\vee T}}{\sqrt{L}}\max_{i,j}\|\pmb{\beta}_{ij}-\widehat{\mathbf{b}}_{ij}\|_2\right),
\end{eqnarray}
 It further yields
\begin{eqnarray*}
\frac{1}{NT}\left\|\sum_{i_2,j}\pmb{\Sigma}_{\mathbf{v},i_1j}^{-1}\pmb{\phi}_{i_1j}^{\bullet\top}\pmb{\Pi}_j^\top \mathbf{W}_{j,6,2}^\top  \mathbf{M}_{\widehat{\mathbf{C}}_{i_1j}} (\mathbf{F}-\widehat{\mathbf{C}}\mathbf{H}^{-1})\pmb{\gamma}_{i_1j}\right\|=O_P\left(\frac{1}{\sqrt{L\wedge T}}\max_{i,j}\|\pmb{\beta}_{ij}-\widehat{\mathbf{b}}_{ij}\| \cdot\frac{1}{\sqrt{T}}\|\mathbf{F}-\widehat{\mathbf{C}}\mathbf{H}^{-1}\|_2\right).
\end{eqnarray*}
Analogously, we have
\begin{eqnarray*}
\frac{1}{NT}\left\|\sum_{i_2,j}\pmb{\Sigma}_{\mathbf{v}, i_1j}^{-1}\pmb{\phi}_{i_1j}^{\bullet\top}\pmb{\Pi}_j^\top \mathbf{W}_{j,6,4}^\top  \mathbf{M}_{\widehat{\mathbf{C}}_{i_1j}} (\mathbf{F}-\widehat{\mathbf{C}}\mathbf{H}^{-1})\pmb{\gamma}_{i_1j}\right\|=O_P\left(\frac{1}{\sqrt{LN\wedge T}}\max_{i,j}\|\pmb{\beta}_{ij}-\widehat{\mathbf{b}}_{ij}\| \cdot\frac{1}{\sqrt{T}}\|\mathbf{F}-\widehat{\mathbf{C}}\mathbf{H}^{-1}\|_2\right).
\end{eqnarray*}
For the term with $\mathbf{W}_{j,6,3}$, by invoking the properties of $\mathbf{M}_{\widehat{\mathbf{C}}}$ and $\mathbf{M}_{\widehat{\mathbf{C}}_{i_1j}}$, we have 
\begin{eqnarray*}
\mathbf{M}_{\widehat{\mathbf{C}}_{i_1j}}\mathbf{W}_{j,6,3}
&=&\frac{1}{LT}\sum_{i_2=1}^L\mathbf{M}_{\widehat{\mathbf{C}}_{i_1j}}\Big(\mathbf{M}_{\widehat{\mathbf{C}}}\mathbf{F}_j^{\bullet}-\widehat{\mathbf{C}}_j^\bullet \mathbf{H}_j^{\bullet -1}\Big)\pmb{\phi}_{i_2j}^\bullet (\pmb{\beta}_{i_2j}-\widehat{\mathbf{b}}_{i_2j}) \pmb{\gamma}_{i_2j}^{\bullet\top}   \mathbf{F}_j^{\bullet\top} \mathbf{M}_{\widehat{\mathbf{C}}} \widehat{\mathbf{C}}_j^\bullet.
\end{eqnarray*}
It follows that
\begin{eqnarray*}
&&\frac{1}{NT}\left\|\sum_{i_2,j}\pmb{\Sigma}_{\mathbf{v},i_1j}^{-1}\pmb{\phi}_{i_1j}^{\bullet\top}\pmb{\Pi}_j^\top \mathbf{W}_{j,6,3}^\top  \mathbf{M}_{\widehat{\mathbf{C}}_{i_1j}} (\mathbf{F}-\widehat{\mathbf{C}}\mathbf{H}^{-1})\pmb{\gamma}_{i_1j}\right\|
\nonumber\\
&=&O_P\left(\max_{i,j}\|\pmb{\beta}_{ij}-\widehat{\mathbf{b}}_{ij}\| \cdot\frac{1}{\sqrt{T}}\|\mathbf{F}-\widehat{\mathbf{C}}\mathbf{H}^{-1}\|_2\cdot \frac{1}{\sqrt{T}}\left\|\mathbf{M}_{\widehat{\mathbf{C}}}\mathbf{F}_j^{\bullet}-\widehat{\mathbf{C}}_j^\bullet \mathbf{H}_j^{\bullet -1}\right\|_2\right).
\end{eqnarray*}
Given the established results for $\mathbf{W}_{j,6}$, we are ready to assert
\begin{equation*}
\sum_{s_2=1,\neq 10}^{25}\mathcal{M}^\ast_{i,31, 6s_2}=o_P\left(\frac{1}{\sqrt{NT}}\right).
\end{equation*}
Analogously, we have $\sum_{s_2=1,\neq 10}^{25}\mathcal{M}^\ast_{i,31, 7s_2}=o_P\left(\frac{1}{\sqrt{NT}}\right)$ as well.

Building on the results established for $\mathbf{W}_{j,8}$, $\cdots$, $\mathbf{W}_{j,12}$, $\mathbf{W}_{j,15}$, $\mathbf{W}_{j,16}$, $\mathbf{W}_{j,20}$, $\cdots$, $\mathbf{W}_{j,25}$ in the equations from  \eqref{mge_e18} to  \eqref{mge_e27}, we can obtain that $\sum_{s_2=1,\neq 10}^{25}\mathcal{M}^\ast_{i,31, s_1s_2}$ is negligible, for $s_1=8,\ldots,12,15,16,20,\ldots,25$. 

We then consider the remaining terms. For $\mathbf{W}_{j,14}$, it is clear to see that
\begin{eqnarray*}
\mathbf{M}_{\widehat{\mathbf{C}}_{i_1j}}\mathbf{W}_{j,14}
&=&\frac{1}{LT}\sum_{i_2=1}^L\mathbf{M}_{\widehat{\mathbf{C}}_{i_1j}} \Big(\mathbf{M}_{\widehat{\mathbf{C}}}\mathbf{F}_j^{\bullet}-\widehat{\mathbf{C}}_j^\bullet \mathbf{H}_j^{\bullet -1}\Big) \pmb{\gamma}_{i_2j}^\bullet \pmb{\gamma}_{i_2j}^{\top}\Big( \mathbf{F}-\mathbf{C}\mathbf{H}^{-1}\Big)^\top \widehat{\mathbf{C}}_j^\bullet,
\end{eqnarray*}
from which we can derive the order for $\mathcal{M}^\ast_{i,31, 14s_2}$:
\begin{eqnarray*}
\sum_{s_2=1,\neq 10}^{25}\mathcal{M}^\ast_{i,31, 14s_2}&=&O_P\left(\frac{1}{\sqrt{T}}\|
\mathbf{M}_{\widehat{\mathbf{C}}}\mathbf{F}_j^{\bullet}-\widehat{\mathbf{C}}_j^\bullet \mathbf{H}_j^{\bullet -1}\|_2\cdot \frac{1}{T}\left\|\mathbf{F}-\widehat{\mathbf{C}}\mathbf{H}^{-1}\right\|^2_2\right).
\end{eqnarray*}
For $\mathbf{W}_{j,17}$,  
\begin{eqnarray*}
\max_j\frac{1}{LT}\left\|\sum_{i=1}^L \mathbf{F}_i^{\circ} \pmb{\gamma}_{ij}^{\circ} \pmb{\gamma}_{ij}^{\circ\top} \mathbf{F}_i^{\circ\top}\right\|_2
&=&\frac{1}{LT}\max_j\left\|\mathbf{F}^{\circ} \text{diag}\big\{ \pmb{\Gamma}_{\centerdot j}^{\circ\top}\big\} \text{diag}\big\{ \pmb{\Gamma}_{\centerdot j}^{\circ\top}\big\}^\top\mathbf{F}^{\circ\top}\right\|_2
\nonumber\\
&\leq&\frac{1}{LT}\left\|\mathbf{F}^{\circ} \right\|_2^2\max_j\left\|\text{diag}\big\{ \pmb{\Gamma}_{\centerdot j}^{\circ\top}\big\}\right\|^2_2
\nonumber\\
&=&O_P\left(\frac{1}{L\wedge T}\right).
\end{eqnarray*}
which immediately leads to the desired order for $\mathcal{M}^\ast_{i,31, 17s_2}$:
\begin{eqnarray*}
\sum_{s_2=1,\neq 10}^{25}\mathcal{M}^\ast_{i,31, 17s_2}=O_P\left(\frac{1}{L\wedge T} \cdot \frac{1}{\sqrt{T}}\left\|\mathbf{F}-\widehat{\mathbf{C}}\mathbf{H}^{-1}\right\|_2\right).
\end{eqnarray*}
Similarly, for the term with $\mathbf{W}_{j,19}$, 
\begin{eqnarray*}
\sum_{s_2=1,\neq 10}^{25}\mathcal{M}^\ast_{i,31, 19s_2}=O_P\left(\frac{1}{\sqrt{L\wedge T}\sqrt{N\wedge T}} \cdot \frac{1}{\sqrt{T}}\left\|\mathbf{F}-\widehat{\mathbf{C}}\mathbf{H}^{-1}\right\|_2\right).
\end{eqnarray*}
Additionally, considering the  arguments that are used in the proof of  \eqref{mge_e12}, we can readily obtain 
\begin{eqnarray}\label{mge_e21}
\frac{1}{NT}\sum_{j=1}^N\pmb{\Sigma}_{\mathbf{v},ij}^{-1}\pmb{\phi}_{ij}^{\bullet\top} \mathcal{M}_{ij,31} \pmb{\gamma}_{ij} &=& \sum_{s_1=13,18}\sum_{s_2=1,\neq 10}^{25}\mathcal{M}^\ast_{i,31, s_1s_2}+o_P\left(\frac{1}{\sqrt{NT}}\right)
\nonumber\\
&=& \sum_{s_1=13,18}\sum_{s_2=12,14}\mathcal{M}^\ast_{i,31, s_1s_2}+o_P\left(\frac{1}{\sqrt{NT}}\right)
\nonumber\\
&=& \frac{1}{NT}\sum_{j=1}^N\pmb{\Sigma}_{\mathbf{v},ij}^{-1}\pmb{\phi}_{ij}^{\bullet\top} \mathcal{R}_{F,j}^{\bullet\top} \mathbf{M}_{\widehat{\mathbf{C}}_{ij}} \mathcal{R}_F \pmb{\gamma}_{ij}+o_P\left(\frac{1}{\sqrt{NT}}\right),
\end{eqnarray}
where $\mathcal{R}_{F,j}^\bullet = \sum_{i=1}^L(\mathbf{F}\pmb{\gamma}_{ij}+\mathbf{F}_{i}^{\circ}\pmb{\gamma}_{ij}^\circ)\pmb{\gamma}_{ij}^{\bullet\top} (\pmb{\Gamma}_{\centerdot j }^{\bullet\top} \pmb{\Gamma}_{\centerdot j }^\bullet)^{-1}$, and  
\begin{eqnarray}\label{mge_e22}
\frac{1}{NT}\left\|\sum_{j=1}^N\pmb{\Sigma}_{\mathbf{v},ij}^{-1}\pmb{\phi}_{ij}^{\bullet\top} \mathcal{R}_{F,j}^{\bullet\top} \mathbf{M}_{\widehat{\mathbf{C}}_{ij}} \mathcal{R}_F \pmb{\gamma}_{ij}\right\|=O_P\left(\frac{1}{L\wedge N \wedge T}\right).
\end{eqnarray}

Using the arguments that are similar to those in the proofs of \eqref{mge_e12} and \eqref{mge_e21}, we can also formulate the leading-order terms in   $\mathcal{M}_{ij,22}$, $\mathcal{M}_{ij,33}$, $\mathcal{M}_{ij,12}$,  $\mathcal{M}_{ij,21}$,   $\mathcal{M}_{ij,13}$ $\mathcal{M}_{ij,23}$ and  $\mathcal{M}_{ij,32}$, and finally obtain
\begin{eqnarray}\label{mge_e24}
\mathbf{D}_{i,1,1}&=&\frac{1}{NT}\sum_{j=1}^N\pmb{\Sigma}_{\mathbf{v},ij}^{-1} \pmb{\phi}_{ij}^{*\top} \mathcal{R}_{ij}^{*\top} \mathbf{M}_{\widehat{\mathbf{C}}_{ij}} \mathcal{R}_{ij}^* \pmb{\gamma}_{ij}^*+o_P\left(\frac{1}{\sqrt{NT}}\right),
\end{eqnarray}
where $\mathcal{R}_{ij}^{*}=(\mathcal{R}_{F}, \mathcal{R}_{F,i}^\circ, \mathcal{R}_{F,j}^\bullet)$ with $\mathcal{R}_{F,i}^\circ=\sum_{j=1}^N(\mathbf{F}\pmb{\gamma}_{ij}+\mathbf{F}_{j}^{\bullet}\pmb{\gamma}_{ij}^\bullet)\pmb{\gamma}_{ij}^{\circ\top} (\pmb{\Gamma}_{i\centerdot  }^{\circ\top} \pmb{\Gamma}_{i\centerdot  }^\circ)^{-1}$. In light of \eqref{mge_e23} and \eqref{mge_e22}, 
\begin{eqnarray}\label{mge_e25}
\frac{1}{NT}\left\|\sum_{j=1}^N\pmb{\Sigma}_{\mathbf{v},ij}^{-1} \pmb{\phi}_{ij}^{*\top} \mathcal{R}_{ij}^{*\top} \mathbf{M}_{\widehat{\mathbf{C}}_{ij}} \mathcal{R}_{ij}^* \pmb{\gamma}_{ij}^*\right\|=O_P\left(\frac{1}{L\wedge N \wedge T}\right).
\end{eqnarray}

We have finished the investigation of $\mathbf{D}_{i,1,1}$ and we now proceed to study  $\mathbf{D}_{i,1,2}$. Similar to the development of \ref{mge_e24}, we only need to investigate the leading-order term in $\mathbf{D}_{i,1,2}$:
\begin{eqnarray*}
\mathbf{D}_{i,1,2}&=&\frac{1}{NT}\sum_{j=1}^N\big(\widehat{\boldsymbol{\Xi}}_{ij}^{-1} -\pmb{\Sigma}_{\mathbf{v},ij}^{-1}\big)\pmb{\phi}_{ij}^{*\top} \mathcal{R}_{ij}^{*\top} \mathbf{M}_{\widehat{\mathbf{C}}_{ij}} \mathcal{R}_{ij}^* \pmb{\gamma}_{ij}^*+o_P\left(\frac{1}{\sqrt{NT}}\right)
\nonumber\\
&=&\mathbf{D}_{i,1,2}^\ast+o_P\left(\frac{1}{\sqrt{NT}}\right)
.
\end{eqnarray*}
It is clear to see that
\begin{eqnarray*}
\mathbf{D}^\ast_{i,1,2}&=&-\frac{1}{NT}\sum_{j=1}^N\widehat{\boldsymbol{\Xi}}_{ij}^{-1}\big(\widehat{\boldsymbol{\Xi}}_{ij}-\pmb{\Sigma}_{\mathbf{v},ij} \big)\pmb{\Sigma}_{\mathbf{v},ij}^{-1}\pmb{\phi}_{ij}^{*\top} \mathcal{R}_{ij}^{*\top} \mathbf{M}_{\widehat{\mathbf{C}}_{ij}} \mathcal{R}_{ij}^* \pmb{\gamma}_{ij}^*,
\end{eqnarray*}

Noteworthily, since $\widehat{\boldsymbol{\Xi}}_{ij}=\pmb{\Sigma}_{\mathbf{v},ij}+o_P(1)$, Assumption \ref{AS2} ensures that $\max_j \left\|\widehat{\boldsymbol{\Xi}}_{ij}^{-1}\right\|_2=(\min_j\lambda_{\rm min}(\widehat{\boldsymbol{\Xi}}_{ij}))^{-1}$ $=O_P(1)$, where $\lambda_{\rm min}(\cdot)$ denotes the the minimum eigenvalue function. 

In connection with \eqref{expandX}, we obtain 
\begin{eqnarray}\label{mge_e28}
\frac{1}{N}\sum_{j=1}^N\big\|\widehat{\boldsymbol{\Xi}}_{ij}^{-1}\big(\widehat{\boldsymbol{\Xi}}_{ij}-\pmb{\Sigma}_{\mathbf{v},ij}\big) \big\|_2^2=O_P\left(\frac{1}{T}\left\|\mathbf{F}-\widehat{\mathbf{C}}\mathbf{H}^{-1}\right\|^2_2\right).
\end{eqnarray}

Using similar arguments in the proof of \eqref{mge_e23}, \eqref{mge_e22}, and \eqref{mge_e25}, we can also establish that 
\begin{eqnarray}\label{mge_e29}
\frac{1}{N}\sum_{j=1}^N\left\|\frac{1}{T}\pmb{\Sigma}_{\mathbf{v},ij}^{-1} \pmb{\phi}_{ij}^{*\top} \mathcal{R}_{ij}^{*\top} \mathbf{M}_{\widehat{\mathbf{C}}_{ij}} \mathcal{R}_{ij}^* \pmb{\gamma}_{ij}^*\right\|^2=O_P\left(\frac{1}{(L\wedge N \wedge T)^2}\right).
\end{eqnarray}
By  \eqref{mge_e28}, \eqref{mge_e29} and directly applying the  Cauchy-Schwarz inequality, we can readily show that
\begin{equation}\label{mge_e30}
\mathbf{D}^\ast_{i,1,2}=O_P\left(\frac{1}{L\wedge N \wedge T}\cdot \frac{1}{\sqrt{T}}\left\|\mathbf{F}-\widehat{\mathbf{C}}\mathbf{H}^{-1}\right\|_2\right).
\end{equation}
By \eqref{mge_e24} and \eqref{mge_e30}, we have
\begin{eqnarray*}
\mathbf{D}_{i,1}&=&\frac{1}{NT}\sum_{j=1}^N\pmb{\Sigma}_{\mathbf{v},ij}^{-1} \pmb{\phi}_{ij}^{*\top} \mathcal{R}_{ij}^{*\top} \mathbf{M}_{\widehat{\mathbf{C}}_{ij}} \mathcal{R}_{ij}^* \pmb{\gamma}_{ij}^*+o_P\left(\frac{1}{\sqrt{NT}}\right).
\end{eqnarray*}

We then proceed to study $\mathbf{D}_{i,2}$. Write
\begin{eqnarray*}
\mathbf{D}_{i,2}&=&\frac{1}{NT}\sum_{j=1}^N\pmb{\Sigma}_{\mathbf{v},ij}^{-1} \mathbf{V}_{ij\centerdot}^\top \mathbf{M}_{\widehat{\mathbf{C}}_{ij}}\big(\mathbf{F}-\widehat{\mathbf{C}}\mathbf{H}^{-1}\big) \pmb{\gamma}_{ij}
+\frac{1}{NT}\sum_{j=1}^N\pmb{\Sigma}_{\mathbf{v},ij}^{-1} \mathbf{V}_{ij\centerdot}^\top \mathbf{M}_{\widehat{\mathbf{C}}_{ij}}\big(\mathbf{M}_{\widehat{\mathbf{C}}}\mathbf{F}_i^\circ-\widehat{\mathbf{C}}_i^\circ\mathbf{H}_i^{\circ -1}\big) \pmb{\gamma}_{ij}^\circ
\nonumber\\
&&+\frac{1}{NT}\sum_{j=1}^N\pmb{\Sigma}_{\mathbf{v},ij}^{-1} \mathbf{V}_{ij\centerdot}^\top \mathbf{M}_{\widehat{\mathbf{C}}_{ij}}\big(\mathbf{M}_{\widehat{\mathbf{C}}}\mathbf{F}_j^\bullet-\widehat{\mathbf{C}}_j^\bullet\mathbf{H}_j^{\bullet -1}\big) \pmb{\gamma}_{ij}^\bullet
+\frac{1}{NT}\sum_{j=1}^N\big(\widehat{\boldsymbol{\Xi}}_{ij}^{-1}-\pmb{\Sigma}_{\mathbf{v},ij}^{-1} \big) \mathbf{V}_{ij\centerdot}^\top\mathbf{M}_{\widehat{\mathbf{C}}_{ij}} \mathbf{F}_{ij}^* \pmb{\gamma}_{ij}^*
\nonumber\\
&:=&\mathbf{D}_{i,2,1}+\cdots+\mathbf{D}_{i,2,4}.
\end{eqnarray*}
For $\mathbf{D}_{i,2,1}$, using arguments that are closely related to those in the proof of \eqref{mge_e12}, we can readily obtain
\begin{eqnarray}
\mathbf{D}_{i,2,1}&=&\frac{1}{NT}\sum_{j=1}^N\pmb{\Sigma}_{\mathbf{v},ij}^{-1} \mathbf{V}_{ij\centerdot}^\top \mathbf{M}_{\widehat{\mathbf{C}}_{ij}}\mathcal{R}_F \pmb{\gamma}_{ij}+o_P\left(\frac{1}{\sqrt{NT}}\right)
\nonumber\\
&=&\frac{1}{NT}\sum_{j=1}^N\pmb{\Sigma}_{\mathbf{v},ij}^{-1} \mathbf{V}_{ij\centerdot}^\top \mathcal{R}_F \pmb{\gamma}_{ij}-\frac{1}{NT^2}\sum_{j=1}^N\pmb{\Sigma}_{\mathbf{v},ij}^{-1} \mathbf{V}_{ij\centerdot}^\top \widehat{\mathbf{C}}\widehat{\mathbf{C}}^\top \mathcal{R}_F \pmb{\gamma}_{ij}
\nonumber\\
&&-\frac{1}{NT^2}\sum_{j=1}^N\pmb{\Sigma}_{\mathbf{v},ij}^{-1} \mathbf{V}_{ij\centerdot}^\top \widehat{\mathbf{C}}_i^{\circ}\widehat{\mathbf{C}}_i^{\circ\top} \mathcal{R}_F \pmb{\gamma}_{ij}-\frac{1}{NT^2}\sum_{j=1}^N\pmb{\Sigma}_{\mathbf{v},ij}^{-1} \mathbf{V}_{ij\centerdot}^\top \widehat{\mathbf{C}}_j^{\bullet}\widehat{\mathbf{C}}_j^{\bullet\top} \mathcal{R}_F \pmb{\gamma}_{ij}.
\end{eqnarray}
Using the $\alpha$-mixing conditions in Assumption \ref{AS2} and the independence of $\mathbf{v}_{ijt}$ with the factors and loadings, we can easily show that 
\begin{eqnarray*}
\frac{1}{NT}\left\|\sum_{j=1}^N\pmb{\Sigma}_{\mathbf{v},ij}^{-1} \mathbf{V}_{ij\centerdot}^\top \mathcal{R}_F \pmb{\gamma}_{ij}\right\|=o_P\left(\frac{1}{NT}\right).
\end{eqnarray*}
For the second term in $\mathbf{D}_{i,2,1}$, 
\begin{eqnarray*}
\frac{1}{NT^2}\left\|\sum_{j=1}^N\pmb{\Sigma}_{\mathbf{v},ij}^{-1} \mathbf{V}_{ij\centerdot}^\top \widehat{\mathbf{C}}\widehat{\mathbf{C}}^\top \mathcal{R}_F \pmb{\gamma}_{ij}\right\|&\leq&\frac{1}{NT^2}\left\|\sum_{j=1}^N\pmb{\Sigma}_{\mathbf{v},ij}^{-1} \mathbf{V}_{ij\centerdot}^\top \big(\widehat{\mathbf{C}}-\mathbf{F}\mathbf{H}\big)\widehat{\mathbf{C}}^\top \mathcal{R}_F \pmb{\gamma}_{ij}\right\|
\nonumber\\
&&+\frac{1}{NT^2}\left\|\sum_{j=1}^N\pmb{\Sigma}_{\mathbf{v},ij}^{-1} \mathbf{V}_{ij\centerdot}^\top \mathbf{F}\mathbf{H}\widehat{\mathbf{C}}^\top \mathcal{R}_F \pmb{\gamma}_{ij}\right\|
\nonumber\\
&=&O_P\left(\frac{1}{\sqrt{N\wedge T}\sqrt{L\wedge N\wedge T}}\cdot \frac{1}{\sqrt{T}}\big\|\widehat{\mathbf{C}}-\mathbf{F}\mathbf{H}\big\|_2\right)+o_P\left(\frac{1}{\sqrt{NT}}\right),
\end{eqnarray*}
where the last equality holds by the fact $\|\mathbf{V}_{i\centerdot\centerdot}\|_2=O_P(\sqrt{N\vee T})$ and $\|\mathcal{R}_F\|_2=O_P\Big(\frac{1}{\sqrt{L\wedge N \wedge T}}\Big)$. Using analogous arguments, we can also show that the third and fourth terms in $\mathbf{D}_{i,2,1}$ are also negligible. Consequently, 
\begin{eqnarray*}
\mathbf{D}_{i,2,1}=o_P\left(\frac{1}{\sqrt{NT}}\right).
\end{eqnarray*}
Similarly, we have 
\begin{eqnarray*}
\mathbf{D}_{i,2,2}=o_P\left(\frac{1}{\sqrt{NT}}\right),\, \mathbf{D}_{i,2,3}=o_P\left(\frac{1}{\sqrt{NT}}\right), \, \mathbf{D}_{i,2,4}=o_P\left(\frac{1}{\sqrt{NT}}\right).
\end{eqnarray*}
Combining these results, we can conclude that $\mathbf{D}_{i,2}$ is negligible. Additionally, since $\mathbf{D}_{i,3}$ has a similar structure with  $\mathbf{D}_{i,2}$, arguments that are analogous to those in the derivation of $\mathbf{D}_{i,2}$ can be applied to show that $\mathbf{D}_{i,3}=o_P\left(\frac{1}{\sqrt{NT}}\right)$.

For $\mathbf{D}_{i,4}$, following the developments for $\mathbf{D}_{i,1}$, $\mathbf{D}_{i,2}$ and $\mathbf{D}_{i,3}$, we only need to study its leading-order term: $\frac{1}{NT}\sum_{j=1}^N\pmb{\Sigma}_{\mathbf{v},ij}^{-1} \mathbf{V}_{ij\centerdot}^\top\mathbf{M}_{\widehat{\mathbf{C}}_{ij}}\pmb{\mathcal{E}}_{ij\centerdot }$. 
Building on the $\alpha$-mixing conditions in Assumption \ref{AS2} and the conditional independence between  $\mathbf{v}_{ijt}$ and $\varepsilon_{ijt}$, we can readily obtain
\begin{eqnarray}\label{asymp_term}
\frac{1}{\sqrt{NT}}\sum_{j=1}^N\pmb{\Sigma}_{\mathbf{v},ij}^{-1} \mathbf{V}_{ij\centerdot}^\top\mathbf{M}_{\widehat{\mathbf{C}}_{ij}}\pmb{\mathcal{E}}_{ij\centerdot }&=&\frac{1}{\sqrt{NT}}\sum_{j=1}^N\pmb{\Sigma}_{\mathbf{v},ij}^{-1} \mathbf{V}_{ij\centerdot}^\top\pmb{\mathcal{E}}_{ij\centerdot }+o_P(1)
\nonumber\\
&\to_D& N(\mathbf{0}, \pmb{\Sigma}_{\mathbf{b},i\centerdot}),
\end{eqnarray} 
where $\pmb{\Sigma}_{\mathbf{b},i\centerdot}=\lim_{N,T\to\infty} \frac{1}{NT}\sum_{j_1,j_2=1}^N \pmb{\Sigma}_{\mathbf{v},ij_1}^{-1} E[\mathbf{V}_{ij_1\centerdot}^\top\pmb{\mathcal{E}}_{ij_1\centerdot } \pmb{\mathcal{E}}_{ij_2\centerdot }^\top \mathbf{V}_{ij_2\centerdot}]\pmb{\Sigma}_{\mathbf{v},ij_2}^{-1}$ and the asymptotic normality can be easily proved  using the standard large block and small block technique (cf., \citealp{Gao}).

(3) Since $\widetilde{\mathbf{b}}_{\centerdot j}$ has a similar structure to $\widetilde{\mathbf{b}}_{i \centerdot}$, we can use analogous arguments to those in the proof of Theorem \ref{M.Thm1}.(2) to prove Theorem \ref{M.Thm1}.(3). Therefore, the detailed proofs are omitted here.

\bigskip

\noindent \textbf{Proof of Theorem \ref{M.Thm2}:}\\
In the following proofs, we denote $E^\ast[\cdot]$ and $\text{Var}^\ast(\cdot)$ as the expectation  and variance conditional on the observed sample, respectively. $\|\cdot\|_{n}$ and $\|\cdot\|^\ast_{n}$ denote the norms associated with $n$: $\|\zeta\|_{n}=E\bigl[\|\zeta\|^{n}\bigr]^{1/n}$ and $\|\zeta\|^\ast_{n}=E^\ast\bigl[\|\zeta\|^{n}\bigr]^{1/n}$, for any random variable $\zeta$ and any positive number $n\geq 1$. Additionally, let $\mathcal{F}_{t}$ and $E_{t}[\cdot]$ be the sigma field generated by $\{\mathbf{v}_{ijs}, \varepsilon_{ijs}\}_{s=t,t-1,\cdots, \,i\in[L], j\in[N]}$ and the expectation conditional on $\mathcal{F}_{t}$, respectively.

Write

\begin{eqnarray}
\widetilde{\mathbf{b}}_{ij}^* - \widetilde{\mathbf{b}}_{ij} &=& \left(\mathbf{X}_{ij\centerdot}^\top \mathbf{M}_{\widehat{\mathbf{C}}_{ij}} \mathbf{X}_{ij\centerdot}\right)^{-1} \mathbf{X}_{ij\centerdot}^\top \mathbf{M}_{\widehat{\mathbf{C}}_{ij}}  (\widehat{\pmb{\mathcal{E}}}_{ij\centerdot}\circ\pmb{\xi} )\nonumber \\
&=&\left(\mathbf{X}_{ij\centerdot}^\top \mathbf{M}_{\widehat{\mathbf{C}}_{ij}} \mathbf{X}_{ij\centerdot}\right)^{-1} \mathbf{X}_{ij\centerdot}^\top \mathbf{M}_{\widehat{\mathbf{C}}_{ij}} ( \pmb{\mathcal{E}}_{ij\centerdot}\circ\pmb{\xi})\nonumber \\
& &  + \left(\mathbf{X}_{ij\centerdot}^\top \mathbf{M}_{\widehat{\mathbf{C}}_{ij}} \mathbf{X}_{ij\centerdot}\right)^{-1} \mathbf{X}_{ij\centerdot}^\top \mathbf{M}_{\widehat{\mathbf{C}}_{ij}} (\mathbf{F}_{ij}^*\pmb{\gamma}_{ij}^*\circ\pmb{\xi})\nonumber \\
& & + \left(\mathbf{X}_{ij\centerdot}^\top \mathbf{M}_{\widehat{\mathbf{C}}_{ij}} \mathbf{X}_{ij\centerdot}\right)^{-1} \mathbf{X}_{ij\centerdot}^\top \mathbf{M}_{\widehat{\mathbf{C}}_{ij}} (\mathbf{X}_{ij} ( \pmb{\beta}_{ij} - \widetilde{\mathbf{b}}_{ij} )\circ\pmb{\xi}).\label{bstar}
\end{eqnarray}

We then investigate the terms on the right hand side of \eqref{bstar} one by one. For the first term, using analogous arguments in the proof of Theorem \ref{M.Thm1}, we can obtain 
\begin{eqnarray*}
\frac{1}{T}\mathbf{X}_{ij\centerdot}^\top \mathbf{M}_{\widehat{\mathbf{C}}_{ij}} ( \pmb{\mathcal{E}}_{ij\centerdot}\circ\pmb{\xi})&=&\frac{1}{T}\mathbf{V}_{ij\centerdot}^\top  ( \pmb{\mathcal{E}}_{ij\centerdot}\circ\pmb{\xi})+o_P(1).
\end{eqnarray*}
In what follows, we show that $\frac{1}{\sqrt{T}}\mathbf{V}_{ij\centerdot}^\top  ( \pmb{\mathcal{E}}_{ij\centerdot}\circ\pmb{\xi})$ can generate the bootstrap distribution. We first show  $\text{Var}^\ast\Bigl( \frac{1}{\sqrt{T}}\mathbf{V}_{ij\centerdot}^\top  ( \pmb{\mathcal{E}}_{ij\centerdot}\circ\pmb{\xi})\Bigr)=\pmb{\Sigma}_{\mathbf{v},\varepsilon,ij}+o_P(1)$. Note that
\begin{eqnarray}\label{btst1}
\text{Var}^\ast\Bigl( \frac{1}{\sqrt{T}}\mathbf{V}_{ij\centerdot}^\top  ( \pmb{\mathcal{E}}_{ij\centerdot}\circ\pmb{\xi})\Bigr)&=&\frac{1}{T}\sum_{s,t=1}^T\mathbf{v}_{ijt}\mathbf{v}_{ijs}^\top\varepsilon_{ijt}\varepsilon_{ijs} E^\ast\left[\xi_{t}\xi_{s}\right]
\nonumber\\
&=&\frac{1}{T}\sum_{s,t=1}^Ta\Bigl(\frac{t-s}{m}\Bigr)\mathbf{v}_{ijt}\mathbf{v}_{ijs}^\top\varepsilon_{ijt}\varepsilon_{ijs}
\nonumber\\
&=&\frac{1}{T}\sum_{t=1}^T\mathbf{v}_{ijt}\mathbf{v}_{ijt}^\top\varepsilon_{ijt}^2+\frac{1}{T}\sum_{t=1}^T\sum_{s=t+1}^Ta\Bigl(\frac{t-s}{m}\Bigr)\mathbf{v}_{ijt}\mathbf{v}_{ijs}^\top\varepsilon_{ijt}\varepsilon_{ijs}
\nonumber\\
&&+\frac{1}{T}\sum_{t=1}^T\sum_{s=1}^{t-1}a\Bigl(\frac{t-s}{m}\Bigr)\mathbf{v}_{ijt}\mathbf{v}_{ijs}^\top\varepsilon_{ijt}\varepsilon_{ijs}
\nonumber\\
&=&\mathbf{B}^\ast_{ij,1}+\mathbf{B}^\ast_{ij,2}+\mathbf{B}^\ast_{ij,3}.
\end{eqnarray}
For the first term on the right-hand side of \eqref{btst1}, it is clear to see that $\mathbf{B}^\ast_{ij,1}\rightarrow_P E[\mathbf{v}_{ijt}\mathbf{v}_{ijt}^\top\varepsilon_{ijt}^2]$. For the second term, 
\begin{eqnarray}\label{btst2}
&&E[\mathbf{B}^\ast_{ij,2}]-\frac{1}{T}\sum_{t=1}^T\sum_{s=t+1}^TE[\mathbf{v}_{ijt}\mathbf{v}_{ijs}^\top\varepsilon_{ijt}\varepsilon_{ijs}] \nonumber \\
&=&\frac{1}{T}\sum_{t=1}^T\sum_{s=t+1}^T\Bigl[a\Bigl(\frac{t-s}{m}\Bigr)-1\Bigr]E[\mathbf{v}_{ijt}\mathbf{v}_{ijs}^\top\varepsilon_{ijt}\varepsilon_{ijs}]
\nonumber\\
&=&\sum_{t=1}^{T-1}\Bigl(1-\frac{t}{T}\Bigr)\Bigl[ a\Bigl(\frac{t}{m}\Bigr)-1\Bigr]E[\mathbf{v}_{ij1}\mathbf{v}_{ij,1+t}^\top  \varepsilon_{ij1}\varepsilon_{ij,1+t}]
\nonumber\\
&=&\sum_{t=1}^{d_T}\Bigl(1-\frac{t}{T}\Bigr)\Bigl[ a\Bigl(\frac{t}{m}\Bigr)-a(0)\Bigr]E[\mathbf{v}_{ij1}\mathbf{v}_{ij,1+t}^\top  \varepsilon_{ij1}\varepsilon_{ij,1+t}]
\nonumber\\
&&+\sum_{t=d_T+1}^{T-1}\Bigl(1-\frac{t}{T}\Bigr)\Bigl[ a\Bigl(\frac{t}{m}\Bigr)-a(0)\Bigr]E[\mathbf{v}_{ij1}\mathbf{v}_{ij,1+t}^\top \varepsilon_{ij1}\varepsilon_{ij,1+t}],
\end{eqnarray}
where $d_T$ is a positive integer that satisfies $d_T^2/m\to 0$ and $d_T\rightarrow\infty$.

For the first term, by the Davydov's inequality for $\alpha$-mixing processes (see \citealp[pp. 19-20]{Bosq1996}) and Assumption \ref{AS2}, 
\begin{eqnarray}\label{btst3}
\|E[\mathbf{v}_{ij1}\mathbf{v}_{ij,1+t}^\top  \varepsilon_{ij1}\varepsilon_{ij,1+t}]\|
&\leq&O(1) \alpha_{ij}(t)^{\kappa_{ij}/(4+\kappa_{ij})}E[|\mathbf{v}_{ij1}|^{2+\kappa_{ij}/2}]^{\frac{4}{4+\kappa_{ij}}} E[|\varepsilon_{ij1}|^{2+\kappa_{ij}/2}]^{\frac{4}{4+\kappa_{ij}}}
\nonumber\\
&=&O(1) \alpha_{ij}(t)^{\kappa_{ij}/(4+\kappa_{ij})}.
\end{eqnarray}
In addition, by the Lipschitz continuity of $a(w)$ on $[-1, 1]$, it is clear to see that the first term in \eqref{btst2} has the order of 
$O(d_T^2/m)$. 

For the second term,  \eqref{btst3} yields that
\begin{eqnarray*}
\Bigl\|\sum_{t=d_T+1}^{T-1}\Bigl(1-\frac{t}{T}\Bigr)\Bigl[ a\Bigl(\frac{t}{m}\Bigr)-a(0)\Bigr]E[\mathbf{v}_{ij1}\mathbf{v}_{ij,1+t}^\top \varepsilon_{ij1}\varepsilon_{ij,1+t}]\Bigr\|
&\leq&O(1)\sum_{t=d_T+1}^{T-1} \alpha_{ij}(t)^{\kappa_{ij}/(4+\kappa_{ij})}.
\end{eqnarray*}
Assumption \ref{AS2} ensures  that $\sum_{t=1}^{T-1}\alpha_{ij}(t)^{\kappa_{ij}/(4+\kappa_{ij})}$ and $\sum_{t=1}^{d_T} \alpha_{ij}(t)^{\kappa_{ij}/(4+\kappa_{ij})}$ have the same limit as $T\rightarrow\infty$ and $d_T\rightarrow\infty$. Therefore, the second term in \eqref{btst2} is also negligible. Then, we proceed to study the order of $\mathbf{B}^\ast_{ij,2}-E[\mathbf{B}^\ast_{ij,2}]$. Define $\kappa_{ij}^\ast$ as
\begin{eqnarray}
\kappa^\ast_{ij}=1/\Bigl(\frac{1}{2+\kappa_{ij}}+\frac{\kappa_{ij}}{2(4+\kappa_{ij})}\Bigr).
\end{eqnarray} 
Simple algebra indicates $\kappa_{ij}^\ast>2$. Then, the following result holds for $\mathbf{B}^\ast_{ij,2}-E[\mathbf{B}^\ast_{ij,2}]$:
\begin{eqnarray*}
\bigl\|\mathbf{B}^\ast_{ij,2}-E[\mathbf{B}^\ast_{ij,2}]\bigr\|_{\kappa_{ij}^\ast/2}&=&\frac{1}{T}\biggl\|\sum_{s=1}^{T-1}\sum_{t=1}^{T-s} a\Bigl(\frac{s}{m}\Bigr)\bigl(\mathbf{v}_{ijt}\mathbf{v}_{ij,t+s}^\top \varepsilon_{ijt}\varepsilon_{ij,t+s}-E[\mathbf{v}_{ijt}\mathbf{v}_{ij,t+s}^\top \varepsilon_{ijt}\varepsilon_{ij,t+s}]\bigr)\biggr\|_{\kappa_{ij}^\ast/2}
\nonumber\\
&\leq&\frac{1}{T}\sum_{s=1}^{T-1} a\Bigl(\frac{s}{m}\Bigr)\biggl\|\sum_{t=1}^{T-s}\bigl(\mathbf{v}_{ijt}\mathbf{v}_{ij,t+s}^\top  \varepsilon_{ijt}\varepsilon_{ij,t+s}-E[\mathbf{v}_{ijt}\mathbf{v}_{ij,t+s}^\top  \varepsilon_{ijt}\varepsilon_{ij,t+s}]\bigr)\biggr\|_{\kappa_{ij}^\ast/2}
\nonumber\\
&=&\frac{1}{T}\sum_{s=1}^{m} a\Bigl(\frac{s}{m}\Bigr)\biggl\|\sum_{t=1}^{T-s}\bigl(\mathbf{v}_{ijt}\mathbf{v}_{ij,t+s}^\top  \varepsilon_{ijt}\varepsilon_{ij,t+s}-E[\mathbf{v}_{ijt}\mathbf{v}_{ij,t+s}^\top  \varepsilon_{ijt}\varepsilon_{ij,t+s}]\bigr)\biggr\|_{\kappa_{ij}^\ast/2}.
\end{eqnarray*}
Moreover, by McLeish's inequality for $\alpha$-mixing processes \citep[see Lemma 2.1 of][]{McLeish1975} and Assumption \ref{AS2}, 
\begin{eqnarray*}
&&\left\|E_{t-t_0}\bigl[\mathbf{v}_{ijt}\mathbf{v}_{ij,t+s}^\top \varepsilon_{ijt}\varepsilon_{ij,t+s}-E[\mathbf{v}_{ijt}\mathbf{v}_{ij,t+s}^\top  \varepsilon_{ijt}\varepsilon_{ij,t+s}]\bigr]\right\|_{\kappa_{ij}^\ast/2}
\nonumber\\
&\leq&6\alpha_{ij}(t_0)^{\kappa_{ij}/(4+\kappa_{ij})}\Bigl\|\mathbf{v}_{ijt}\mathbf{v}_{ij,t+s}^\top \varepsilon_{ijt}\varepsilon_{ij,t+s}-E[\mathbf{v}_{ijt}\mathbf{v}_{ij,t+s}^\top  \varepsilon_{ijt}\varepsilon_{ij,t+s}]\Bigr\|_{(2+\kappa_{ij})/2}
\nonumber\\
&\leq&12\alpha_{ij}(t_0)^{\kappa_{ij}/(4+\kappa_{ij})}\bigl\|\mathbf{v}_{ij1}\mathbf{v}_{ij,1+s}^\top\bigr\|_{(2+\kappa_{ij})/2} \bigl\|\varepsilon_{ij1}\varepsilon_{ij,1+s}\bigr\|_{(2+\kappa_{ij})/2}
\nonumber\\
&\leq&12\alpha_{ij}(t_0)^{\kappa_{ij}/(4+\kappa_{ij})}\bigl\|\mathbf{v}_{ij1} \bigr\|_{2+\kappa_{ij}} \bigl\|\varepsilon_{ij1}\bigr\|_{2+\kappa_{ij}}
,
\end{eqnarray*}
for any positive integer $t_0$. Together with Lemma A of \cite{Hansen1992}, it gives
\begin{eqnarray*}
&&\left\|\sum_{t=1}^{T-s}\bigl(\mathbf{v}_{ijt}\mathbf{v}_{ij,t+s}^\top \varepsilon_{ijt}\varepsilon_{ij,t+s}-E[\mathbf{v}_{ijt}\mathbf{v}_{ij,t+s}^\top  \varepsilon_{ijt}\varepsilon_{ij,t+s}]\bigr)\right\|_{\kappa^\ast_{ij}/2}
\nonumber\\
&\leq& 36c_{\kappa,ij} \Bigl(\kappa^\ast_{ij}/(\kappa^\ast_{ij}-2)\Bigr)^{3/2}\sum_{t_0=1}^\infty \alpha_{ij}(t_0)^{\kappa_{ij}/(4+\kappa_{ij})}\left(T-s \right)^{2/\underline{\kappa}_{ij}},
\end{eqnarray*}
where $c_{\kappa,ij}=12 \|\mathbf{v}_{ij1}  \|_{2+\kappa_{ij}}  \|\varepsilon_{ij1} \|_{2+\kappa_{ij}}$ and $\underline{\kappa}_{ij}=\min(\kappa^\ast_{ij},4)$. It follows that

\begin{eqnarray*}
\bigl\|\mathbf{B}^\ast_{ij,2}-E[\mathbf{B}^\ast_{ij,2}]
\bigr\|_{\kappa^\ast_{ij}/2}&=&O(m T^{2/\underline{\kappa}_{ij}-1})=o(1),
\end{eqnarray*}
which immediately yields $\mathbf{B}^\ast_{ij,2}-E[\mathbf{B}^\ast_{ij,2}]=o_P(1)$. Therefore, we obtain
\begin{eqnarray*}
\mathbf{B}^\ast_{ij,2}-\frac{1}{T}\sum_{t=1}^T\sum_{s=t+1}^TE[\mathbf{v}_{ijt}\mathbf{v}_{ijs}^\top\varepsilon_{ijt}\varepsilon_{ijs}]=o_P(1).
\end{eqnarray*}
Analogously, for $\mathbf{B}^\ast_{ij,3}$, 
\begin{eqnarray*}
\mathbf{B}^\ast_{ij,3}-\frac{1}{T}\sum_{t=1}^T\sum_{s=1}^{t-1}E[\mathbf{v}_{ijt}\mathbf{v}_{ijs}^\top\varepsilon_{ijt}\varepsilon_{ijs}]=o_P(1).
\end{eqnarray*}
In summary of these results, we have
\begin{equation}\label{btst4}
\text{Var}^\ast\Bigl( \frac{1}{\sqrt{T}}\mathbf{V}_{ij\centerdot}^\top  ( \pmb{\mathcal{E}}_{ij\centerdot}\circ\pmb{\xi})\Bigr)=\pmb{\Sigma}_{\mathbf{v},\varepsilon,ij}+o_P(1).
\end{equation}

Since $\xi_{t}$ is $m$-dependent time series, we can follow the Theorem 3.1 of \cite{shao2010} and adopt the large-block and small-block argument to prove the central limit theorem for $\frac{1}{\sqrt{T}}\mathbf{V}_{ij\centerdot}^\top  ( \pmb{\mathcal{E}}_{ij\centerdot}\circ\pmb{\xi})$ conditional on the observed sample. Define $l_T$ and $s_T$ as the lengths for the large and small blocks and $k_T=\lfloor T/(l_T+s_T)\rfloor$ such that $l_T,s_T\rightarrow\infty$ and they satisfy $\frac{m}{s_T},\,\frac{k_Ts_T}{T},\Bigl(\frac{m l_T}{T}\Bigr)^{1+\kappa_{ij}/4}k_T\rightarrow 0$.

For $k=1,\ldots, k_T$, we define a series of variables for the large and small blocks, respectively:

\begin{eqnarray*}
&&\boldsymbol{\xi}^\ast_{ijk,1} = \sum_{t=(k-1)(l_T+s_T)+1}^{jl_T+(k-1)s_T} \mathbf{v}_{ijt}\varepsilon_{ijt}\xi_{t},\quad \boldsymbol{\xi}^\ast_{ijk,2} = \sum_{t=jl_T+ (k-1)s_T+1}^{j(l_T+s_T)} \mathbf{v}_{ijt}\varepsilon_{ijt}\xi_{t}.
\end{eqnarray*}
Moreover, we define $\boldsymbol{\xi}^\ast_{ij,0} =\sum_{t=k_T(l_T+s_T)+1}^T \mathbf{v}_{ijt}\varepsilon_{ijt}\xi_{t}$ to contain the reminder terms. 

 Without loss of generality, we assume $l_T,s_T>m$. Then, it is clear to see that $\{\boldsymbol{\xi}^\ast_{ij1,1},\cdots,\boldsymbol{\xi}^\ast_{k_T,1}\}$ are independent conditional on the observed data, as are $\{\boldsymbol{\xi}^\ast_{ij1,2},\cdots,\boldsymbol{\xi}^\ast_{ijk_T,2}\}$. Therefore,
\begin{eqnarray*}
&&\frac{1}{T}E\Biggl[E^\ast\Biggl[\Biggl(\sum_{k=1}^{k_T} \boldsymbol{\xi}^\ast_{ijk,2}\Biggr)\Biggl(\sum_{k=1}^{k_T} \boldsymbol{\xi}^\ast_{ijk,2}\Biggr)^\top\Biggr]\Biggr]\nonumber \\
&=&\frac{1}{T}\sum_{k=1}^{k_T}E\left[E^\ast\left[ \boldsymbol{\xi}^\ast _{ijk,2}\boldsymbol{\xi}^{\ast \top}_{ijk,2}\right]\right]
\nonumber\\
&=&\frac{1}{T}\sum_{k=1}^{k_T}\sum_{t,s=kl_T+ (k-1)s_T+1}^{k(l_T+s_T)}  E\left[E^\ast\left[ \mathbf{v}_{ijt}\mathbf{v}_{ijs}^\top\varepsilon_{ijt}\varepsilon_{ijs}\xi_{t}\xi_{s}\right]\right]
\nonumber\\
&\leq&\frac{1}{T}\sum_{k=1}^{k_T}\sum_{t=kl_T+ (k-1)s_T+1}^{k(l_T+s_T)}\sum_{s=-s_T+1}^{s_T-1}a\Bigl(\frac{s}{m}\Bigr)  \left|E\left[\mathbf{v}_{ijt}\mathbf{v}_{ijs}^\top\varepsilon_{ijt}\varepsilon_{ijs}\right]\right|
\nonumber\\
&=& O\left(\frac{k_Ts_T}{T}\right).
\end{eqnarray*}
Under the condition $\frac{k_Ts_T}{T}\rightarrow0$,  we have $\frac{1}{\sqrt{T}}\sum_{k=1}^{k_T} \boldsymbol{\xi}^\ast_{ijk,2}=o_P(1)$. Analogously, we can show that $\frac{1}{\sqrt{T}} \boldsymbol{\xi}^\ast_{ij,0}=o_P(1)$. 

In what follows, we establish the asymptotic normality of $\frac{1}{\sqrt{T}}\sum_{k=1}^{k_T} \boldsymbol{\xi}^\ast_{ijk,1}$ (conditional on the observed sample) using the Lindeberg CLT. Similarly to \eqref{btst4}, we can also show that 
\[
\frac{1}{T}\text{Var}^\ast\Bigl(\sum_{k=1}^{k_T} \boldsymbol{\xi}^\ast_{ijk,1}\Bigr)=\pmb{\Sigma}_{\mathbf{v},\varepsilon,ij}+o_P(1).
\]

For any $\epsilon>0$, we have

\begin{eqnarray}\label{btst5}
\frac{1}{T}\sum_{k=1}^{k_T}E^\ast\bigl[\big\|\boldsymbol{\xi}^{\ast }_{ijk,1}\bigr\|^2\cdot I(\|\boldsymbol{\xi}^\ast_{ijk,1}\|\geq \epsilon \sqrt{T}) \bigr]&\leq&\frac{1}{T}\sum_{k=1}^{k_T}\{E^\ast\|\boldsymbol{\xi}^\ast_{ijk,1}\|^{2+\kappa_{ij}/2}\}^{\frac{4}{4+\kappa_{ij}}} \left\{ \frac{E^\ast\|\boldsymbol{\xi}^\ast_{ijk,1}\|^{2+\kappa_{ij}/2}}{\epsilon^{2+\kappa_{ij}/2} T^{\frac{4+\kappa_{ij}}{4}}}\right\}^{\frac{\kappa_{ij}}{4+\kappa_{ij}}}
\nonumber\\
&\leq&\frac{1}{\epsilon^{\kappa_{ij}/2} T^{1+\kappa_{ij}/4}}\sum_{k=1}^{k_T}E^\ast\big\|\boldsymbol{\xi}^\ast_{ijk,1}\big\|^{ 2+\kappa_{ij}/2}.
\end{eqnarray}

 Without loss of generality, we now use the Rosenthal inequality to study the order of $\|\boldsymbol{\xi}^\ast_{ij1,1}\|^\ast_{2+\kappa_{ij}/2}$. Noteworthily, Rosenthal inequality is designed for the independent random variables. Therefore, we consider the following decomposition for $\boldsymbol{\xi}^\ast_{ij1,1}$: $\boldsymbol{\xi}^\ast_{ij1,1}=\sum_{k=1}^{m+1} \boldsymbol{\xi}^\ast_{ij1,1k}$,
where 
\begin{equation*}
\boldsymbol{\xi}^\ast_{ij1,1k}=\sum_{s=1}^{\lfloor (l_T-k)/(m+1) \rfloor}\mathbf{v}_{ij,k+(s-1)(m+1)}\varepsilon_{ij,k+(s-1)(m+1)}\xi_{k+(s-1)(m+1)}.
\end{equation*}
By this definition, it is clear to see that the elements involved in $\boldsymbol{\xi}^\ast_{ij1,1k}$ are independent conditional on the observed sample.  Directly applying the triangle inequality and Rosenthal inequality sequentially, we obtain
\begin{eqnarray}\label{btst6}
&&\|\boldsymbol{\xi}^\ast_{ij1,1}\|^\ast_{2+\kappa_{ij}/2}\leq \sum_{k=1}^{m+1} \|\boldsymbol{\xi}^\ast_{ij1,1k}\|^\ast_{2+\kappa_{ij}/2}
\nonumber\\
&\leq&O(1)\sum_{k=1}^{m+1} \Biggl[\Biggl\|\sum_{s=1}^{\lfloor (l_T-k)/(m+1) \rfloor}\bigl\|\mathbf{v}_{ij,k+(s-1)(m+1)}\bigr\|^2\varepsilon_{ij,k+(s-1)(m+1)}^2\xi^2_{k+(s-1)(m+1)}\Biggr\|^\ast_{1+\kappa_{ij}/4}\Biggr]^{1/2}
\nonumber\\
&\leq&O(1)\sum_{k=1}^{m+1} \Biggl[\sum_{s=1}^{\lfloor (l_T-k)/(m+1) \rfloor}\bigl\|\mathbf{v}_{ij,k+(s-1)(m+1)}\bigr\|^2\varepsilon_{ij,k+(s-1)(m+1)}^2\Biggr]^{1/2}
\nonumber\\
&\leq&O(\sqrt{m})\Biggl[\sum_{k=1}^{m+1} \sum_{s=1}^{\lfloor (l_T-k)/(m+1) \rfloor}\bigl\|\mathbf{v}_{ij,k+(s-1)(m+1)}\bigr\|^2\varepsilon_{ij,k+(s-1)(m+1)}^2\Biggr]^{1/2}
\nonumber\\
&=&O(\sqrt{m})\Biggl[\sum_{t=1}^{l_T}\bigl\|\mathbf{v}_{ijt}\bigr\|^2\varepsilon_{ijt}^2\Biggr]^{1/2}. 
\end{eqnarray}
By \eqref{btst5} and \eqref{btst6}, 
\begin{eqnarray*}
&&\frac{1}{T}\sum_{k=1}^{k_T}E \bigl[E^\ast\bigl[\bigl\|\boldsymbol{\xi}^{\ast }_{ijk,1}\bigr\|^2\cdot I(\bigl\|\boldsymbol{\xi}^\ast_{ijk,1}\bigr\|\geq \epsilon \sqrt{T}) \bigr]\bigr]\nonumber \\
&\leq&O(1)\left(\frac{m}{T}\right)^{1+\kappa_{ij}/4}\sum_{k=1}^{k_T}E\Biggl[\Biggl(\sum_{t=(k-1)(l_T+s_T)+1}^{kl_T+(k-1)s_T} \bigl\|\mathbf{v}_{ijt}\bigr\|^2\varepsilon_{ijt}^2\Biggr)^{1+\kappa_{ij}/4}\Biggr]
\nonumber\\
&=&O\Bigl(\Bigl(\frac{m l_T}{T}\Bigr)^{1+\kappa_{ij}/4}k_T\Bigr)=o(1),
\end{eqnarray*}
which implies that $\frac{1}{T}\sum_{k=1}^{k_T}E^\ast\bigl[\big\|\boldsymbol{\xi}^{\ast }_{ijk,1}\bigr\|^2\cdot I(\|\boldsymbol{\xi}^\ast_{ijk,1}\|\geq \epsilon \sqrt{T}) \bigr]=o_P(1)$. In summary of the results that we have established,  the Lindeberg condition can be satisfied and
\begin{equation*}
\frac{1}{\sqrt{T}}\mathbf{V}_{ij\centerdot}^\top  ( \pmb{\mathcal{E}}_{ij\centerdot}\circ\pmb{\xi})\rightarrow_{D^\ast} N(\mathbf{0}, \pmb{\Sigma}_{\mathbf{v},\varepsilon,ij}).
\end{equation*}

Additionally, using analogous arguments to those in the proof of Theorem \ref{M.Thm1}, we can readily obtain that $\frac{1}{T}\mathbf{X}_{ij\centerdot}^\top \mathbf{M}_{\widehat{\mathbf{C}}_{ij}} \mathbf{X}_{ij\centerdot}=\pmb{\Sigma}_{\mathbf{v},ij}^{-1}+o_P(1)$ and the second and third terms in \eqref{bstar} are asymptotically negligible. Therefore, we are ready to complete the proof of Theorem \ref{M.Thm2}. \hspace*{\fill}{$\blacksquare$}

\bigskip

\noindent \textbf{Proof of Lemma  \ref{LemA3}:}

(1).  Consider the first $\ell$ eigenvalues of $\widehat{\pmb{\Sigma}}$, and note that by PCA, we have

\begin{eqnarray*}
\widehat{\mathbf{V}}= \widehat{\mathbf{C}}^\top\widehat{\pmb{\Sigma}}\widehat{\mathbf{C}}.
\end{eqnarray*}
We focus on the first $\ell$ columns of $\widehat{\mathbf{C}}$, which is denoted as $\widehat{\mathbf{C}}^\sharp$ in the description of above this lemma.  By \eqref{ff1}, we know that
 
\begin{eqnarray}\label{EQA.19}
\frac{1}{\sqrt{T}}\|\widehat{\mathbf{C}}^\sharp  - \mathbf{F}  \mathbf{H}^\sharp\|_2 = O_P\left( \frac{1}{\sqrt{LN}}\|\pmb{\beta}_{\centerdot \centerdot}-\widetilde{\mathbf{b}}_{\centerdot\centerdot}\| +\frac{1}{\sqrt{L\wedge N\wedge T}}\right).
\end{eqnarray}
where $\mathbf{H}^\sharp$ is also defined in the description of above this lemma. By the development of $\mathbf{J}_1$ to $\mathbf{J}_{25}$ in Lemma \ref{M.LM1}, it is easy to see  that

\begin{eqnarray}\label{EQA.20}
\|\widehat{\pmb{\Sigma}}  - \pmb{\Sigma} \|_2=O_P\left( \frac{1}{\sqrt{LN}}\|\pmb{\beta}_{\centerdot \centerdot}-\widetilde{\mathbf{b}}_{\centerdot\centerdot}\| +\frac{1}{\sqrt{L\wedge N\wedge T}}\right).
\end{eqnarray}
In the context of this lemma, we have also defined $\widehat{\lambda}_s$ and $\lambda_s$. These notations and results will be repeatedly used below.

\medskip

Let's now consider $ \widehat{\lambda}_s-\lambda_s$, and write

\begin{eqnarray*}
&&\widehat{\lambda}_s -\lambda_s \\
&=&\frac{1}{\sqrt{T}}(\widehat{\mathbf{C}}_s - \mathbf{F} \mathbf{H}_s^{\sharp} +\mathbf{F} \mathbf{H}_s^{\sharp} )^\top (\widehat{\pmb{\Sigma}} -\pmb{\Sigma} +\pmb{\Sigma} )\frac{1}{\sqrt{T}}(\widehat{\mathbf{C}}_s - \mathbf{F} \mathbf{H}_s^{\sharp} +\mathbf{F} \mathbf{H}_s^{\sharp} )\nonumber \\
&&-\frac{1}{T}\mathbf{H}_s^{\sharp\top} \mathbf{F}^{\top} \pmb{\Sigma} \mathbf{F} \mathbf{H}_s^{\sharp} \nonumber \\
&=&\frac{1}{\sqrt{T}}(\widehat{\mathbf{C}}_s - \mathbf{F} \mathbf{H}_s^{\sharp} )^\top (\widehat{\pmb{\Sigma}} -\pmb{\Sigma} )\frac{1}{\sqrt{T}}(\widehat{\mathbf{C}}_s - \mathbf{F} \mathbf{H}_s^{\sharp} ) \nonumber \\
&&+\frac{2}{\sqrt{T}}(\widehat{\mathbf{C}}_s - \mathbf{F} \mathbf{H}_s^{\sharp} )^\top (\widehat{\pmb{\Sigma}}  -\pmb{\Sigma} )\frac{1}{\sqrt{T}} \mathbf{F} \mathbf{H}_s^{\sharp} \nonumber \\
&&+\frac{1}{\sqrt{T}}(\widehat{\mathbf{C}}_s - \mathbf{F} \mathbf{H}_s^{\sharp} )^\top \pmb{\Sigma} \frac{1}{\sqrt{T}}(\widehat{\mathbf{C}}_s - \mathbf{F} \mathbf{H}_s^{\sharp} )\nonumber \\
&&+\frac{2}{\sqrt{T}}(\widehat{\mathbf{C}}_s - \mathbf{F} \mathbf{H}_s^{\sharp} )^\top \pmb{\Sigma} \frac{1}{\sqrt{T}}\mathbf{F} \mathbf{H}_s^{\sharp} +\frac{1}{\sqrt{T}}( \mathbf{F} \mathbf{H}_s^{\sharp}   )^\top (\widehat{\pmb{\Sigma}} -\pmb{\Sigma} )\frac{1}{\sqrt{T}}  \mathbf{F}  \mathbf{H}_s^{\sharp} \\
&:=& \mathbf{A}_1+2\mathbf{A}_2+\mathbf{A}_3+2\mathbf{A}_4+\mathbf{A}_5,
\end{eqnarray*}
where the definitions of $\mathbf{A}_1$ to $\mathbf{A}_5$ are obvious. 

By \eqref{EQA.19} and \eqref{EQA.20}, we can immediately conclude that $|\mathbf{A}_1| =o_P(|\mathbf{A}_5|)$, $|\mathbf{A}_2| =o_P(|\mathbf{A}_5|)$, and $|\mathbf{A}_3|=o_P(|\mathbf{A}_4|)$. Thus, we focus on $\mathbf{A}_4$ and $\mathbf{A}_5$ below. For $\mathbf{A}_4$, we write

\begin{eqnarray*}
|\mathbf{A}_4|&\le & \frac{1}{\sqrt{T}}\| \widehat{\mathbf{C}}_s - \mathbf{F} \mathbf{H}_s^{\sharp}  \|_2\cdot \| \pmb{\Sigma} \|_2 \cdot \frac{1}{\sqrt{T}} \|\mathbf{F} \mathbf{H}_s^{\sharp} \|_2 \nonumber \\
&=&O_P\left( \frac{1}{\sqrt{LN}}\|\pmb{\beta}_{\centerdot \centerdot}-\widetilde{\mathbf{b}}_{\centerdot\centerdot}\| +\frac{1}{\sqrt{L\wedge N\wedge T}}\right),
\end{eqnarray*}
where the last equality follows from \eqref{EQA.19} and the fact that $ \| \pmb{\Sigma} \|_2 =O_P(1)$ and $\frac{1}{\sqrt{T}} \| \mathbf{F} \mathbf{H}_\ell^{\sharp} \|_2=O_P(1)$ by the construction. For $\mathbf{A}_5$, write

\begin{eqnarray*}
|\mathbf{A}_5|&=&\left|\frac{1}{\sqrt{T}}( \mathbf{F}  \mathbf{H}_s^{\sharp}   )^\top (\widehat{\pmb{\Sigma}} -\pmb{\Sigma} )\frac{1}{\sqrt{T}} \mathbf{F} \mathbf{H}_s^{\sharp}   \right| \le  \|\widehat{\pmb{\Sigma}} - \pmb{\Sigma} \|_2\cdot \frac{1}{T}\| \mathbf{F}  \mathbf{H}_s^{\sharp}  \|_2^2\nonumber \\
&=&O_P\left( \frac{1}{\sqrt{LN}}\|\pmb{\beta}_{\centerdot \centerdot}-\widetilde{\mathbf{b}}_{\centerdot\centerdot}\| +\frac{1}{\sqrt{L\wedge N\wedge T}}\right),
\end{eqnarray*}
where the last step follows from \eqref{EQA.20}.

This concludes the proof of the first result of this lemma.

\medskip

(2). To investigate the second result, we start the proof by introducing some notations. We denote $\mathbf{F}^{\perp}$ as a $T\times (T - \ell)$ matrix such that $\frac{1}{T}(\mathbf{F}^{\perp}, \mathbf{F} \mathbf{R})^\top (\mathbf{F}^{\perp}, \mathbf{F} \mathbf{R})= \mathbf{I}_{T}$, where $\mathbf{R}$ is an $\ell\times \ell$ rotation matrix. The matrices $\frac{1}{\sqrt{T}} \mathbf{F}^{\perp}$, $\frac{1}{\sqrt{T}} \mathbf{F} \mathbf{R}$, $\pmb{\Sigma}$, and $\widehat{\pmb{\Sigma}} -\pmb{\Sigma}$ correspond to $\mathbf{Q}_1$, $\mathbf{Q}_2$, $\mathbf{A}$, and $\mathbf{E}$ of Lemma \ref{LemA1}. Thus, the counterpart of the matrix $\mathbf{Q}_1^0$ becomes 

\begin{eqnarray*}
\widehat{\mathbf{C}}^{\perp}= \frac{1}{\sqrt{T}} (\mathbf{F}^{\perp} + \mathbf{F} \mathbf{R}  \mathbf{P})( \mathbf{I}_{T - \ell} + \mathbf{P}^\top  \mathbf{P})^{-1/2},
\end{eqnarray*}
in which 
{\footnotesize
\begin{eqnarray}\label{EQA.21}
\| \mathbf{P}\|_2 &\le & \frac{4}{\text{sep}(0,\frac{1}{T} \mathbf{R}^\top   \mathbf{F}^{\top} \pmb{\Sigma}\mathbf{F} \mathbf{R} )} \|\widehat{\pmb{\Sigma}} - \pmb{\Sigma} \|_2 =O_P\left( \frac{1}{\sqrt{LN}}\|\pmb{\beta}_{\centerdot \centerdot}-\widetilde{\mathbf{b}}_{\centerdot\centerdot}\| +\frac{1}{\sqrt{L\wedge N\wedge T}}\right).
\end{eqnarray}}

Moreover, $\widehat{\mathbf{C}}^{\perp}$ is an orthonormal basis for a subspace that is invariant for $\widehat{\pmb{\Sigma}}$. In addition, note that

\begin{eqnarray*}
&&\left\|\widehat{\mathbf{C}}^{\perp} - \frac{1}{\sqrt{T}} \mathbf{F}^{\perp}\right\|_2 \nonumber \\
&=& \frac{1}{\sqrt{T}}\left\| \left[  (\mathbf{F}^{\perp} + \mathbf{F}\mathbf{R} \mathbf{P}) - \mathbf{F}^{\perp} (\mathbf{I}_{T - \ell } +\mathbf{P}^\top \mathbf{P})^{1/2} \right](\mathbf{I}_{T - \ell} +\mathbf{P}^\top \mathbf{P})^{-1/2}\right\|_2 \nonumber \\
&\le & \frac{1}{\sqrt{T}}\left\| \mathbf{F}^{\perp} \left[ \mathbf{I}_{T-\ell}  -  (\mathbf{I}_{T - \ell} +\mathbf{P}^\top \mathbf{P})^{1/2} \right](\mathbf{I}_{T- \ell} +\mathbf{P}^\top \mathbf{P})^{-1/2}\right\|_2  \nonumber \\
&&+ \frac{1}{\sqrt{T}}\left\| \mathbf{F} \mathbf{R} \mathbf{P} (\mathbf{I}_{T - \ell} +\mathbf{P}^\top \mathbf{P})^{-1/2}\right\|_2\nonumber \\
&\le &O_P(1) \left\| \left[ \mathbf{I}_{T-\ell}  -  (\mathbf{I}_{T - \ell} +\mathbf{P}^\top \mathbf{P})^{1/2} \right](\mathbf{I}_{T - \ell} +\mathbf{P}^\top \mathbf{P})^{-1/2}\right\|_2\nonumber \\
&&+\left\| \mathbf{P} (\mathbf{I}_{T - \ell} +\mathbf{P}^\top \mathbf{P})^{-1/2}\right\|_2 \nonumber\\
&\le &O_P(1) \left\| \mathbf{I}_{T-\ell}  -  (\mathbf{I}_{T - \ell} +\mathbf{P}^\top \mathbf{P})^{1/2} \right\|_2+O_P(1)\left\| \mathbf{P} \right\|_2 \nonumber \\
&=& O_P(\| \mathbf{P}\|_2) = O_P\left( \frac{1}{\sqrt{LN}}\|\pmb{\beta}_{\centerdot \centerdot}-\widetilde{\mathbf{b}}_{\centerdot\centerdot}\| +\frac{1}{\sqrt{L\wedge N\wedge T}}\right),
\end{eqnarray*}
where the last equality follows from \eqref{EQA.21}.

Now, let $\widehat{ \mathbf{C}}_s^{\perp} $ and $\mathbf{F}_s^{\perp}$ be the $s^{th}$ columns of $\widehat{ \mathbf{C}}^{\perp} $ and $\mathbf{F}^\perp$ respectively. Since $\widehat{ \mathbf{C}}^{\perp}$ is an orthonormal basis for a subspace that is invariant for $\widehat{\pmb{\Sigma}}$, for $\ell=1,\ldots, T-\ell$ we write

\begin{eqnarray*}
\widehat{\lambda}_{\ell+s} &= & \left(\widehat{ \mathbf{C}}_s^{\perp} - \frac{1}{\sqrt{T}}  \mathbf{F}_s^{\perp}+\frac{1}{\sqrt{T}}  \mathbf{F}_s^{\perp}\right)^\top (\widehat{\pmb{\Sigma}} - \pmb{\Sigma} +\pmb{\Sigma} )  \left(\widehat{ \mathbf{C}}_s^{\perp} - \frac{1}{\sqrt{T}}  \mathbf{F}_s^{\perp}+\frac{1}{\sqrt{T}}  \mathbf{F}_s^{\perp}\right) \nonumber \\
&\le &\left\|\widehat{\mathbf{C}}^{\perp} - \frac{1}{\sqrt{T}} \mathbf{F}^{\perp}\right\|_2^2 \cdot \|\widehat{\pmb{\Sigma}} -\pmb{\Sigma}  \|_2\\
&&+2\left\|\widehat{\mathbf{C}}^{\perp} - \frac{1}{\sqrt{T}} \mathbf{F}^{\perp}\right\|_2  \cdot \|\widehat{\pmb{\Sigma}} - \pmb{\Sigma} \|_2 \cdot  \frac{1}{\sqrt{T}}  \|\mathbf{F}^{\perp}\|_2\nonumber \\
&&+ \left\|\widehat{\mathbf{C}}^{\perp} - \frac{1}{\sqrt{T}} \mathbf{F}^{\perp}\right\|_2^2 \cdot \|\pmb{\Sigma}\|_2\nonumber \\
&=& O_P\left( \frac{1}{LN}\|\pmb{\beta}_{\centerdot \centerdot}-\widetilde{\mathbf{b}}_{\centerdot\centerdot}\|^2 +\frac{1}{ L\wedge N\wedge T}\right).
\end{eqnarray*}

The proof of the second result of this lemma is now completed. \hspace*{\fill}{$\blacksquare$}

\bigskip

\noindent \textbf{Proof of Lemma \ref{LemA4}:}

The proof is almost identical to Lemma \ref{LemA3}, so we omit the details. \hspace*{\fill}{$\blacksquare$}

\bigskip

\noindent \textbf{Proof of Theorem \ref{Theorem2.3}:}

Note that 

\begin{eqnarray*}
\Pr(\widehat{\ell} =\ell, \widehat{\pmb{\ell}}^\circ= \pmb{\ell}^\circ, \widehat{\pmb{\ell}}^\bullet= \pmb{\ell}^\bullet) = \Pr( \widehat{\pmb{\ell}}^\circ= \pmb{\ell}^\circ, \widehat{\pmb{\ell}}^\bullet= \pmb{\ell}^\bullet |\widehat{\ell} =\ell) \Pr(\widehat{\ell} =\ell).
\end{eqnarray*}
Therefore, in what follows, we first show that $\Pr(\widehat{\ell} =\ell) \to 1$, and then prove that $ \Pr( \widehat{\pmb{\ell}}^\circ= \pmb{\ell}^\circ, \widehat{\pmb{\ell}}^\bullet= \pmb{\ell}^\bullet |\widehat{\ell} =\ell) \to 1$ in the second step.

\medskip

Step 1. First, consider the case when $\ell =0$. By Lemma \ref{LemA3}, we have

\begin{eqnarray*}
\widehat{\lambda}_{\ell+s} =\widehat{\lambda}_{s}  = O_P\left( \frac{1}{\sqrt{LN}}\|\pmb{\beta}_{\centerdot \centerdot}-\widetilde{\mathbf{b}}_{\centerdot\centerdot}\| +\frac{1}{\sqrt{L\wedge N\wedge T}}\right),
\end{eqnarray*}
for $s =1,\ldots, d_{\max}$, which is less than $\omega$ with a probability approaching one. By the construction of the mock eigenvalue, we immediately obtain that $\Pr(\ell=0)\to 1$.

\medskip

Next, we consider the case with $\ell>0$. Note that for $s =1,\ldots, \ell$,

\begin{eqnarray*}
\lambda_s &=& \frac{1}{T}\mathbf{H}_s^{\top} \mathbf{F}^{\top} \pmb{\Sigma}  \mathbf{F}  \mathbf{H}_s \nonumber \\
&=& \frac{1}{T} \mathbf{H}_s^{\top} \mathbf{F}^{\top} \mathbf{F}\cdot \frac{1}{LN}\pmb{\Gamma}_{\centerdot \centerdot}^{\top} \pmb{\Gamma}_{\centerdot \centerdot} \cdot \frac{1}{T} \mathbf{F}^{\top}\mathbf{F}  \mathbf{H}_s \nonumber \\
&\asymp &  \frac{1}{T} \widehat{\mathbf{C}}_s^{\top} \mathbf{F} \cdot \frac{1}{LN}\pmb{\Gamma}_{\centerdot \centerdot}^{\top} \pmb{\Gamma}_{\centerdot \centerdot} \cdot \frac{1}{T} \mathbf{F}^{\top}\widehat{\mathbf{C}}_s  \asymp 1,
\end{eqnarray*}
where the first $\asymp$ follows from \eqref{EQA.19}.

By Lemma \ref{LemA3}, $ \widehat{\lambda}_s  \asymp \lambda_s $ for $s=1,\ldots,\ell$, which are larger than $\omega $ with a probability approaching one. Thus, for $s=1,\ldots,\ell-1$ we can conclude that

\begin{eqnarray*}
\frac{  \widehat{\lambda}_{s+1}}{  \widehat{\lambda}_{s} }I(\widehat{\lambda}_s\ge \omega) + I(\widehat{\lambda}_s<\omega) \asymp 1.
\end{eqnarray*}
For $s=\ell+1,\ldots,d_{\max}$, by Lemma \ref{LemA3}, $\widehat{\lambda}_s=O_P\left( \frac{1}{LN}\|\pmb{\beta}_{\centerdot \centerdot}-\widetilde{\mathbf{b}}_{\centerdot\centerdot}\|^2 +\frac{1}{ L\wedge N\wedge T}\right)$, which is less than $\omega$ with a probability approaching one. Thus,

\begin{eqnarray*}
\frac{  \widehat{\lambda}_{s+1}}{  \widehat{\lambda}_{s} }I(\widehat{\lambda}_s\ge \omega) + I(\widehat{\lambda}_s<\omega)  = 1
\end{eqnarray*}
for $s=\ell+1,\ldots,d_{\max}$ by construction. In addition, for $s =\ell$, it is straightforward to obtain that 

\begin{eqnarray*}
\frac{ \widehat{\lambda}_{\ell+1}}{\widehat{\lambda}_{\ell}} =  O_P\left( \frac{1}{LN}\|\pmb{\beta}_{\centerdot \centerdot}-\widetilde{\mathbf{b}}_{\centerdot\centerdot}\|^2 +\frac{1}{ L\wedge N\wedge T}\right)
\end{eqnarray*}
using the facts that $ \widehat{\lambda}_\ell \asymp 1$ and $\widehat{\lambda}_{\ell+1} = O_P\left( \frac{1}{LN}\|\pmb{\beta}_{\centerdot \centerdot}-\widetilde{\mathbf{b}}_{\centerdot\centerdot}\|^2 +\frac{1}{ L\wedge N\wedge T}\right)$. Thus, we are ready to conclude that $P(\widehat{\ell} = \ell) \to 1$.  

\medskip

Step 2. Note that $i$ and $j$ are symmetric. Thus, without loss of generality, we consider two cases: (i) there is at least one $\widehat{\ell}_i^\circ<\ell_i^\circ$ in $\widehat{\pmb{\ell}}^\circ$, and (ii) there is at least one $\widehat{\ell}_i^\circ > \ell_i^\circ$ in $\widehat{\pmb{\ell}}^\circ$. Note that case (i) does not rule out the possibility that other estimated numbers of factors may be larger than the true value. Similarly, case (ii) does not rule out the possibility that other estimated numbers of factors may be less than the true value. If we can rule out both cases with a probability approaching one, then $ \Pr( \widehat{\pmb{\ell}}^\circ= \pmb{\ell}^\circ, \widehat{\pmb{\ell}}^\bullet= \pmb{\ell}^\bullet |\widehat{\ell} =\ell) \to 1$. 

We now consider case (i), and suppose that $\widehat{\ell}_i^\circ<\ell_i^\circ$. By Lemma \ref{LemA4}, we can show that $$\frac{\widehat{\lambda}_{i,\widehat{\ell}_i^\circ+1}^\circ}{\widehat{\lambda}_{i, \widehat{\ell}_i^\circ}^\circ}  I(\widehat{\lambda}_{i,\widehat{\ell}_i^\circ}^\circ \ge \omega)+ I(\widehat{\lambda}_{i, \widehat{\ell}_i^\circ}^\circ < \omega) \asymp 1.$$

By replacing $\widehat{\ell}_i^\circ$ of $\widehat{\pmb{\ell}}^\circ$ with $\ell_i^\circ$, we find another $\widetilde{\pmb{\ell}}^S=(\widehat{\ell}_1^\circ,\ldots,\widehat{\ell}_{i-1}^\circ, \ell_i^\circ,\widehat{\ell}_{i+1}^\circ ,\ldots\widehat{\ell}_L^\circ)$, which yields a smaller value for the objective function considered in \eqref{EQ2.9} with a probability approaching one. However, this is contradictory to the definition of $\widehat{\pmb{\ell}}^\circ$.  

Next, we consider case (ii), and suppose that $\widehat{\ell}_i^\circ>\ell_i^\circ$. Again, Lemma \ref{LemA4} yields that $$\frac{\widehat{\lambda}_{i,\widehat{\ell}_i^\circ+1}^\circ}{\widehat{\lambda}_{i, \widehat{\ell}_i^\circ}^\circ}  I(\widehat{\lambda}_{i,\widehat{\ell}_i^\circ}^\circ \ge \omega)+ I(\widehat{\lambda}_{i, \widehat{\ell}_i^\circ}^\circ < \omega)  =1.$$

By replacing $\widehat{\ell}_i^\circ$ of $\widehat{\pmb{\ell}}^\circ$ with $\ell_i^\circ$, we find another $\widetilde{\pmb{\ell}}^\circ=(\widehat{\ell}_1^\circ,\ldots,\widehat{\ell}_{i-1}^S, \ell_i^\circ,\widehat{\ell}_{i+1}^\circ ,\ldots\widehat{\ell}_L^\circ)$, which yields a smaller value for the objective function considered in \eqref{EQ2.9} with a probability approaching one. However, it is contradictory to the definition of $\widehat{\pmb{\ell}}^\circ$. Based on the above development, we conclude that $ \Pr( \widehat{\pmb{\ell}}^\circ= \pmb{\ell}^\circ, \widehat{\pmb{\ell}}^\bullet= \pmb{\ell}^\bullet |\widehat{\ell} =\ell) \to 1$. 

In view of Steps 1 and 2, the proof is now completed.  \hspace*{\fill}{$\blacksquare$}

}

\end{document}